\documentclass[12pt]{article}
\usepackage[cp1251]{inputenc}
\usepackage[english,russian]{babel}
\usepackage{latexsym}
\usepackage{amssymb}
\usepackage{multirow}
\usepackage{enumerate}
\usepackage{hhline}
\usepackage{amsmath}
\usepackage{amsthm}
\usepackage{graphicx,color}
\usepackage{tocloft}
\usepackage{minitoc}
\usepackage{url}
\usepackage[unicode,bookmarksopen,bookmarksopenlevel=1,bookmarksnumbered]{hyperref}
\hypersetup{pdfstartview={XYZ null null 1.25}}
\newtheorem{theorem}{Theorem}[section]
\newtheorem{note}{Note}[section]

\newtheorem{corollary}{Corollary}[section]
\newtheorem{example}{Example}[section]
\newtheorem{definition}{Definition}[section]
\newtheorem{algorithm}{Algorithm}[section]

\def\hcorrection{\hspace{-0.3em}}
\def\UDK#1{\rightline{UDK {#1}}}
\def\Author#1{\vspace{4.0ex plus 0.2ex minus 0.2ex}\centerline{\Large{#1}}}
\def\Title#1{\section*{\hcorrection{#1}}}
\def\Abstract#1{{\footnotesize\baselineskip=12pt\begin{quotation}\noindent\hcorrection{#1}\end{quotation}}}
\def\References#1{{\footnotesize\baselineskip=12pt\begin{thebibliography}{99}{#1}\end{thebibliography}}}

\def\myf#1{\mathit{\tilde{#1}}}
\def\myff#1{\mathit{\tilde{\tilde {#1}}}}
\def\sps{, \ \ }
\def\spsd{. \ \ }
\numberwithin{equation}{section}
\numberwithin{theorem}{section}
\numberwithin{note}{section}
\numberwithin{definition}{section}
\numberwithin{task}{section}
\begin{document}
\renewcommand{\proofname}{Proof.}
\renewcommand{\refname}{References}
\renewcommand{\figurename}{Fig.}
\renewcommand{\contentsname}{Contents}
\renewcommand{\partname}{}
\renewcommand{\tablename}{Table}
\renewcommand{\contentsname}{Contents/Содержание}
\renewcommand{\listtablename}{Tables/Список таблиц}
\renewcommand{\listfigurename}{Figures/Список рисунков}
\renewcommand{\tanh}{\mathrm{th}}
\doparttoc
\dopartlof
\dopartlot
\setcounter{parttocdepth}{4}
\addtocontents{toc}{\cftpagenumbersoff{part}}
\makeatletter
\@addtoreset{section}{part}
\@addtoreset{table}{part}
\@addtoreset{figure}{part}
\makeatother
\foreignlanguage{english}{
\thispagestyle{empty}
\begin{center}
MINISTRY OF EDUCATION AND SCIENCE\\
RUSSIAN FEDERATION\\
STATE EDUCATIONAL INSTITUTION\\
OF HIGHER PROFESSIONAL EDUCATION\\
BASHKIR STATE UNIVERSITY\\
\end{center}
\vspace{7cm}

\UDK{514.744.2}
\vspace{1cm}
\Author{K.V. Andreev}
\vspace{1cm}
\begin{center}
\Title{On the spinor formalism for even n.}

\vspace{6cm}
UFA-2011
\end{center}
 
\part{English edition}
\parttoc
\partlot
\partlof
\newpage
\section{Introduction}
In this article the answers to the following questions are considered for any even $n\ge 4$:
\begin{enumerate}
\item What is a Clifford algebra and how to construct it?
\item What is the real and complex representations?
\item What is an involution, and how it helps in a transition to real inclusions?
\item How to construct the complex and real representations of the Clifford generators (the connecting operators)?
\item How does the involution act on the Clifford generators (the connecting operators)?
\item How to lead a complex orthogonal matrix to a block diagonal form?
\item How to construct the basic isomorphisms (including the double coverings) and other relations in the explicit form with the help of the connecting operators?
\item How to construct a partial solution of the Clifford equation for the connecting operators for even $n\ge 4$?
\item How to construct the structure constants of the hypercomplex (sedenion) algebra (without division for n$>$8) by means of the connecting operators for $n\mod \ 8 = 0$?
\item How to enter and to coordinate the connection in the tangent and spinor bundles?
\item How to construct the Lie operator analogues and what are the conditions of their construction?
\item How to construct the curvature spinors?
\item What is the communication between the twistor equation, the derivational equation of the normalized Grassmannian and the conformal Killing equation?
\item What is the difference between the spinor formalism for $n \le 8$ and the one for $n>8$, and how to construct it for small dimensions?
\item How to construct the geometric representation of spinors (twistors) for $\mathbb R^6_{(2,4)}$?
\item How to construct a generalization of the Cartan triality principle to the Klein correspondence and what is the geometry of such the generalization for $n=8$?
\item How to construct (one-to-one) the spin (spin-pair) analogues of the Lie operators for $n=4$ in an explicit form?
\end{enumerate}
The conclusion of all results of this article is made on the basis \cite{Andreev0}.\footnote{The Russian edition (on the pp. \pageref*{originb}-\pageref*{origine})
contains the original variant of the article \textbf{with the corrected errors} and the original numbering of the pages (pp. 1-138)}

\section{Clifford algebra}
\Abstract{
$\phantom{ff}$In this section, it is told about how to construct the real 2n-dimensional Clifford algebra with any functional G according to
\cite[lecture 13, pp. 258-299]{Postnicov1}. In the case when the functional G in a suitable basis has the diagonal form with the identical quantity $\ll+\gg$ and $\ll-\gg$ on the main diagonal, the generators of such the Clifford algebra create the generators of the complex n-dimensional Clifford algebra which is isomorphic to the algebra of complex matrixes $\mathbb C(2^{\frac{n}{2}})$. The conclusion of all results of this section is made on the basis \cite{Postnicov1}.
}

Let $\mathcal V$ be a vector space over $\mathbb R$. Then $\mathcal V$ will be a module over $\mathbb R$: 1). a (x+y) =ax+ay,
2). (a+b) x=ax+bx, 3). ~(ab) x=a (bx), 4). 1$\cdot$ x=x, where x, y $\in\ \mathcal V$, a, b $\in\ \mathbb R$.
Let's consider the ideal
\begin{equation}
J(G):=\{T\otimes (x\otimes x - \frac{1}{2}G(x));\ T\in T_0(\mathcal V), x\in T_0{}^1(\mathcal V)\}
\end{equation}
in the algebra $T_0 (\mathcal V) =T_0 {} ^0 (\mathcal V) \oplus...\oplus T_0 {} ^q (\mathcal V) \oplus... $. Also we will define the Clifford algebra as $CL (G^{2n}): =T_0 (\mathcal V)/J(G)$. We will construct a representation of such the algebra into an polynomial algebra
\begin{equation}
\begin{array}{ll}
\alpha :\mathcal V\longmapsto \mathcal A\sps & \tilde\alpha : T_0(\mathcal V)\longmapsto \mathcal A\sps \\
\alpha(\mathcal V)=B^\myf\Lambda x_\myf\Lambda\sps &
\tilde\alpha(T_0(\mathcal V))=A+B^\myf\Lambda x_\myf\Lambda+C^{\myf\Lambda\myf\Psi}x_\myf\Lambda x_\myf\Psi +...\ \spsd\\
\end{array}
\end{equation}
Hereinafter $\myf\Lambda\sps\myf\Psi\sps... = \overline {1,2n}$. Suppose that the relation $\hat\alpha (J (G)) =0$ is executed for some mapping
$\hat\alpha:CL (G) \longmapsto C\mathcal A$. It will mean that
\begin{equation}
\label{e2.1}
x_\myf\Lambda x_\myf\Psi + x_\myf\Psi x_\myf\Lambda=G(x_\myf\Lambda , x_\myf\Psi)\spsd
\end{equation}
Therefore, the form $G (x_\myf\Lambda, x_\myf\Psi) $ in a suitable basis has the diagonal form. Suppose that for such the basis, the parities
\begin{equation}
x_\myf\Lambda x_\myf\Psi + x_\myf\Psi x_\myf\Lambda =0\sps (\myf\Lambda\ne \myf\Psi);\ \ x_\myf\Lambda{}^2=\pm 1
\end{equation}
are carried out that will define the mapping
\begin{equation}
\hat\alpha (Cl(G^{2n}))=A+B^\myf\Lambda x_\myf\Lambda+C^{[\myf\Lambda\myf\Psi ]}x_\myf\Lambda x_\myf\Psi +...
\end{equation}
In turn, it means that the Clifford algebra is \emph{finite-dimensional} and $dim\ CL(G^{2n})=2^{2n}\sps dim\ \mathcal V =2n$. Let's construct now the complex Clifford algebra $CL(g^n)$. For this purpose, we will consider the real Clifford algebra $CL(G^{2n}_{(n,n)})$. We will demand that $(x_{2\Lambda})^2=1\sps (x_{2\Lambda-1})^2=-1$ and set
\begin{equation}
\sqrt 2z_\Lambda=x_{2\Lambda}+ix_{2\Lambda -1}\sps \sqrt 2\bar z_\Lambda=x_{2\Lambda}-ix_{2\Lambda -1}\spsd
\end{equation}
Hereinafter $\Lambda\sps\Psi\sps ... =\overline{1,n}$. Then
\begin{equation}
\begin{array}{lcll}
2(z_\Lambda z_\Psi +z_\Psi z_\Lambda)&=&
(x_{2\Lambda}+ix_{2\Lambda -1})(x_{2\Psi}+ix_{2\Psi -1})+(x_{2\Psi}+ix_{2\Psi -1})(x_{2\Lambda}+ix_{2\Lambda -1})=&\\
&=&x_{2\Lambda}x_{2\Psi}-x_{2\Lambda -1}x_{2\Psi -1}+i(x_{2\Lambda -1}x_{2\Psi}+x_{2\Lambda}x_{2\Psi -1})+&\\
&+&x_{2\Psi}x_{2\Lambda}-x_{2\Psi -1}x_{2\Lambda -1}+i(x_{2\Psi -1}x_{2\Lambda}+x_{2\Psi}x_{2\Lambda -1})=&\\
& &
=\begin{cases}
4,&\Lambda =\Psi\sps\\
0,&\Lambda \ne\Psi\sps\\
\end{cases}\\
2(z_\Lambda \bar z_\Psi +\bar z_\Psi z_\Lambda)&=&
(x_{2\Lambda}+ix_{2\Lambda -1})(x_{2\Psi}-ix_{2\Psi -1})+(x_{2\Psi}-ix_{2\Psi -1})(x_{2\Lambda}+ix_{2\Lambda -1})=&\\
&=&x_{2\Lambda}x_{2\Psi}+x_{2\Lambda -1}x_{2\Psi -1}-i(-x_{2\Lambda -1}x_{2\Psi}+x_{2\Lambda}x_{2\Psi -1})+&\\
&+&x_{2\Psi}x_{2\Lambda}+x_{2\Psi -1}x_{2\Lambda -1}-i(x_{2\Psi -1}x_{2\Lambda}-x_{2\Psi}x_{2\Lambda -1})=0\sps&\\
\end{array}
\end{equation}
\begin{equation}
\label{e2.2}
\begin{array}{l}
z_\Lambda z_\Psi +z_\Psi z_\Lambda =g(z_\Lambda ,z_\Psi)\spsd
\end{array}
\end{equation}
Therefore, the factorization of the complex algebra $T_0(\mathcal V^\mathbb C)= T_0(\mathcal V^\mathbb C)(z)$ by the ideal
$J(g)=\{T\otimes (z\otimes z-\frac{1}{2}g(z));\ \ T\in T_0(\mathcal V^\mathbb C)\sps z \in T_0{}^1(\mathcal V^\mathbb C)\}$ will define the complex algebra $CL(g^n)\cong T_0(\mathcal V^\mathbb C)/J(g)$.
\begin{theorem}
\label{theorem2.1}
For the Clifford algebra, there is the decomposition
\begin{equation}
\begin{array}{c}
CL(G_{(n+1,n+1)}^{(2n+2)})=CL(G_{(n,n)}^{2n})\otimes CL(G_{(1,1)}^2)\sps\\
CL(G_{(2n+2,0)}^{(2n+2)})=CL(G_{(2n,0)}^{2n})\otimes CL(G_{(2,0)}^2)\sps
CL(g^n)=CL(g^{n-2})\otimes CL(g^2)
\end{array}
\end{equation}
with the multiplication $a\cdot b=(\tilde a\sps\tilde{\tilde a})\cdot (\tilde b\sps\tilde{\tilde b})=(\tilde a\tilde b\sps\tilde{\tilde a}\tilde{\tilde b})$.
\end{theorem}
\begin{proof}
Let $x_\myf\Lambda$ be a basis in $\mathcal V^{2n}$, and let $x_{2n+1},x_{2n+2}$ be a basis in $\mathcal V^2$
\begin{equation}
\begin{array}{ll}
x_\myf\Lambda x_\myf\Psi + x_\myf\Psi x_\myf\Lambda =0\sps  (\myf\Lambda\ne \myf\Psi);&
(x_{2\Lambda}){}^2=1;\ \ (x_{2\Lambda-1}){}^2=-1\sps\\
x_{2n+1} x_{2n+2} + x_{2n+2} x_{2n+1} =0;\ \  &
(x_{2n+2}){}^2=-1;\ \ (x_{2n+1}){}^2=1\spsd
\end{array}
\end{equation}
It will give a chance to define the basis in $\mathcal V^{2n+2}$ as
\begin{equation}
X_\myf\Lambda=x_\myf\Lambda x_{2n+1}x_{2n+2}\sps
X_{2n+1}=x_{2n+2}\sps X_{2n+2}=x_{2n+1}
\end{equation}
with the conditions
\begin{equation}
%
\end{equation}
Then for any $a\in CL(G_{(n+1,n+1)}^{2n+2})$, the decomposition
\begin{equation}
\label{e2.7}
\begin{array}{c}
a=a_1X_{2n+2}X_{2n+1}+a_2X_{2n+2}+a_3X_{2n+1}+a_4=\\
=a_1\otimes x_{2n+1}x_{2n+2}+a_2\otimes x_{2n+1}+a_3\otimes x_{2n+2}+a_4\otimes 1
\end{array}
\end{equation}
is executed. Let's spend the replacement of the basis
\begin{equation}
\begin{array}{c}
y_{2n+1}=(x_{2n+1}+x_{2n+2})/2\sps y_{2n+2}=(x_{2n+1}-x_{2n+2})/2\sps\\
(y_{2n+1}){}^2=0\sps (y_{2n+2}){}^2=0\sps
2y_{2n+1}y_{2n+2}=1-x_{2n+1}x_{2n+2}\spsd
\end{array}
\end{equation}
Then, the decomposition (\ref{e2.7}) will have the form
\begin{equation}
a=(-a_1+a_4)y_{2n+1}y_{2n+2}+(a_2+a_3)y_{2n+1}+(a_2-a_3)y_{2n+2}+(a_1+a_4)y_{2n+2}y_{2n+1}\spsd
\end{equation}
Hence, any element $a\in CL(G_{(1,1)}^2)$ is decomposed as
\begin{equation}
%
\end{equation}
and for any two elements $a,b\in CL(G_{(n+1,n+1)}^{2n+2})$, the relation
$$
\begin{array}{ccccccccl}
c:=a\cdot b&=&     (a_1\otimes x_{2n+1}x_{2n+2}&+&a_2\otimes x_{2n+1}&+&a_3\otimes x_{2n+2}&+&a_4\otimes 1)\cdot\\
           & &\cdot(b_1\otimes x_{2n+1}x_{2n+2}&+&b_2\otimes x_{2n+1}&+&b_3\otimes x_{2n+2}&+&b_4\otimes 1)=\\
\end{array}
$$
\begin{equation}
\begin{array}{lll}
=\underbrace{(a_2b_3-a_3b_2+a_1b_4+a_4b_1)}_{:=c_1}\otimes x_{2n+1}x_{2n+2}&+&
 \underbrace{(a_2b_4+a_4b_2-a_1b_3+a_3b_1)}_{:=c_2}\otimes x_{2n+1}+\\
+\underbrace{(a_3b_4+a_4b_3-a_1b_2+a_2b_1)}_{:=c_3}\otimes x_{2n+2}&+&
 \underbrace{(a_1b_1+a_4b_4+a_2b_2-a_3b_3)}_{:=c_4}
\end{array}
\end{equation}
is executed. At the same time, for any two elements $a,b\in CL(G_{(1,1)}^2)$, the identity
$$
c:=a\cdot b=
\left(
%
\right)\spsd
\end{equation}
is true. That will allow to define the correspondence between the generators of the Clifford algebra $CL(G_{(1,1)}^{2})$ and their matrix representations
\begin{equation}
x_{2n+1}=
\left(
\begin{array}{rr}
0 & 1\\
1 & 0
\end{array}
\right)\sps
x_{2n+2}=
\left(
\begin{array}{rr}
 0 & 1\\
-1 & 0
\end{array}
\right)\spsd
\end{equation}
 This means that for any two elements $a,b\in CL(G_{(n+1,n+1)}^{2n+2})$, for the similar matrix representation, $c_1,c_2,c_3,c_4\in CL(G_{(n,n)}^{2n})$. Other two representations are similarly proved also.
\end{proof}
\begin{theorem}
\label{theorem2.2}
The complex Clifford algebra for even n is isomorphic to the algebra of complex matrixes of kind
\begin{equation}
CL(g)\cong \mathbb C(2^{n/2})\spsd
\end{equation}
\end{theorem}
\begin{proof}
Let's determine according to the proof of Theorem \ref{theorem2.1} and replacing $X$ on $x$
$$
\sqrt 2z_{n+1}=x_{2n+2}+ix_{2n+1}\sps
\sqrt 2z_{n+2}=x_{2n+4}+ix_{2n+3}\sps
$$
\begin{equation}
\begin{array}{c}
z_{n+1}z_{n+1}=1\sps z_{n+1}\bar z_{n+1}=ix_{2n+1}x_{2n+2}=-\bar z_{n+1}z_{n+1}\sps\\
z_{n+2}z_{n+2}=1\sps z_{n+2}\bar z_{n+2}=ix_{2n+3}x_{2n+4}=-\bar z_{n+2}z_{n+2}\sps
\end{array}
\end{equation}
$$
z_{n+1}z_{n+2}+z_{n+2}z_{n+1}=0\spsd
$$
This will allow to define the correspondence between the generators of the Clifford algebra $CL(G_{(2,2)}^{4})$ and their matrix representations
\begin{equation}
\label{e2.4}
\begin{array}{c}
x_{2n+2}=
\left(
\begin{array}{rrrr}
0 & 0 & 1 & 0\\
0 & 0 & 0 & 1\\
1 & 0 & 0 & 0\\
0 & 1 & 0 & 0
\end{array}
\right)\sps
x_{2n+1}=
\left(
\begin{array}{rrrr}
0 & 0 & 1 & 0\\
0 & 0 & 0 &-1\\
-1& 0 & 0 & 0\\
0 & 1 & 0 & 0
\end{array}
\right)\sps\\
x_{2n+4}=
\left(
\begin{array}{rrrr}
0 & 0 & 0 & 1\\
0 & 0 &-1 & 0\\
0 &-1 & 0 & 0\\
1 & 0 & 0 & 0
\end{array}
\right)\sps
x_{2n+3}=
\left(
\begin{array}{rrrr}
 0 & 0 & 0 &-1\\
 0 & 0 &-1 & 0\\
 0 & 1 & 0 & 0\\
 1 & 0 & 0 & 0
\end{array}
\right)\spsd
\end{array}
\end{equation}
Then for the generators $CL(g^2)$ in the special basis similar to the one of Example \ref{ex2.2}\footnote{In the example, $x_\myf\Lambda\longmapsto \frac{1}{\sqrt{2}}x_\myf\Lambda$, $z_\Lambda\longmapsto \frac{1}{\sqrt{2}}z_\Lambda$.}, we will obtain
\begin{equation}
\label{e2.5}
\sqrt{2}z_{n+1}=
\left(
\begin{array}{rr}
  0  & 1+i\\
 1-i & 0
\end{array}
\right)\sps
\sqrt{2}z_{n+2}=
\left(
\begin{array}{rr}
   0 & 1-i\\
 1+i & 0
\end{array}
\right)\sps
\end{equation}
as will prove our theorem. Strictly to construct the specified conformity in the common view, the knowledge of some additional facts, which are worthy to devote them the following section, is necessary.
\end{proof}

Thus, the real Clifford algebra $CL(G_{(n,n)}^{2n})$ is isomorphic to the real algebra of n-dimension matrixes
\begin{equation}
CL(G^{2n}_{(n,n)})\cong \mathbb R(2^n)\spsd
\end{equation}
Therefore, there is a mapping
\begin{equation}
\label{e2.6}
\gamma:\ \ \mathbb R(2^n)\longmapsto (\mathcal V^\mathbb R)^{2n}\spsd
\end{equation}
On the basis (\ref{e2.1}), the operators $\gamma_\myf\Lambda$  should satisfy the equation
\begin{equation}
\label{e2.8}
\gamma_\myf\Lambda\gamma_\myf\Psi+\gamma_\myf\Psi\gamma_\myf\Lambda=G_{\myf\Lambda\myf\Psi}\sps
\end{equation}
which is called \emph{Clifford equation}. At the same time, the complex Clifford algebra is represented by the matrix algebra $\mathbb C(2^{n/2})$. The multiplication in the complex Clifford algebra delegates the matrix multiplication in $\mathbb C(2^{n/2})$. Therefore, one can construct mapping
\begin{equation}
\label{e2.9}
\gamma:\ \ \mathbb C(2^{n/2})\longmapsto (\mathcal V^\mathbb C)^n\spsd
\end{equation}
From (\ref{e2.2}), it follows that the operators $\gamma_\Lambda$ should satisfy the complex Clifford equation
\begin{equation}
\label{e2.10}
\gamma_\Lambda\gamma_\Psi+\gamma_\Psi\gamma_\Lambda=g_{\Lambda\Psi}\spsd
\end{equation}

\section{Complex and real representations}
\Abstract {
$\phantom{ff}$In this section, it is told about how locally to construct the n-dimensional complex representation on the real one of the 2n-dimensional pseudo-Riemannian manifold with tangent bundle fibers isomorphic to $\mathbb R^{2n}_{(n, n)}$. For this purpose, we construct the corresponding complex reparametrization of an atlas in a neighborhood of some point that induces the Neifeld operators \cite{Newfield1} connected to the Norden affinor \cite{Norden1} in the tangent bundle. Accordingly, by means of the connecting operators $\gamma_{\tilde\Lambda}$, the Neifeld operator analogues are induced to within a sign in the spinor bundle. It will give a chance to pass to the complex representation of the connecting operators $\gamma_\Lambda$. The conclusion of all results of this section is made on the basis  \cite{Newfield1}, \cite{Norden1}.
}

Under \emph{complex analytical Riemannian space} $\mathbb CV_n$, we will further understand an analytical complex bundle supplied with \emph{analytical quadratic metric}, i.e., a metric defined by means of a symmetric nonsingular tensor $g_{\Lambda\Psi}$ (here $\Lambda , \Psi ,... = \overline {1, n} $; $\myf\Lambda ,\myf\Psi ,... = \overline {1,2n}$), which the coordinates are analytical functions of the point coordinates. To this tensor, there corresponds \emph{complex Riemannian torsion-free connection}, the coefficients of which are defined by the Christoffel symbols, and hence these coefficients are analytical functions. The tangent bundle to this manifold $\tau^\mathbb C (\mathbb CV_n)$ has fibers $\tau_x^\mathbb C\cong \mathbb C\mathbb R^n $ that is fibers isomorphic to the n-dimensional complex Euclidian space, the metric of which is defined by the value of the metric tensor at the given point. The real representation $V_{2n}$ of $\mathbb CV_n$ has the tangent bundle $\tau^\mathbb R(V_{2n})$ with fibers isomorphic to $\mathbb R^{2n}_{(n, n)}$. Let on $V_{2n}$, an atlas $(U; x^\myf\Lambda)$ be set. We will consider \emph{reparametrization} of this atlas $(U; w^\Lambda)$ such that $w^\Lambda=\frac{1}{\sqrt{2}}(u^\Lambda(x^\myf\Lambda)+iv^\Lambda (x^\myf\Lambda))$, which is locally solvable as $x^\myf\Lambda= x^\myf\Lambda (u^\Lambda, v^\Lambda)$. Set
\begin{equation}
\label{e1.1}
 m^\Lambda{}_\myf\Lambda:=\frac{1}{\sqrt{2}}(
 \frac{\partial u^\Lambda}{\partial x^\myf\Lambda}+i
 \frac{\partial v^\Lambda}{\partial x^\myf\Lambda})=:\frac{\partial w^\Lambda}{\partial x^\myf\Lambda}\sps
 m_\Lambda{}^\myf\Lambda:=\frac{1}{\sqrt{2}}(
 \frac{\partial x^\myf\Lambda}{\partial u^\Lambda}-i
 \frac{\partial x^\myf\Lambda}{\partial v^\Lambda})=:\frac{\partial x^\myf\Lambda}{\partial w^\Lambda}\spsd
\end{equation}
Then
\begin{equation}
 \Delta_\myf\Lambda{}^\myf\Psi:= m^\Lambda{}_\myf\Lambda m_\Lambda{}^\myf\Psi=\frac{1}{2}((
 \underbrace{
 \frac{\partial u^\Lambda}{\partial x^\myf\Lambda}\ \frac{\partial x^\myf\Psi}{\partial u^\Lambda}+
 \frac{\partial v^\Lambda}{\partial x^\myf\Lambda}\ \frac{\partial x^\myf\Psi}{\partial v^\Lambda}}
 _{=\delta_\myf\Lambda{}^\myf\Psi})+
 i(\underbrace{
   \frac{\partial v^\Lambda}{\partial x^\myf\Lambda}\ \frac{\partial x^\myf\Psi}{\partial u^\Lambda}-
   \frac{\partial u^\Lambda}{\partial x^\myf\Lambda}\ \frac{\partial x^\myf\Psi}{\partial v^\Lambda}}
   _{=:f_\myf\Lambda{}^\myf\Psi}))\sps
\end{equation}
that will define the complex structure f on $V_{2n} $, where $\Delta_\Psi{}^\Lambda$ is \emph{Norden affinor}. We will define \emph{Neifeld operators} $\bar m^{\Psi '}{}_\myf\Psi:=\overline{m^\Psi{}_\myf\Psi}$ as
\begin{equation}
\begin{array}{c}
 0=:m^\Lambda{}_\myf\Lambda \bar m_{\Psi '}{}^\myf\Lambda=
 \frac{1}{2}((
 \underbrace{
 \frac{\partial u^\Lambda}{\partial x^\myf\Lambda}\ \frac{\partial x^\myf\Lambda}{\partial u^{\Psi '}}-
 \frac{\partial v^\Lambda}{\partial x^\myf\Lambda}\ \frac{\partial x^\myf\Lambda}{\partial v^{\Psi '}}}
 _{\equiv 0})+
 i\ (\frac{\partial u^\Lambda}{\partial x^\myf\Lambda}\ \frac{\partial x^\myf\Lambda}{\partial v^{\Psi '}}+
    \frac{\partial v^\Lambda}{\partial x^\myf\Lambda}\ \frac{\partial x^\myf\Lambda}{\partial u^{\Psi '}}
   ))\sps\\
\end{array}
\end{equation}
\begin{equation}
\begin{array}{c}
 \delta_\Psi{}^\Lambda=:m^\Lambda{}_\myf\Lambda m_\Psi{}^\myf\Lambda=
 \frac{1}{2}((
 \underbrace{
 \frac{\partial u^\Lambda}{\partial x^\myf\Lambda}\ \frac{\partial x^\myf\Lambda}{\partial u^\Psi}+
 \frac{\partial v^\Lambda}{\partial x^\myf\Lambda}\ \frac{\partial x^\myf\Lambda}{\partial v^\Psi}}
 _{=2\delta_\Psi{}^\Lambda})-
 i\ (\frac{\partial u^\Lambda}{\partial x^\myf\Lambda}\ \frac{\partial x^\myf\Lambda}{\partial v^\Psi}-
    \frac{\partial v^\Lambda}{\partial x^\myf\Lambda}\ \frac{\partial x^\myf\Lambda}{\partial u^\Psi}
   ))\sps
\end{array}
\end{equation}
 Then, $\frac{\partial u^\Lambda}{\partial x^\myf\Lambda}\ \frac{\partial x^\myf\Lambda}{\partial v^\Psi}\equiv 0$ (where $v^{\Psi '}=\delta_\Psi{}^{\Psi '} v^\Psi$ and $u^{\Psi '}=\delta_\Psi{}^{\Psi '} u^\Psi$, and the tensor $\delta_\Psi{}^{\Psi '}$ has the identity matrix). Hence,
\begin{equation}
 \frac{\partial u^\Lambda}{\partial x^\myf\Lambda} f_\myf\Psi{}^\myf\Lambda=
 \frac{\partial u^\Lambda}{\partial x^\myf\Lambda}\
 (\frac{\partial v^\Psi}{\partial x^\myf\Psi}\ \frac{\partial x^\myf\Lambda}{\partial u^\Psi}-
  \frac{\partial u^\Psi}{\partial x^\myf\Psi}\ \frac{\partial x^\myf\Lambda}{\partial v^\Psi})=
 \frac{\partial v^\Lambda}{\partial x^\myf\Psi}
\end{equation}
are \emph{Cauchy-Riemann conditions} if the complex structure has the canonical form in the given basis
\begin{equation}
f_\myf\Lambda{}^\myf\Psi=\left(
\begin{array}{cc}
  0 & E \\
 -E & 0
\end{array}\right)\sps
\end{equation}
where E is the identity matrix of dimension $n\times n$. It is easy to check up that $f^2 =-E $. If now to demand that maps of the atlas on $\mathbb CV_n$ have been coordinated by means of holomorphic transformations then by such a way, one can identify the two manifolds: $\mathbb CV_n $ and $V_{2n}$. Thus, it is possible to construct the isomorphism between fibers of the tangent bundles: $\tau^\mathbb C (\mathbb CV_n) $ and $\tau^\mathbb C (V_{2n})$ as well as between fibers of their real representations: $\tau^\mathbb R(\mathbb CV_n)$ and $\tau^\mathbb R(V_{2n})$.\\

Let in the same basis in which the complex structure has the canonical form, our metric be
\begin{equation}
 G_{\myf\Lambda\myf\Psi}=
 \left(
 \begin{array}{cc}
  E &  0\\
  0 & -E
 \end{array}
 \right)\sps
\end{equation}
where E is the identity matrix of the dimension $n\times n$. The special operators $m_\Lambda{}^\myf\Lambda\sps m^\Lambda{}_\myf\Lambda$ which by definition, should satisfy the following system of the equations
\begin{equation}
\label{e1.2}
 \left\{
 \begin{array}{l}
  m_\Lambda{}^\myf\Lambda m^\Psi{}_\myf\Lambda=\delta_\Lambda{}^\Psi\sps
  \\ \\
  m_\Lambda{}^\myf\Lambda \bar m^{\Psi '}{}_\myf\Lambda=0\sps
  \\ \\
  m_\Lambda{}^\myf\Lambda m^\Lambda{}_\myf\Psi=\frac{1}{2}(\delta_\myf\Psi{}^\myf\Lambda+if_\myf\Psi{}^\myf\Lambda):=
  \triangle_\myf\Psi{}^\myf\Lambda\sps
  \\
 \end{array}
 \right.
\end{equation}
answer for the conformity $V_{2n} \longleftrightarrow \mathbb CV_{n}$. Then from the average equation, the parities
\begin{equation}
 m^\Psi{}_\myf\Omega=m^\Lambda{}_\myf\Omega (m_\Lambda{}^\myf\Lambda m^\Psi{}_\myf\Lambda)=(m^\Lambda{}_\myf\Omega m_\Lambda{}^\myf\Lambda)
 m^\Psi{}_\myf\Lambda=\frac{1}{2}m^\Psi{}_\myf\Omega+\frac{i}{2} f_\myf\Omega{}^\myf\Lambda m^\Psi{}_\myf\Lambda\spsd
\end{equation}
will follow. Whence, the relations
\begin{equation}
\label{e1.3}
 m^\Psi{}_\myf\Omega=i\ f_\myf\Omega{}^\myf\Lambda m^\Psi{}_\myf\Lambda
\end{equation}
 are the Cauchy-Riemann conditions again. Therefore, in the presence of the metric $G_{\myf\Lambda\myf\Psi}$ for $\mathbb R^{2n}_{(n,n)}$,  the metric
\begin{equation}
\label{e1.4}
\begin{array}{c}
 g_{\Lambda\Psi}:=G_{\myf\Lambda\myf\Psi} m_\Lambda{}^\myf\Lambda m_\Psi{}^\myf\Psi\sps
 \bar g_{\Lambda ' \Psi '}:=G_{\myf\Lambda\myf\Psi} \bar m_{\Lambda '}{}^\myf\Lambda \bar m_{\Psi '}{}^\myf\Psi\sps\\
 g_{\Lambda ' \Psi }:=G_{\myf\Lambda\myf\Psi} \bar m_{\Lambda '}{}^\myf\Lambda m_\Psi{}^\myf\Psi\sps
 \bar g_{\Lambda \Psi '}:=G_{\myf\Lambda\myf\Psi} \bar m_\Lambda{}^\myf\Lambda \bar m_{\Psi '}{}^\myf\Psi
\end{array}
\end{equation}
is induced in the complex representation. In the standard complex representation (where instead of the space $\mathbb R^{2n}_{(n, n)}$, the space $\mathbb R^{2n}$ undertakes), this means that $g_{\Lambda '\Psi '}=\bar g_{\Lambda ' \Psi '}=0$ and leads to the Hermitian metric $g_{\Lambda '\Psi}$. However, in our case, it is necessary to take advantage of the conditions
\begin{equation}
\label{e1.5}
 g_{\Lambda '\Psi  }=\bar g_{\Lambda \Psi '}=0\spsd
\end{equation}
It will lead to some restrictions on the metric tensor
\begin{equation}
\begin{array}{c}
 0=g_{\Lambda ' \Psi} \bar m^{\Lambda '}{}_\myf\Omega m^\Psi{}_\myf\Theta=
 (G_{\myf\Lambda\myf\Psi} \bar m_{\Lambda '}{}^\myf\Lambda m_\Psi{}^\myf\Psi) \bar m^{\Lambda '}{}_\myf\Omega m^\Psi{}_\myf\Theta=
 G_{\myf\Lambda\myf\Psi}(\bar m_{\Lambda '} {}^\myf\Lambda \bar m^{\Lambda '}{}_\myf\Omega )( m_\Psi{}^\myf\Psi m^\Psi{}_\myf\Theta)=\\
 =G_{\myf\Lambda\myf\Psi}\cdot \frac{1}{2}(\delta_\myf\Omega{}^\myf\Lambda - if_\myf\Omega{}^\myf\Lambda ) \cdot
 \frac{1}{2}(\delta_\myf\Theta{}^\myf\Psi + if_\myf\Theta{}^\myf\Psi)=\\
 =\frac{1}{4}(\underbrace{G_{\myf\Omega\myf\Theta}+(f_\myf\Omega{}^\myf\Lambda f_\myf\Theta{}^\myf\Psi G_{\myf\Lambda\myf\Psi})}_{=0} +
  i(\underbrace{(f_\myf\Omega{}^\myf\Lambda G_{\myf\Lambda\myf\Theta})-(f_\myf\Theta{}^\myf\Psi G_{\myf\Omega\myf\Psi}))}_{=0})=0\spsd
\end{array}
\end{equation}
Therefore, our pseudo-Euclidian metric with the canonical complex structure satisfies the equation
\begin{equation}
\label{e1.6}
 f_\myf\Omega{}^\myf\Lambda G_{\myf\Lambda\myf\Psi}f_\myf\Theta{}^\myf\Psi=-G_{\myf\Omega\myf\Theta}\sps
\end{equation}
$$
\left(
\begin{array}{cc}
  0 & E \\
 -E & 0
\end{array}
\right)
\left(
\begin{array}{cc}
  E & 0 \\
  0 & -E
\end{array}
\right)
\left(
\begin{array}{cc}
  0 & -E \\
  E &  0
\end{array}
\right)
=
-\left(
\begin{array}{cc}
  E &  0 \\
  0 & -E
\end{array}
\right)\spsd
$$
Besides, it will be executed
\begin{equation}
\label{e1.7}
 G_{\myf\Lambda\myf\Psi}=
 m^\Lambda{}_\myf\Lambda m^\Psi{}_\myf\Psi g_{\Lambda\Psi}+
 \bar m^{\Lambda '}{}_\myf\Lambda \bar m^{\Psi '}{}_\myf\Psi \bar g_{\Lambda '\Psi '}\spsd
\end{equation}
\vspace{4mm}
\begin{example} At n=1 for $\mathbb R^2_{(1,1)}$
\begin{equation}
%
\end{equation}
\end{example}
Now the operators (\ref{e1.2}) can be applied to (\ref{e2.6}). For this purpose, we will consider the decomposition for n=2 (N=1)
\begin{equation}
\label{e1.8}
 f_\myf\Psi{}^\myf\Lambda\gamma_\myf\Lambda=\tilde F\gamma_\myf\Psi \tilde {\tilde F}\sps \tilde F{}^4=\tilde {\tilde F}{}^4=E\spsd
\end{equation}
The proof of this formula will be given in the following section (\ref{e3.9}). From (\ref {e1.8}), it will follow
\begin{equation}
\begin{array}{c}
 \triangle_\myf\Psi{}^\myf\Lambda \gamma_\myf\Lambda=
 \frac{1}{2}(\delta_\myf\Psi{}^\myf\Lambda+if_\myf\Psi{}^\myf\Lambda)\gamma_\myf\Lambda=\\ \\
=\underbrace{\frac{1}{2}(\delta +\frac{(1+i)}{2}\tilde F+\frac{(1-i)}{2}\tilde  F^3)}_{:=\tilde\triangle_+}\gamma_\myf\Psi
 \underbrace{\frac{1}{2}(\delta +\frac{(1+i)}{2}\tilde {\tilde F}+\frac{(1-i)}{2}\tilde {\tilde F}^3)}_{:=\tilde {\tilde\triangle}_+}+\\
+\underbrace{\frac{1}{2}(\delta -\frac{(1+i)}{2}\tilde F-\frac{(1-i)}{2}\tilde F^3)}_{:=\tilde\triangle_-}\gamma_\myf\Psi
 \underbrace{\frac{1}{2}(\delta -\frac{(1+i)}{2}\tilde {\tilde F}-\frac{(1-i)}{2}\tilde {\tilde F}^3)}_{:=\tilde{\tilde\triangle}_-}=\\
=\tilde\triangle_+\gamma_\myf\Psi\tilde {\tilde\triangle}_+ +\tilde\triangle_-\gamma_\myf\Psi\tilde{\tilde\triangle}_-\spsd
\end{array}
\end{equation}
This means that one can set for n=2 (N=1)
\begin{equation}
\label{e1.9}
 %
 \right.  \\
 \end{array}
\end{equation}
Here $\myf A,\myf B,\myf C,\myf D, ... =\overline{1,(2N)^2}$, $\hat A,\hat B,\hat C,\hat D, ... =\overline{1,2N}$. Note that we can introduce the operation of \emph{conjugation} (not complex!) such that the operators $\tilde m_{\pm}\ (\tilde{\tilde m}_{\pm})$ and $\tilde m_{\mp}\ (\tilde{\tilde m}_{\mp})$ are conjugate to each other for such the operation. Hence, as the operators $\gamma_\Lambda$, one can take either $\gamma_+{}_\Lambda$ or $\gamma_-{}_\Lambda$
\begin{equation}
\label{e1.10}
\gamma_\pm{}_\Lambda{}_{\hat A}{}^{\hat B}:=
m{}_\Lambda{}^\myf\Lambda\gamma_\myf\Lambda{}_\myf C{}^\myf D
\tilde m_\pm{}_{\hat A}{}^\myf C \tilde {\tilde m}_\pm{}^{\hat B}{}_\myf D\spsd
\end{equation}

\vspace{4mm}
\begin{example} At n=2 for $\mathbb R^4_{(2,2)}$
\label{ex2.2}
\begin{equation}
\begin{array}{c}
 m_\Lambda{}^\myf\Lambda=\frac{1}{\sqrt{2}}
\left(
\begin{array}{cccc}
  1 & i & 0 & 0 \\
  0 & 0 & 1 & i
\end{array}
\right)\sps
 (m^T)_\myf\Lambda{}^\Lambda=\frac{1}{{\sqrt{2}}}
\left(
\begin{array}{cc}
  1 & 0 \\
 -i & 0 \\
  0 & 1 \\
  0 &-i
\end{array}
\right)\sps
 g_{\Lambda\Psi}=
\left(
\begin{array}{cc}
  1 &  0 \\
  0 &  1
\end{array}
\right)\sps\\
 G_{\myf\Lambda\myf\Psi}=
\left(
\begin{array}{cccc}
  1 &  0 & 0 &  0 \\
  0 & -1 & 0 &  0 \\
  0 &  0 & 1 &  0 \\
  0 &  0 & 0 & -1 \\
\end{array}
\right)\sps
 f_\myf\Lambda{}^\myf\Psi=
\left(
\begin{array}{cccc}
  0 & 1 & 0 &  0 \\
 -1 & 0 & 0 &  0 \\
  0 & 0 & 0 &  1 \\
  0 & 0 &-1 &  0 \\
\end{array}\right)\spsd
\end{array}
\end{equation}
$\gamma_\myf\Lambda$ are defined according to (\ref{e2.8}), $\gamma_\Lambda$ are defined according to (\ref{e2.10}).
\begin{equation}
\begin{array}{c}
 m_{\hat A}{}^\myf B:=\tilde m_+{}_{\hat A}{}^\myf B=\tilde {\tilde m}_+{}_{\hat A}{}^\myf B=
 \frac{1}{{\sqrt{2}}}\left(
\begin{array}{cccc}
  1 & i & 0 & 0 \\
  0 & 0 & 1 & 1
\end{array}
\right)\sps\\
 (m^T)_\myf B{}^{\hat A}=(\tilde m_+^T){}_\myf B{}^{\hat A}=(\tilde {\tilde m}_+^T){}_\myf B{}^{\hat A}=
 \frac{1}{{\sqrt{2}}}\left(
\begin{array}{cc}
  1 & 0 \\
 -i & 0 \\
  0 & 1 \\
  0 & 1
\end{array}
\right)\sps\\
 \tilde m_-{}_{\hat A}{}^\myf B=\tilde {\tilde m}_-{}_{\hat A}{}^\myf B=
 \frac{1}{{\sqrt{2}}}\left(
\begin{array}{cccc}
  1 &-i & 0 & 0 \\
  0 & 0 & 1 &-1
\end{array}
\right)\sps\\
 (\tilde m_-^T){}_\myf B{}^{\hat A}=(\tilde {\tilde m}_-^T){}_\myf B{}^{\hat A}=
 \frac{1}{{\sqrt{2}}}\left(
\begin{array}{cc}
  1 & 0 \\
  i & 0 \\
  0 & 1 \\
  0 &-1
\end{array}
\right)\sps\\
F^\myf A{}_\myf B:=\tilde F^\myf A{}_\myf B=\tilde {\tilde F}^\myf A{}_\myf B=
\left(
\begin{array}{cccc}
  0 & 1 & 0 &  0 \\
 -1 & 0 & 0 &  0 \\
  0 & 0 & 0 &  1 \\
  0 & 0 & 1 &  0 \\
\end{array}\right)\spsd
\end{array}
\end{equation}
By the way, it once again says that the operator $F$ is a matrix fourth root from the matrix identity. The square $F^2$ corresponds to the trivial transformation multiplied by -1 in the space $\mathbb R^4_{(2,2)}$.
\end{example}

In the case of arbitrary n, the algorithm for the reduction of the spinor dimension is as follows.
\begin{algorithm}
\label{a2.1}
The complex structure can be represented as a product of elementary transformations of the dimension $4 \times 4$. Each I-th elementary transformation generates the operator $(m_I)_\Lambda{}^{\tilde\Lambda}$ which is responsible for the transition from the real representation of the subspace $\mathbb (R_I)^4_{(2,2)}\subset\mathbb R^{2n}_{(2n,2n)}$ to the complex representation  $\mathbb (C_I)^2\subset R^{2n-4}_{(2n-2,2n-2)}\oplus\mathbb (C_I)^2$, and hence this lowers the dimension of the space on 2. Accordingly, in the spinor space, the operators $\tilde{\tilde m}_{I_\pm} ,\ \tilde m_{I_\pm}$ which can reduce the dimension of the spinor space by 2 times are generated. Obviously, the $n/2$ steps are required $(I =\overline{1, \frac{n}{2}})$. This would reduce the spinor dimension by $2^{\frac{n}{2}}=2N$ times and would lead to the complex matrix representation of the dimension $2N \times 2N$. The operators $\tilde{\tilde m}_{I_\pm} ,\ \tilde m_{I_\pm}$ will satisfy the following relations
\begin{equation}
\label{e1.11}
 \begin{array}{c}
 \left\{
 \begin{array}{l}
  \tilde m_{I_\pm}{}_{\hat A}{}^\myf B \tilde m_{I_\pm}{}^{\hat C}{}_\myf B=\delta_{\hat A}{}^{\hat C}\sps
  \\ \\
  \tilde m_{I_\pm}{}_{\hat A}{}^\myf B \tilde m_{I_\mp}{}^{\hat C}{}_\myf B=0\sps
  \\ \\
  \tilde m_{I_\pm}{}_{\hat A}{}^\myf B \tilde m_{I_\pm}{}^{\hat A}{}_\myf C=\tilde\triangle_{I_\pm}{}_\myf C{}^\myf B\sps
  \\
 \end{array}
 \right.
 \left\{
 \begin{array}{l}
  \tilde{\tilde m}_{I_\pm}{}_{\hat A}{}^\myf B \tilde{\tilde m}_{I_\pm}{}^{\hat C}{}_\myf B=\delta_{\hat A}{}^{\hat C}\sps
  \\ \\
  \tilde{\tilde m}_{I_\pm}{}_{\hat A}{}^\myf B \tilde{\tilde m}_{I_\mp}{}^{\hat C}{}_\myf B=0\sps
  \\ \\
  \tilde{\tilde m}_{I_\pm}{}_{\hat A}{}^\myf B \tilde{\tilde m}_{I_\pm}{}^{\hat A}{}_\myf C=\tilde{\tilde\triangle}_{I_\pm}{}_\myf C{}^\myf B\spsd
  \\
 \end{array}
 \right.  \\
 \end{array}
\end{equation}
Here for I step $\myf A\sps\myf B\sps\myf C\sps\myf D, ... =\overline{1,\frac{(2N)^2}{2^{I-1}}}$, $\hat A,\hat B,\hat C,\hat D, ... =\overline{1,\frac{(2N)^2}{2^I}}$. Then we can construct the operators ($\tilde m_{J_{z_J}}:=\tilde m_{J_\pm}{}_{\hat A}{}^\myf B$,
$\tilde{\tilde m}^*_{J_{z_J}}:=\tilde{\tilde m}_{J_\pm}{}^{\hat A}{}_\myf B$)
\begin{equation}
\label{e1.12}
\begin{array}{l}
\tilde M_K:=\tilde m_{\frac{n}{2}_{z_{\frac{n}{2}}}}\tilde m_{{\frac{n}{2}-1}_{z_{\frac{n}{2}-1}}}\cdot ...\cdot\tilde m_{2_{z_2}}\tilde m_{1_{z_1}}\sps
\tilde{\tilde M}_K:=\tilde{\tilde m}^*_{\frac{n}{2}_{z_{\frac{n}{2}}}}\tilde{\tilde m}^*_{{\frac{n}{2}-1}_{z_{\frac{n}{2}-1}}}\cdot ...\cdot\tilde{\tilde m}^*_{2_{z_2}}\tilde{\tilde m}^*_{1_{z_1}}\sps\\
\end{array}
\end{equation}
where $z_J\ (J=\overline{1,\frac{n}{2}})$ is equal to 0 for the sign $\ll-\gg$ or 1 for the sign $\ll+\gg$ then $K=\sum\limits_{J=1}^{\frac{n}{2}}z_J\cdot 2^{J-1}+1$. If it does not matter which of the K-th operators will be selected then the number K will be omitted. Using these operators, one can define
\begin{equation}
\label{e1.13}
(\gamma_K)_\Lambda:=\tilde M_K (m_\Lambda{}^\myf\Lambda\gamma_\myf\Lambda)(\tilde{\tilde M}_K)^T\spsd
\end{equation}
\end{algorithm}
The corresponding example will be analyzed below (Algorithm \ref{a6.1}).

\section{Real inclusion. Involution}
\Abstract{
$\phantom{ff}$In this section it is told about how locally to enclose a real n-dimensional \discretionary{(pseudo-)}{}{(pseudo-)} Riemannian space into the complex representation $\mathbb CV_n$ of the 2n-dimensional pseudo-Riemannian space. For this purpose, a real surface in $\mathbb CV_n$ with a real parametrization is  considered. This enclosure is induced in the tangent bundle by means of an inclusion operator $H_i{}^\Lambda$ \cite{Newfield1} with which the help one can obtain the fiber of a real tangent bundle equipped with the (pseudo-)Euclidian metric. The index of such the metric will be significantly depend on the kind of the inclusion operator. The conclusion of all results of this section is made on the basis \cite{Newfield1}.
}

We will consider the real (pseudo-) Riemannian space $V_n$ as a surface of the real dimension n in the space $\mathbb CV_n$, i.e., a surface defined by means of the parametrical equation
\begin{equation}
\label{e5.1}
    w^\Lambda=w^\Lambda(u^i)\sps (\Lambda, \Psi ,... ,i,j,g,h=\overline{1,n})\sps
\end{equation}
where $w^\Lambda$ are the complex coordinates of a point $x$ of the base, and $u^i$ are parameters: the local coordinates of a point of the space $V_n$. The partial derivatives $(\partial_i w^\Lambda =:H_i{}^\Lambda)$ locally define \emph{inclusion} of the real tangent spaces (\ref {e5.1}) in the complex tangent space $\tau_x^\mathbb C$ as follows
\begin{equation}
    H:\tau_x^\mathbb R\longmapsto\tau_x^\mathbb C\sps
\end{equation}
\begin{equation}
\begin{array}{c}
    w^\Lambda=w^\Lambda(u^i(t))\sps r^\Lambda:=\frac{dw^\Lambda}{dt}=
    H_i{}^\Lambda\frac{du^i}{dt}=:H_i{}^\Lambda r^i\sps\\[2ex]
    \frac{du^i}{dt}\in\tau_x^\mathbb R\longmapsto \frac{dw^\Lambda}{dt}\in\tau_x^\mathbb C\sps
\end{array}
\end{equation}
where the differentiation is conducted along a real curve $\gamma(t)$ of the surface (\ref {e5.1}). Since $\parallel H_i{}^\Lambda \parallel$ is a nonsingular Jacobian matrix then there is the operator $H^i{}_\Lambda $ such that
\begin{equation}
\label{e5.2}
    \left\{
\begin{array}{c}
    H^i{}_\Lambda H_i{}^\Psi=\delta_\Lambda{}^\Psi\sps\\
    H^i{}_\Lambda H_j{}^\Lambda=\delta_j{}^i\spsd
\end{array}
    \right.
\end{equation}
From here, it follows that the operator $H_i{}^\Lambda$ defines \emph{involution}
\begin{equation}
\label{e5.3}
    S_\Lambda {}^{\Psi '}=H^i{}_\Lambda\bar  H_i{}^{\Psi '}
\end{equation}
in the complex space, where the coordinates $\bar H_i{}^{\Psi '}$ are in \emph{complex conjugate} to the coordinates $H_i{}^\Psi$. Therefore,
\begin{equation}
    r^i=H^i{}_\Lambda r^\Lambda=\overline{H^i{}_\Lambda r^\Lambda}
    \ \ \Rightarrow \ \ S_\Lambda{}^{\Psi '}r^\Lambda=\bar r^{\Psi '}\spsd
\end{equation}
It is a necessary and sufficient condition for the vector $r^\Lambda\in \tau^\mathbb C_x$ to be real. Thus,
\begin{equation}
    S_\Lambda{}^{\Psi '}\bar S_{\Psi '}{}^\Phi=\delta_\Lambda{}^\Phi\spsd
\end{equation}
We will define the metric of $V_n$ (the real (pseudo-) Riemannian space)  by the condition
\begin{equation}
\label{e5.4}
    g_{\Lambda\Psi}r^\Lambda r^\Psi=\overline{g_{\Lambda\Psi}r^\Lambda r^\Psi}\sps
    \forall \bar r^{\Psi '}=S_\Lambda{}^{\Psi '} r^\Lambda\spsd
\end{equation}
This means that a real tensor of the space $V_n$ is defined as the tensor self-conjugated under the specified Hermitian involution
\begin{equation}
    g_{\Lambda\Psi}=S_\Lambda{}^{\Phi '}S_\Psi{}^{\Theta '}\bar g_{\Phi '\Theta '}\spsd
\end{equation}
Therefore, the tensor
\begin{equation}
\label{e5.5}
    g_{ij}:=H_i{}^\Lambda H_j{}^\Psi g_{\Lambda\Psi}=\overline{H_i{}^\Lambda H_j{}^\Psi g_{\Lambda\Psi}}
\end{equation}
will be the metric tensor of $V_n\subset \mathbb CV_n$. The kind of the metric $g_{ij}$ significantly depends on the structure of the operator $H_i{}^\Lambda$ and hence the involution $S_\Lambda{}^{\Psi '}$.
\begin{example}
Let there be the complex Euclidian space $\mathbb C\mathbb R^4$. An inclusion of a real space in the complex can be obtain by means of one of the three various ways
\begin{equation}
\begin{array}{c}
\begin{array}{rrr}
H_i{}^\Lambda=
\left(
\begin{array}{cccc}
 1 & 0 & 0 & 0 \\
 0 & 1 & 0 & 0 \\
 0 & 0 & 1 & 0 \\
 0 & 0 & 0 & 1
\end{array}
\right)\sps &
H_i{}^\Lambda=
\left(
\begin{array}{cccc}
 1 & 0 & 0 & 0 \\
 0 & 1 & 0 & 0 \\
 0 & 0 & i & 0 \\
 0 & 0 & 0 & i
\end{array}
\right)\sps &
H_i{}^\Lambda=
\left(
\begin{array}{cccc}
 1 & 0 & 0 & 0 \\
 0 & i & 0 & 0 \\
 0 & 0 & i & 0 \\
 0 & 0 & 0 & i
\end{array}
\right)\spsd
\end{array}\\ \\
G_{\Lambda\Psi}=
\left(
\begin{array}{cccc}
 1 & 0 & 0 & 0 \\
 0 & 1 & 0 & 0 \\
 0 & 0 & 1 & 0 \\
 0 & 0 & 0 & 1
\end{array}
\right)\sps\\ \\
\begin{array}{rrr}
g_{ij}=
\left(
\begin{array}{cccc}
 1 & 0 & 0 & 0 \\
 0 & 1 & 0 & 0 \\
 0 & 0 & 1 & 0 \\
 0 & 0 & 0 & 1
\end{array}
\right)\sps &
g_{ij}=
\left(
\begin{array}{cccc}
 1 & 0 & 0 & 0 \\
 0 & 1 & 0 & 0 \\
 0 & 0 &-1 & 0 \\
 0 & 0 & 0 &-1
\end{array}
\right)\sps &
g_{ij}=
\left(
\begin{array}{cccc}
 1 & 0 & 0 & 0 \\
 0 &-1 & 0 & 0 \\
 0 & 0 &-1 & 0 \\
 0 & 0 & 0 &-1
\end{array}
\right)\spsd
\end{array}
\end{array}
\end{equation}
\end{example}

\section{Elementary transformations of the orthogonal group}
\Abstract{
$\phantom{ff}$In this section, it is told about how to lead a pseudo-orthogonal transformation to a block-diagonal form, and then to pass to the complex representation of such the transformation. The conclusion of all results of this section is made on the basis \cite{Berger1}, \cite{Rozenfeld2}.
}

Let's consider an orthogonal transformation in the space $\mathbb R^{2n}_{(n, n)}$ which is set by the formula
\begin{equation}
\label{e4.2}
G_{\myf\Lambda\myf\Psi}S_\myf\Omega{}^\myf\Lambda S_\myf\Gamma{}^\myf\Psi=G_{\myf\Omega\myf\Gamma}\spsd
\end{equation}
Thus, a basis is chosen so that $\parallel G_{\myf\Lambda\myf\Psi} \parallel$ will have the diagonal form.
\begin {theorem}
\label {theorem4.1}
Conformal transformations of the space $\mathbb R^{2n-2}_{(n-1, n-1)}$ form the group $O(n,n)$ consisting of \\
\begin{tabular}{lll}
1. \texttt{rotations} from $O(n-1,n-1)$, & 3. \texttt{translations}, & 5. \texttt{superpositions} 1-4. \\
2. \texttt{dilations}, & 4. \texttt{inversions},
\end{tabular}\\
Then any transformation of $O(n,n)$ can be represented as a product of \texttt{elementary transformations}
\begin{equation}
\label{e4.1}
S_{\myf\Lambda_1}{}^{\myf\Lambda_{J+1}}=\pm\prod_{I=1}^J(\pm (r_I)_{\myf\Lambda_I}(r_I)^{\myf\Lambda_{I+1}}-
\delta_{\myf\Lambda_I}{}^{\myf\Lambda_{I+1}})\sps (r_I)_{\myf\Lambda_I}(r_I)^{\myf\Lambda_I}=\pm 2\spsd
\end{equation}
\end{theorem}

\begin{proof}
The elementary transformations (\ref{e4.1}) are really orthogonal
\begin{equation}
(\pm r_\myf\Lambda r^\myf\Psi-\delta_\myf\Lambda{}^\myf\Psi)
(\pm r_\myf\Psi r^\myf\Theta-\delta_\myf\Psi{}^\myf\Theta)=
(\pm 2)r_\myf\Lambda r^\myf\Theta\mp r_\myf\Lambda r^\myf\Theta\mp r_\myf\Lambda r^\myf\Theta+\delta_\myf\Lambda{}^\myf\Theta=
\delta_\myf\Lambda{}^\myf\Theta\spsd
\end{equation}
Usual rotations from the group $O(2)$ can be represented as (\ref{e4.1}). At the same time, $r^\myf\Lambda=\sqrt{2}(\cos\frac{\alpha}{2}\sps\sin\frac{\alpha}{2}$), and there are the two disconnected classes of rotations
\begin{equation}
\begin{array}{rr}
a).\left(
\begin{array}{rr}
 \cos\alpha & \sin\alpha\\
 \sin\alpha &-\cos\alpha
\end{array}
\right) \sps
b).\left(
\begin{array}{rr}
 \cos\alpha & \sin\alpha\\
-\sin\alpha & \cos\alpha
\end{array}
\right) \spsd
\end{array}
\end{equation}
However, the transformation from b). is obtained by a superposition of the two transformations from a).
\begin{enumerate}
\item
A transformation from the group $O(1,1)$ is \emph{boost} of one of the 4 forms
\begin{equation}
\begin{array}{cccc}
a).&\left(
\begin{array}{rr}
 \ch\theta & \sh\theta\\
 \sh\theta & \ch\theta
\end{array}
\right) \sps &
b).&\left(
\begin{array}{rr}
 \ch\theta &-\sh\theta\\
 \sh\theta &-\ch\theta
\end{array}
\right) \sps
\\
c).&\left(
\begin{array}{rr}
-\ch\theta & \sh\theta\\
-\sh\theta & \ch\theta
\end{array}
\right) \sps&
d).&\left(
\begin{array}{rr}
-\ch\theta &-\sh\theta\\
-\sh\theta &-\ch\theta
\end{array}
\right) \spsd
\end{array}
\end{equation}
 The transformation from b). is represented as in the equation (\ref{e4.1}), where $r^\myf\Lambda=\sqrt{2}(\ch\frac{\theta}{2}\sps\sh \frac{\theta}{2})$. The transformation from a). is a superposition of the two transformations from b). The transformations from c). and d). differ only in the sign from the transformations from b). and a). respectively. This describes the four disconnected components of $O(1,1)$.
\item
Let \emph{one-dimensional dilation} in $\mathbb R^1$ has the form $\tilde x=\lambda x$. Consider the light cone in $\mathbb R^3_{(1,2)}$ defined as $T^2-Z^2-X^2=0$. We cut it by the plane $T+Z=1$ and perform the stereographic projection of the cross section onto the line $T = 1$, $Z=0$ that induces the one-dimensional space $\mathbb R^1$ with the single coordinate $x=\frac{X}{T-Z}$. Then $\tilde T+\tilde Z=\lambda(T+Z)$, $\tilde T-\tilde Z=\lambda^{-1}(T-Z)$, $\tilde X=X$.
If $\lambda>0$ then this transformation will be the boost from a).; if $\lambda<0$, this transformation will the boost from d).
\item
Consider \emph{translation} $\tilde x = x + a$ in the space $\mathbb R^1$. This will lead to the relations: $\tilde T-\tilde Z=T-Z$, $\tilde T+\tilde Z=T+Z+2aX+a^2(T-Z)$, $\tilde X=X+a(T-Z)$. The fixed vector of this transformation will be an isotropic vector of the form $(b,0,b)$.  It is impossible to translate an isotropic vector to a non-isotropic vector by any pseudo-orthogonal transformation since such the transformations keep the constant value $T^2-X^2-Z^2$. This means that we can not apply such the transformations to the basis to direct an isotropic vector along a non-isotropic axis. However, this does not prevent for the dilation to be decomposed into elementary transformations: \emph{rotation}, \emph{boost}, and one more such \emph{rotation}. To do this, as the rotation we take the elementary transformation
\begin{equation}
\left(
\begin{array}{rrr}
 1 &          0 &          0 \\
 0 & \cos\alpha &-\sin\alpha \\
 0 & \sin\alpha & \cos\alpha
\end{array}
\right)
\end{equation}
with $\tg\alpha =\frac{a}{2}$, and as the boost from a)., we take the elementary transformation
\begin{equation}
\left(
\begin{array}{rrr}
 \ch\theta & \sh\theta & 0\\
 \sh\theta & \ch\theta & 0\\
         0 &         0 & 1
\end{array}
\right)
\end{equation}
with $\ch\theta =\frac{a^2}{2}+1$. Then, this composition will have the form
\begin{equation}
\begin{array}{c}
\left(
\begin{array}{ccc}
 1 &          0 &          0 \\
 0 & \frac{1}{\sqrt{1+\frac{a^2}{4}}} &-\frac{\frac{a}{2}}{\sqrt{1+\frac{a^2}{4}}} \\
 0 & \frac{\frac{a}{2}}{\sqrt{1+\frac{a^2}{4}}} & \frac{1}{\sqrt{1+\frac{a^2}{4}}}
\end{array}
\right)
\left(
\begin{array}{ccc}
 1+\frac{a^2}{2}         & a\sqrt{1+\frac{a^2}{4}} & 0\\
 a\sqrt{1+\frac{a^2}{4}} & 1+\frac{a^2}{2}         & 0\\
                       0 &                       0 & 1
\end{array}
\right)\cdot\\
\cdot\left(
\begin{array}{ccc}
 1 &          0 &          0 \\
 0 & \frac{1}{\sqrt{1+\frac{a^2}{4}}} &-\frac{\frac{a}{2}}{\sqrt{1+\frac{a^2}{4}}} \\
 0 & \frac{\frac{a}{2}}{\sqrt{1+\frac{a^2}{4}}} & \frac{1}{\sqrt{1+\frac{a^2}{4}}}
\end{array}
\right)=
\left(
\begin{array}{ccc}
 1+\frac{a^2}{2} & a &  -\frac{a^2}{2} \\
               a & 1 &              -a \\
   \frac{a^2}{2} & a & 1-\frac{a^2}{2}
\end{array}
\right)\spsd
\end{array}
\end{equation}
\item
\emph{Inversion} $\tilde x =\frac{1}{x}$ in $\mathbb R^1$ induces the rotation $\tilde T=T$,  $\tilde Z=-Z$, $\tilde X=X$.
\item
\emph{Superpositions} of all one-dimensional transformations of the form 1-4 and \emph{rotations} from $O(n-1,n-1)$ represent the conformal transformation group of the space $\mathbb R^{2n-2}_{(n-1, n-1)}$ which will be isomorphic to the group $O(n,n)$.
\end{enumerate}
\end{proof}

\begin{example}
Let's consider the group of pseudo-orthogonal transformations $O(1,2)$ of the space $\mathbb R^3_{(1,2)}$. As is known, eigenvalues of any such a transformation are roots of a polynomial of third degree. One of such the values should be obligatory a real number. Since pseudo-orthogonal transformations satisfy the parity (\ref{e4.2}) then the square of such the real number is equal to 1. Therefore, any transformation from the group $O(1,2)$ possesses a fixed axis. However, not always this axis can be combined with a coordinate axis using pseudo-orthogonal transformations of the basis only.
\begin{enumerate}
\item
Let's consider the composition of a rotation and a boost for the space $\mathbb R^3_{(1,2)}$
\begin{equation}
\left(
\begin{array}{rrr}
 \ch\theta &-\sh\theta & 0\\
 \sh\theta &-\ch\theta & 0\\
         0 &         0 & 1
\end{array}
\right)
\left(
\begin{array}{rrr}
 1 &          0 &          0 \\
 0 & \cos\alpha & \sin\alpha \\
 0 & \sin\alpha &-\cos\alpha
\end{array}
\right) =
\left(
\begin{array}{rrr}
 \ch\theta &-\cos\alpha\sh\theta &-\sin\alpha\sh\theta \\
 \sh\theta &-\cos\alpha\ch\theta &-\sin\alpha\ch\theta \\
         0 &          \sin\alpha &         -\cos\alpha
\end{array}
\right) \spsd
\end{equation}
Let this composition will leave fixed the vector $(a,b,c)$
\begin{equation}
\left\{
\begin{array}{rrr}
 a\ch\theta -b\cos\alpha\sh\theta -c\sin\alpha\sh\theta &=&a\sps\\
 a\sh\theta -b\cos\alpha\ch\theta -c\sin\alpha\ch\theta &=&b\sps\\
             b\sin\alpha          -c\cos\alpha          &=&c\sps
\end{array}
\right.
\begin{array}{l}
\frac{c}{b}=\tg\frac{\alpha}{2}\sps \frac{a}{b}=\cth\frac{\theta}{2}\sps \\ \\
r=(b\cth\frac{\theta}{2},b,b\tg\frac{\alpha}{2})\spsd
\end{array}
\end{equation}
Let's spend the basis transformation
\begin{equation}
\left(
\begin{array}{rrr}
 \ch\frac{\theta}{2} &-\sh\frac{\theta}{2} & 0\\
-\sh\frac{\theta}{2} & \ch\frac{\theta}{2} & 0\\
                     0 &                 0 & 1
\end{array}
\right)
\left(
\begin{array}{c}
 b\cth\frac{\theta}{2} \\
 b \\
 b\tg\frac{\alpha}{2}
\end{array}
\right)=
\left(
\begin{array}{c}
 b/\sh\frac{\theta}{2} \\
 0 \\
 b\tg\frac{\alpha}{2}
\end{array}
\right)\spsd
\end{equation}
As now $|r|^2=b^2(1/\sh^2\frac{\theta}{2}-\tg^2\frac{\alpha}{2})$, we will set $b:=\pm\frac{1}{\sqrt{|1/\sh^2\frac{\theta}{2}-\tg^2\frac{\alpha}{2}}|}$. Then there is the basis transformation
\begin{equation}
\left(
\begin{array}{rrr}
 \ch\phi & 0 & \sh\phi\\
       0 & 1 &       0\\
 \sh\phi & 0 & \ch\phi
\end{array}
\right)
\left(
\begin{array}{c}
 b/\sh\frac{\theta}{2} \\
 0 \\
 b\tg\frac{\alpha}{2}
\end{array}
\right)\sps
\begin{array}{ll}
\mbox{a). or } \ch\phi=\ \frac{b}{\sh\frac{\theta}{2}}\sps\sh\phi=-b\tg\frac{\alpha}{2}\sps\\
\mbox{b). or } \sh\phi=-\frac{b}{\sh\frac{\theta}{2}}\sps\ch\phi=\ b\tg\frac{\alpha}{2}\spsd
\end{array}
\end{equation}
In the case a)., the coordinate axis (1,0,0) is fixed, and in the case b)., the axis (0,0,1) is fixed. If $1/\sh\frac{\theta}{2}=\pm\tg\frac{\alpha}{2}$ then it is impossible to simplify the initial superposition with the help of any pseudo-orthogonal rotation of the basis only.
\item
Let's consider the composition of two boosts for the space $\mathbb R^3_{(1,2)}$
\begin{equation}
\left(
\begin{array}{rrr}
 \ch\theta &-\sh\theta & 0\\
 \sh\theta &-\ch\theta & 0\\
         0 &         0 & 1
\end{array}
\right)
\left(
\begin{array}{rrr}
 \ch\psi & 0 &-\sh\psi \\
       0 & 1 &       0 \\
 \sh\psi & 0 &-\ch\psi
\end{array}
\right) =
\left(
\begin{array}{rrr}
 \ch\theta\ch\psi &-\sh\theta &-\sh\psi\ch\theta \\
 \sh\theta\ch\psi &-\ch\theta &-\sh\psi\sh\theta \\
          \sh\psi &         0 &         -\ch\psi
\end{array}
\right) \spsd
\end{equation}
Let this composition will leave motionless the vector $(a,b,c)$
\begin{equation}
\left\{
\begin{array}{rrr}
 a\ch\theta\ch\psi -b\sh\theta -c\sh\psi\ch\theta &=&a\\
 a\sh\theta\ch\psi -b\ch\theta -c\sh\psi\sh\theta &=&b\\
          a\sh\psi             -c\ch\psi          &=&c
\end{array}
\right.\sps
\begin{array}{l}
\frac{c}{a}=\tanh\frac{\psi}{2}\sps\frac{b}{a}=\tanh\frac{\theta}{2}\sps\\ \\
r=(a,a\tanh\frac{\theta}{2},a\tanh\frac{\psi}{2})\spsd
\end{array}
\end{equation}
Let's spend the basis transformation
\begin{equation}
\left(
\begin{array}{rrr}
 \ch\frac{\theta}{2} &-\sh\frac{\theta}{2} & 0\\
-\sh\frac{\theta}{2} & \ch\frac{\theta}{2} & 0\\
                   0 &                   0 & 1
\end{array}
\right)
\left(
\begin{array}{c}
 a\\
 a\tanh\frac{\theta}{2} \\
 a\tanh\frac{\psi}{2}
\end{array}
\right)=
\left(
\begin{array}{c}
 a/\ch\frac{\theta}{2} \\
 0 \\
 a\tanh\frac{\psi}{2}
\end{array}
\right)\spsd
\end{equation}
As now $|r|^2=a^2(1/\ch^2\frac{\theta}{2}-\tanh^2\frac{\psi}{2})$, we will set $a:=\pm\frac{1}{\sqrt{|1/\ch^2\frac{\theta}{2}-\tanh^2\frac{\psi}{2}}|}$. Then there is the basis transformation
\begin{equation}
\left(
\begin{array}{rrr}
 \ch\phi & 0 & \sh\phi\\
       0 & 1 &       0\\
 \sh\phi & 0 & \ch\phi
\end{array}
\right)
\left(
\begin{array}{c}
 a/\ch\frac{\theta}{2} \\
 0 \\
 a\tanh\frac{\psi}{2}
\end{array}
\right)\sps
\begin{array}{ll}
\mbox{a). or } \ch\phi=\ \frac{a}{\ch\frac{\theta}{2}}\sps\sh\phi=-a\tanh\frac{\psi}{2}\sps\\
\mbox{b). or } \sh\phi=-\frac{a}{\ch\frac{\theta}{2}}\sps\ch\phi=\ a\tanh\frac{\psi}{2}\spsd
\end{array}
\end{equation}
In the case a)., the coordinate axis (1,0,0) is fixed, and in the case b)., the axis (0,0,1) is fixed. If $1/\ch^2\frac{\theta}{2}=\pm\tanh^2\frac{\psi}{2}$, then it is impossible to simplify the initial superposition with the help of any pseudo-orthogonal rotation of the basis only.
\end{enumerate}
\end{example}

\begin{example}
Let's consider a superposition of elementary transformations from the group $O(1,2)$
\begin{equation}
%
\end{equation}
The solution of this system will be
\begin{equation}
\begin{array}{c}
\sin\alpha =\ch\tilde\psi\sin\tilde\alpha\sps\sh\psi =\frac{\sh\tilde\psi}{\sqrt{1-\ch^2\tilde\psi\sin^2\tilde\alpha}}\sps
\sh(\phi-\tilde\phi)=\frac{\sh\tilde\psi\sin\tilde\alpha}{\sqrt{1-\ch^2\tilde\psi\sin^2\tilde\alpha}}\spsd\\ \\
\sin\tilde\alpha =\frac{\sin\alpha}{\sqrt{1+\sh^2\psi\cos^2\alpha}}\sps\sh\tilde\psi =\sh\psi\cos\alpha\sps
\sh(\phi-\tilde\phi)=\frac{\sh\psi\sin\alpha}{\sqrt{1+\sh^2\psi\cos^2\alpha}}\spsd
\end{array}
\end{equation}
Besides, an unusual case is $\ch^2\tilde\psi\sin^2\tilde\alpha\geq 1$. Therefore, it is not always one can rearrange elementary transformations in a superposition unlike the orthogonal group $O(3)$.
\end{example}

\newpage
\begin{example}
Let's consider transformations from the group $O(2,2)$. Any transformation from this group is presented by means of four transformations (\ref{e4.1}) according to the  Cartan-Dieudonne theorem \cite[v. 2, p. 33 (rus)]{Berger1}
\begin{equation}
\pm r_\myf\Lambda r^{\myf\Psi}-\delta_\myf\Lambda{}^\myf\Psi\sps r_\myf\Lambda r^\myf\Lambda=\pm 2\spsd
\end{equation}
However, always it is possible to pick up the decomposition so that two of such the transformations will have a diagonal matrix. Elements of the main diagonal of the matrix $I_\myf\Lambda{}^\myf\Psi $ on the module are equal to 1. Therefore, (\ref{e4.1}) can be copied as
\begin{equation}
\begin{array}{c}
(I_1)_\myf\Phi{}^\myf\Lambda(\pm (r_1)_\myf\Lambda (r_1)^{\myf\Psi}-\delta_\myf\Lambda{}^\myf\Psi)
(I_2)_\myf\Psi{}^\myf\Theta(\pm (r_2)_\myf\Theta (r_2)^{\myf\Omega}-\delta_\myf\Theta{}^\myf\Omega)\sps\\
(r_1)^\myf\Psi (I_2)_\myf\Psi{}^\myf\Theta (r_2)_\myf\Theta =0\sps (r_2)^\myf\Psi (I_1)_\myf\Psi{}^\myf\Theta (r_1)_\myf\Theta =0\spsd
\end{array}
\end{equation}
There are significantly the various 3 mutually exclusive variants.
\begin{enumerate}
\item
In this case, a pseudo-orthogonal transformation can have the block-diagonal structure
\begin{equation}
%
\end{equation}
\item
In this case, a pseudo-orthogonal transformation can have the block-diagonal structure
\begin{equation}
I_1
\left(
\begin{array}{rrrr}
 \cos\theta & \sin\theta  & 0        &        0\\
 \sin\theta & -\cos\theta & 0        &        0\\
          0 &           0 & 1        &        0\\
          0 &           0 & 0        &        1
\end{array}
\right)
I_2
\left(
\begin{array}{rrrr}
          1 &           0 & 0        &        0\\
          0 &           1 & 0        &        0\\
          0 &           0 &-\cos\phi & -\sin\phi\\
          0 &           0 &-\sin\phi &  \cos\phi
\end{array}
\right)=
\end{equation}
$$
I_1\left(\underbrace{\left(
\begin{array}{c}
 \sqrt{2}\cos\frac{\theta}{2}\\
 \sqrt{2}\sin\frac{\theta}{2}\\
                            0\\
                            0
\end{array}
\right)}_{(r_1)_\myf\Theta}
\underbrace{\left(
\begin{array}{rrrr}
\sqrt{2}\cos\frac{\theta}{2} & \sqrt{2}\sin\frac{\theta}{2} &         0 &            0
\end{array}
\right)}_{(r_1)^\myf\Omega}-
\underbrace{\left(
\begin{array}{rrrr}
          1 &                  0 &        0 &                0\\
          0 &                  1 &        0 &                0\\
          0 &                  0 &        1 &                0\\
          0 &                  0 &        0 &                1
\end{array}
\right)}_{\delta_\myf\Theta{}^\myf\Omega}
\right)
$$
$$
I_2\left(\underbrace{\left(
\begin{array}{c}
                           0\\
                           0\\
-\sqrt{2}\cos\frac{\phi}{2} \\
-\sqrt{2}\sin\frac{\phi}{2}
\end{array}
\right)}_{(r_2)_\myf\Theta}
\underbrace{\left(
\begin{array}{rrrr}
         0 &            0 & \sqrt{2}\cos\frac{\phi}{2} & \sqrt{2}\sin\frac{\phi}{2}
\end{array}
\right)}_{(r_2)^\myf\Omega}+
\underbrace{\left(
\begin{array}{rrrr}
          1 &                  0 &        0 &                0\\
          0 &                  1 &        0 &                0\\
          0 &                  0 &        1 &                0\\
          0 &                  0 &        0 &                1
\end{array}
\right)}_{\delta_\myf\Theta{}^\myf\Omega}
\right)\sps
$$
$$
(r_1)^\myf\Psi (r_1)_\myf\Psi=2\sps (r_2)^\myf\Psi (r_2)_\myf\Psi=-2\spsd
$$
\item
In this case, a pseudo-orthogonal transformation will not have a block-diagonal structure, but can be presented as
\begin{equation}
%
\end{equation}
\end{enumerate}
\end{example}

According to the same Cartan-Dieudonne theorem, any transformation from the group $O(2n,2n)$ in some basis will have the block-diagonal structure consisting of blocks $O(2,2)$. This means, that (\ref{e4.1}) is executed
\begin{equation}
\label{e4.3}
\begin{array}{c}
S_{\myf\Lambda_1}{}^{\myf\Lambda_{J+1}}=\pm\prod\limits_{I=1}^J
(\pm (r_I)_{\myf\Lambda_I}(r_I)^{\myf\Lambda_{I+1}}-\delta_{\myf\Lambda_I}{}^{\myf\Lambda_{I+1}})\spsd
\end{array}
\end{equation}
Let now  the canonical complex structure be set in the space $\mathbb R^{2n}_{(n, n)}$ then in the decomposition (\ref{e4.1}), one can allocate $J/2$ factors so that the relations
\begin{equation}
\begin{array}{c}
f_{\myf\Lambda_{2I}}{}^{\myf\Lambda_I}((r_I)_{\myf\Lambda_I}(r_I)^{\myf\Lambda_{I+1}}-\delta_{\myf\Lambda_I}{}^{\myf\Lambda_{I+1}})
f_{\myf\Lambda_{I+1}}{}^{\myf\Lambda_{2I+1}}=
(r_{2I})_{\myf\Lambda_{2I}}(r_{2I})^{\myf\Lambda_{2I+1}}+\delta_{\myf\Lambda_{2I}}{}^{\myf\Lambda_{2I+1}}\sps\\
(r_I)_{\myf\Lambda}(r_I)^{\myf\Lambda}=2\sps
(r_{2I})_{\myf\Lambda}(r_{2I})^{\myf\Lambda}=-2\sps \forall I=\overline{1,J/2}
\end{array}
\end{equation}
will be executed. For any pseudo-orthogonal transformation $S_\myf\Lambda{}^\myf\Psi$, one can define the parities
\begin{equation}
\label{e4.4}
\begin{array}{c}
S_\Lambda{}^\Psi:=m_\Lambda{}^\myf\Lambda S_\myf\Lambda{}^\myf\Psi m^\Psi{}_\myf\Psi\sps
S_\Lambda{}^{\Psi '}:=m_\Lambda{}^\myf\Lambda S_\myf\Lambda{}^\myf\Psi \bar m^{\Psi '}{}_\myf\Psi=0\sps\\
\bar S_{\Lambda '}{}^\Psi:=\bar m_{\Lambda '}{}^\myf\Lambda S_\myf\Lambda{}^\myf\Psi m^\Psi{}_\myf\Psi=0\sps
\bar S_{\Lambda '}{}^{\Psi '}:=\bar m_{\Lambda '}{}^\myf\Lambda S_\myf\Lambda{}^\myf\Psi \bar m^{\Psi '}{}_\myf\Psi\sps\\
-f_{\myf\Omega}{}^\myf\Lambda S_\myf\Lambda{}^\myf\Psi f_{\myf\Psi}{}^\myf\Theta=S_\myf\Omega{}^\myf\Theta\spsd
\end{array}
\end{equation}
Thus, a complex symmetry will be defined as
\begin{equation}
\begin{array}{c}
m_\Theta{}^\myf\Theta (r_\myf\Theta r^\myf\Sigma-\delta_\myf\Theta{}^\myf\Sigma)
f_\myf\Sigma{}^\myf\Psi (r_\myf\Psi r^\myf\Omega-\delta_\myf\Psi{}^\myf\Omega)f_\myf\Omega{}^\myf\Xi m^\Xi{}_\myf\Xi=\\
=m_\Theta{}^\myf\Theta
(-r_\myf\Theta r^\myf\Sigma f_\myf\Sigma{}^\myf\Psi f_\myf\Psi{}^\myf\Xi
 -f_\myf\Theta{}^\myf\Psi   r_\myf\Psi r^\myf\Omega f_\myf\Omega{}^\myf\Xi+
  f_\myf\Theta{}^\myf\Psi   f_\myf\Psi{}^\myf\Xi)
m^\Xi{}_\myf\Xi=\\
=(2r_\Theta r^\Xi-\delta_\Theta{}^\Xi)\sps
\end{array}
\end{equation}
Besides,
\begin{equation}
r_\Lambda r^\Lambda=m_\Lambda{}^\myf\Psi r_\myf\Psi m^\Lambda{}_\myf\Phi r^\myf\Phi=
\frac{1}{2}(\delta_\myf\Phi{}^\myf\Psi+if_\myf\Phi{}^\myf\Psi)r_\myf\Psi r^\myf\Phi=
\frac{1}{2}r_\myf\Psi r^\myf\Psi=1\sps
\end{equation}
that definitively confirms the existence of the decomposition
\begin{equation}
\label{e4.5}
\begin{array}{c}
S_{\Lambda_1}{}^{\Lambda_{J+1}}=\pm\prod\limits_{I=1}^J
((r_I)_{\Lambda_I}(r_I)^{\Lambda_{I+1}}-\delta_{\Lambda_I}{}^{\Lambda_{I+1}})\sps
(r_I)_{\Lambda_I}(r_I)^{\Lambda_I}=2\spsd
\end{array}
\end{equation}
However, it is necessary to notice that not always $(r_I)_{\Lambda}(r_J)^{\Lambda}=0\sps (I\ne J)$ in view of the existence of isotropic vectors as this has been shown in the previous example. As to the block structure of transformations from the group $O(n,\mathbb C)$ that the dimension of the blocks will be equal to 4 too.

\begin{example}
Let the three-dimensional pseudo-Euclidian space $\mathbb R^3_{(1,2)}$ be given. We will consider the transformation
\begin{equation}
S_i{}^j=
\left(
\begin{array}{ccc}
1+\frac{a^2}{2} &        a &   -\frac{a^2}{2}\\
              a &        1 &               -a\\
  \frac{a^2}{2} &        a &   1-\frac{a^2}{2}
\end{array}
\right)\spsd
\end{equation}
Let's define the inclusion operator
\begin{equation}
H^i{}_\Lambda=
\left(
\begin{array}{rrrr}
               1 &        0 &                0\\
               0 &       -i &                0\\
               0 &        0 &               -i
\end{array}
\right)
\sps
H_i{}^\Lambda=
\left(
\begin{array}{rrrr}
               1 &        0 &                0\\
               0 &        i &                0\\
               0 &        0 &                i
\end{array}
\right)
\spsd
\end{equation}
It will define the complexification of our transformation
\begin{equation}
S_\Lambda{}^\Psi=H^i{}_\Lambda S_i{}^j H_j{}^\Psi=
\left(
\begin{array}{ccc}
1+\frac{a^2}{2} &       ia &  -i\frac{a^2}{2}\\
            -ia &        1 &               -a\\
-i\frac{a^2}{2} &        a &   1-\frac{a^2}{2}
\end{array}
\right)
\end{equation}
with the complex Euclidian metric. To pass to the real representation, the operators
\begin{equation}
(m^T)_\myf\Lambda{}^\Lambda=
\frac{1}{\sqrt{2}}\left(
\begin{array}{rrr}
 1 & 0 & 0 \\
 0 & i & 0 \\
 0 & 0 & i \\
-i & 0 & 0 \\
 0 & 1 & 0 \\
 0 & 0 & 1
\end{array}
\right)\sps
m_\Lambda{}^\myf\Lambda=
\frac{1}{\sqrt{2}}\left(
\begin{array}{rrrrrr}
 1 & 0 & 0 & i & 0 & 0 \\
 0 &-i & 0 & 0 & 1 & 0 \\
 0 & 0 &-i & 0 & 0 & 1
\end{array}
\right)\sps
\end{equation}
will be required to us. Accordingly, the real representation of our transformation will look like
\begin{equation}
S_\myf\Lambda{}^\myf\Psi=m^\Lambda{}_\myf\Lambda S_\Lambda{}^\Psi m_\Psi{}^\myf\Psi+
\bar m^{\Lambda '}{}_\myf\Lambda S_{\Lambda '}{}^{\Psi '}\bar m_{\Psi '}{}^\myf\Psi=
\left(
\begin{array}{cccccc}
1+\frac{a^2}{2} &        a &    -\frac{a^2}{2}  &               0 &        0 &                0 \\
              a &        1 &                -a  &               0 &        0 &                0 \\
  \frac{a^2}{2} &        a &   1-\frac{a^2}{2}  &               0 &        0 &                0 \\
              0 &        0 &                 0  & 1+\frac{a^2}{2} &        a &   -\frac{a^2}{2} \\
              0 &        0 &                 0  &              a  &        1 &               -a \\
              0 &        0 &                 0  &  \frac{a^2}{2}  &        a &   1-\frac{a^2}{2}
\end{array}
\right)\spsd
\end{equation}
It is visible that such the representation has the block-diagonal structure, however, the complex representation has no such a structure.
\end{example}

Along with the transformations (\ref{e4.4}), it makes sense to consider the pseudo-orthogonal transformations
\begin{equation}
\label{e4.6}
\begin{array}{c}
S_\Lambda{}^\Psi:=m_\Lambda{}^\myf\Lambda S_\myf\Lambda{}^\myf\Psi m^\Psi{}_\myf\Psi=0\sps
S_\Lambda{}^{\Psi '}:=m_\Lambda{}^\myf\Lambda S_\myf\Lambda{}^\myf\Psi \bar m^{\Psi '}{}_\myf\Psi\sps\\
\bar S_{\Lambda '}{}^\Psi:=\bar m_{\Lambda '}{}^\myf\Lambda S_\myf\Lambda{}^\myf\Psi m^\Psi{}_\myf\Psi\sps
S_{\Lambda '}{}^{\Psi '}:=\bar m_{\Lambda '}{}^\myf\Lambda S_\myf\Lambda{}^\myf\Psi \bar m^{\Psi '}{}_\myf\Psi=0\sps\\
f_{\myf\Omega}{}^\myf\Lambda S_\myf\Lambda{}^\myf\Psi f_{\myf\Psi}{}^\myf\Theta=S_\myf\Omega{}^\myf\Theta\sps
S_\myf\Lambda{}^\myf\Psi S_\myf\Psi{}^\myf\Phi=\delta_\myf\Lambda{}^\myf\Phi\sps
S_\Lambda{}^{\Psi '}\bar S_{\Psi '}{}^\Phi=\delta_\Lambda{}^\Phi\sps
S_\Lambda{}^{\Psi '}S_\Phi {}^{\Omega '}\bar g_{\Psi '\Omega '}=g_{\Lambda\Phi}\spsd
\end{array}
\end{equation}
The complex representation of such the transformations will be an involution on the space $\mathbb C^n$ responsible for a real inclusion, therefore in the decomposition (\ref{e4.1}), one can allocate $J/2$ factors so that the relations
\begin{equation}
\begin{array}{c}
f_{\myf\Lambda_{2I}}{}^{\myf\Psi_I}((r_I)_{\myf\Psi_I}(r_I)^{\myf\Lambda_{I+1}}-\delta_{\myf\Psi_I}{}^{\myf\Lambda_{I+1}})
f_{\myf\Lambda_{I+1}}{}^{\myf\Lambda_{2I+1}}=
-((r_{2I})_{\myf\Lambda_{2I}}(r_{2I})^{\myf\Lambda_{2I+1}}+\delta_{\myf\Lambda_{2I}}{}^{\myf\Lambda_{2I+1}})\sps\\
(r_I)_{\myf\Lambda}(r_I)^{\myf\Lambda}=2\sps
(r_{2I})_{\myf\Lambda}(r_{2I})^{\myf\Lambda}=-2\sps \forall I=\overline{1,J/2}\spsd
\end{array}
\end{equation}
 will be executed. Thus, eigenvalues of the transformation $S_\Lambda{}^{\Psi '}$ are equal to $\pm 1$. This automatically means, that such the transformation is orthogonally diagonalized in the complex representation, and the real representation in some basis has the block-diagonal structure
\begin{equation}
\left(
\begin{array}{rrrrr}
 \cos\theta & \sin\theta  & 0          &          0 & ......\\
 \sin\theta & -\cos\theta & 0          &          0 & ......\\
          0 &           0 &-\cos\theta &-\sin\theta & ......\\
          0 &           0 &-\sin\theta & \cos\theta & ......\\
............&.............&............&............&.......
\end{array}
\right)
\end{equation}
and also is diagonalized. It is necessary to address for more detailed statement to \cite[pp. 79-126]{Rozenfeld2}, \cite[v. 2]{Berger1}.
\newpage

\section{Clifford equation. Double covering}
\Abstract{
$\phantom{ff}$In this section, it is told about how to pass to the reduced Cartan spinors. This will allow to construct the one-to-one correspondence between
vectors of the space $\mathbb C^n$ and bivectors of the spinor space $\mathbb C^N$ that will lead to the construction of the double covering $Spin(n,\mathbb C)/\{\pm 1\}\cong SO(n,\mathbb C)$. The conclusion of all results of this section is made on the basis \cite{Rozenfeld2}, \cite{Penrouz1}.
}

Suppose that a pseudo-Riemannian space $V_{2n}$ is the real representation of the complex space $\mathbb CV_n$, where n is even. Let a set of non-degenerate operators $\gamma_\myf\Lambda$ ($\myf\Lambda,\ \myf\Psi,\ ...=\overline{1,2n}$) be given at each point of $V_{2n}$. Moreover, we assume that each operator $\gamma_\myf\Lambda:=m_\myf\Lambda{}^\Lambda\gamma_\Lambda+\bar m_\myf\Lambda{}^{\Lambda '}\gamma_{\Lambda '}$ can be represented by a real matrix $(2N)^2 \times (2N)^2$, where $N=2^{n/2-1}$. In addition, the operators must satisfy the condition (\ref{e2.1})
\begin{equation}
\label{e3.0}
 \gamma_\myf\Lambda\gamma_\myf\Psi+\gamma_\myf\Psi\gamma_\myf\Lambda=G_{\myf\Lambda\myf\Psi}\sps
\end{equation}
where $G_{\myf\Lambda\myf\Psi}=G_{\myf\Psi\myf\Lambda}$  is a non-degenerate metric tensor from (\ref{e1.4})-(\ref{e1.5}) defined in the real representation of the tangent bundle $V_{2n}$. The operators (\ref{e1.13}) (if it does not matter which of the K-th operators will be selected then the number K will be omitted)
\begin{equation}
\label{e3.1}
\gamma_\Lambda:= \tilde M(m_\Lambda{}^\myf\Lambda\gamma_\myf\Lambda) \tilde {\tilde M}^T
\end{equation}
are the complex representation of the operators $\gamma_\myf\Lambda$. They will satisfy the equation
\begin{equation}
\label{e3.2}
 \gamma_\Lambda\gamma_\Psi+\gamma_\Psi\gamma_\Lambda=g_{\Lambda\Psi}\sps
 \bar \gamma_{\Lambda '}\bar \gamma_{\Psi '}+\bar \gamma_{\Psi '}\bar \gamma_{\Lambda '}=g_{\Lambda '\Psi '}\spsd
\end{equation}
Let $r^\Lambda$ be a vector of the complex tangent bundle $\tau^\mathbb C$. As the base can be regarded both the complex representation $\mathbb CV_n$ and the real representation $V_{2n}$. A fiber of this bundle is isomorphic to the space $\mathbb C^n$, the real representation of which will be the pseudo-Euclidean space $R_{(n,n)}^{2n}$, where n is the index of the pseudo-Euclidean metric (the number of $"+"$ on the main diagonal). Thus, a vector of the space $\mathbb C^n$ ($\mathbb R^{2n}_{(n, n)})$ can be uniquely associated to an operator from the space $\mathbb C^{2N}$ by the rule
\begin{equation}
\label{e3.3}
 R:=r^\Phi\gamma_\Phi\spsd
\end{equation}
Because from the Clifford equation, the relation
\begin{equation}
tr(\gamma_\Lambda\gamma_\Psi)=N g_{\Lambda\Psi}
\end{equation}
follows then each operator R can be uniquely associated to the vector of the space $\mathbb C^n$($R^{2n}_{(n,n)})$ by means of the taking of the trace from the both parts of the conformity (\ref{e3.3}) with $\gamma^\Lambda:=g^{\Lambda\Psi}\gamma_\Psi$
\begin{equation}
 g^{\Lambda\Psi} tr(\gamma_\Psi R)=r^\Phi g^{\Lambda\Psi}tr(\gamma_\Psi\gamma_\Phi)\sps
\end{equation}
and then
\begin{equation}
 \frac{1}{N} tr(\gamma^\Psi R)=r^\Psi \spsd
\end{equation}
Thus, it is obtained the one-to-one correspondence between the vectors $r^\Lambda $ of the space $\mathbb C^n$ ($\mathbb R^{2n}_{(n, n)})$ and the operators R (special kind, of course) acting on the space $\mathbb C^{2N}$. Using the Clifford equation (\ref{e3.2}), we can write down the identity
\begin{equation}
\label{e3.4}
 R\gamma_\Psi R=
 r^\Lambda\gamma_\Lambda \gamma_\Psi \gamma_\Phi r^\Phi=
 r^\Lambda (g_{\Lambda\Psi}-\gamma_\Psi\gamma_\Lambda)
 \gamma_\Phi r^\Phi= (r_\Psi r^\Phi-\frac{1}{2}(r_\Lambda r^\Lambda)\delta_\Psi{}^\Phi)
 \gamma_\Phi\spsd
\end{equation}
Let $r_\Lambda r^\Lambda:=\pm 2$. It is known that any (pseudo-) orthogonal transformation $S_\Psi{}^\Phi$ can be represented as
the finite product of the elementary transformations $S_I:=(r_I){}_\Psi (r_I)^\Phi-\frac{1}{2}((r_I)_\Lambda (r_I)^\Lambda)\delta_\Psi{}^\Phi$ (I runs over the finite values from 1 to any finite J, for example). The previous section is devoted to the proof of this fact. Define
\begin{equation}
 R_{I}:=(r_{I}){}^\Lambda\gamma_\Lambda\sps
 S=\prod\limits_{I=1}^{J} S_I=S_1S_2\ldots S_J\sps
\end{equation}
\begin{equation}
 \tilde S=\prod\limits_{I=1}^J R_I=R_1 R_2 \ldots R_J \sps
 \tilde{\tilde S}:=\prod\limits_{I=J}^1 R_I = R_J\ldots R_2 R_1\spsd
\end{equation}
Here $\tilde S$ is the right product of the operators $R_I$, and $\tilde {\tilde S}$ is the left product of the operators $R_I$. Therefore, the matrixes of the operators $\tilde S$ and $\tilde {\tilde S}$ are significantly various and in any way with each other are not connected. It will allow to copy the conformity (\ref {e3.4}) for the orthogonal transformation $S_\Lambda{}^\Psi$ as
\begin{equation}
\label{e3.5}
 S_\Lambda{}^\Psi\gamma_\Psi=\tilde S\gamma_\Lambda\tilde{\tilde S}\sps  \tilde{\tilde S}=\tilde S^{-1}\spsd
\end{equation}
The equation (\ref{e3.4}) is true for the real representation that will give
\begin{equation}
\label{e3.6}
 S_\myf\Lambda{}^\myf\Psi\gamma_\myf\Psi=\tilde S\gamma_\myf\Lambda\tilde{\tilde S}\sps
 f_\myf\Theta{}^\myf\Lambda S_\myf\Lambda{}^\myf\Psi f_\myf\Psi{}^\myf\Phi=-S_\myf\Theta{}^\myf\Phi\spsd
\end{equation}
Accordingly, for the involution, the equation (\ref{e3.6}) can be rewritten as
\begin{equation}
\label{e3.7}
 S_\myf\Lambda{}^\myf\Psi\gamma_\myf\Psi=\tilde S\gamma_\myf\Lambda\tilde{\tilde S}\sps
 f_\myf\Theta{}^\myf\Lambda S_\myf\Lambda{}^\myf\Psi f_\myf\Psi{}^\myf\Phi=S_\myf\Theta{}^\myf\Phi\sps
 S_\myf\Lambda{}^\myf\Psi S_\myf\Psi{}^\myf\Phi = \delta_\myf\Lambda{}^\myf\Phi\spsd
\end{equation}
Then for the complex representation, we obtain the identity
\begin{equation}
\label{e3.8}
 S_\Lambda{}^{\Psi '}\bar \gamma_{\Psi '}=\bar{\tilde S}\gamma_\Lambda \tilde {\tilde S}\sps
 S_\Lambda{}^{\Psi '}\bar S_{\Psi '}{}^\Phi =\delta_\Lambda{}^\Phi\sps \tilde S\bar{\tilde S} = \pm E\sps
 \tilde {\tilde S}\bar{\tilde {\tilde S}} = \pm E\spsd
\end{equation}
It is now possible to complexify the real representation $\mathbb R^{2n}_{(n, n)}$ considering the vector $r^\myf\Lambda$ in the same basis, but with complex coefficients. In this case, $\gamma_\myf\Lambda$ will remain the same and $S_\myf\Lambda{}^\myf\Psi$ from (\ref{e3.6}) will be a complex transformation. In particular, as such a transformation, we can consider the transformation $if_\myf\Psi{}^\myf\Phi$ (see (\ref{e1.6})). This means that the complex structure  can be expanded  as
\begin{equation}
\label{e3.9}
 if_\myf\Lambda{}^\myf\Psi\gamma_\myf\Psi=\tilde I\tilde F\gamma_\myf\Lambda \tilde {\tilde F}\tilde {\tilde I}\spsd
\end{equation}
For further investigations of this correspondence, it is necessary to study the structure of the operators $\gamma_\myf\Lambda$. Without loss of generality, we can consider a basis in which the metric tensor $G_{\myf\Lambda\myf\Psi} $ on the main diagonal has $\pm 1$, and all remaining components are equal to 0
\begin{equation}
\label{e3.10}
 \gamma_\myf\Lambda\gamma_\myf\Psi=-\gamma_\myf\Psi\gamma_\myf\Lambda\sps \myf\Lambda\neq\myf\Psi\spsd
\end{equation}
Construct the operator
\begin{equation}
 \gamma_0:=\prod_{\myf\Lambda =1}^{2n} \gamma_\myf\Lambda
\end{equation}
according to \cite[v. 2, Appendix, p. 442 (eng), (B.8)]{Penrouz1}. Then from (\ref{e3.10}), the relations
\begin{equation}
\label{e3.11}
 \gamma_0\gamma_\myf\Lambda=-\gamma_\myf\Lambda\gamma_0 \sps
\end{equation}
$$
 (\gamma_0-\lambda E)\gamma_\myf\Lambda=-\gamma_\myf\Lambda(\gamma_0+\lambda E)
$$
will follow. Therefor,
\begin{equation}
 det(\gamma_0-\lambda E)=0=det(\gamma_0+\lambda E)
\end{equation}
will be the equation for the eigenvalues of $\gamma_0$. It is seen that these values are the paired: $\pm\lambda$. Then positive and negative eigenvalues correspond to the two different eigenspaces. Then there is a basis in which a block representation of the operator $\gamma_0$ (the blocks have the dimension $2N^2\times 2N^2$)
\begin{equation}
\gamma_0=\left(
\begin{array}{cc}
\zeta_0   & 0 \\
 0         & \xi_0
\end{array}\right)
\end{equation}
can exist, where $\zeta_0$ responsible for the positive eigenvalues, and $\xi_0$ responsible for the negative eigenvalues. Define
\begin{equation}
\gamma_\myf\Lambda=\left(
\begin{array}{cc}
\zeta_\myf\Lambda   & \sigma_\myf\Lambda \\
\eta_\myf\Lambda & \xi_\myf\Lambda
\end{array}\right)\spsd
\end{equation}
Taking into account this definition, the condition (\ref{e3.10}) will lead to the relations
\begin{equation}
\zeta_0\zeta_\myf\Lambda=-\zeta_\myf\Lambda\zeta_0\sps \xi_0\xi_\myf\Lambda=-\xi_\myf\Lambda\xi_0\spsd
\end{equation}
This means that $\zeta_0$ and $\xi_0$ should have pairs of the positive and negative values (even if $\zeta_\myf\Lambda$ and $\xi_\myf\Lambda$ are degenerate, but not equal to 0). Therefore, $\zeta_\myf\Lambda=\xi_\myf\Lambda\equiv 0$. Thus,
\begin{equation}
\gamma_\myf\Lambda=\left(
\begin{array}{cc}
0              & \sigma_\myf\Lambda \\
\eta_\myf\Lambda  & 0
\end{array}\right)\sps
\end{equation}
and the equation (\ref{e3.0}) is equivalent to the system
\begin{equation}
\label{e3.12} \left\{
\begin{array}{c}
 \eta_\myf\Lambda\sigma_\myf\Psi+\eta_\myf\Psi\sigma_\myf\Lambda=G_{\myf\Lambda\myf\Psi}\sps\\
 \sigma_\myf\Lambda\eta_\myf\Psi+\sigma_\myf\Psi\eta_\myf\Lambda=G_{\myf\Lambda\myf\Psi}\spsd
\end{array}\right.
\end{equation}
Note that if we pass to an other basis in the space $\mathbb R^{(2N)^2}$ with a non-degenerate operator S from the general linear group $\tilde{\gamma}_\myf\Lambda: = S\gamma_\myf\Lambda S^{-1}$ then the Clifford equation (\ref{e3.0}), generally speaking, will not save its form.

We now turn to the complex structure. Its square is equal to $-E$; this matrix is orthogonal. According to (\ref{e3.12}), for n=2, N=1, such the transformation is represented by means of the formula (\ref{e3.6}) as
\begin{equation}
{\tilde F}^2=\tilde{\tilde F}^2=
\left(
\begin{array}{rr}
-E & 0 \\
 0 & E
\end{array}
\right)\spsd
\end{equation}
If the complex structure $f_\myf\Lambda{}^\myf\Psi$ has the canonical form in the given basis then in the decomposition (\ref{e3.9}), $F:=\tilde F=\tilde {\tilde F}$, $I:=\tilde I=\tilde {\tilde I}$ has a block structure too (due to $\gamma_\myf\Lambda$ having a block structure and the conformity (\ref{e3.4}) written for the real representation)
\begin{equation}
F=
\left(
\begin{array}{rr}
\hat F & 0 \\
0 & \hat {\hat F}^T
\end{array}
\right)\sps
\hat F^2=-E\sps \hat {\hat F}^2=E\sps
I=
\left(
\begin{array}{rr}
iE & 0 \\
 0 & E
\end{array}
\right)\spsd
\end{equation}
With regard to the transformation I, it is determined from the relationship
\begin{equation}
i\gamma_\Lambda=\frac{1}{2}(i\gamma_\Lambda+(-i)(-\gamma_\Lambda))=
\underbrace{\frac{1}{2}((1+i)E+(1-i){\tilde F}^2)}_{:=\tilde I}\gamma_\Lambda\underbrace{\frac{1}{2}((1+i)E+(1-i){\tilde{\tilde F}}^2)}_{:=\tilde{\tilde I}}\spsd
\end{equation}
Whereas, the quantity of the elementary transformations, making the complex structure $f_\myf\Lambda{}^\myf\Psi$, is always multiple to 2 (since the dimension 2n of the real representation for even n is multiple to 4) then the block structure will be on the main diagonal. Hence, for the block components of $\gamma_\myf\Lambda$, the relation
\begin{equation}
f_\myf\Lambda{}^\myf\Psi\sigma_\myf\Psi=\hat F \sigma_\myf\Lambda\hat{\hat F}^T\sps
f_\myf\Lambda{}^\myf\Psi\eta_\myf\Psi=\hat {\hat F}^T\eta_\myf\Lambda\hat F
\end{equation}
will be true that defines the positive representation of (\ref{e1.9}) for the top sign and the negative representation of (\ref {e1.9}) for the lower sign. If we will set ($\myf A,\myf B,\myf C,\myf D, ... =\overline{1,4}$, $\hat A,\hat B,\hat C,\hat D, ... =\overline{1,2}$,
$\tilde{\tilde A},\tilde{\tilde B},\tilde{\tilde C},\tilde{\tilde D}, ... =\overline{1,2}$, $A,B,C,D ... =\overline{1,1}$)
\begin{equation}
\tilde m_{\pm}{}_{\hat A}{}^{\tilde B}:=
\left(
\begin{array}{ll}
\tilde m_{\pm}{}_{A_1}{}^{\tilde{\tilde B}_1} &                                             0\\
                                            0 & \tilde{\tilde m}_{\pm}{}^{A_2}{}_{\tilde{\tilde B}_2}
\end{array}
\right)\sps
\tilde{\tilde m}_{\pm}{}_{\hat A}{}^{\tilde B}:=
\left(
\begin{array}{ll}
\tilde m_{\pm}{}_{A_1}{}^{\tilde{\tilde B}_1} &                                                     0\\
                                            0 & \tilde{\tilde m}_{\pm}{}^{A_2}{}_{\tilde{\tilde B}_2}
\end{array}
\right)
\end{equation}
then the condition (\ref{e1.9}) will take the form
\begin{equation}
\label{e3.12/1}
 %
\end{equation}

\begin{algorithm}
To construct the step operators (\ref{e3.12/1}) for an arbitrary even n, we use the algorithm ~\ref{a2.1}
($N=2^{n/2-1}$, $I=\overline{1,\frac{n}{2}}$, $\myf A,\myf B,\myf C,\myf D, ... =\overline{1,\frac{(2N)^2}{2^{I-1}}}$,  $\hat A,\hat B,\hat C,\hat D, ... =\overline{1,\frac{(2N)^2}{2^I}}$, $\tilde{\tilde A},\tilde{\tilde B},\tilde{\tilde C},\tilde{\tilde D}, ... =\overline{1,\frac{2N^2}{2^{I-1}}}$, $A,B,C,D ... =\overline{1,\frac{2N^2}{2^I}}$)
\begin{equation}
\label{e3.12/2}
 %
 \right.  \\
 \end{array}
\end{equation}
Thus,
\begin{equation}
(\tilde\triangle_{I_\pm}){}_{\tilde{\tilde C}}{}^{\tilde{\tilde A}}:=
\frac{1}{2}(\delta_{\tilde{\tilde C}}{}^{\tilde{\tilde A}}\pm\hat I_{\tilde{\tilde C}}{}^{\tilde{\tilde K}}\hat F_{\tilde{\tilde K}}{}^{\tilde{\tilde A}})\sps
(\tilde{\tilde\triangle}_{I_\pm}){}_{\tilde{\tilde C}}{}^{\tilde{\tilde A}}:=
\frac{1}{2}(\delta_{\tilde{\tilde C}}{}^{\tilde{\tilde A}}\pm\hat{\hat I}_{\tilde{\tilde C}}{}^{\tilde{\tilde K}}\hat{\hat F}_{\tilde{\tilde K}}{}^{\tilde{\tilde A}})\spsd
\end{equation}
It is now possible to define the operators $(\tilde{\tilde A},\tilde{\tilde B},\tilde{\tilde C},\tilde{\tilde D}, ... =\overline{1,2N^2}$, $A,B,C,D ... =\overline{1,N}$, $K\ne J)$
\begin{equation}
\label{e3.13/1}
\begin{array}{cc}
\tilde M_K:=\tilde m_{\frac{n}{2}_{z_{\frac{n}{2}}}}\tilde m_{{\frac{n}{2}-1}_{z_{\frac{n}{2}-1}}}\cdot ...\cdot\tilde m_{2_{z_2}}\tilde m_{1_{z_1}}\sps &
\tilde{\tilde M}_K:=\tilde{\tilde m}_{\frac{n}{2}_{z_{\frac{n}{2}}}}\tilde{\tilde m}_{{\frac{n}{2}-1}_{z_{\frac{n}{2}-1}}}\cdot ...\cdot\tilde{\tilde m}_{2_{z_2}}\tilde{\tilde m}_{1_{z_1}}\sps\\
(\tilde M_K)_C{}^\myff A(\tilde M^*_K)^B{}_\myff A=\delta_C{}^B\sps &
(\tilde{\tilde M}_K)_C{}^\myff A(\tilde{\tilde M}^*_K)^B{}_\myff A=\delta_C{}^B\sps\\
(\tilde M_K)_C{}^\myff A(\tilde M^*_J)^B{}_\myff A=0\sps &
(\tilde{\tilde M}_K)_C{}^\myff A(\tilde{\tilde M}^*_J)^B{}_\myff A=0\sps\\
\sum\limits_{K=1}^{2N}(\tilde M_K)_A{}^\myff C(\tilde M^*_K)^A{}_\myff B=\delta_\myff B{}^\myff C\sps &
\sum\limits_{K=1}^{2N}(\tilde{\tilde M}_K)_A{}^\myff C(\tilde{\tilde M}^*_K)^A{}_\myff B=\delta_\myff B{}^\myff C\sps\\
\end{array}
\end{equation}
where $z_J\ (J=\overline{1,\frac{n}{2}})$ is equal to 0 for the sign $\ll-\gg$ or 1 for the sign $\ll+\gg$ then $K=\sum\limits_{J=1}^{\frac{n}{2}}z_J\cdot 2^{J-1}+1$. If it does not matter which of the K-th operators will be selected then the number K will be omitted. Using these operators, it is possible to define ($\tilde M:=\tilde M_C{}^\myff A$, $\tilde M^*:=\tilde M^{*C}{}_\myff A$, $\tilde{\tilde M}:=\tilde{\tilde M}_C{}^\myff A$, $\tilde{\tilde M}^*:=\tilde{\tilde M}^{*C}{}_\myff A$)
\begin{equation}
\label{e3.13/2}
\sigma_\Lambda:=\tilde M (m_\Lambda{}^\myf\Lambda\sigma_\myf\Lambda)\tilde{\tilde M}^T\sps
\eta_\Lambda:=\tilde{\tilde M}^* (m_\Lambda{}^\myf\Lambda\eta_\myf\Lambda)(\tilde M^*){}^T\spsd
\end{equation}
\end{algorithm}

For further calculations, we must impose the condition
\begin{equation}
tr(\eta_\myf\Psi \sigma_\myf\Phi)=\frac{N}{2}g_{\myf\Psi\myf\Phi}\spsd
\end{equation}
Going to the complex representation in (\ref{e3.12}) with the help of the operators (\ref{e3.13/1}), we can obtain
\begin{equation}
\label{e3.13} \left\{
\begin{array}{c}
 \eta_\Lambda\sigma_\Psi+\eta_\Psi\sigma_\Lambda=g_{\Lambda\Psi}\sps\\
 \sigma_\Lambda\eta_\Psi+\sigma_\Psi\eta_\Lambda=g_{\Lambda\Psi}\sps\\
tr(\eta_\Psi \sigma_\Lambda)=\frac{N}{2}g_{\Psi\Lambda}\spsd
\end{array}\right.
\end{equation}
According to (\ref{e3.3}), this will lead to
\begin{equation}
\label{e3.14}
 R_\eta:=r^\Lambda\eta_\Lambda\sps
 R_\sigma:=r^\Lambda\sigma_\Lambda\sps
 \frac{2}{N}tr((R_\sigma)\eta_\Lambda)=r_\Lambda\sps
 \frac{2}{N}tr((R_\eta)\sigma_\Lambda)=r_\Lambda\sps
\end{equation}
\begin{equation}
 R=\left(
 \begin{array}{cc}
   0       & R_\sigma \\
   R_\eta  & 0
 \end{array}\right)\sps
\end{equation}
where the tensors $ R_\eta, \ R_\sigma$ have the dimension $N \times N$. Hence, (\ref{e3.4}) is separated into the pair of the equations
\begin{equation}
 R_\sigma\eta_\Psi R_\sigma=
 (r_\Psi r^\Phi-\frac{1}{2}(r_\Omega r^\Omega)\delta_\Psi{}^\Phi) \sigma_\Phi\sps
 R_\eta\sigma_\Psi R_\eta=
 (r_\Psi r^\Phi-\frac{1}{2}(r_\Omega r^\Omega)\delta_\Psi{}^\Phi) \eta_\Phi\spsd
\end{equation}
Since
\begin{equation}
R_\sigma R_\eta=\frac{1}{2}(r_\Omega r^\Omega)E\sps
\end{equation}
then one can determine the formalization of the operators $\eta_\Lambda\ , \sigma_\Lambda$ as follows
\begin{equation}
\eta_\Lambda:=\eta_\Lambda{}^{AB}\sps
\sigma_\Lambda:=\sigma_\Lambda{}_{AB}=\eta_\Lambda{}_{BA}\sps
(A,B,... =\overline{1,N})\spsd
\end{equation}
Then (\ref{e3.5}) can be rewritten as
\begin{equation}
\label{e3.15}
 I).\  S_\Psi{}^\Lambda\eta_\Lambda={\tilde S}^T\eta_\Psi\tilde{\tilde S}\sps
 II).\ S_\Psi{}^\Lambda\eta_\Lambda=\tilde{\tilde S}^T\sigma_\Psi\tilde S\sps
\end{equation}
but here the operators $\tilde {\tilde S}$ and $\tilde S$ have the dimension $N \times N$. However, the decomposition of these operators for the special transformation $S_\Lambda{}^\Psi\ (\det\parallel S_\Lambda{}^\Psi\parallel =1)$ (case I) will have the form
\begin{equation}
 \tilde S^T=\prod\limits_{I=1}^{J}
 (R_\eta)_{2I-1}(R_\sigma)_{2I}={(R_\eta)}_1{(R_\sigma)}_2{(R_\eta)}_3{(R_\sigma)}_4 \ldots {(R_\eta)}_{2J-1}{(R_\sigma)}_{2J}\sps
\end{equation}
\begin{equation}
 \tilde {\tilde S}=\prod\limits_{I=J}^{1} (R_\sigma)_{2I}(R_\eta)_{2I-1}=
 {(R_\sigma)}_{2J}{(R_\eta)}_{2J-1}\ldots {(R_\sigma)}_4{(R_\eta)}_3{(R_\sigma)}_2{(R_\eta)}_1 \sps
\end{equation}
and for the non-special transformation $S_\Lambda{}^\Psi \ (\det\parallel S_\Lambda{}^\Psi\parallel =-1)$ (case II) will have the form
\begin{equation}
 \tilde {\tilde S}^T=(R_\eta)_{1}\prod\limits_{I=1}^{J} (R_\sigma)_{2I}(R_\eta)_{2I+1}={(R_\eta)}_1{(R_\sigma)}_2{(R_\eta)}_3
 \ldots {(R_\sigma)}_{2J}{(R_\eta)}_{2J+1}\sps\\
\end{equation}
\begin{equation}
 \tilde S=(\prod\limits_{I=J}^{1} (R_\eta)_{2I+1}(R_\sigma)_{2I})(R_\eta)_1=
 {(R_\eta)}_{2J+1}{(R_\sigma)}_{2J}\ldots {(R_\eta)}_3{(R_\sigma)}_2{(R_\eta)}_1 \spsd
\end{equation}
Thus, the equation (\ref{e3.15}) defines the algebraic realization of the double covering $Spin(n,\mathbb C)/\{\pm 1\}\cong SO(n,\mathbb C)$. Accordingly, for the involution, the condition (\ref{e3.8}) determines
\begin{equation}
\label{e3.16}
\begin{array}{llll}
 I).&\  S_\Psi{}^{\Lambda '}\eta_{\Lambda '}=\tilde S{}^T\eta_\Psi\tilde{\tilde S}\sps &
 II).&\ S_\Psi{}^{\Lambda '}\eta_{\Lambda '}=\tilde{\tilde S}^T\sigma_\Psi\tilde S\sps\\
 &\tilde S \bar{\tilde S}=\pm E\sps \tilde{\tilde S}\bar{\tilde{\tilde S}}=\pm E\spsd &
 &\tilde S =\pm \bar{\tilde S}\sps \tilde{\tilde S}=\pm \bar{\tilde{\tilde S}}\spsd
\end{array}
\end{equation}
The bottom line of the parities are inherited by the real representation of the involution in the tangent and spinor spaces. But the real representation of an involution in the tangent space is an orthogonal transformation, the square of which has the identity matrix. Accordingly, in the real representation of the spinor space, this property inherits, and hence the square of the spinor involution in the real representation (case I) up to sign is the identity transformation too. It remains to pass to the complex representation on the rule (\ref{e1.10}).

\vspace{12mm}
\begin{example} 2n=2, N=1.
\label{ex3.1}
if $g_{\myf\Lambda\myf\Psi}=\left(

\end{equation}
Thus, the identity $tr(\eta_\myf\Lambda\sigma_\myf\Psi)=\frac{1}{2}g_{\myf\Lambda\myf\Psi}$ is false. Nevertheless, we can use
(\ref{e3.5}) and (\ref{e3.6}). Define
\begin{equation}
%
\end{equation}
Then for any vector $x^\myf\Lambda=(T,X)$, the conformity $X=x^\myf\Lambda\gamma_\myf\Lambda\sps (S_1)_\myf\Lambda{}^\myf\Psi x_\myf\Psi \longleftrightarrow R_1 X R_1$ will have the form
\begin{equation}
 \left(%
\end{equation}
Then for any vector $x^\myf\Lambda=(T,X)$, the conformity $X=x^\myf\Lambda\gamma_\myf\Lambda\sps (S_2)_\myf\Lambda{}^\myf\Psi x_\myf\Psi \longleftrightarrow R_2 X R_2$ will have the form
\begin{equation}
\left(\begin{array}{cc}
  \cos 2\phi & -\sin 2\phi \\
 -\sin 2\phi & -\cos 2\phi
\end{array}\right)
\left(
\begin{array}{c}
  T \\
  X
\end{array}\right)\longleftrightarrow
\left(
\begin{array}{cc}
  0         & e^{-i\phi} \\
  e^{i\phi} & 0
\end{array}
\right) \left(
\begin{array}{cc}
  0    & T+iX \\
  T-iX & 0
\end{array}
\right) \left(
\begin{array}{cc}
  0         & e^{-i\phi} \\
  e^{i\phi} & 0
\end{array}
\right)\spsd
\end{equation}
It is now possible to construct the special transformation $S:=S_1S_2$ for which $S_\myf\Lambda{}^\myf\Psi x_\myf\Psi\longleftrightarrow R_1 R_2X R_2 R_1$ and $\tilde S=R_1 R_2\sps \tilde{\tilde S}=R_2 R_1$. The explicit record of the conformity will have the form
\begin{equation}
 \left(\begin{array}{cc}
  \cos 4\phi & -\sin 4\phi \\
  \sin 4\phi &  \cos 4\phi
 \end{array}\right)
 \left(
 \begin{array}{c}
  T \\
  X
 \end{array}\right)\longleftrightarrow
 \left(
 \begin{array}{cc}
  e^{2i\phi} & 0 \\
  0          & e^{-2i\phi}
 \end{array}
 \right) \left(
 \begin{array}{cc}
  0    & T+iX \\
  T-iX & 0
 \end{array}
 \right) \left(
 \begin{array}{cc}
  e^{-2i\phi} & 0 \\
  0           & e^{2i\phi}
 \end{array}
 \right)\spsd
\end{equation}
It is obvious that $SO(2)$ is formed by such the transformations $S_\myf\Lambda{}^\myf\Psi$ then $\left(\begin{array}{cc}  e^{2i\phi} &
0 \\ 0 & e^{-2i\phi} \end{array}\right)$  represents the group, double covering the group $\left(\begin{array}{cc}  e^{4i\phi}
& 0 \\ 0 & e^{-4i\phi} \end{array}\right)\cong U(1)$. At the same time, $\tilde{\tilde S}=\overline{\tilde S}$ (\emph{complex conjugate}). This proved that $SO (2)\cong U(1)$.
\end{example}

\vspace{12mm}

\begin{example} 2n=2, N=1.
\label{ex3.2}
If $g_{\myf\Lambda\myf\Psi}=\left(%
\end{equation}
Thus, the identity $tr(\eta_\myf\Lambda\sigma_\myf\Psi)=\frac{1}{2}g_{ij}$ is false. The explicit record of the conformity $S_\myf\Lambda{}^\myf\Psi  x_\myf\Psi \longleftrightarrow \tilde S X \tilde {\tilde S}$  will have the form
\begin{equation}
 \left(\begin{array}{cc}
  \ch\ 4\phi & \sh\ 4\phi \\
  \sh\ 4\phi & \ch\ 4\phi
 \end{array}\right)
 \left(
 \begin{array}{c}
  T \\
  X
 \end{array}\right)\longleftrightarrow
 \left(
 \begin{array}{cc}
  e^{2\phi} & 0 \\
  0          & e^{-2\phi}
 \end{array}
 \right) \left(
 \begin{array}{cc}
  0    & T+X \\
  T-X & 0
 \end{array}
 \right) \left(
 \begin{array}{cc}
  e^{-2\phi} & 0 \\
  0           & e^{2\phi}
 \end{array}
 \right)\sps
\end{equation}
that determines the conformity $SO^+(1,1)\cong R$ (\emph{additive}), where $SO^+(1,1) $ is the connected identity component with $S_1{}^1>0$.
\end{example}

\vspace{12mm}

\begin{example} 2n=4, N=2.
\label{ex3.3}
If $g_{ij}=\left(\begin{array}{cccc}
  1 & 0 &  0 &  0 \\
  0 & 1 &  0 &  0 \\
  0 & 0 & -1 &  0 \\
  0 & 0 &  0 & -1
\end{array}\right)$ then we can construct
\begin{equation}
%
\end{equation}
The operators $\gamma_i$ constructed in accordance with $\gamma_i=\left(\begin{array}{cc} 0 & \sigma_i \\ \eta_i  & 0 \end{array}\right)$. It is easy to verify that $tr(\eta_i\sigma_j)=tr(\sigma_i\eta_j)=g_{ij}$. The calculations in this case are more cumbersome, but similar to the case n = 2 (and in part, will be carried out below). Just note that the fact that here the covering $SO^+(2,2)\cong SL(2)\times SL(2)/{\pm 1}$ acts. However, it would be desirable to know how to pass from the case n=4 to the case n=2.
\end{example}

\vspace{12mm}

\begin{example} $n=2 \subset 2n=4,\ N=2.$
\begin{equation}
%
\right)\sps\\
 x^\myf\Lambda=(T,Y,Z,X)\sps x^\Lambda= m^\Lambda{}_\myf\Lambda x^\myf\Lambda =\frac{1}{2}(T-Y+i(Z-X),-i(T+Y)+Z+X)
\end{array}
\end{equation}
then each special orthogonal transformation $S_\myf\Lambda{}^\myf\Psi \in SO^+(2,2)$ $(\det\parallel S_\Lambda{}^\Psi\parallel =1$, and the determinant of the upper left minor $2\times 2$ is positive) can be represented as a finite product of the elementary transformations. Since $fSf =-S$ due to the presence of the complex structure in the tangent bundle then such the transformation will be of the two kinds
\begin{equation}

\end{equation}
The complex representation $S_\myf\Lambda{}^\myf\Psi $ can be defined as $S_\Lambda{}^\Psi=m_\Lambda{}^\myf\Lambda S_\myf\Lambda{}^\myf\Psi m^\Psi{}_\myf\Psi$
\begin{equation}
 (S_1)_\Lambda{}^\Psi=\left(
 \begin{array}{cccc}
  -\sin 2\varphi & -i\cos 2\varphi\\
  i\cos 2\varphi &   \sin 2\varphi
 \end{array}\right)\sps
 (S_2)_\Lambda{}^\Psi=\left(
 \begin{array}{cccc}
  i\sh\ 2\psi & -i\ch\ 2\psi \\
  i\ch\ 2\psi & -i\sh\ 2\psi \\
 \end{array}\right)\spsd
\end{equation}
Therefore, $S=S_2S_1$ describes the group $SO(2,\mathbb C)$ completely. Now let
\begin{equation}
 \begin{array}{lcl}
  (r_1){}^\myf\Lambda =\sqrt{2}(\cos\phi ,\sin\phi ,0,0) & , & (r_2){}^\myf\Lambda=\sqrt{2}(0,0, \cos\phi , \sin\phi )\sps \\
  (r_1){}_\myf\Lambda =\sqrt{2}(\cos\phi ,\sin\phi ,0,0) & , & (r_2){}_\myf\Lambda=\sqrt{2}(0,0,-\cos\phi ,-\sin\phi )\spsd
 \end{array}
\end{equation}
Construct $(S_1)_\myf\Lambda{}^\myf\Theta=((r_1)_\myf\Lambda (r_1)^\myf\Psi -\delta_\myf\Lambda{}^\myf\Psi)
((r_2)_\myf\Psi (r_2)^\myf\Theta +\delta_\myf\Psi{}^\myf\Theta)$. At the same time, $(r_1)_\myf\Lambda (r_1)^\myf\Lambda=
-(r_2)_\myf\Lambda (r_2)^\myf\Lambda=2$. Let $\eta_\myf\Lambda$ and $\sigma_\myf\Lambda$ be the same as in the previous example
\begin{equation}
%
\end{equation}
Construct $(S_2)_\myf\Lambda{}^\myf\Theta=((r_3)_\myf\Lambda (r_3)^\myf\Psi -\delta_\myf\Lambda{}^\myf\Psi)
((r_4)_\myf\Psi (r_4)^\myf\Theta +\delta_\myf\Psi{}^\myf\Theta)$. At the same time, $(r_3)_\myf\Lambda (r_3)^\myf\Lambda=
-(r_4)_\myf\Lambda (r_4)^\myf\Lambda=2$.
\begin{equation}
%
\right)$ form the group $U(1)$ and $\left(
\begin{array}{cc}
-\ch\ 2\psi & \sh\ 2\psi \\
 \sh\ 2\psi &-\ch\ 2\psi
\end{array}\right)$ form the additive group R.  If to make the substitution $\eta$ on $\sigma$ in the relevant equations then
\begin{equation}
%
\end{equation}
\end{example}

\vspace{12mm}

\begin{example} $n=2 \subset 2n=4,\ N=2.$
Let the involution tensor in the tangent fiber for the real representation  $V_{4}$ be the tensor (the left inclusion of the metric of index equal to 2, the right inclusion of the metric of index equal to 1)
\begin{equation}
%
\end{equation}
and all the identities of the previous examples are executed. Then for the complex representation $\mathbb CV_{2}$, it will have the form $S_\Lambda{}^{\Psi '}=m_\Lambda{}^\myf\Lambda S_\myf\Lambda{}^\myf\Psi\bar m ^{\Lambda '}{}_\myf\Psi$
\begin{equation}
%
\end{equation}
On the other hand, there is an image of the involution (as letters T denotes the transposition), and in some basis
$S_\myf\Lambda{}^\myf\Psi \eta_\myf\Psi=\tilde S^T\eta_\myf\Psi\tilde S$
\begin{equation}
%
\end{equation}
In the first case (index equal to 2)
\begin{equation}
%
\right)\sps
\end{equation}
and we again come to the isomorphism $SO(2) \cong U(1)$
\begin{equation}
%
\end{equation}
In the second case (index 1),
\begin{equation}
%
\right)\sps
\end{equation}
and we again come to the isomorphism $SO^+(1,1)\cong R(additive)$
\begin{equation}
%
\end{equation}
\end{example}
The following example is worthy of a separate section.

\section{Quaternions}
\Abstract{
$\phantom{ff}$In this section, it is told about how to contact the quaternion algebra with the Clifford algebra and how to construct explicitly the double covering corresponding to this case. The conclusion of all results of this section is made on the basis \cite{Postnicov1}.
}

It is known that the Clifford algebra $CL(g^n)$ can be represented as
\begin{equation}
%
\end{equation}
 that is called \emph{$Z_2$ grading}. In this case, $CL^0(g^n)$ is a subalgebra of $CL(g^n)$. If we consider n=4 with the 8-dimensional Clifford algebra $CL^0(g^n)$ then the basis of this algebra has the form
\begin{equation}
 \begin{array}{l}
  1\sps e_1e_2\sps e_1e_3 \sps  e_2e_3 \sps
  e_1e_4\sps e_2e_4 \sps e_3e_4\sps e_1e_2e_3e_4\spsd
 \end{array}
\end{equation}
From the Clifford equation (\ref{e2.2}), the relation
\begin{equation}
 e_\Lambda e_\Psi=1\sps\Lambda = \Psi;\ \ \
 e_\Lambda e_\Psi=-e_\Psi e_\Lambda \sps \Lambda \neq \Psi
\end{equation}
will follow in some basis
\begin{equation}
 \begin{array}{ll}
 I=1\sps & i:=e_1e_2\sps \\ j=:e_1e_3\sps & k:=e_2e_3\sps \\
 e:=e_3e_4\sps & f:=ie=e_1e_2e_3e_4\sps \\ g:=je=e_1e_3e_3e_4=e_1e_4 \sps & h:=ke=e_2e_3e_3e_4=e_2e_4\sps
 \end{array}
\end{equation}
 Consider the inclusion $\mathbb R^4\subset\mathbb C^4$. Then we can find the representation of this basis with the operators $\gamma_i =H_i{}^\Lambda \gamma_\Lambda$ constructed above. But the basis, in turn, is the direct sum of two quaternionic bases; for this, the isomorphism $CL^0(g^n_{(n,0)})\cong \mathbb H\oplus \mathbb H$ is responsible. In order to construct an appropriate representation of the specified basis, it is necessary to demand the performance of the condition $tr(\eta_i\sigma_j)=\frac{N}{2}g_{ij}$ for the parts of the operators $\gamma_i$. And then, indeed, it may be restricted by the quaternion basis (just having 2 generators i and j (k = ij) and the quaternion identity I)
\begin{equation}

\end{equation}
Now suppose that in the space $\mathbb R^4\subset \mathbb C^4$, the vectors
\begin{equation}
\begin{array}{c}
 r_1=(\cos\phi_1 ,\sin\phi_1,0,0)\sps
 r_2=(\cos\phi_2 ,0,\sin\phi_2,0)\sps
 r_3=(0,\cos\phi_3 ,\sin\phi_3,0)\sps\\
 r_4=(\cos\phi_4 ,0,0,\sin\phi_4)\sps
 r_5=(0,\cos\phi_5 ,0,\sin\phi_5)\sps
 r_6=(0,0,\cos\phi_6 ,\sin\phi_6)\sps
\end{array}
\end{equation}
generating the elementary orthogonal transformations, are defined. Here, as before, the transformations $(r_I)^i (r_I)_j-\delta_j{}^i $  $(I=\overline{1,6})$ generate all rotations of the group O(4). Accordingly, on the space $\mathbb C^2$ ($N=2^{2/2-1}=2$), their images
\begin{equation}
%
\end{equation}
will be induced. To simplify calculations, we will be restricted to the case when $\phi_1 =\phi_2 =\phi_4 =\frac{\pi}{2}$. Then
\begin{equation}
\label{e7.1}
%
\end{equation}
By this, the double covering $SO(3)\cong SU(2)/\{\pm 1\}$ is constructed, and therefore $Spin(3)\cong SU(2)$
\begin{equation}
 {S_1}_i{}^j \eta_j= A_1{}^T\eta_i A_1^*\spsd
\end{equation}
If we now set $\phi_3 =\phi_5 =\phi_6 =\frac{\pi}{2}$ and carry out similar calculations, we obtain
\begin{equation}
%
 \right) =:A_2{}^{\sharp} \spsd
\end{array}
\end{equation}
Therefor,
\begin{equation}
 {S_2}_i{}^j \eta_j= A_2{}^T\eta_i A_2^{\sharp}\spsd
\end{equation}
The transformations $A_1,A_2$ are the representatives of $SU(2)$ and fully describe the group. Gathering the two equations together, we find that
\begin{equation}
 S_i{}^j \eta_j:= (S_2S_1)_i{}^j \eta_j=
 A_1{}^TA_2{}^T \eta_j A_2^\sharp A_1^*:=\tilde S {}^T\eta_j \tilde{\tilde S}\spsd
\end{equation}
Generally speaking, knowing the transformation $A_1{}^TA_2{}^T$, we can say nothing about $A_2^\sharp A_1^*$. Therefore, the transformations $\tilde S{}^T$ and $\tilde{\tilde S}$ are distinct. This means that there is the double covering  $SO(4)\cong SU(2) \times SU(2)/\{\pm 1\}$. And so $Spin(4)\cong SU(2) \times SU(2)$.

\section{Particular solutions of the Clifford equation}
\Abstract{
$\phantom{ff}$In this section it is told about how to construct particular solutions of the Clifford equation with the help of which can uniquely construct the double covering $Spin(n,\mathbb C)/\{\pm~1\}\cong SO(n,\mathbb C)$. Such the solutions will give the chance uniquely to prolong the Riemannian torsion-free connection into the spinor bundle. The conclusion of all results of this section is made on the basis of previous calculations.
}

In order to find some particular solutions of the Clifford equation (\ref{e3.13}), we rewrite it as
\begin{equation}
\label{e6.1}
\eta_\Lambda{}^{AB}\eta_\Psi{}_{CB}+\eta_\Psi{}^{AB}\eta_\Lambda{}_{CB}=g_{\Lambda\Psi}\delta_C{}^A\spsd
\end{equation}
\begin{example}
\label{ex6.1}
n=4, N=2.\\
In this case, we use the isomorphism $\mathbb C^4 \cong \mathbb C(2)$ which is constructed by means of the connecting operators $\eta^\Lambda{}_{AB}$ (\ref{e3.14})
\begin{equation}
r^\Lambda=\frac{1}{2}\eta^\Lambda{}_{AB} R^{AB}\spsd
\end{equation}
Let the metric in $\mathbb C^4$ have the form
\begin{equation}
g_{\Lambda\Psi}=
\left(
\begin{array}{rrrr}
 1 & 0 & 0 & 0\\
 0 & 1 & 0 & 0\\
 0 & 0 & 1 & 0\\
 0 & 0 & 0 & 1
\end{array}
\right)\spsd
\end{equation}
It is obvious that any tensor $R^{AB}$ can be decomposed into the three symmetrical and one antisymmetrical parts
\begin{equation}
%
\end{equation}
with the metric spin-tensor
$\varepsilon_{AB}=
\left(
%
\end{equation}
where $(\varepsilon_I)^{ab}$ are taken from the previous example (n=4, N=2). It is one of possible variants (the 4 symmetrical and 2 antisymmetric parts). If all received $(\varepsilon_I)^{AB}$ ($I =\overline{1,2,3,4,\tilde 1,\tilde 2}$) multiply by
\begin{equation}
\left(
\begin{array}{rrrr}
 0 & 0 & 0 & i\\
 0 & 0 &-i & 0\\
 0 & i & 0 & 0\\
-i & 0 & 0 & 0
\end{array}
\right)
\end{equation}
on the right side then one can construct the new tensors (the 6 antisymmetric parts)
\begin{equation}
%
\right)\sps
\end{array}
\end{equation}
which will enable to define
\begin{equation}
R^{AB}=-R^{BA}=R_{\tilde 1}(\varepsilon_{\tilde 1})^{AB}+R_{\tilde 2}(\varepsilon_{\tilde 2})^{AB}+
R_{\tilde 3}(\varepsilon_{\tilde 3})^{AB}+R_{\tilde 4}(\varepsilon_{\tilde 4}^{AB})+
R_{\tilde 5}(\varepsilon_{\tilde 5})^{AB}+R_{\tilde 6}(\varepsilon_{\tilde 6}^{AB})\spsd
\end{equation}
\end{example}

In order to obtain a similar construction for any even n, we need in the following definition.
\begin{definition}
\label{d6.1}
Let in the space $\mathbb C^n$ $(n\ge 4)$, n orthogonal vectors $(\eta_I){}^\Lambda$
\begin{equation}
(\eta_I){}^\Lambda (\eta_J){}_\Lambda=2\delta_{IJ}\sps \Lambda ,\ \Psi , \ ... = \overline{1,n}
\end{equation}
be given. Then for the spinor space, the vectors are uniquely associated to the n tensors
\begin{equation}
(\varepsilon_I)_{AB}:=(\eta_I)^\Lambda \eta_\Lambda{}_{AB}\spsd
\end{equation}
Therefore, among the n vectors $(\eta_I){}^\Lambda$, some $\frac{n}{2}+q$ vectors
\begin{equation}
\eta^\Psi{}_{(AB)}=\frac{1}{2}\sum_{Q=1}^{\frac{n}{2}+q}(\eta_Q){}^\Psi(\varepsilon_Q)_{AB}\sps
(\varepsilon_Q)_{AB}=(\varepsilon_Q)_{BA}
\end{equation}
and some $\frac{n}{2}+\tilde q $ vectors
\begin{equation}
\eta^\Psi{}_{[AB]}=\frac{1}{2}\sum_{\tilde Q =1}^{\frac{n}{2}+\tilde q}(\eta_{\tilde Q}){}^\Psi(\varepsilon_{\tilde Q}){}_{AB}\sps
(\varepsilon_{\tilde Q}){}_{AB}=-(\varepsilon_{\tilde Q}){}_{BA}\sps q+\tilde q=0
\end{equation}
must be found. In addition, we require that the real representation of these operators satisfy similar relations.
\end{definition}

\begin{corollary}
From the relation
\begin{equation}
\label{e6.3/1}
\tilde T_A{}^B(\varepsilon_I)_{BC}+\tilde{\tilde T}_C{}^B(\varepsilon_I)_{AB}=0\sps
\end{equation}
performed for all $I=\overline{1,n}$, the parities  $\tilde T_A{}^B=\tilde{\tilde T}_A{}^B=0$ follow.
\end{corollary}

\begin{proof}
We use the identity
\begin{equation}
\begin{array}{c}
\eta_{\left[\right.\Gamma}{}^{CN}\eta_{\Omega\left.\right] MN}\varepsilon_{CD}{}^{ML}=
\frac{1}{2}\sum\limits_I \eta_{\left[\right.\Gamma}{}^{CN}\eta_{\Omega\left.\right] MN}(\varepsilon_I)_{CD}{}(\varepsilon_I)^{ML}=\\
=\frac{1}{2}\sum\limits_I(-\eta_{\left[\right.\Gamma}{}_{|CD}(\varepsilon_{I|})^{CN}+(\eta_I)_{\left[\right.\Gamma}\delta_{|D|}{}^N)
\eta_{\Omega\left.\right] MN}(\varepsilon_I)^{ML}=\\
=\frac{1}{2}\sum\limits_I (-\eta_{\left[\right.\Gamma}{}_{|CD|})(-\eta_{\Omega\left.\right]}{}^{CL}+
(\eta_{|I|})_{\Omega\left.\right]}(\varepsilon_I)^{CL})+(\eta_I)_{\left[\right.\Gamma}\eta_{\Omega\left.\right] MD}(\varepsilon_I)^{ML}=
\frac{n-4}{2}\eta_{\left[\right.\Gamma}{}_{|CD|}\eta_{\Omega\left.\right]}{}^{CL}\spsd
\end{array}%
\end{equation}
Consequently, $\tilde{\tilde T}_A{}^D=\tilde T_A{}^D=0$.
\end{proof}

Thus, on the generating tensors $(\varepsilon_I)_{AB}$, the condition $(I\ne J)$
\begin{equation}
\label{e6.2}
\begin{array}{c}
(\varepsilon_J)_{AB}=-(\varepsilon_I)_{CB}(\varepsilon_J)^{CD}(\varepsilon_I)_{AD}\sps
\eta_\Lambda{}^{AB}(\varepsilon_I)_{AD}(\varepsilon_I)_{CB}=-\eta_\Lambda{}_{CD}+(\eta_I)_\Lambda(\varepsilon_I)_{CD}\sps\\
(\eta_I){}^\Lambda=\frac{2}{N}\eta^\Lambda{}^{AB}(\varepsilon_I)_{AB}\sps
(\varepsilon_I)_{AB}=(\eta_I)^\Lambda \eta_\Lambda{}_{AB}\sps
(\varepsilon_I)_{AB}(\varepsilon_I)^{AC}=\delta_B{}^C
\end{array}
\end{equation}
is imposed by the Clifford equation (\ref{e6.1}).

\begin{definition}
\label{d6.2}
In the bivector space, the 4-valent tensor
\begin{equation}
\label{e6.5}
\begin{array}{ll}
\varepsilon_{ABCD}:=\eta_\Lambda{}_{AB}\eta^\Lambda{}_{CD}\sps &
\bar \varepsilon_{A'B'C'D'}:=\bar \eta_{\Lambda '}{}_{A'B'}\bar \eta^{\Lambda '}{}_{C'D'}\sps\\ \\
g_{\Lambda\Psi}=
\left(\frac{2}{N}\right)^2\eta^\Lambda{}^{AB}\eta^\Psi{}^{CD}\varepsilon_{ABCD}\sps &
\bar g_{\Lambda '\Psi '}=
\left(\frac{2}{N}\right)^2\bar \eta^{\Lambda '}{}^{A'B'}\bar \eta^{\Psi '}{}^{C'D'}\bar \varepsilon_{A'B'C'D'}
\end{array}
\end{equation}
with the help of which we can raise and lower pair indices is defined.
\end{definition}

\begin{note}
It should be noted that for a real inclusion the condition $(\eta_I){}^i(\eta_J){}_i=2 \delta_{IJ}$ is not always feasible. It should be modified to $(\eta_I){}^i(\eta_J){}_i=2g_{IJ}$, where $g_{IJ}=0\sps I\ne J$ and $g_{II}=\pm 1$, while the sign is chosen in depending on the metric index and the number of the vector. This literally means that the metric in some basis has the diagonal form  with $\pm 1$ on the main diagonal. Then the equation (\ref{e6.2}) will take the form $(I\ne J)$
\begin{equation}
\label{e6.2a}
\tag{\ref{e6.2}$'$}
\begin{array}{c}
(\varepsilon_J)_{AB}=-g_{II}(\varepsilon_I)_{CB}(\varepsilon_J)^{CD}(\varepsilon_I)_{AD}\sps
g_{II}\eta_i{}^{AB}(\varepsilon_I)_{AD}(\varepsilon_I)_{CB}=-\eta_i{}_{CD}+(\eta_I)_i(\varepsilon_I)_{CD}\sps\\
(\eta_I){}^i=\frac{2}{N}\eta^i{}^{AB}(\varepsilon_I)_{AB}\sps
g_{II}(\varepsilon_I)_{AB}=(\eta_I)^i \eta_i{}_{AB}\sps
g_{II}(\varepsilon_I)_{AB}(\varepsilon_I)^{AC}=\delta_B{}^C\spsd
\end{array}
\end{equation}
In the case of the complex space, we can always restrict ourselves to $+1$ with the Euclidean metric $\delta_{IJ}$. This will be enough to prove the main identities. In addition, one can pass to the real spaces using the corresponding inclusion operator.
\end{note}

Therefore, it is possible the two significantly different ways depending on multiplicity n
\begin{equation}
\begin{array}{lclc}
I).& n\ mod\ 4 =0\sps & II).& n\ mod\ 4 =2
\end{array}
\end{equation}
which determine either the pair of the metric tensors $\varepsilon_{AB}=-\varepsilon_{BA}$, $\tilde\varepsilon_{AB}=\tilde\varepsilon_{BA}$  in Case I). or the pair of the involutions $E_A{}^B$, $\tilde E_A$ in Case II). (the proof is given in Appendix)
\begin{equation}
\label{e6.3}
%
\end{equation}

\begin{note}
For real inclusion, the tensors from (\ref{e6.3}) multiply by $g:=(\prod\limits_{P=1}^{(n/2+q)}\sqrt{g_{PP}})$, where either of the two versions $\pm i$ for $g_{\tilde P\tilde P}=-1$ can be chosen as the root $\sqrt{g_{\tilde P\tilde P}}$.
\end{note}

We will show that such the operators exist. For n = 4, one can construct the representation as described in Example \ref{ex6.1}. Suppose that we have constructed some representation $\eta_\alpha{}^{ab}\ (\alpha\ ,\beta\ , ...=\overline{1,n-2};a,\ b,\  ...=\overline{1,2^{(n-2)/2-1}})$. The operators $\eta_\Lambda{}^{AB}\ (\Lambda\ ,\Psi\ , ...=\overline{1,n};A,\ B,\  ...=\overline{1,2^{n/2-1}})$ are constructed as follows. Let the tensor $g_{\alpha \beta}$ contain only +1 on the main diagonal then
\begin{equation}
\label{e6.6}
\begin{array}{lcclcc}
\eta_\alpha{}^{AB}&=&
\left(
\begin{array}{cc}
   \eta_\alpha{}^{ab} & 0                    \\
                    0 & -(\eta^T)_\alpha{}_{cd}
\end{array}
\right)\sps &
\eta^\alpha{}_{AB}&=&
\left(
\begin{array}{cc}
  \eta^\alpha{}_{pk} & 0                    \\
                   0 & -(\eta^T)^\alpha{}^{lm}
\end{array}
\right)\sps\\ \\
\eta_{n-1}{}^{AB}&=&
\frac{1}{\sqrt{2}}\left(
\begin{array}{cc}
             0 & i\delta^a{}_d\\
-i\delta_c{}^b & 0
\end{array}
\right)\sps &
\eta^{n-1}{}_{AB}&=&
\frac{1}{\sqrt{2}}\left(
\begin{array}{cc}
             0 & -i\delta_p{}^m\\
 i\delta^l{}_k & 0
\end{array}
\right)\sps\\ \\
\eta_{n}{}^{AB}&=&
\frac{1}{\sqrt{2}}\left(
\begin{array}{cc}
              0 & \delta^a{}_d\\
 \delta_c{}^b & 0
\end{array}
\right)\sps &
\eta^{n}{}_{AB}&=&
\frac{1}{\sqrt{2}}\left(
\begin{array}{cc}
              0 & \delta_p{}^m\\
 \delta^l{}_k & 0
\end{array}
\right)\sps
\end{array}
\end{equation}
Under the general scheme of the constructing (\ref{e6.6}) for n = 4, the equation
\begin{equation}
\label{e6.3/1}
\tilde T_A{}^B(\varepsilon_I)_{BC}+\tilde{\tilde T}_C{}^B(\varepsilon_I)_{AB}=0
\end{equation}
leads to $\tilde T_A{}^B=\tilde{\tilde T}_A{}^B=0$ that can be verified directly. Suppose that this equation holds for some even n-2. Then for n, the equation splits into the parts
\begin{enumerate}
\item
\begin{equation}
\tilde T_a{}^b(\varepsilon_I)_{bc}+\tilde{\tilde T}_c{}^b(\varepsilon_I)_{ab}=0\sps
\tilde T^a{}_b(\varepsilon_I)^{bc}+\tilde{\tilde T}^c{}_b(\varepsilon_I)^{ab}=0\sps I=\overline{1,n-2}\sps
\end{equation}
which immediately give $\tilde T_a{}^b=\tilde{\tilde T}_a{}^b=0\sps \tilde T^a{}_b=\tilde{\tilde T}^a{}_b=0$.
\item
\begin{equation}
\begin{array}{c}
\tilde T_{ab}(\varepsilon_{n+1})^b{}_c+\tilde{\tilde T}_{cb}(\varepsilon_{n+1})_a{}^b=0\sps\Rightarrow \tilde T_{ac}-\tilde{\tilde T}_{ca}=0\sps\\
\tilde T^{ab}(\varepsilon_{n+1})_b{}^c+\tilde{\tilde T}^{cb}(\varepsilon_{n+1})^a{}_b=0\sps\Rightarrow -\tilde T^{ac}+\tilde{\tilde T}^{ca}=0\spsd
\end{array}
\end{equation}
\item
\begin{equation}
\begin{array}{c}
\tilde T_{ab}(\varepsilon_{n+2})^b{}_c+\tilde{\tilde T}_{cb}(\varepsilon_{n+2})_a{}^b=0\sps\Rightarrow\tilde T_{ac}+\tilde{\tilde T}_{ca}=0\sps\\
\tilde T^{ab}(\varepsilon_{n+2})_b{}^c+\tilde{\tilde T}^{cb}(\varepsilon_{n+2})^a{}_b=0\sps\Rightarrow\tilde T^{ac}+\tilde{\tilde T}^{ca}=0\sps
\end{array}
\end{equation}
which immediately give $\tilde T_{ab}=\tilde{\tilde T}_{ab}=0\sps \tilde T^{ab}=\tilde{\tilde T}^{ab}=0$. Thus, again $\tilde T_A{}^B=\tilde{\tilde T}_A{}^B=0$.
\end{enumerate}
Now suppose that there are the real connecting operators $\eta_\myf\Lambda{}^{\myff A\myff B}$. In this case, there is a set of orthogonal vectors such that a part $(\varepsilon_Q)^{\myff A\myff B}$ is symmetric and the remaining part $(\varepsilon_{\tilde Q})^{\myff A\myff B}$ is antisymmetric. We will show that one can move towards the complex representation of such the operators which preserve this property and therefore satisfy Definition \ref{d6.1}.

\begin{algorithm}$ $\\
\label{a6.1}
For the first step, the operators
\begin{equation}

\end{equation}
are used as the operators (\ref{e3.12}) and (\ref{e3.12/1}). As the operators $(\tilde{\tilde m}_{1_-})^A{}_{\myff A}$, we can take
$
\left(
\begin{array}{cccc}
 \delta^a{}_{\myff a} & 0                    & 0                     &-i\delta^a{}_{\myff q}\\
 0                    & \delta_c{}^{\myff c} &-\delta_c{}^{\myff p}  & 0
\end{array}
\right)\spsd
$
If we now use the operators $m_\Lambda{}^\myf\Lambda$ translating the subspace $\mathbb R^4_{(2,2)}$ in the subspace  $\mathbb C^2$ and leaving unchanged the subspace $\mathbb R^{2n-4}_{(n-2,n-2)}$ then we can construct the connecting operators  $\eta_\Lambda{}^{ab}$ of the corresponding space $\mathbb C^2\oplus \mathbb R^{2n-4}_{(n-2,n-2)}$, while the dimension of the spinor space will be reduced by 2 times. \\
2. For the second step, we will use the operators
$$
\underbrace{\frac{1}{\sqrt 2}\left(
\begin{array}{cccccccc}
 \delta^a{}_{\myff a}       & 0                          & 0                           & i\delta^a{}_{\myff q}       & 0                          & 0                        & 0                           & 0                           \\
 0                          & \delta_c{}^{\myff c}       & \delta_c{}^{\myff p}        & 0                           & 0                          & 0                        & 0                           &                             \\
 0                          & 0                          & 0                           & 0                           & \delta_{a_1}{}^{\myff a_1} & 0                          & 0                           & i\delta_{a_1}{}^{\myff p_1} \\
 0                          & 0                          & 0                           & 0                           & 0                          & \delta^{c_1}{}_{\myff c_1} & \delta^{c_1}{}_{\myff q_1}  & 0
\end{array}
\right)}_{:=(\tilde{\tilde m}_2)^A{}_{\myff A}}\cdot
$$
$$
\underbrace{\left(
\begin{array}{cccccccc}
 \eta_\alpha{}^{\myff a\myff b}         & \alpha\delta^{\myff a}{}_{\myff d}     & \gamma\delta^{\myff a}{}_{\myff k}     & 0                                      & \phi_1\delta{}^{\myff a}{}_{\myff b_1}  & 0                                      & 0                                      & 0                                       \\
 \beta\delta_{\myff c}{}^{\myff b}      &-(\eta^T)_\alpha{}_{\myff c\myff d}     & 0                                      & \gamma\delta_{\myff c}{}^{\myff l}     & 0                                       & \phi_1\delta{}_{\myff c}{}^{\myff d_1} & 0                                      & 0                                       \\
 \delta\delta_{\myff p}{}^{\myff b}     & 0                                      &-(\eta^T)_\alpha{}_{\myff p\myff k}     &-\alpha\delta_{\myff p}{}^{\myff l}     & 0                                       & 0                                      & \phi_1\delta{}_{\myff p}{}^{\myff k_1} & 0                                       \\
 0                                      & \delta\delta^{\myff q}{}_{\myff d}     &-\beta\delta^{\myff q}{}_{\myff k}      & \eta_\alpha{}^{\myff q\myff l}         & 0                                       & 0                                      & 0                                      & \phi_1\delta{}^{\myff q}{}_{\myff l_1}  \\
 \psi_1\delta{}_{\myff a_1}{}^{\myff b} & 0                                      & 0                                      & 0                                      &-(\eta^T)_\alpha{}_{\myff a_1\myff b_1}  &-\alpha\delta_{\myff a_1}{}^{\myff d_1} &-\gamma\delta_{\myff a_1}{}^{\myff k_1} & 0                                       \\
\end{array}
\right)}_{\eta_\myf\Lambda{}^{\myff A\myff B}}
$$
\begin{equation}
\cdot\underbrace{\frac{1}{\sqrt 2}\left(
\begin{array}{cccc}
 \delta_{\myff b}{}^b       & 0                        & 0                            & 0                          \\
 0                          & \delta^{\myff d}{}_d     & 0                            & 0                          \\
 0                          & \delta^{\myff k}{}_d     & 0                            & 0                          \\
-i\delta_{\myff l}{}^b      & 0                        & 0                            & 0                          \\
 0                          & 0                        & \delta^{\myff b_1}{}_{d_2}   & 0                          \\
 0                          & 0                        & 0                            & \delta^{\myff d_1}{}_{b_2} \\
 0                          & 0                        & 0                            & \delta^{\myff k_1}{}_{b_2} \\
 0                          & 0                        & -i\delta^{\myff l_1}{}_{d_2} & 0
\end{array}
\right)}_{:=(\tilde m^T_2){}_{\myff B}{}^B}\spsd
\end{equation}
If we now use the operators $m_\Lambda{}^\myf\Lambda$ translating the subspace $\mathbb R^4_{(2,2)}$ in the subspace  $\mathbb C^2$ and leaving unchanged the subspace $\mathbb C^2\oplus \mathbb R^{2n-8}_{(n-4,n-4)}$  then we can construct the connecting operators  $\eta_\Lambda{}^{ab}$ of the corresponding space $\mathbb C^4\oplus \mathbb R^{2n-8}_{(n-4,n-4)}$, while the dimension of the spinor space will be reduced by 2 times. \\
3. Proceeding similarly, we can obtain the space $\mathbb C^{n-2}\oplus \mathbb R^{4}_{(2,2)}$ and the corresponding connecting operators. It is obvious that at each step, only those components of the connecting operators which responsible for the transition from the real representation of the subspace $\mathbb R^4_{(2,2)}$ to its complex analog $\mathbb C^2$ are modified. \\
4. The final step is carried out using the operators discussed above in Example \ref{ex2.2}. \\
\end{algorithm}

\begin{corollary}
\label{c6.1}
The connecting operators from Definition \ref{d6.1} satisfy the identity (the proof is given in Appendix)
\begin{equation}
\label{e6.4}
\eta_\Lambda{}^{AB}\eta_\Psi{}_{AD}\eta^\Omega{}^{CD}\eta^\Theta{}_{CB}=
\frac{N}{4}(g_{\Lambda\Psi}g^{\Omega\Theta}+\delta_\Lambda{}^\Theta\delta_\Psi{}^\Omega-\delta_\Psi{}^\Theta\delta_\Lambda{}^\Omega)\spsd
\end{equation}
\end{corollary}

\begin{corollary}
\label{c6.2}
For the orthogonal transformation (\ref{e3.15}), there are the two variants of a decomposition for each case:
\begin{enumerate}
\renewcommand{\theenumi}{\Roman{enumi})}
\item Special rotation.
\begin{equation}
S_\Lambda{}^\Psi\eta_\Psi{}^{AB}=\eta_\Lambda{}^{CD}\tilde S_C{}^A\tilde{\tilde S}_D{}^B\spsd
\end{equation}
\begin{enumerate}[1.]
\item n mod 4 =2\\
\begin{equation}
\tilde{\tilde S}_A{}^B=-E_A{}^C\tilde S_C{}^DE_D{}^B\sps
\tilde{\tilde S}_A{}^B=-\tilde E_A{}^C\tilde S_C{}^D\tilde E_D{}^B\spsd
\end{equation}
\item n mod 4 =0\\
\begin{equation}
\begin{array}{c}
\tilde{\tilde S}_A{}^B\tilde{\tilde S}_C{}^D\varepsilon_{BD}=\varepsilon_{AC}\sps
\tilde S_A{}^B\tilde S_C{}^D\varepsilon_{BD}=\varepsilon_{AC}\sps\\
\tilde{\tilde S}_A{}^B\tilde{\tilde S}_C{}^D\tilde\varepsilon_{BD}=\tilde\varepsilon_{AC}\sps
\tilde S_A{}^B\tilde S_C{}^D\tilde\varepsilon_{BD}=\tilde\varepsilon_{AC}\spsd
\end{array}
\end{equation}
\end{enumerate}
\item Non-special rotation.
\begin{equation}
S_\Lambda{}^\Psi\eta_\Psi{}^{BA}=\eta_\Lambda{}_{CD}\tilde S^{CA}\tilde{\tilde S}^{DB}\spsd
\end{equation}
\begin{enumerate}[1.]
\item n mod 4 =2\\
\begin{equation}
\tilde{\tilde S}^{CA}=-E_K{}^A\tilde S^{KL}E_L{}^C\sps
\tilde{\tilde S}^{CA}=\tilde E_K{}^A\tilde S^{KL}\tilde E_L{}^C\spsd
\end{equation}
\item n mod 4 =0\\
\begin{equation}
\begin{array}{c}
\tilde{\tilde S}^{AB}\tilde{\tilde S}^{CD}\varepsilon_{BD}=\varepsilon^{AC}\sps
\tilde S^{AB}\tilde S^{CD}\varepsilon_{BD}=\varepsilon^{AC}\sps\\
\tilde{\tilde S}^{AB}\tilde{\tilde S}^{CD}\tilde\varepsilon_{BD}=\tilde\varepsilon^{AC}\sps
\tilde S^{AB}\tilde S^{CD}\tilde\varepsilon_{BD}=\tilde\varepsilon^{AC}\spsd
\end{array}
\end{equation}
\end{enumerate}
\end{enumerate}
\end{corollary}

\begin{corollary}
\label{c6.3}
For the involution (\ref{e3.16}), there are the two variants of a decomposition:
\begin{enumerate}
\renewcommand{\theenumi}{\Roman{enumi})}
\item Analog of the special rotation.
\begin{equation}
S_\Lambda{}^{\Psi '}\bar\eta_{\Psi '}{}^{A'B'}=\eta_\Lambda{}^{CD}\tilde S_C{}^{A'}\tilde{\tilde S}_D{}^{B '}\spsd
\end{equation}
\begin{equation}
\tilde{\tilde S}_A{}^{B'}\bar{\tilde{\tilde S}}_{B'}{}^D=\pm\delta_A{}^D\sps
\tilde S_A{}^{B'}\bar{\tilde S}_{B'}{}^D=\pm\delta_A{}^D\spsd
\end{equation}
\item Analog of the non-special rotation.
\begin{equation}
S_\Lambda{}^{\Psi'}\bar\eta_{\Psi'}{}^{B'A'}=\eta_\Lambda{}_{CD}\tilde S^{CA'}\tilde{\tilde S}^{DB'}\spsd\\
\end{equation}
\begin{equation}
\tilde{\tilde S}^{AB'}=\pm\bar{\tilde{\tilde S}}{}^{B'A}\sps
\tilde S{}^{AB'}=\pm\bar{\tilde S}{}^{B'A}\spsd
\end{equation}
\end{enumerate}
\end{corollary}

\begin{table}[p]
\caption{The matrix form of the tensor S for the real inclusions of some 2-dimensional spaces in the complex space for $r^i=(T,Z)$.}
\label{t6.1}
\begin{center}
\begin{tabular}{|c|c|c|c|c|c|c|}
      \hline
      N & Space & Form of S &
      S & Isomorphism & Form of $\sqrt{2}r^i\eta_i$ & Form of $\sqrt{2}r^i\sigma_i$\\
      \hline
      1a  &
      $
      \begin{array}{c}
      R^2   \\
      \ll ++ \gg
      \end{array}
      $  &
      $S_{AA'}$ &
      $\left(
      \begin{array}{c}
      1
      \end{array}
      \right)
      $&
      $\small{U(1) \cong SO^+(2)}
      $&
      $T+iZ
      $&
      $T-iZ
      $\\
      \hline
      1b  &
      $
      \begin{array}{c}
      R^2   \\
      \ll -- \gg
      \end{array}
      $  &
      $S_{AA'}$ &
      $\left(
      \begin{array}{c}
      i
      \end{array}
      \right)
      $&
      $\small{U(1) \cong SO^+(2)}
      $&
      $-iT-Z
      $&
      $-iT+Z
      $\\
      \hline
      2a  &
      $
      \begin{array}{c}
      R^2_{(1,1)}   \\
      \ll +- \gg
      \end{array}
      $  &
      $S_A{}^{A'}$ &
      $\left(
      \begin{array}{cccc}
      1
      \end{array}
      \right)
      $&
      $\small{R_+ \cong SO^+(1,1)}
      $&
      $T+Z
      $&
      $T-Z
      $\\
      \hline
      2b  &
      $
      \begin{array}{c}
      R^2_{(1,1)}   \\
      \ll -+ \gg
      \end{array}
      $  &
      $S_A{}^{A'}$ &
      $\left(
      \begin{array}{cccc}
      1
      \end{array}
      \right)
      $&
      $\small{R_+ \cong SO^+(1,1)}
      $&
      $-T-Z
      $&
      $T-Z
      $\\
      \hline
\end{tabular}
\end{center}
\end{table}

\begin{table}[p]
\caption{The matrix form of the tensor S for the real inclusions of some 4-dimensional spaces in the complex space for $r^i=(T,X,Y,Z)$.}
\label{t6.2}
\begin{center}
\begin{tabular}{|c|c|c|c|c|}
       \hline
       N & Space & S &  Form of $\sqrt{2}r^i\eta_i$/$\sqrt{2}r^i\sigma_i$\\
       \hline
       1a  &
       $
       \begin{array}{c}
       R^4   \\
       \ll ---- \gg\\
       {}_{SU(2)\times SU(2)/\pm 1 \cong SO^+(4)}
       \end{array}
       $  &
       $\begin{array}{c}
       \tilde S_A{}^{A'}=\tilde{\tilde S}_A{}^{A'}\\
       \left(
       \begin{array}{cc}
        0 & 1 \\
       -1 & 0
       \end{array}
       \right)
       \end{array}
       $&
       $\begin{array}{c}
       \left(
       \begin{array}{cc}
       {}_{-iT-Z}  & {}_{iX+Y} \\
       {}_{iX-Y}   & {}_{iT-Z}
       \end{array}
       \right)
       \\
       \left(
       \begin{array}{cc}
       {}_{-iT+Z} &   {}_{iX+Y} \\
       {}_{iX-Y}  &   {}_{iT+Z}
       \end{array}
       \right)
       \end{array}
       $\\
       \hline
       1b  &
       $
       \begin{array}{c}
       R^4   \\
       \ll ++++ \gg\\
       {}_{SU(2)\times SU(2)/\pm 1 \cong SO^+(4)}
       \end{array}
       $  &
       $\begin{array}{c}
       \tilde S_A{}^{A'}=\tilde{\tilde S}_A{}^{A'}\\
       \left(
       \begin{array}{cc}
        0 & i \\
       -i & 0
       \end{array}
       \right)
       \end{array}
       $&
       $\begin{array}{c}
       \left(
       \begin{array}{cc}
       {}_{T+iZ}   & {}_{iY+X} \\
       {}_{-iY+X}   & {}_{-T+iZ}
       \end{array}
       \right)\\
       \left(
       \begin{array}{cc}
       {}_{T-iZ} &   {}_{iY+X} \\
       {}_{-iY+X} &   {}_{-T-iZ}
       \end{array}
       \right)
       \end{array}
       $\\
\hline
\end{tabular}
\end{center}
\end{table}

\renewcommand{\multirowsetup}{\centering}
\begin{table}[p]
\caption{The matrix form of the tensor S for the real inclusions of some 6-dimensional spaces in the complex space for $r^i=(T,X,V,W,Y,Z)$.}
\label{t6.3}
\begin{center}
\begin{tabular}{|c|c|c|c|}
      \hline
      \multirow{3}{5mm}{N} &\multirow{3}{5mm}{Space} & \multirow{3}{14mm}{Form of S} &
      Isomorphism \\\cline{4-4} &&& Form of $\sqrt{2}r^i\eta_i$\\\cline{4-4} &&& Form of $\sqrt {2}r^i\sigma_i$\\
      \hline
      \hline
      \multirow{3}{14mm}{\\[8ex]1}&
      \multirow{3}{14mm}{\\[8ex]$R^6$}&
      $S_{AA'}$ & $SU(4)/\pm 1 \cong SO^+(6)$\\\cline{3-4}&&
      \multirow{2}{35mm}
      {\\$
      \left(
      \begin{array}{cccc}
      _0    & _0   &   _0 & _1\\
      _0    & _0   &_{-1} & _0\\
      _0    &_1    &   _0 & _0\\
      _{-1} & _0   &   _0 & _0
      \end{array}
      \right)
      $}&
      $\left(
      \begin{array}{cccc}
      _{T+iZ}  &  _{X+iY} & _{V+iW}    &_{0}       \\
      _{X-iY}  &  _{-T+iZ}& _{0}       &_{V+iW}    \\
      _{V-iW}  &  _{0   } & _{-T+iZ}   &_{-X-iY} \\
      _{0}     &  _{V-iW} & _{-X+iY}   &_{T+iZ}    \\
      \end{array}
      \right)$
      \\\cline{4-4}&&&
      $\left(
      \begin{array}{cccc}
      _{T-iZ}     & _{X+iY}    & _{V+iW}   & _{0}       \\
      _{X-iY}     & _{-T-iZ}   & _{0}      & _{V+iW}   \\
      _{V-iW}     & _{0}       & _{-T-iZ}  & _{-X-iY}   \\
      _{0}        & _{V-iW}    & _{-X+iY}  & _{T-iZ}    \\
      \end{array}
      \right)
      $\\
      \hline
      \hline
      \multirow{3}{14mm}{\\[8ex]2}&
      \multirow{3}{14mm}{\\[8ex]$R^6_{(1,5)}$}&
      $S_A{}^{A'}$ & $SL(2,\mathbb H)/\pm 1 \cong SO^+(1,5)$\\
\hline
      \hline
\end{tabular}
\end{center}
\end{table}

We assume that the operators $\eta_\Lambda{}^{AB}(\sigma_\Lambda{}_{AB})$ are constructed either by the scheme (\ref{e6.6})($\alpha=\overline{1,n-2}$)
\begin{equation}
\label{e6.7}
%
\end{equation}
or by the scheme
\begin{equation}
\label{e6.7a}
\tag{\ref{e6.7}$'$}
%
\end{equation}
This means that from the previous dimension, the conjugating spin-tensors(\ref{e3.16})
\begin{equation}
\label{e6.8}
%
\right)\sps
\end{array}
\end{equation}
defined to within a sign, are inherited.

\begin{example}
In the table \ref{t6.1}, the various options for the real inclusion for n=2 are given (see Example \ref{ex3.1} and Example \ref{ex3.2}).
\end{example}

\begin{example}
In the table \ref{t6.2}, the various options for the real inclusion for n=4 are given (see Example \ref{ex3.3}).
\end{example}

\begin{example}
In the table \ref{t6.3}, the various options for the real inclusion for n=6 are given. See Example \ref{ex6.2} in the case, when $\tilde\eta_\Lambda{}^{AC}:=\eta_\Lambda{}^{AB}\varepsilon_B{}^C$ is an antisymmetric tensor.
Recall that $\eta_i:=H_i{}^\Lambda\eta_\Lambda$ according to (\ref{e5.2}).
\end{example}

\section{On the structural constants of the non-division algebra. Sedenions}
\Abstract{
$\phantom{ff}$In this section it is told about how on the connecting operators which satisfy the Clifford equation to construct the structural constants of the sedenion algebra. Such the algebras are constructed by the inductive transition for n mod 8 = 0 based on the Bott periodicity. We consider the axioms of such the algebras. The conclusion of all results of this section is made on the basis \cite{Baez1}, \cite[pp. 192-244 (rus)]{Husemoller1}.
}

We use the fact that for n mod 8 = 0, there are the connecting operators $\eta_\Lambda{}^{AB}$, the inductive construction of which is not the doubling procedure, but is based on the Bott periodicity. This is means by Definition \ref{d6.1} that
\begin{equation}
\eta_\Lambda{}^{AB}+\eta_\Lambda{}^{BA}=\eta_\Lambda\varepsilon^{AB}\sps \eta_\Lambda:=(\eta_{n})_\Lambda
\end{equation}
and $\varepsilon^{AB}$ is a symmetric metric tensor in the spinor space. This will give the opportunity to set
\begin{equation}
\label{e13.0}
\begin{array}{cccc}
I).  & P_\Lambda{}^A:=\eta_\Lambda{}^{BA}X_B\sps P_\Lambda{}^AP_\Psi{}_A=g_{\Lambda\Psi}\sps &
II). & P_\Lambda{}^A:=\eta_\Lambda{}^{AB}X_B\sps P_\Lambda{}^AP_\Psi{}_A=g_{\Lambda\Psi}
\end{array}
\end{equation}
for some $X_B\ (X^AX_A=2)$. Then we can define some structural constants as follows
\begin{equation}
\begin{array}{cccc}
I).  & \eta_{\Lambda\Psi}{}^\Theta:=\sqrt{2}\eta_\Lambda{}^{AB}P_\Psi{}_AP^\Theta{}_B\sps
II). & \eta_{\Lambda\Psi}{}^\Theta:=\sqrt{2}\eta_\Lambda{}^{BA}P_\Psi{}_AP^\Theta{}_B\spsd
\end{array}
\end{equation}
Both variants are identical. Therefore, we will further consider Case I). Will they define \emph{group algebra}? The answer is not quite unequivocally defined, and that's why.

1. \emph{First axiom of multiplication} is executed. For any two elements $(r_1)^\Lambda\sps (r_2)^\Psi$, it can always only be one way to construct their \emph{product}
\begin{equation}
\label{e13.1}
(r_1,r_2)^\Theta:=(r_1)^\Lambda(r_2)^\Psi\eta_{\Lambda\Psi}{}^\Theta\spsd
\end{equation}

2. \emph{Second axiom of multiplication} is executed. There is uniquely defined \emph{identity element for multiplication} $e:=\frac{1}{\sqrt{2}}\eta^\Lambda$.
\begin{equation}
\label{e13.2}
\begin{array}{c}
\frac{1}{\sqrt{2}}\eta^\Lambda\eta_{\Lambda\Psi}{}^\Theta=\varepsilon^{AB}P_\Psi{}_AP^\Theta{}_B=\delta_\Psi{}^\Theta\sps\\
\frac{1}{\sqrt{2}}\eta^\Psi\eta_{\Lambda\Psi}{}^\Theta=\eta_\Lambda{}^{AB}\varepsilon_{XA}X^XP^\Theta{}_B=P_\Lambda{}^BP^\Theta{}_B=\delta_\Lambda{}^\Theta\spsd
\end{array}
\end{equation}

3. For any non-isotropic element, there is uniquely defined \emph{inverse element for multiplication}. For the inclusion $\mathbb R^n\subset \mathbb C^n\ (n\ mod\ 8=0)$, \emph{Third axiom of multiplication} is executed for any nonzero element. Accordingly, with the help of the inclusion operators $H_i{}^\Lambda$, the structure constants $\eta_{ij}{}^k$,  which are real not for all spinors $X^A$, are defined.
\begin{equation}
\label{e13.3}
r^{-1}:=\frac{1}{<r,r>}(2<r,e>e-r)\sps <r_1,r_2>:=g_{\Lambda\Psi}(r_1)^\Lambda(r_2)^\Psi\spsd
\end{equation}
\begin{equation}
\begin{array}{c}
(r^{-1})^\Psi=\frac{1}{r^\Phi r_\Phi}r^\Omega(\eta_\Omega\eta^\Psi-\delta_\Omega{}^\Psi)\sps\\
r^\Lambda \frac{1}{r^\Phi r_\Phi}r^\Omega(\eta_\Omega\eta^\Psi-\delta_\Omega{}^\Psi)\eta_{\Lambda\Psi}{}^\Theta=
\frac{\sqrt{2}}{r^\Phi r_\Phi}R^{AB}r^\Omega(\eta_\Omega\eta^\Psi-\delta_\Omega{}^\Psi)\eta_\Psi{}_{XA}X^XP^\Theta{}_B=\\
=\frac{\sqrt{2}}{r^\Phi r_\Phi}R^{AB}R_{AX}X^XP^\Theta{}_B=\frac{1}{\sqrt{2}}X^BP^\Theta{}_B=\frac{1}{\sqrt{2}}\eta^\Theta\sps
\\
\frac{1}{r^\Phi r_\Phi}r^\Omega(\eta_\Omega\eta^\Lambda-\delta_\Omega{}^\Lambda)r^\Psi \eta_{\Lambda\Psi}{}^\Theta=
\frac{\sqrt{2}}{r^\Phi r_\Phi}R^{BA}R_{XA}X^XP^\Theta{}_B=\frac{1}{\sqrt{2}}\eta^\Theta\spsd
\end{array}
\end{equation}
These identities, executed for any non-isotropic r, are equivalent to the relation
\begin{equation}
\eta_{\Lambda\Psi}{}^\Omega+\eta_\Lambda{}^\Omega{}_\Psi=\sqrt{2}\eta_\Lambda\delta_{\Psi}{}^\Omega\spsd
\end{equation}

4. \emph{Commutative axiom of multiplication} is not executed. However, there is the identity
\begin{equation}
\label{e13.4}
\begin{array}{c}
\frac{1}{2}((r_1,r_2)+(r_2,r_1))=-<r_1,r_2>e+<r_1,e>r_2+<r_2,e>r_1\spsd
\end{array}
\end{equation}
\begin{equation}
\begin{array}{c}
\eta_{(\Lambda\Psi)}{}^\Theta=
\sqrt{2}\eta_{\left(\right.\Lambda}{}^{AB}P_{\Psi\left.\right)A}P^\Theta{}_B=
\sqrt{2}(-\eta_{\left(\right.\Lambda}{}^{AB}\eta_{\Psi\left.\right)AX}+2\eta_{\left(\right.\Lambda}{}^{AB}\eta_{\Psi\left.\right)(AX)})X^XP^\Theta{}_B=\\
=-g_{\Lambda\Psi}\big(\frac{1}{\sqrt{2}}\eta^\Theta\big)+2\big(\frac{1}{\sqrt{2}}\eta_{\left(\right.\Psi}\big)\delta_{\Lambda\left.\right)}{}^\Theta\spsd
\end{array}
\end{equation}
The special case of this identity
\begin{equation}
\label{e13.4a}
\tag{\ref{e13.4}$'$}
r^2=-<r,r>e+2<r,e>r
\end{equation}
leads to (\ref{e13.3}) and means that there is \emph{power-associativity}. Therefore, \emph{Fourth axiom of multiplication} is not executed. But a replacement can be found.

5. \emph{Associative axiom of multiplication} is not executed. However, there is \emph{alternative-elastic identity}
\begin{equation}
\label{e13.5}
\begin{array}{c}
((r_1,r_1),r_2)-(r_1,(r_1,r_2))-(r_1,(r_2,r_1))=((r_2,(r_1,r_1))-((r_2,r_1),r_1)-((r_1,r_2),r_1)\spsd
\end{array}
\end{equation}
\begin{equation}
\begin{array}{c}
\label{e13.5/1}
((r_1,r_1),r_2)-(r_1,(r_1,r_2))-(r_1,(r_2,r_1))=
-<r_1,r_1>r_2+2<r_1,e>(r_1,r_2)-\\
-2(r_1,-<r_1,r_2>e+<r_1,e>r_2+<r_2,e>r_1)=\\=
-<r_1,r_1>r_2+2<r_1,r_2>r_1-2<r_2,e>(-<r_1,r_1>e+2<r_1,e>r_1)=\\
=2<r_2,e><r_1,r_1>e+(2<r_1,r_2>-4<r_2,e><r_1,e>)r_1-<r_1,r_1>r_2\sps\\ \\
(r_2,(r_1,r_1))-((r_2,r_1),r_1))-((r_1,r_2),r_1)=
-<r_1,r_1>r_2+2<r_1,e>(r_2,r_1)-\\
-2(-<r_1,r_2>e+<r_1,e>r_2+<r_2,e>r_1,r_1)=\\=
-<r_1,r_1>r_2+2<r_1,r_2>r_1-2<r_2,e>(-<r_1,r_1>e+2<r_1,e>r_1)=\\
=2<r_2,e><r_1,r_1>e+(2<r_1,r_2>-4<r_2,e><r_1,e>)r_1-<r_1,r_1>r_2\sps\\ \\
\end{array}
\end{equation}
Therefore, \emph{Fifth axiom of multiplication} is not executed. But a replacement can be found.

4${}^*$. The Clifford equation gives another identity: \emph{flexible identity} \cite{Albert1} (according to Definitions \ref{d6.1}-\ref{d6.2}, $\varepsilon^{[A(BC)D]}=0$)
\begin{equation}
\tag{\ref{e13.5}$'$}
(r_1r_2)r_1=r_1(r_2r_1)\spsd
\end{equation}
\begin{equation}
\tag{\ref{e13.5/1}$'$}
\begin{array}{c}
\frac{1}{2}\eta_{\left(\right.\Lambda|\Psi}{}^\Theta\eta_{\Theta|\Omega\left.\right)}{}^\Gamma=
\eta_{\left(\right.\Lambda}{}^{AB}P_{\Omega\left.\right) C}\eta_\Theta{}^{CD}P_{\Psi A}P^\Theta{}_BP^\Gamma{}_D=\\
=\eta_{\left(\right.\Lambda}{}^{CD} P_{\Omega\left.\right) C}\eta_\Psi P^\Gamma{}_D-
\varepsilon^{ABCD}P_{\left(\right.\Omega |C|}P_{\Lambda\left.\right)B} P_{\Psi A} P^\Gamma{}_D=\\
=\eta_{\left(\right.\Lambda}\delta_{\Omega\left.\right)}{}^\Gamma\eta_\Psi-\frac{1}{2}g_{\Omega\Lambda}\eta_\Psi\eta_\Gamma
-\varepsilon^{ABCD}P_{\left(\right.\Omega |C|}P_{\Lambda\left.\right)B} P_{\Psi A} P^\Gamma{}_D\sps\\ \\
\frac{1}{2}\eta_{\left(\right.\Lambda|\Theta}{}^\Gamma\eta_{\Psi|\Omega\left.\right)}{}^\Theta=
\eta_{\left(\right.\Lambda}{}^{AB}P_{\Omega\left.\right) C}\eta_\Psi{}^{CD}P_{\Theta A}P^\Theta{}_DP^\Gamma{}_B=\\
=\eta_{\left(\right.\Lambda}P_{\Omega\left.\right) C}\eta_\Psi{}^{CD}P^\Gamma{}_D-
P_{\left(\right.\Lambda |D|}P_{\Omega\left.\right) C}\eta_\Psi{}^{CD}\eta^\Gamma
+\eta_\Theta{}^{BA}\eta_\Psi{}^{CD}P_{\left(\right.\Lambda |A|}P_{\Omega\left.\right) C}P^\Theta{}_DP^\Gamma{}_B=\\
=\eta_{\left(\right.\Lambda}P_{\Omega\left.\right) C}\eta_\Psi{}^{CD}P^\Gamma{}_D-
P_{\left(\right.\Lambda |D|}P_{\Omega\left.\right) C}\eta_\Psi{}^{CD}\eta^\Gamma+
\eta_\Psi{}^{BA}P_{\left(\right.\Lambda |A|}\eta_{\Omega\left.\right)}P^\Gamma{}_B-\\
-\varepsilon^{BACD}P_{\left(\right.\Lambda |A|}P_{\Omega\left.\right) C}P^\Psi{}_DP^\Gamma{}_B=\\
=\eta_{\left(\right.\Lambda}\delta_{\Omega\left.\right)}{}^\Gamma\eta_\Psi-\frac{1}{2}g_{\Omega\Lambda}\eta_\Psi\eta_\Gamma
-\varepsilon^{ABCD}P_{\left(\right.\Omega |C|}P_{\Lambda\left.\right)B} P_{\Psi A} P^\Gamma{}_D\spsd\\ \\
\end{array}
\end{equation}
From (\ref{e13.4a}), it follows that \emph{Jordan identity} $(r_1)^k(r_2r_1)=((r_1)^kr_2)r_1,\ \forall k \in \mathbb Z$ is executed too.

Such the algebra is normalized in the sense of
\begin{equation}
\tag{\ref{e13.5/1}$''$}
g_{\Theta\Gamma}\eta_{\left(\right.\Lambda|\Psi|}{}^\Theta\eta_{\Omega\left.\right)\Phi}{}^\Gamma=
g_{\Theta\Gamma}\eta^\Theta{}_{\Psi\left(\right.\Lambda}\eta^\Gamma{}_{|\Phi|\Omega\left.\right)}
\end{equation}
that coincides with the condition of \emph{normalizability} (\ref{e12.9/1}) for n = 8 .

According to the general theory, the sedenion algebra has \emph{0-divisors} for $n\ne 8$.
\begin{equation}
(r_1,r_2)=0\sps\\
(r_1)^\Lambda(r_2)^\Psi\eta_{\Lambda\Psi}{}^\Omega=0\sps
(R_1)^{AB}(R_2)_{DA}X^DP^\Omega{}_BP_\Omega{}^C=0\spsd
\end{equation}
0-divisors appear because the operator $P_\Omega{}^C$ is degenerate for $n>8$. However, there exists a sedenion basis for which
\begin{equation}
\label{e13.6}
\begin{array}{c}
(r_i,e)=(e,r_i)\sps e^2=e\sps\\
(r_i,r_j)=-(r_j,r_i)\sps r_i\ne e\sps r_j\ne e\sps i\ne j\sps\\
(r_i)^2=-e\sps r_i\ne e\spsd\\
\end{array}
\end{equation}
The algebras for which we can construct the basis (\ref{e13.6}) whose elements are not 0-divisors and satisfy \emph{alternative identities}
\begin{equation}
(r_i(r_i,r_j))=((r_i,r_i),r_j)\sps ((r_j,r_i),r_i)=(r_j,(r_i,r_i))\spsd
\end{equation}
 are the most interesting for the study. Such the algebras are called \emph{sedenion algebras}. Of course, not for all hypercomplex algebras, such the basis exists. Therefore, the hypercomplex algebras will be classified by the maximum number of basic elements, which are not 0-divisors. The inductive construction of the higher dimension hypercomplex metric group alternative-elastic algebra looks as follows.

\begin{algorithm}
\begin{enumerate}
\item Let us know antisymmetric operators $\eta_\alpha{}^{ab}$ for n mod 8 = 6. Then the connecting operators $\eta_\Lambda{}^{AB}$ for n mod 8 = 0 are constructed according to the scheme
\begin{equation}
\label{e8.7}

\end{equation}
that would provide the existence of the symmetric metric spin-tensor $\tilde\varepsilon_{AB}\equiv\varepsilon_{AB}$ for n mod 8 = 0.
\item Let us know the connecting operators $\eta_\alpha{}^{ab}$ for n mod 8 = 0. Then the connecting operators $\eta_\Lambda{}^{AB}$ for n mod 8 = 2 are constructed according to the scheme
\begin{equation}
\label{e8.8}
%
\end{equation}

\item Let us know the connecting operators $\eta_\alpha{}^{ab}$ for n mod 8 = 2. Then the connecting operators $\eta_\Lambda{}^{AB}$ for n mod 8 = 4 are constructed according to the scheme
\begin{equation}
\label{e8.9}
%
\end{equation}
providing for n mod 8 = 4 the existence of the antisymmetric metric spin-tensor
\begin{equation}
\varepsilon_{AB}:=(\varepsilon_1)_{AC}(\varepsilon_2)^{DK}(\varepsilon_3)_{BK}=
\left(
\begin{array}{cccc}
 0                     & 0                     & 0                     & i(\varepsilon_1)_{ab}\\
 0                     & 0                     &-i(\varepsilon_1)^{cd} & 0                    \\
 0                     & i(\varepsilon_1)^{kl} & 0                     & 0                    \\
-i(\varepsilon_1)_{mn} & 0                     & 0                     & 0                    \\
\end{array}
\right)\sps
\end{equation}
where $(\varepsilon_1)_{ab}$ is the symmetric metric spin-tensor for n mod 8 = 0.

\item Let us know the connecting operators $\eta_\alpha{}^{ab}$ for n mod 8 = 4. Then the connecting operators $\eta_\Lambda{}^{AB}$ for n mod 8 = 6 are constructed according to the scheme
\begin{equation}
\label{e8.10}
%
\end{equation}
Let's, in addition, spend the transformation of the connecting operators according to the scheme
\begin{equation}
\label{e8.10a}
\tag{\ref{e8.10}$'$}
\tilde\eta_\Lambda{}^{AC}:=\eta_\Lambda{}^{AB}\varepsilon_B{}^C\sps
\varepsilon_B{}^C:=
\left(
\begin{array}{ll}
                 0 &-i\varepsilon_{bm}\\
 i\varepsilon^{dk} & 0
\end{array}
\right)\sps
\end{equation}
where $\varepsilon_{ab}$ $(n/2+q=3)$ for n mod 8 = 4 will be the metric antisymmetric spin-tensor. Then for the next implementation n mod 8 = 6 the operators $\tilde\eta_\Lambda{}^{AC}=-\tilde\eta_\Lambda{}^{CA}$ will be completely antisymmetric $(n/2+q)=0$. It is easy to verify that the Clifford equation (\ref{e6.1}) is executed for such the operators. If the further construction for n mod 8 = 0 to lead on the basis of such the operators then there will exist the symmetric metric spin-tensor $\varepsilon_{AB}\equiv(\varepsilon_1)_{AB}$, $n/2+q=1$. For n mod 8 = 2, there will exist the involution at $n/2+q=2$. And again, for n mod 8 = 4, we obtain $n/2+q=3$. The circle has become isolated. The beginning has been made in Examples \ref{ex6.1}-\ref{ex6.2}.

\item Suppose there is an algebra over $\mathbb R^n$ with the structural constants generated from the connecting operators $\eta_\Lambda{}^{AB}$ with the metric spin-tensor $\varepsilon^{XZ}$ and the inclusion operator $H_i{}^\Lambda$. We assume that the metric tensor $g_{\Lambda\Psi}$ on the main diagonal contains $\ll+\gg$ only. Then we can construct the antisymmetric operators for the space $\mathbb C^{n +6}$
\begin{equation}

\end{equation}

\item The transition to the connecting operators of the space $\mathbb C^{n +8}$ is carried out as follows
\begin{equation}
%
\end{equation}
with the metric spin-trnsor
$
\varepsilon^{XZ}:=
\left(
\begin{array}{cc}
 0              &  \delta^a{}_d  \\
 \delta_c{}^b   &  0
\end{array}
\right)\sps
$
$
\varepsilon_{XZ}:=
\left(
\begin{array}{cc}
 0              &  \delta_a{}^d  \\
 \delta^c{}_b   &  0
\end{array}
\right)\spsd
$
Then we pass to the connecting operators of the space $\mathbb R^{n+8}\subset\mathbb C^{n+8}$ using the corresponding inclusion operator. And such the operators generate the structural constants of the algebra with dimension equal to $n+8$.
\end{enumerate}
\end{algorithm}

\section{Infinitesimal transformations}
\Abstract{
$\phantom{ff}$In this section, it is told about how on the connecting operators which satisfies the Clifford equation, to construct the infinitesimal transformations. The conclusion of all results of this section is made on the basis \cite[v. 1, pp. 175-177 (eng)]{Penrouz1}.
}

Using Corollary \ref{c6.2} for the special orthogonal transformations, the decomposition
\begin{equation}
\label{e8.1}
\begin{array}{ll}
1). & S_\Lambda{}^\Psi\eta_\Psi{}^{AB}=-\eta_\Lambda{}^{CD}S_C{}^A E_D{}^K S_K{}^M E_M{}^B\sps \\
2). & S_\Lambda{}^\Psi\eta_\Psi{}^{AB}=\eta_\Lambda{}^{CD}S_C{}^A (\eta_I)^\Omega S_\Omega{}^\Phi\eta_\Phi{}^{KB}\varepsilon_{KL}S_X{}^L \varepsilon^{MX}(\varepsilon_I)_{MD}\spsd
\end{array}
\end{equation}
is obtained. Suppose that there is \emph{one-parameter family of orthogonal transformations}
\begin{equation}
S_\Lambda{}^\Psi (\lambda) S_\Omega{}^\Phi (\lambda) g_{\Psi\Phi}=g_{\Lambda\Omega}\sps
S_\Lambda{}^\Psi (0):=\delta_\Lambda{}^\Psi
\end{equation}
that induces the one-parameter transformations in the spinor space
\begin{equation}
S_C{}^A(\lambda)\sps S_C{}^A(0)=\delta_C{}^A\spsd
\end{equation}
\emph{Infinitesimal transformation} is defined as
\begin{equation}
\begin{array}{ll}
T_\Lambda{}^\Psi=\left.\left[\frac{d}{d\lambda} S_\Lambda{}^\Psi(\lambda)\right]\right|_{\lambda =0}\sps &
\begin{array}{lll}
&& T_{\Lambda\Psi}=-T_{\Psi\Lambda}\sps
\end{array}\\
T_C{}^A=\left.\left[\frac{d}{d\lambda}S_C{}^A(\lambda)\right]\right|_{\lambda =0}\sps &
\begin{array}{ll}
1).& T_C{}^A=-E_C{}^B T_B{}^D E_D{}^A\sps\\
2).& T_C{}^A=-T^A{}_C\spsd\\
\end{array}
\end{array}
\end{equation}
Differentiating in zero  (\ref{e8.1}), we will obtain the identity (the calculations are given in Appendix)
\begin{equation}
\label{e8.2}
T_\Lambda{}^\Psi\eta_\Psi{}^{AB}=\eta_\Lambda{}^{CB}T_C{}^A-\eta_\Lambda{}^{AD}(\varepsilon_I){}^{KB}T_K{}^M(\varepsilon_I){}_{MD}+
\eta_\Lambda{}^{AD}(\varepsilon_I){}_{MD}(\eta_I)^\Omega T_\Omega{}^\Phi\eta_\Phi{}^{MB}\spsd
\end{equation}
\emph{Particular solution} of this equation will have the form
\begin{equation}
\label{e8.3}
T_C{}^A=\frac{1}{2}T^{\Theta\Phi}\eta_\Phi{}^{AB}\eta_\Theta{}_{CB}\sps
\end{equation}
that is verified by the direct substitution. \emph{Homogeneous solution} of the equation
\begin{equation}
\eta_\Lambda{}^{CB}T_C{}^A-\eta_\Lambda{}^{AD}(\varepsilon_I){}^{KB}T_K{}^M(\varepsilon_I){}_{MD}=0
\end{equation}
is rewritten as (\ref{e6.3/1}) that means triviality of such the solution.

\section{Complex and real connections}
\Abstract{
$\phantom{ff}$In this section, it is told about how using some particular solutions of the Clifford equation, to construct a prolongation of the Riemannian torsion-free connection into the spinor bundle. We consider the two different variants. This will give the opportunity to construct the spinor analogues of the Lie operators. The conclusion of all results of this section is made on the basis \cite{Bilyalov1}, \cite[v. 1, pp. 228-231 (eng)]{Penrouz1}.
}

For the real representation of the tangent bundle $\tau^\mathbb R (V_{2n})$, the equation of the geodesics take the form
\begin{equation}
\frac{d^2x^\myf\Omega}{ds^2}+\Gamma_{\myf\Phi\myf\Psi}{}^\myf\Omega\ \frac{dx^\myf\Phi}{ds}\frac{dx^\myf\Psi}{ds}=0\spsd
\end{equation}
For the parameterization of the atlas $w^\Theta=\frac{1}{\sqrt{2}}(u^\Theta(x^\myf\Theta)+iv^\Theta(x^\myf\Theta))$, we obtain
\begin{equation}
\begin{array}{c}
\frac{d^2x^\myf\Omega}{ds^2}=\frac{d}{ds}(\frac{\partial x^\myf\Omega}{\partial w^\Theta}\ \frac{dw^\Theta}{ds}+
\frac{\partial x^\myf\Omega}{\partial w^{\Theta '}}\ \frac{dw^{\Theta '}}{ds})=\\
\frac{\partial ^2x^\myf\Omega}{\partial w^\Theta ds}\ \frac{dw^\Theta}{ds}+
\frac{\partial x^\myf\Omega}{\partial w^\Theta}\ \frac{d^2w^\Theta}{ds^2}+
\frac{\partial ^2x^\myf\Omega}{\partial w^{\Theta '} ds}\ \frac{dw^{\Theta '}}{ds}+
\frac{\partial x^\myf\Omega}{\partial w^{\Theta '}}\ \frac{d^2w^{\Theta '}}{ds^2}\sps\\
\frac{d^2w^\Theta}{ds^2}\ \frac{\partial x^\myf\Omega}{\partial w^\Theta}+
\frac{d^2w^{\Theta '}}{ds^2}\ \frac{\partial x^\myf\Omega}{\partial w^{\Theta'}}+
\frac{\partial ^2x^\myf\Omega}{\partial w^\Theta ds}\ \frac{dw^\Theta}{ds}+
\frac{\partial ^2x^\myf\Omega}{\partial w^{\Theta '} ds}\ \frac{dw^{\Theta '}}{ds}+\\
+\Gamma_{\myf\Phi\myf\Psi}{}^\myf\Omega\frac{dx^\myf\Phi}{ds}\ \frac{\partial x^\myf\Psi}{\partial w^\Theta}\ \frac{dw^\Theta}{ds}+
\Gamma_{\myf\Phi\myf\Psi}{}^\myf\Omega\frac{dx^\myf\Phi}{ds}\ \frac{\partial x^\myf\Psi}{\partial w^{\Theta '}}\ \frac{dw^{\Theta '}}{ds}=0\spsd
\end{array}
\end{equation}
We multiply the both sides by $\frac{dw^\Lambda}{dx^\myf\Omega}=m^\Lambda{}_\myf\Omega$. Considering the identities (\ref{e1.2}) for the operators $m^\Lambda{}_\myf\Omega$ and $m^{\Lambda '}{}_\myf\Omega$, we will obtain
\begin{equation}
\begin{array}{c}
\frac{d^2w^\Lambda}{ds^2}+(\frac{\partial w^\Lambda}{\partial x^\myf\Omega}\ \frac{\partial^2x^\myf\Omega}{\partial w^\Theta ds}\ \frac{dw^\Theta}{ds}+
\Gamma_{\myf\Phi\myf\Psi}{}^\myf\Omega \frac{\partial x^\myf\Psi}{\partial w^\Theta}\frac{dx^\myf\Phi}{ds}\ \frac{dw^\Theta}{ds}\ \frac{\partial w^\Lambda}{\partial x^\myf\Omega})+\\
+(\frac{\partial w^\Lambda}{\partial x^\myf\Omega}\ \frac{\partial^2x^\myf\Omega}{\partial w^{\Theta '} ds}\ \frac{dw^{\Theta '}}{ds}+
\Gamma_{\myf\Phi\myf\Psi}{}^\myf\Omega \frac{\partial x^\myf\Psi}{\partial w^{\Theta '}}\frac{dx^\myf\Phi}{ds}\ \frac{dw^{\Theta '}}{ds}
\frac{\partial w^\Lambda}{\partial x^\myf\Omega})=0\spsd
\end{array}
\end{equation}
For the complex representation of the tangent bundle $\tau^\mathbb C(V_{2n})$, the geodesic equations take the form
\begin{equation}
\frac{d^2w^\Lambda}{ds^2}+(\Gamma_{\myf\Phi\Theta}{}^\Lambda\ \frac{dw^\Theta}{ds}+
\Gamma_{\myf\Phi\Theta '}{}^\Lambda\ \frac{dw^{\Theta '}}{ds})\ \frac{dx^\myf\Phi}{ds}=0\spsd
\end{equation}
Comparing these two equations with Definition \ref{e1.2} of the operators $m^\Lambda{}_\myf\Omega$, $m_\Lambda{}^\myf\Omega$, we obtain
\begin{equation}
\left\{
\begin{array}{l}
m^\Lambda{}_\myf\Omega\ (\partial_{\myf\Phi}m_\Theta{}^\myf\Omega+\Gamma_{\myf\Phi\myf\Psi}{}^\myf\Omega m_\Theta{}^\myf\Psi)=\Gamma_{\myf\Phi\Theta}{}^\Lambda\sps\\
m^\Lambda{}_\myf\Omega\ (\partial_{\myf\Phi}\bar m_{\Lambda '}{}^\myf\Omega+\Gamma_{\myf\Phi\myf\Psi}{}^\myf\Omega \bar m_{\Theta '}{}^\myf\Psi)=\Gamma_{\myf\Phi\Theta '}{}^\Lambda
\end{array}
\right.
\end{equation}
taking into account the identity
\begin{equation}
\frac{\partial^2x^\myf\Omega}{\partial w^\Theta ds}=\frac{\partial^2x^\myf\Omega}{\partial w^\Theta  \partial x^\myf\Phi}\frac{dx^\myf\Phi}{ds}=
\frac{dx^\myf\Phi}{ds}\partial_\myf\Phi\frac{\partial x^\myf\Omega}{\partial w^\Theta}=\frac{dx^\myf\Phi}{ds}\partial_\myf\Phi m_\Theta{}^\myf\Omega\spsd
\end{equation}
Other record of the system has the form
\begin{equation}
%
\end{equation}
Let's add these two equations
\begin{equation}
\partial_\myf\Phi m^\Lambda{}_\myf\Lambda-\Gamma_{\myf\Phi\myf\Lambda}{}^\myf\Omega m^\Lambda{}_\myf\Omega+
\Gamma_{\myf\Phi\Xi}{}^\Lambda m^\Xi{}_\myf\Lambda+\Gamma_{\myf\Phi\Xi '}{}^\Lambda \bar m^{\Xi '}{}_\myf\Lambda=0\spsd
\end{equation}
We will contract it with $\bar m_{\Psi '}{}^\myf\Lambda$
\begin{equation}
\begin{array}{c}
\bar m_{\Psi '}{}^\myf\Lambda\partial_\myf\Phi m^\Lambda{}_\myf\Lambda-
\Gamma_{\myf\Phi\myf\Lambda}{}^\myf\Omega m^\Lambda{}_\myf\Omega \bar m_{\Psi '}{}^\myf\Lambda+\Gamma_{\myf\Phi\Psi '}{}^\Lambda=0\sps\\
\Gamma_{\myf\Phi\Psi '}{}^\Lambda=\Gamma_{\myf\Phi\myf\Lambda}{}^\myf\Omega m^\Lambda{}_\myf\Omega \bar m_{\Psi '}{}^\myf\Lambda-
    \frac{i}{2}\bar m_{\Psi '}{}^\myf\Lambda m^\Lambda{}_\myf\Theta\partial_\myf\Phi f_\myf\Lambda{}^\myf\Theta\spsd
\end{array}
\end{equation}
Therefore, the connection in the tangent bundle should be predetermined by the system
\begin{equation}
\left\{
\begin{array}{l}
\nabla_{\myf\Phi}m^\Lambda{}_\myf\Lambda=0\sps\\[2ex]
\nabla_{\myf\Phi}\bar m^{\Lambda '}{}_\myf\Lambda=0\spsd
\end{array}
\right.
\end{equation}
Indeed, since $m^\Psi{}_\myf\Omega=i\ f_\myf\Omega{}^\myf\Lambda m^\Psi{}_\myf\Lambda$ then
\begin{equation}
\begin{array}{c}
\nabla_\myf\Phi f_\myf\Omega{}^\myf\Lambda=0\sps\\
\partial_\myf\Phi f_\myf\Omega{}^\myf\Lambda=(\Gamma_{\myf\Phi\myf\Omega}{}^\myf\Theta f_\myf\Theta{}^\myf\Lambda-
\Gamma_{\myf\Phi\myf\Theta}{}^\myf\Lambda f_\myf\Omega{}^\myf\Theta)\spsd
\end{array}
\end{equation}
We will multiply this equation by $\frac{i}{2}\bar m_{\Psi '}{}^\myf\Omega m^\Lambda{}_\myf\Lambda$
\begin{equation}
\frac{i}{2}\bar m_{\Psi '}{}^\myf\Omega m^\Lambda{}_\myf\Lambda\partial_\myf\Phi f_\myf\Omega{}^\myf\Lambda=
\bar m_{\Psi '}{}^\myf\Omega m^\Lambda{}_\myf\Lambda\Gamma_{\myf\Phi\myf\Omega}{}^\myf\Lambda \spsd
\end{equation}
Hence,
\begin{equation}
\Gamma_{\myf\Phi\Psi '}{}^\Lambda=0\spsd
\end{equation}
Therefore, we can write down
\begin{equation}
\nabla_\myf\Phi m^\Lambda{}_\myf\Lambda:=\partial_\myf\Phi m^\Lambda{}_\myf\Lambda-
\Gamma_{\myf\Phi\myf\Lambda}{}^\myf\Omega m^\Lambda{}_\myf\Omega+\Gamma_{\myf\Phi\Psi}{}^\Lambda m^\Psi{}_\myf\Lambda=0\spsd
\end{equation}
Thus, if we know the coefficients of the connection in one of the representation (real or complex) then the connection coefficients in the other representation (complex or real, respectively) are uniquely restored from this equation. If now in the real representation of the tangent bundle, the torsion-free connection, compatible with the metric
\begin{equation}
\label{e9.0}
\nabla_\myf\Lambda G_{\myf\Psi\myf\Omega}=0\sps
\end{equation}
is given then it can be prolonged into the spinor bundle with the help of the condition
\begin{equation}
\label{e9.2}
\nabla_\myf\Lambda\eta_\myf\Psi{}^{\myff A\myff B}=0\spsd
\end{equation}
Now let I). n mod 4 = 0, n$>$ 2, II). n mod 4 = 2, n$>$ 4, then
\begin{equation}
\begin{array}{cc}
I). & \nabla_\myf\Psi \eta_\myf\Lambda{}_{\myff C \myff D}=0\sps \\
& \nabla_\myf\Psi \varepsilon_{\myff A\myff C}\varepsilon_{\myff B\myff D}=0\sps\\
& \varepsilon_{\myff A\myff C}\nabla_\myf\Psi\varepsilon_{\myff B\myff D}+\varepsilon_{\myff B\myff D}\nabla_\myf\Psi\varepsilon_{\myff A\myff C}=0\sps\\
& \eta_\myf\Lambda{}^{\myff A\myff B}(\delta_\myff A{}^\myff K\varepsilon^{\myff L\myff D}\nabla_\myf\Psi\varepsilon_{\myff B\myff D}+\delta_{\myff B}{}^{\myff L}\varepsilon^{\myff K\myff C}\nabla_\myf\Psi\varepsilon_{\myff A\myff C})=0\sps\\
& \eta_\myf\Lambda{}^{\myff K\myff B}\varepsilon^{\myff L\myff D}\nabla_\myf\Psi\varepsilon_{\myff B\myff D}+\eta_\myf\Lambda{}^{\myff B\myff L}\varepsilon^{\myff K\myff C}\nabla_\myf\Psi\varepsilon_{\myff B\myff C}=0\sps\\ \\
II). & \nabla_\myf\Psi \eta_\myf\Lambda{}_{\myff C\myff D}=0\sps\\
& \myf\nabla_\myf\Psi E_{\myff D}{}^{\myff A} E_{\myff C}{}^{\myff B}=0\sps\\
& E_{\myff D}{}^{\myff A} \myf\nabla_\myf\Psi E_{\myff C}{}^{\myff B}+E_{\myff C}{}^{\myff B}\myf\nabla_\myf\Psi E_{\myff D}{}^{\myff A}=0\sps\\
& \eta_\myf\Lambda{}_{\myff A\myff B}(\delta_{\myff K}{}^{\myff A} E_{\myff L}{}^{\myff C}\nabla_\myf\Psi E_{\myff C}{}^{\myff B}+\delta_{\myff L}{}^{\myff B} E_{\myff K}{}^{\myff D}\nabla_\myf\Psi E_{\myff D}{}^{\myff A})=0\sps\\
& \eta_\myf\Lambda{}_{\myff K\myff B} E_{\myff L}{}^{\myff C}\nabla_\myf\Psi E_{\myff C}{}^{\myff B}+\eta_\myf\Lambda{}_{\myff A\myff L}E_{\myff K}{}^{\myff D}\nabla_\myf\Psi E_{\myff D}{}^{\myff A}=0\spsd
\end{array}
\end{equation}
Turn to the equation (\ref{e6.3/1}). This will give
\begin{equation}
\begin{array}{cccc}
I). & \varepsilon^{\myff L\myff D}\nabla_\myf\Psi\varepsilon_{\myff B\myff D}=0\sps & II). & E_{\myff L}{}^{\myff C}\nabla_\myf\Psi E_{\myff C}{}^{\myff B}=0\spsd\\
\end{array}
\end{equation}
Thus, we see that
\begin{equation}
\label{e9.1}
\begin{array}{cccc}
I). &\nabla_\myf\Psi\varepsilon_{\myff B\myff D}=0\sps       &II).& \nabla_\myf\Psi E_{\myff C}{}^{\myff B}=0\sps\\
    &\nabla_\myf\Psi\tilde\varepsilon_{\myff B\myff D}=0\sps &    & \nabla_\myf\Psi\tilde E_{\myff C}{}^{\myff B}=0\spsd
\end{array}
\end{equation}
Therefore, the condition (\ref{e9.2}) imposes the restrictions
\begin{equation}
\begin{array}{cccc}
I). &\nabla_\myf\Psi\eta_\myf\Lambda{}^{\myff A\myff B}\varepsilon_{\myff A\myff B}=0\sps            &II).& \nabla_\myf\Psi\eta_\myf\Lambda{}^{\myff A\myff B}\eta_\myf\Phi{}_{\myff A\myff C}E_{\myff B}{}^{\myff C}=0\sps        \\
    &\nabla_\myf\Psi\eta_\myf\Lambda{}^{\myff A\myff B}\tilde\varepsilon_{\myff A\myff B}=0\sps      &    & \nabla_\myf\Psi\eta_\myf\Lambda{}^{\myff A\myff B}\eta_\myf\Phi{}_{\myff A\myff C}\tilde E_{\myff B}{}^{\myff C}=0\spsd
\end{array}
\end{equation}
This kind of the connection is possible with the normalization of the spinor basis according to \cite[v. 1, p. 230 (eng)]{Penrouz1}. However, we can relax the conditions (\ref{e9.0}) and (\ref{e9.2}) remembering that we consider the connection in the tangent bundle. Therefore, it is necessary to go back
\begin{equation}
\label{e9.3}
\eta^\myf\Psi{}_{\myff A\myff B}\nabla_\myf\Lambda\eta_\myf\Omega{}^{\myff A\myff B}=0
\end{equation}
which can be rewritten as
\begin{equation}
\label{e9.4}
\eta^\myf\Psi{}_{\myff A\myff B}\partial_\myf\Lambda \eta_\myf\Omega{}^{\myff A\myff B}-
\Gamma_{\myf\Lambda\myf\Omega}{}^\myf\Theta\eta^\myf\Psi{}_{\myff A\myff B}\eta_\myf\Theta{}^{\myff A\myff B}+
\tilde{\tilde\Gamma}_{\myf\Lambda\myff D}{}^{\myff A}\eta^\myf\Psi{}_{\myff A\myff B}\eta_\myf\Omega{}^{\myff D\myff B}+
\tilde\Gamma_{\myf\Lambda\myff D}{}^{\myff B}\eta^\myf\Psi{}_{\myff A\myff B}\eta_\myf\Omega{}^{\myff A\myff D}=0\spsd
\end{equation}
Therefore, in order to prolong uniquely the constructed connection into the spinor bundle, it is necessary to demand
\begin{equation}
\begin{array}{c}
\tilde{\tilde\Gamma}_{\myf\Lambda \myff L}{}^{\myff M}:=\frac{1}{N^2}\tilde{\tilde\Gamma}_{\myf\Lambda \myff K}{}^{\myff A}(
\eta_\myf\Psi{}_{\myff L\myff B}\eta_\myf\Omega{}^{\myff M\myff B} \eta^\myf\Psi{}_{\myff A\myff C}\eta^\myf\Omega{}^{\myff K\myff C}
-\frac{(2n-2)}{4}\delta_{\myff L}{}^{\myff M}\delta_{\myff K}{}^{\myff A})\sps\\ \\
\tilde\Gamma_{\myf\Lambda \myff K}{}^{\myff A}:=-\frac{2}{2n-4}\tilde{\tilde\Gamma}_{\myf\Lambda \myff C}{}^{\myff D}(\varepsilon_{\myff D\myff K}{}^{\myff C\myff A}-
\frac{2n-2}{2N^2}\delta_{\myff K}{}^{\myff A}\delta_{\myff D}{}^{\myff C})\spsd
\end{array}
\end{equation}
Indeed,
\begin{equation}
\begin{array}{c}
\eta_\myf\Lambda{}^{\myff A\myff B}\eta_\myf\Psi{}_{\myff K\myff B} \cdot (-\frac{2}{2n-4})(\varepsilon_{\myff D\myff A}{}^{\myff C\myff K}-
\frac{2n-2}{2N^2}\delta_{\myff A}{}^{\myff K}\delta_{\myff D}{}^{\myff C})=\eta_{\left[\right.\myf\Lambda}{}^{\myff B\myff C}\eta_{\myf\Psi\left.\right]}{}_{\myff B\myff D}-\\
-\frac{1}{2n-4}(\frac{2n}{2}-\frac{2n-2}{2N^2}\cdot 2N^2)g_{\myf\Lambda\myf\Psi}\delta_{\myff D}{}^{\myff C}=
\eta_\myf\Lambda{}^{\myff B\myff C}\eta_\myf\Psi{}_{\myff B\myff D}\spsd
\end{array}
\end{equation}

\begin{note}
\label{n9.1}
It should be noted that for n = 4, all conclusions remain valid and the last identity will have the form
\begin{equation}
\begin{array}{c}
\eta_\myf\Lambda{}^{\myff A\myff C}\eta_\myf\Psi{}_{\myff K\myff C}=(\eta_\myf\Lambda{}^{\myff C\myff A}+\eta_\myf\Lambda\varepsilon^{\myff A\myff C})\eta_\myf\Psi{}_{\myff K\myff C}=
\eta_\myf\Lambda{}^{\myff C\myff A}(\eta_\myf\Psi{}_{\myff C\myff K}+\eta_\myf\Psi\varepsilon_{\myff K\myff C})+\eta_\myf\Lambda\eta_\myf\Psi{}_{\myff K}{}^{\myff A}=\\
=\eta_\myf\Lambda{}^{\myff C\myff A}\eta_\myf\Psi{}_{\myff C\myff K}-\eta_\myf\Psi\eta_\myf\Lambda{}_{\myff K}{}^{\myff A}+\eta_\myf\Lambda\eta_\myf\Psi{}_{\myff K}{}^{\myff A}\spsd
\end{array}
\end{equation}
Thus, the contraction of (\ref{e9.4}) with $\eta_\myf\Psi\eta^\myf\Omega$ determines the expression $\Gamma_{\myf\Lambda\myff A}{}^{\myff A}$, after substituting  of which in (\ref{e9.4}), the coefficients $\Gamma_{\myf\Lambda\myff A}{}^{\myff D}$ are uniquely determined. For this purpose, it is enough to contract (\ref{e9.4}) with $\eta_\myf\Psi{}_{\myff K\myff L}\eta^\myf\Omega{}^{\myff M\myff L}$. It is caused by the identity $\eta_\Psi{}_{\myff A\myff B}\eta^\Psi{}_{\myff C\myff D}=\varepsilon_{\myff A\myff B\myff C\myff D}=\varepsilon_{\myff A\myff C}\varepsilon_{\myff B\myff D}$, where $\varepsilon_{\myff B\myff D}$, as before, is the antisymmetric metric spin-tensor (see Example \ref{ex6.1} which is easily adaptable for the inclusion $\mathbb R^4\subset\mathbb C^4$). But in this case, the condition (\ref{e9.3}) is equivalent to (\ref{e9.2}).
\end{note}

If we perform the symmetrization of the expression (\ref{e9.4}) on $\Omega$ and omitted $\Psi$ then from the analogue of (\ref{e6.5}), we obtain
\begin{equation}
\begin{array}{c}
\underbrace{\frac{N^2}{2}(\partial_\myf\Lambda G_{\myf\Psi\myf\Omega}-
2\Gamma_{\myf\Lambda(\myf\Omega\myf\Psi)})}_{=\frac{N^2}{2}\nabla_\Lambda G_{\myf\Psi\myf\Omega}}+
\underbrace{(\frac{1}{2N^2}\eta_\myf\Psi{}_{\myff A\myff B}\eta_\myf\Omega{}_{\myff C\myff D}\partial_\Lambda\varepsilon^{\myff A\myff B\myff C\myff D}+
\tilde\Gamma_{\myf\Lambda\myff B}{}^{\myff B}G_{\myf\Psi\myf\Omega})}_
{=\frac{1}{2N^2}\eta_{\myf\Psi}{}_{\myff A\myff B}\eta_\myf\Omega{}_{\myff C\myff D}\nabla_\Lambda \varepsilon^{\myff A\myff B\myff C\myff D}}=0\spsd
\end{array}
\end{equation}
If the metric tensor is covariantly constant then
\begin{equation}
\eta_\myf\Psi{}_{\myff A\myff B} \eta_\myf\Omega{}_{\myff C\myff D}\nabla_\myf\Lambda\varepsilon^{\myff A\myff B\myff C\myff D}=0\spsd
\end{equation}
Accordingly, the complex connection is determined by the relations
\begin{equation}
\nabla_\Lambda:=m_\Lambda{}^\myf\Lambda\nabla_\myf\Lambda\sps
\bar\nabla_{\Lambda '}:=\bar m_{\Lambda '}{}^\myf\Lambda\nabla_\myf\Lambda
\end{equation}
that in the absence of torsion, will lead to the connection
\begin{equation}
\nabla_\Lambda g_{\Psi\Phi}=0\sps  \bar\nabla_{\Lambda '} \bar g_{\Psi '\Phi '}=0
\end{equation}
which prolongs into the complex spinor bundle by the conditions (see (\ref{e3.13/1}))
\begin{equation}
\label{e9.8}
\begin{array}{c}
\nabla_\Lambda (\tilde M_K)_A{}^{\myff B}=0\sps \nabla_\Lambda (\tilde{\tilde M}_K)_A{}^{\myff B}=0\sps
\bar\nabla_{\Lambda '} (\tilde M_K)_A{}^{\myff B}=0\sps \bar\nabla_{\Lambda '} (\tilde{\tilde M}_K)_A{}^{\myff B}=0\sps\\
\sum\limits_{K=1}^{2N}(\eta_K)^\Psi{}_{AB}\bar\nabla_{\Lambda '}(\eta_K)_\Omega{}^{AB}=0\sps
\sum\limits_{K=1}^{2N}(\eta_K)^\Psi{}_{AB}\nabla_\Lambda(\eta_K)_\Omega{}^{AB}=0\spsd\\
\end{array}
\end{equation}
It should be noted that the upper relations
\begin{equation}
\begin{array}{c}
\partial_\Lambda (\tilde M_K)_A{}^{\myff B}+\tilde\Gamma_\Lambda{}_{\myff C}{}^{\myff B}(\tilde M_K)_A{}^{\myff C}-(\tilde\Gamma_K)_\Lambda{}_A{}^C(\tilde M_K)_C{}^{\myff B}=0\sps\\
\partial_\Lambda (\tilde{\tilde M}_K)_A{}^{\myff B}+\tilde{\tilde\Gamma}_\Lambda{}_{\myff C}{}^{\myff B}(\tilde{\tilde M}_K)_A{}^{\myff C}-(\tilde{\tilde\Gamma}_K)_\Lambda{}_A{}^C(\tilde{\tilde M}_K)_C{}^{\myff B}=0
\end{array}
\end{equation}
are define the different coefficients of the connection: $(\tilde\Gamma_K)_\Lambda{}_A{}^C$ and $(\tilde{\tilde\Gamma}_K)_\Lambda{}_A{}^C$; the bottom relations are rewritten as
\begin{equation}
\begin{array}{c}
\sum\limits_{K=1}^{2N}((\eta_K)^\Psi{}_{AB}\partial_\Lambda (\eta_K)_\Omega{}^{AB}-\frac{N}{2}\cdot\Gamma_{\Lambda\Omega}{}^\Psi+\\
+(\tilde\Gamma_K)_{\Lambda D}{}^A(\eta_K)^\Psi{}_{AB}(\eta_K)_\Omega{}^{DB}+(\tilde{\tilde\Gamma}_K)_{\Lambda D}{}^B(\eta_K)^\Psi{}_{AB}(\eta_K)_\Omega{}^{AD})=0\spsd\\
\end{array}
\end{equation}
Therefore, in order to prolong uniquely the connection into the spinor bundle, it is necessary to demand
\begin{equation}
\label{e9.9}
\begin{array}{c}
(\tilde{\tilde\Gamma}_K)_{\Lambda L}{}^M:=\frac{2}{N}(\tilde{\tilde\Gamma}_K)_{\Lambda D}{}^A(
(\eta_K)_\Psi{}_{LB}(\eta_K)_\Omega{}^{MB} (\eta_K)^\Psi{}_{AC}(\eta_K)^\Omega{}^{DC}
-\frac{(n-2)}{4}\delta_L{}^M\delta_D{}^A)\sps\\ \\
(\tilde\Gamma_K)_\Lambda{}_A{}^C:=-\frac{2}{n-4}(\tilde{\tilde\Gamma}_K)_\Lambda{}_K{}^L((\varepsilon_K)_{LA}{}^{KC}-\frac{n-2}{N}\delta_L{}^K\delta_A{}^C)\spsd\\
\end{array}
\end{equation}
Therefore, from the covariant constancy of the metric tensor
\begin{equation}
\begin{array}{c}
\nabla_\Lambda g_{\Psi\Omega}=(\frac{2}{N})^2\nabla_\Lambda((\eta_K){}_\Psi{}_{AB} (\eta_K){}_\Omega{}_{CD}(\varepsilon_K)^{ABCD})=\\
=\frac{2}{N}((\eta_K){}_\Psi{}^{AB}\nabla_\Lambda(\eta_K){}_\Omega{}_{AB}+(\eta_K){}_\Omega{}^{AB}\nabla_\Lambda(\eta_K){}_\Psi{}_{AB})+\\
+(\frac{2}{N})^2(\eta_K){}_\Psi{}_{AB} (\eta_K){}_\Omega{}_{CD}\nabla_\Lambda(\varepsilon_K)^{ABCD}=0\sps
\end{array}
\end{equation}
the relation
\begin{equation}
\label{e9.5}
\sum\limits_{K=1}^{2N}(\eta_K)_\Psi{}_{AB} (\eta_K)_\Omega{}_{CD}\nabla_\Lambda(\varepsilon_K)^{ABCD}=0
\end{equation}
will follow. We now turn to a real inclusion. The equation
\begin{equation}
\nabla_\Lambda H_i{}^\Psi=0
\end{equation}
can be rewritten as
\begin{equation}
\partial_\Lambda H_i{}^\Psi -\Gamma_{\Lambda i}{}^j H_j{}^\Psi+\Gamma_{\Lambda\Phi}{}^\Psi H_i{}^\Phi=0\spsd
\end{equation}
Accordingly, in the spinor bundle, the real connection can be prolonged by the condition
\begin{equation}
\sum\limits_{I=1}^{2N}(\eta_K)^i{}_{AB}\nabla_k(\eta_K)_j{}^{AB}=0\spsd
\end{equation}
This means that on the real space, the connection compatible with the metric takes the form
\begin{equation}
\nabla_i g_{ik}=0\sps
\end{equation}
where
\begin{equation}
\nabla_i:=H_i{}^\Lambda\nabla_\Lambda\spsd
\end{equation}

\begin{example}
n=4.  Metric index is equal to 1. \\
Let
\begin{equation}
g^i{}_{AB'}:=(\eta_1)^i{}_{AB}\bar S_{B'}{}^B\sps g_i{}^{AB'}:=(\eta_1)_i{}^{AB}S_B{}^{B'}\sps \overline{g_j{}^{A'B}}=\bar g_j{}^{B'A}=g_j{}^{AB'}\spsd
\end{equation}
Then the special case of the coordination for the connections can look like
\begin{equation}
g^i{}_{AB'}\nabla_k g_j{}^{AB'}+\bar g^i{}_{B'A}\nabla_k \bar g_j{}^{B'A}=0\spsd
\end{equation}
The condition can be rewritten as
\begin{equation}
\begin{array}{c}
g^i{}_{AB'}(\partial_k g_j{}^{AB'}-\Gamma_{kj}{}^lg_l{}^{AB'}+\tilde{\tilde\Gamma}_{kC}{}^Ag_j{}^{CB'}+\tilde\Gamma_{kC'}{}^{B'}g_j{}^{AC'}+
\partial_k g_j{}^{AB'}-\Gamma_{kj}{}^lg_l{}^{AB'}+\\+\overline{\tilde{\tilde\Gamma}}_{kC'}{}^{B'}g_j{}^{AC'}+\overline{\tilde\Gamma}_{kC}{}^Ag_j{}^{CB'})=0\spsd
\end{array}
\end{equation}
Then due to Note \ref{n9.1}, this relation can be rewritten as
\begin{equation}
\partial_k g_j{}^{AB'}-\Gamma_{kj}{}^lg_l{}^{AB'}+\frac{1}{2}(\tilde{\tilde\Gamma}_{kC}{}^A+\overline{\tilde\Gamma}_{kC}{}^A)g_j{}^{CB'}+
\frac{1}{2}\overline{(\tilde{\tilde\Gamma}_{kC}{}^B+\overline{\tilde\Gamma}_{kC}{}^B)}g_j{}^{AC'}=0\spsd
\end{equation}
Define $\Gamma_{kC}{}^A:=\frac{1}{2}(\tilde{\tilde\Gamma}_{kC}{}^A+\overline{\tilde\Gamma}_{kC}{}^A)$ which will lead to the connection from \cite[v. 1, p. 230 (eng)]{Penrouz1}
\begin{equation}
\nabla_k g_j{}^{AB'}=0\spsd
\end{equation}
\end{example}
This proves the theorem.
\begin{theorem}
\label{theorem9.0}
Let the pseudo-Riemannian manifold $V_{2n}$ ($n\ge 4$) be given. Let in the tangent bundle with fibers isomorphic to $R^{2n}_{(n, n)}$, a torsion-free connection be introduced. Then the compatibility condition of the connections between the spinor and tangent bundles can have the form
\begin{equation}
\eta^\myf\Psi{}_{\myff A\myff B}\nabla_\myf\Lambda \eta_\myf\Omega{}^{\myff A\myff B}=0\spsd
\end{equation}
In order to prolong uniquely the constructed connection into the spinor bundle, it is necessary to demand
\begin{equation}
\begin{array}{c}
\tilde{\tilde\Gamma}_{\myf\Lambda \myff L}{}^{\myff M}:=\frac{1}{N^2}\tilde{\tilde\Gamma}_{\myf\Lambda \myff K}{}^{\myff A}(
\eta_\myf\Psi{}_{\myff L\myff B}\eta_\myf\Omega{}^{\myff M\myff B} \eta^\myf\Psi{}_{\myff A\myff C}\eta^\myf\Omega{}^{\myff K\myff C}
-\frac{(2n-2)}{4}\delta_{\myff L}{}^{\myff M}\delta_{\myff K}{}^{\myff A})\sps\\ \\
\tilde\Gamma_{\myf\Lambda \myff K}{}^{\myff A}:=-\frac{2}{2n-4}\tilde{\tilde\Gamma}_{\myf\Lambda \myff C}{}^{\myff D}(\varepsilon_{\myff D\myff K}{}^{\myff C\myff A}-
\frac{2n-2}{2N^2}\delta_{\myff K}{}^{\myff A}\delta_{\myff D}{}^{\myff C})\spsd
\end{array}
\end{equation}
In order to prolong uniquely the connection onto the complex representation $\mathbb CV_{n}$ with fibers isomorphic $\mathbb C^n$, it is necessary to demand
\begin{equation}
\nabla_\myf\Lambda m_\myf\Psi{}^\Psi=0\sps\nabla_\myf\Lambda \bar m_\myf\Psi{}^{\Psi '}=0\sps
\nabla_\Lambda:= m^\myf\Lambda{}_\Lambda\nabla_\myf\Lambda\sps
\bar\nabla_{\Lambda '}:= \bar m^\myf\Lambda{}_{\Lambda '}\nabla_\myf\Lambda\spsd
\end{equation}
Then the compatibility condition of the connections will have the form
\begin{equation}
\label{e9.12}
\begin{array}{c}
\nabla_\Lambda (\tilde M_K)_A{}^{\myff B}=0\sps \nabla_\Lambda (\tilde{\tilde M}_K)_A{}^{\myff B}=0\sps
\bar\nabla_{\Lambda '} (\tilde M_K)_A{}^{\myff B}=0\sps \bar\nabla_{\Lambda '} (\tilde{\tilde M}_K)_A{}^{\myff B}=0\sps\\
\sum\limits_{K=1}^{2N}(\eta_K)^\Psi{}_{AB}\bar\nabla_{\Lambda '}(\eta_K)_\Omega{}^{AB}=0\sps
\sum\limits_{K=1}^{2N}(\eta_K)^\Psi{}_{AB}\nabla_\Lambda(\eta_K)_\Omega{}^{AB}=0\spsd\\
\end{array}
\end{equation}
In order to prolong uniquely the connection, we will demand
\begin{equation}
\begin{array}{c}
(\tilde{\tilde\Gamma}_K)_{\Lambda L}{}^M:=\frac{2}{N}(\tilde{\tilde\Gamma}_K)_{\Lambda D}{}^A(
(\eta_K)_\Psi{}_{LB}(\eta_K)_\Omega{}^{MB} (\eta_K)^\Psi{}_{AC}(\eta_K)^\Omega{}^{DC}
-\frac{(n-2)}{4}\delta_L{}^M\delta_D{}^A)\sps\\ \\
(\tilde\Gamma_K)_\Lambda{}_A{}^C:=-\frac{2}{n-4}(\tilde{\tilde\Gamma}_K)_\Lambda{}_K{}^L((\varepsilon_K)_{LA}{}^{KC}-\frac{n-2}{N}\delta_L{}^K\delta_A{}^C)\spsd\\
\end{array}
\end{equation}
In the case of an inclusion of a real space into the complex space, it is necessary to satisfy the relations
\begin{equation}
\nabla_\Lambda H_i{}^\Psi=0\sps
\bar\nabla_{\Lambda '}\bar H_i{}^{\Psi '}=0\sps \nabla_i:=H_i{}^\Lambda \nabla_\Lambda\spsd
\end{equation}
Then the compatibility condition will have the form
\begin{equation}
\sum\limits_{I=1}^{2N}(\eta_K)^i{}_{AB}\nabla_k(\eta_K)_j{}^{AB}=0\spsd
\end{equation}
\end{theorem}

\begin{corollary}
If in the tangent bundle to the manifold $V_{2n}$, a torsion-free connection compatible with the metric
\begin{equation}
\nabla_\myf\Lambda G_{\myf\Psi\myf\Omega}=0
\end{equation}
is given then in the tangent bundle of the complex representation $\mathbb CV_{n}$, the connection
\begin{equation}
\nabla_\Lambda g_{\Omega\Psi}=0\sps \bar\nabla_{\Lambda '}\bar g_{\Omega '\Psi '}=0
\end{equation}
is induced. For an inclusion $V_n\subset\mathbb CV_{n}$, the connection
\begin{equation}
\nabla_i g_{ik}=0
\end{equation}
is induced.
\end{corollary}

We construct the operators $P^\myf\Psi{}_{\myff A}:=\eta^\myf\Psi{}_{\myff B\myff A}X^{\myff B}$ and $P_\myf\Psi{}^{\myff A}:=\eta_\myf\Psi{}^{\myff B\myff A}Y_{\myff B}$ such that $X^{\myff A}Y_{\myff A}=2$ and $P_\myf\Lambda{}^{\myff A}P_\myf\Psi{}_{\myff A}=G_{\myf\Lambda\myf\Psi}$. This is always possible for $n\ge 8$. Then the compatibility condition of the connections for the real representation has the form
\begin{equation}
\nabla_\myf\Lambda P_\myf\Psi{}^{\myff A}=0
\end{equation}
which can be rewritten as
\begin{equation}
\partial_\myf\Lambda P_\myf\Psi{}^{\myff A}-\Gamma_{\myf\Lambda\myf\Psi}{}^\myf\Phi P_\myf\Phi{}^{\myff A}+\Gamma_\myf\Lambda{}_{\myff C}{}^{\myff A}P_\myf\Psi{}^{\myff C}=0\spsd
\end{equation}
Therefore, we can uniquely prolong the connection into the spinor bundle as
\begin{equation}
\label{e9.6}
\Gamma_\myf\Lambda{}_{\myff C}{}^{\myff A}:=-P^\myf\Psi{}_{\myff C}\partial_\myf\Lambda P_\myf\Psi{}^{\myff A}+\Gamma_{\myf\Lambda\myf\Psi}{}^\myf\Phi P^\myf\Psi{}_{\myff C}P_\myf\Phi{}^{\myff A}\spsd
\end{equation}
For the complex representation, the operators
\begin{equation}
\label{e9.10}
\begin{array}{c}
(P_K)^\Psi{}_A:=m_\myf\Psi{}^\Psi P^\myf\Psi{}_{\myff A}(\tilde M_K)_A{}^{\myff A}\sps
(P_K)^{\Psi '}{}_A:=\bar m_\myf\Psi{}^{\Psi '} P^\myf\Psi{}_{\myff A}(\tilde M_K)_A{}^{\myff A}\sps\\ \\
(P^*_K)_\Psi{}^A:=m^\myf\Psi{}_\Psi P_\myf\Psi{}^{\myff A}(\tilde M^*_K)^A{}_{\myff A}\sps
(P^*_K)_{\Psi '}{}^A:=\bar m^\myf\Psi{}_{\Psi '} P_\myf\Psi{}^{\myff A}(\tilde M^*_K)^A{}_{\myff A}\spsd\\
\end{array}
\end{equation}
are defined.

\begin{theorem}
\label{theorem9.1}
Let the pseudo-Riemannian manifold $V_{2n}$ $(n\ge 4)$ be given. Let in the tangent bundle with fibers isomorphic to $R^{2n}_{(n, n)}$, a torsion-free connection be introduced. Then the compatibility condition of the connections between the spinor and tangent bundles can have the form
\begin{equation}
\nabla_\myf\Lambda P_\myf\Psi{}^{\myff A}=0\spsd
\end{equation}
In order to prolong uniquely the constructed connection into the spinor bundle, it is necessary to demand
\begin{equation}
\Gamma_\myf\Lambda{}_{\myff C}{}^{\myff A}:=-P^\myf\Psi{}_{\myff C}\partial_\myf\Lambda P_\myf\Psi{}^{\myff A}+\Gamma_{\myf\Lambda\myf\Psi}{}^\myf\Phi P^\myf\Psi{}_{\myff C}P_\myf\Phi{}^{\myff A}\spsd
\end{equation}
In order to prolong uniquely the connection onto the complex representation $\mathbb CV_{n}$ with fibers isomorphic $\mathbb C^n$, it is necessary to demand
\begin{equation}
\nabla_\myf\Lambda m_\myf\Psi{}^\Psi=0\sps\nabla_\myf\Lambda \bar m_\myf\Psi{}^{\Psi '}=0\sps
\nabla_\Lambda:= m^\myf\Lambda{}_\Lambda\nabla_\myf\Lambda\sps
\bar\nabla_{\Lambda '}:= \bar m^\myf\Lambda{}_{\Lambda '}\nabla_\myf\Lambda\spsd
\end{equation}
Then the compatibility condition of the connections has the form
\begin{equation}
\begin{array}{c}
\nabla_\Lambda (\tilde M_K)_A{}^{\myff B}=0\sps \bar\nabla_{\Lambda '} (\tilde M_K)_A{}^{\myff B}=0\sps\\
\nabla_\Lambda (P_K)^\Psi{}_A=0\sps  \bar\nabla_{\Lambda '}(P_K)^\Psi{}_A=0\sps
\nabla_\Lambda (\bar P_K)^{\Psi '}{}_A=0\sps  \bar\nabla_{\Lambda '}(\bar P_K)^{\Psi '}{}_A=0\spsd
\end{array}
\end{equation}
In the case of an inclusion of a real space in the complex space, it is necessary to satisfy the relations
\begin{equation}
\nabla_\Lambda H_i{}^\Psi=0\sps
\bar\nabla_{\Lambda '}\bar H_i{}^{\Psi '}=0\sps \nabla_i:=H_i{}^\Lambda \nabla_\Lambda\spsd
\end{equation}
Then the compatibility condition of the connections has the form
\begin{equation}
\nabla_i (P_K)^j{}_A=0\sps\nabla_i (\bar P_K)^j{}_{A'}=0\spsd\\
\end{equation}
\end{theorem}

Now one can construct the operators
\begin{equation}
%
\end{equation}
These operators are invariants of the connection only if
\begin{equation}
\begin{array}{c}
\partial_\Lambda (P_K)^\Psi{}_B=0\sps \bar\partial_{\Lambda '} (P_K)^\Psi{}_B=0\sps
\partial_\Lambda (P^*_K)_\Psi{}^B=0\sps \bar\partial_{\Lambda '} (P^*_K)_\Psi{}^B=0\sps\\
\partial_\Lambda (\bar P_K)^{\Psi '}{}_B=0\sps \bar\partial_{\Lambda '} (\bar P_K)^{\Psi '}{}_B=0\sps
\partial_\Lambda (\bar P^*_K)_{\Psi '}{}^B=0\sps \bar\partial_{\Lambda '} (\bar P^*_K)_{\Psi '}{}^B=0
\end{array}
\end{equation}
under the condition
\begin{equation}
\partial_\myf\Lambda P^\myf\Psi{}_\myff B=0\sps
\partial_\myf\Lambda P_\myf\Psi{}^\myff B=0\sps
\partial_\myf\Lambda m^\myf\Psi{}_\Psi=0\sps
\partial_\myf\Lambda m_\myf\Psi{}^\Psi=0\sps
\partial_\myf\Lambda \bar m^\myf\Psi{}_{\Psi '}=0\sps
\partial_\myf\Lambda \bar m_\myf\Psi{}^{\Psi '}=0\spsd
\end{equation}
Then from (\ref{e9.6}) and (\ref{e3.13/1}), the parities
\begin{equation}
\begin{array}{c}
(\Gamma_K)_\Lambda{}_C{}^A=\Gamma_{\Lambda\Psi}{}^\Phi \underbrace{(P_K)^\Psi{}_C (P^*_K)_\Phi{}^A}_{:=(P_K)_\Phi{}^\Psi{}{}_C{}^A}\sps
(\bar\Gamma_K)_{\Lambda '}{}_C{}^A=\Gamma_{\Lambda'\Psi}{}^\Phi \underbrace{(P_K)^\Psi{}_C (P^*_K)_\Phi{}^A}_{:=(P_K)_\Phi{}^\Psi{}_C{}^A}=0\\
\end{array}
\end{equation}
will follow that will ensure the invariance of the operators. Then the following identities will be executed (the proof is given in Appendix)
\begin{equation}
\label{e9.7}
%
\end{equation}
For n = 4, such the connections were considered in \cite{Bilyalov1}. However, in \cite{Bilyalov1}, the expression of the form $(\eta_K)^\Psi{}_{DC} (\eta_K)_\Phi{}^{DA}$ unlike the expression $(P_K)_\Phi{}^\Psi{}_C{}^A$ of this article was considered, and this would cause the torsion of such the operators.

\begin{theorem}
\label{theorem9.2}
The operators
\begin{equation}
%
\end{equation}
will be invariants of the complex connection from Theorem \ref{theorem9.1} under the condition of the constancy of the operators
\begin{equation}
\label{e9.11}
\begin{array}{c}
\partial_\Lambda (P_K)^\Psi{}_B=0\sps \bar\partial_{\Lambda '} (P_K)^\Psi{}_B=0\sps
\partial_\Lambda (P^*_K)_\Psi{}^B=0\sps \bar\partial_{\Lambda '} (P^*_K)_\Psi{}^B=0\sps\\
\partial_\Lambda (\bar P_K)^{\Psi '}{}_B=0\sps \bar\partial_{\Lambda '} (\bar P_K)^{\Psi '}{}_B=0\sps
\partial_\Lambda (\bar P^*_K)_{\Psi '}{}^B=0\sps \bar\partial_{\Lambda '} (\bar P^*_K)_{\Psi '}{}^B=0\sps\\
\partial_\myf\Lambda P^\myf\Psi{}_\myff B=0\sps
\partial_\myf\Lambda P_\myf\Psi{}^\myff B=0\sps\\
\partial_\myf\Lambda m^\myf\Psi{}_\Psi=0\sps
\partial_\myf\Lambda m_\myf\Psi{}^\Psi=0\sps
\partial_\myf\Lambda \bar m^\myf\Psi{}_{\Psi '}=0\sps
\partial_\myf\Lambda \bar m_\myf\Psi{}^{\Psi '}=0\sps\\
\sum\limits_{K=1}^{2N}(P_K)^\Psi{}_B(P^*_K)_\Phi{}^B=\delta_\Psi{}^\Phi\sps
\sum\limits_{K=1}^{2N}(\bar P_K)^{\Psi '}{}_B(\bar P^*_K)_{\Phi '}{}^B=\delta_{\Psi '}{}^{\Phi '}\spsd
\end{array}
\end{equation}
These operators satisfy the identities
\begin{equation}
%
\end{equation}
and are the natural prolongation of the operators $L_x(y)^\Psi=x^\Omega\partial_\Omega y^\Psi-y^\Omega\partial_\Omega x^\Psi$,
$\bar L_{\bar x}(\bar y)^{\bar\Psi}=\bar x^{\Omega '}\bar \partial_{\Omega '}\bar y^{\Psi '}-\bar y^{\Omega '}\bar\partial_{\Omega '}\bar x^{\Psi '}$ into the spinor bundle
\begin{equation}
\begin{array}{c}
L_x(y)^\Psi=\sum\limits_{K=1}^{2N}L_x(Y_K)^A(P_K)^\Psi{}_A\sps \bar L_{\bar x}(\bar y)^{\Psi '}=\sum\limits_{K=1}^{2N}\bar L_{\bar x}(\bar Y_K)^A(\bar P_K)^{\Psi '}{}_A\sps\\
L_x(Y_K)^A=L_x(y)^\Psi (P^*_K)_\Psi{}^A\sps\bar L_{\bar x}(\bar Y_K)^A=\bar L_{\bar x}(\bar y)^{\Psi '}(\bar P^*_K)_{\Psi '}{}^A\spsd
\end{array}
\end{equation}
\end{theorem}

\section{On the classification of tensors with the symmetries of the Riemann curvature tensor. Curvature spinors}
\Abstract{
$\phantom{ff}$In this section, it is told about how to associate tensors possessing the symmetries of the curvature tensor to their spinor analogues. This simplifies the classification of such the tensors for small dimensions. The conclusion of all results of this section is made on the basis \cite[v. 1, p. 231-246 (eng)]{Penrouz1}.
}

After the introduction of the connection, we can pass to the study of the curvature tensor properties with the restriction $(\Gamma_K)_{\Lambda A}{}^B=(\tilde\Gamma_K)_{\Lambda A}{}^B=(\tilde{\tilde\Gamma}_K)_{\Lambda A}{}^B$ in the corresponding bundles. The curvature tensors can be calculated by the formulas
\begin{equation}
\nabla_{\left[\right.\Lambda}\nabla_{\Psi\left.\right]}r^\Omega=R_{\Lambda\Psi\Phi}{}^\Omega r^\Phi\sps
\nabla_{\left[\right.\Lambda}\nabla_{\Psi\left.\right]}X^A=\sum\limits_{K=1}^{2N}(\mathcal R_K)_{\Lambda\Psi C}{}^A X^C
\end{equation}
Consider the coordination of the complex connections induced by the formulas (\ref{e9.8}), (\ref{e9.9}) with the covariant constancy of the metric tensor. We require that
\begin{equation}
\nabla_\Lambda (\varepsilon_K)_{AB}{}^{CD}=\partial_\Lambda (\varepsilon_K)_{AB}{}^{CD}=0\spsd
\end{equation}
Then
\begin{equation}
\sum\limits_{K=1}^{2N}(\nabla_\Theta(\eta_K)^\Psi{}_{AB})(\nabla_\Lambda(\eta_K)_\Omega{}^{AB})=0\spsd
\end{equation}
Therefore, the integrability condition of (\ref{e9.12}) from Theorem \ref{theorem9.0} takes the form
\begin{equation}
\begin{array}{c}
\sum\limits_{K=1}^{2N}(\eta_K)^\Theta{}_{AB}\nabla_{\left[\right.\Lambda}\nabla_{\Psi\left.\right]}(\eta_K)_\Omega{}^{AB}=0\sps\\
N^2R_{\Lambda\Psi\Omega}{}^\Theta=\sum\limits_{K=1}^{2N}
((\mathcal R_K)_{\Lambda\Psi C}{}^A(\eta_K)^\Theta{}_{AB}(\eta_K)_\Omega{}^{CB}+
(\mathcal R_K)_{\Lambda\Psi C}{}^B(\eta_K)^\Theta{}_{AB}(\eta_K)_\Omega{}^{AC})\spsd
\end{array}
\end{equation}
Define
\begin{equation}
\label{e10.1}
(A_K)_{\Theta\Phi L}{}^X:=\frac{1}{2}(\eta_K)_{\left[\right.\Theta}{}^{MX}(\eta_K)_{\Phi\left.\right]}{}_{ML}\spsd
\end{equation}
Then from (\ref{e6.4}), the identity
\begin{equation}
(A_K)_{\Phi\Theta L}{}^X (A_K)_{\Omega\Gamma X}{}^L=\frac{N}{8}g_{\Phi\left[\right.\Gamma}g_{\Omega\left.\right]\Theta}
\end{equation}
will follow. Then we can set
\begin{equation}
\label{e10.2}
\begin{array}{c}
(R_K)_{\Lambda\Psi C}{}^N:=-R_{\Lambda\Psi \Theta \Phi}(A_K)^{\Theta\Phi}{}_C{}^N\sps
R_{\Lambda\Psi \Theta \Phi}=\frac{8}{N}(A_K)_{\Theta\Phi}{}_C{}^N (R_K)_{\Lambda\Psi N}{}^C\sps\\[2ex]
(R_K)_{\Lambda\Psi C}{}^M=-\frac{2}{n-4}(R_K)_{\Lambda\Psi L}{}^D\varepsilon_{CD}{}^{ML}
\end{array}
\end{equation}
(the proof of the last equation is given in Appendix). It is obvious that $(R_K)_{\Lambda\Psi C}{}^N=(\mathcal R_K)_{\Lambda\Psi C}{}^N$, for example, if the coordination condition of the connections has the form
\begin{equation}
\label{e10.2/1}
\nabla_\Lambda(\eta_K)_\Psi{}^{AB}=0\spsd
\end{equation}
For further calculations, we should use the identities (the number K will be omitted since all the other calculations do not depend on this number)
\begin{equation}
\label{e10.3}
\eta_{\left[\right. \Lambda_1}{}^{A_1A_2}\eta_{\Lambda_2\left.\right]}{}_{A_1A_3}
\eta_{\left[\right. \Lambda_3}{}^{A_4A_3}\eta_{\Lambda_4\left.\right]}{}_{A_4A_2}=
\frac{N}{2} g_{\Lambda_1 \left[\right. \Lambda_4}
            g_{\Lambda_3 \left.\right] \Lambda_2}\spsd
\end{equation}
\begin{equation}
\label{e10.4}
\eta_{\left[\right. \Lambda_1}{}^{A_1A_2}\eta_{\Lambda_2\left.\right]}{}_{A_1A_3}
\eta_{\left[\right. \Lambda_3}{}^{A_4A_3}\eta_{\Lambda_4\left.\right]}{}_{A_4A_5}
\eta_{\left[\right. \Lambda_5}{}^{A_6A_5}\eta_{\Lambda_6\left.\right]}{}_{A_6A_2}=
N g_{\left[\right. \Lambda_3 | \left[\right. \Lambda_2}
  g_{\Lambda_1 \left.\right] | \left[\right. \Lambda_6}
  g_{\Lambda_5 \left.\right] | \Lambda_4 \left]\right.}\spsd
\end{equation}
\begin{equation}
\label{e10.5}
\begin{array}{c}
\eta_{\left[\right. \Lambda_1}{}^{A_1A_2}\eta_{\Lambda_2\left.\right]}{}_{A_1A_3}
\eta_{\left[\right. \Lambda_3}{}^{A_4A_3}\eta_{\Lambda_4\left.\right]}{}_{A_1A_5}
\eta_{\left[\right. \Lambda_5}{}^{A_6A_5}\eta_{\Lambda_6\left.\right]}{}_{A_6A_7}
\eta_{\left[\right. \Lambda_7}{}^{A_8A_7}\eta_{\Lambda_8\left.\right]}{}_{A_8A_2}=\\
=\frac{N}{4}
(g_{\Lambda_1\left[\right. \Lambda_8} g_{\Lambda_7 \left.\right]\Lambda_2}g_{\Lambda_3\left[\right. \Lambda_6} g_{\Lambda_5 \left.\right]\Lambda_4}-
g_{\Lambda_1\left[\right. \Lambda_6} g_{\Lambda_5 \left.\right]\Lambda_2}g_{\Lambda_4\left[\right. \Lambda_8} g_{\Lambda_7 \left.\right]\Lambda_3}+\\
+g_{\Lambda_1\left[\right. \Lambda_4} g_{\Lambda_3 \left.\right]\Lambda_2}g_{\Lambda_5\left[\right. \Lambda_8} g_{\Lambda_7 \left.\right]\Lambda_6})+
N(
g_{\left[\right. \Lambda_3 | \left[\right. \Lambda_2}
g_{\Lambda_1 \left.\right]   \left[\right. \Lambda_8}
g_{\Lambda_7 \left.\right]   \left[\right. \Lambda_6}
g_{\Lambda_5 \left.\right] | \Lambda_4 \left.\right]}+\\
+g_{\left[\right. \Lambda_5 | \left[\right. \Lambda_4}
g_{\Lambda_3 \left.\right]   \left[\right. \Lambda_7}
g_{\Lambda_8 \left.\right]   \left[\right. \Lambda_2}
g_{\Lambda_1 \left.\right] | \Lambda_6 \left.\right]}+
g_{\left[\right. \Lambda_3 | \left[\right. \Lambda_7}
g_{\Lambda_8 \left.\right]   \left[\right. \Lambda_5}
g_{\Lambda_6 \left.\right]   \left[\right. \Lambda_1}
g_{\Lambda_2 \left.\right] | \Lambda_4 \left.\right]})
\end{array}
\end{equation}
(the proof is given in Appendix).
\begin{theorem}
The classification of a bitensor, possessing  the properties ($\Lambda ,\Psi ,...=\overline{1,n}$, $A,B ,...=\overline{1,N}$)
\begin{equation}
    R_{\Lambda \Psi \Phi \Theta}=R_{\left[ \Lambda \Psi\right]
    \left[\Phi \Theta] \right.} \sps
    R_{\Lambda \Psi  \Phi \Theta}=R_{\Phi \Theta \Lambda \Psi}
\end{equation}
and belonging to the tangent bundle $\tau(\mathbb CV_n)$ over the analytic Riemannian space $\mathbb CV_n$, can be reduced to the classification of a tensor $R_A{}^B{}_C{}^D$ of the N-dimensional complex spinor space such that
\begin{equation}
    R_C{}^D{}_S{}^R:=R_{\Lambda \Psi \Phi \Theta}A^{\Lambda \Psi}{}_C{}^D
    A^{\Phi \Theta}{}_S{}^R\sps A_{\Lambda \Psi C}{}^D:=\frac{1}{2}\eta_{\left[\Lambda\right.}{}^{AD}\eta_{\left.\Psi\right]}{}_{AC}\spsd
\end{equation}
Besides, the relations
\begin{equation}
    R_K{}^K{}_S{}^R=R_S{}^R{}_K{}^K=0 \sps  R_C{}^D{}_S{}^R=
    R_S{}^R{}_C{}^D
\end{equation}
will be executed. The decomposition
\begin{equation}
\begin{array}{c}
    R_C{}^K{}_M{}^A=C_C{}^K{}_M{}^A-\frac{4}{N(n-2)}R_G{}^D{}_N{}^P\varepsilon^{ABKL}\varepsilon^{GN}{}_{ML}\varepsilon_{DPCB}-\\
    -\big(\frac{1}{2(n-1)(n-2)}-\frac{1}{4(n-2)}\big)R\varepsilon^{ABKL}\varepsilon_{CBML}-
    \big(\frac{1}{4(n-2)}-\frac{n}{8(n-1)(n-2)}\big)R\delta_C{}^A\delta_M{}^K
\end{array}
\end{equation}
corresponds to the decomposition of the tensor $R_{\Lambda \Psi}{}^{\Phi \Theta}$
\begin{equation}
    R_{\Lambda \Psi}{}^{\Phi \Theta}=C_{\Lambda \Psi}{}^{\Phi \Theta}+
    \frac{4}{n-2} R_{\left[\Lambda \right.}{}^{\left[ \Phi \right.}
    g_{\left.\Psi \right]}{}^{\left. \Theta \right]}-
    \frac{2}{(n-1)(n-2)}Rg_{\left[\Lambda \right.}{}^{\left[ \Phi \right.}
    g_{\left.\Psi \right]}{}^{\left. \Theta \right]}
\end{equation}
into the irreducible components not resulted by orthogonal transformations. At the same time, the Bianchi identity
\begin{equation}
    R_{\Lambda \Psi  \Phi \Theta}+R_{\Lambda \Theta \Psi \Phi}+
    R_{\Lambda \Phi \Theta \Psi}=0
\end{equation}
will have the form
\begin{equation}
    R_L{}^D{}_S{}^L=-\frac{1}{8}\cdot R\delta_S{}^D\spsd
\end{equation}
\end{theorem}

\begin{proof}$ $\\
\begin{enumerate}
\item The first step. The Bianchi identity $R_{[\Lambda \Psi \Phi] \Theta}=0$.
\begin{equation}
\begin{array}{c}
R_{[\Lambda \Psi \Phi] \Theta}\eta^\Lambda{}_{AB}\eta^\Psi{}^{AC}\eta^\Phi{}_{DC}\eta^\Theta{}^{DL}=\\
=R_{\Lambda \Psi \Phi \Theta}\frac{1}{3}(\eta^\Lambda{}_{AB}\eta^\Psi{}^{AC}\eta^\Phi{}_{DC}+
\eta^\Phi{}_{AB}\eta^\Lambda{}^{AC}\eta^\Psi{}_{DC}+\eta^\Psi{}_{AB}\eta^\Phi{}^{AC}\eta^\Lambda{}_{DC})\eta^\Theta{}^{DL}=\\
=R_{\Lambda \Psi \Phi \Theta }\frac{1}{3}(\eta^\Lambda{}_{AB}\eta^\Psi{}^{AC}\eta^\Phi{}_{DC}+g^{\Phi\Lambda}\eta^\Psi{}_{DB}-\\
-\eta^\Lambda{}_{AB}\eta^\Phi{}^{AC}\eta^\Psi{}_{DC}+\eta^\Lambda{}_{AB}\eta^\Psi{}^{AC}\eta^\Phi{}_{DC}-g^{\Phi\Psi}\eta^\Lambda{}_{DB})\eta^\Theta{}^{DL}=\\
=R_{\Lambda \Psi \Phi \Theta }(\eta^\Lambda{}_{AB}\eta^\Psi{}^{AC}\eta^\Phi{}_{DC}+
g^{\Phi\Lambda}\eta^\Psi{}_{DB})\eta^\Theta{}^{DL}=4R_B{}^C{}_C{}^L+\frac{1}{2}R\delta_B{}^L=0\spsd
\end{array}
\end{equation}
Inversely,
\begin{equation}
\begin{array}{c}
(4R_{A_1}{}^C{}_C{}^{A_5}+\frac{1}{2}R\delta_{A_1}{}^{A_5})
\eta_{\left[\right. \Lambda_5}{}_{|A_5A_6|}\eta_{\Lambda_6\left.\right]}{}^{A_7A_6}
\eta_{\left[\right. \Lambda_7}{}_{|A_7A_8|}\eta_{\Lambda_8\left.\right]}{}^{A_1A_8}=\\
=R_{\Lambda_1\Lambda_2\Lambda_3\Lambda_4}
\eta^{\Lambda_1}{}_{A_1A_2}\eta^{\Lambda_2}{}^{A_3A_2}
\eta^{\Lambda_3}{}_{A_3A_4}\eta^{\Lambda_4}{}^{A_5A_4}
\eta_{\left[\right. \Lambda_5}{}_{|A_5A_6|}\eta_{\Lambda_6\left.\right]}{}^{A_7A_6}
\eta_{\left[\right. \Lambda_7}{}_{|A_7A_8|}\eta_{\Lambda_8\left.\right]}{}^{A_1A_8}+\\
+\frac{1}{2}R
\eta_{\left[\right. \Lambda_5}{}_{|A_1A_6|}\eta_{\Lambda_6\left.\right]}{}^{A_7A_6}
\eta_{\left[\right. \Lambda_7}{}_{|A_7A_8|}\eta_{\Lambda_8\left.\right]}{}^{A_1A_8}=
\frac{N}{4}(R_{\Lambda_8\Lambda_7\Lambda_6\Lambda_5}-R_{\Lambda_6\Lambda_5\Lambda_7\Lambda_8}-\\
-Rg_{\Lambda_5 \left[\right. \Lambda_8} g_{\Lambda_7\left.\right]\Lambda_6})+
N(-R_{\left[\right. \Lambda_8|\left[\right. \Lambda_5} g_{\Lambda_6\left.\right]\Lambda_7\left.\right]}+
R_{[\Lambda_6|[\Lambda_8\Lambda_7]|\Lambda_5]}+\\
+R_{\left[\right. \Lambda_6|\left[\right. \Lambda_7} g_{\Lambda_8\left.\right]|\Lambda_5\left.\right]})+
\frac{1}{2}\cdot\frac{N}{2}Rg_{\Lambda_5 \left[\right. \Lambda_8} g_{\Lambda_7\left.\right]\Lambda_6}=
\frac{N}{2}(R_{\Lambda_8\Lambda_7\Lambda_6\Lambda_5}+2R_{\Lambda_6[\Lambda_8\Lambda_7]\Lambda_5})=0\spsd
\end{array}
\end{equation}
\item Second step. The Ricci tensor $R_{\Lambda\Psi}$.\\
We use the two auxiliary identities.
\begin{equation}
\begin{array}{c}
R_{\Lambda\Psi C}{}^M=-\frac{2}{n-4}R_{\Lambda\Psi L}{}^D\varepsilon_{CD}{}^{ML}\sps\\
\eta_{\left[\right.\Gamma}{}^{CL}\eta_{\Omega\left.\right]}{}_{CD}=
-\frac{2}{n-4}\eta_{\left[\right.\Gamma}{}^{CN}\eta_{\Omega\left.\right] MN}\varepsilon_{CD}{}^{ML}\spsd
\end{array}
\end{equation}
This can determine the identity
\begin{equation}
\eta_{\left[\right.\Gamma}{}_{|CD|}\eta_{\Omega\left.\right]}{}^{CL}R_{\Lambda\Psi L}{}^D=
-\eta_{\left[\right.\Gamma}{}^{CN}\eta_{\Omega\left.\right] MN}R_{\Lambda\Psi C}{}^M\spsd
\end{equation}
Therefor,
\begin{equation}
%
\end{equation}
\item Third step. The scalar curvature R.\\
\begin{equation}
\begin{array}{c}
g^{\left[\right.\Delta |\left[\right.\Psi }g^{\Phi \left.\right] |\Lambda\left.\right]}
\eta_{\left[\right.\Delta}{}^{AB}\eta_{\Lambda\left.\right]}{}_{CB}\eta_{\left[\right.\Psi}{}^{KL}\eta_{\Phi\left.\right]}{}_{ML}=\\
=g^{\Delta\Psi}g^{\Phi\Lambda}\eta_\Delta{}^{AB}\eta_\Lambda{}_{CB}(\eta_\Psi{}^{KL}\eta_\Phi{}_{ML}-\frac{1}{2}g_{\Psi\Phi}\delta_M{}^K)=
\varepsilon^{ABKL}\varepsilon_{CBML}-\frac{n}{4}\delta_C{}^A\delta_M{}^K\spsd
\end{array}
\end{equation}
\end{enumerate}
Now it is necessary to gather all results and to receive the analogue of the decomposition on irreducible components not resulted by orthogonal transformations.
\end{proof}

\begin{corollary}$ $
\label{c10.1}
\begin{enumerate}
    \item
    The simplicity condition of an bivector of the n-dimensional space $\mathbb C^n$ can be written down as
\begin{equation}
\label{e10.6}
    p^{\left[\Lambda\Psi\right.}p^{\Phi\Omega\left.\right]}=0\spsd
\end{equation}
    The coordinates of the bivector can be associated to the traceless complex matrix
\begin{equation}
\label{e10.7}
    p_L{}^Dp_S{}^L-\frac{1}{N}(p_L{}^Kp_K{}^L)\delta_S{}^D=0\spsd
\end{equation}
    \item
    A simple bivector of the space $\mathbb C^n$ with the condition $p^{\Lambda\Psi}p_{\Lambda\Psi}=0$ can be associated to a nilpotent operator with index equal to 2: $p_L{}^Dp_S{}^L=0$.\\
\end{enumerate}
\end{corollary}
\begin{proof}$ $\\
1). A bivector is simple if and only if when there is the decomposition
\begin{equation}
    p^{\Lambda\Psi}=x^\Lambda y^\Psi-y^\Lambda x^\Psi\spsd
\end{equation}
Therefore, the conditions (\ref{e10.6}) are satisfied automatically. Conversely, if the conditions (\ref{e10.6}) are executed then they can be written down as
\begin{equation}
    p^{\Lambda\Psi}p^{\Phi\Omega}-
    p^{\Lambda\Phi}p^{\Psi\Omega}+
    p^{\Psi\Phi}p^{\Lambda\Omega}=0\spsd
\end{equation}
Contract this equation with the nonzero covectors $t_\Omega$ and $z_\Phi$, that $p^{\Phi\Omega}z_\Phi t_\Omega \ne 0$
\begin{equation}
    p^{\Lambda\Psi}=\frac{1}{p^{\Theta\Xi}z_\Theta t_\Xi}
    (p^{\Lambda\Phi}z_\Phi p^{\Psi\Omega}t_\Omega-
     p^{\Psi\Phi}z_\Phi p^{\Lambda\Omega}t_\Omega)\spsd
\end{equation}
Set
\begin{equation}
    x^\Lambda:=\frac{1}{p^{\Theta\Xi}z_\Theta t_\Xi}p^{\Lambda\Phi}z_\Phi\sps
    y^\Psi:=\frac{1}{p^{\Theta\Xi}z_\Theta t_\Xi}p^{\Psi\Omega}t_\Omega
\end{equation}
from which the simplicity condition will imply. Since the tensor $R_{\Lambda\Psi\Phi\Omega}=p_{\Lambda\Psi}p_{\Phi\Omega}$ satisfies the conditions of the classification theorem, the formula (\ref{e10.7}) is a direct consequence of the Bianchi identity.

2). In the condition of the first paragraph, we add the condition $p^{\Lambda\Psi}p_{\Lambda\Psi}=0$ which takes the form $p_L{}^Kp_K{}^L=0$. From here, the existence of the nilpotent operator with index equal to 2 will follow.
\end{proof}
We construct the analogue of the differential Bianchi identity
\begin{equation}
\nabla_{\left[\right.\Lambda}R_{\Psi\Phi\left.\right]\Theta\Omega}=0
\end{equation}
under the covariant constancy of the connecting operators. For this purpose, we will contract it with the expression $\eta^\Lambda{}_{AB}A^{\Psi\Phi}{}_C{}^D A^{\Theta\Omega}{}_K{}^L$
\begin{equation}
\begin{array}{c}
\eta^\Lambda{}_{AB}A^{\Psi\Phi}{}_C{}^D(\eta_{\Lambda}{}^{MN} A_{\Psi\Phi}{}_P{}^Q-\eta_{\Psi}{}^{MN} \eta_{\Lambda}{}^{XQ}\eta_\Phi{}_{XP})\nabla_{MN}R_Q{}^P{}_K{}^L=\\
=-\frac{N^2}{16}\nabla_{AB}R_C{}^D{}_K{}^L-\frac{N}{8}(\varepsilon_{YCXP}\varepsilon_{AB}{}^{XQ}\nabla^{YD}-
\varepsilon^{YD}{}_{XP}\varepsilon_{AB}{}^{XQ}\nabla_{YC})R_Q{}^P{}_K{}^L=0\spsd
\end{array}
\end{equation}
This will lead to the differential spinor Bianchi identity
\begin{equation}
\label{e10.8}
\nabla_{AB}R_C{}^D{}_K{}^L=\frac{4}{N}(\varepsilon^{YD}{}_{XP}\varepsilon_{AB}{}^{XQ}\nabla_{YC}-\frac{1}{2}\delta_C{}^D\varepsilon_{AB}{}^{XQ}\nabla_{XP})R_Q{}^P{}_K{}^L\spsd
\end{equation}

\section{Twistor equation}
\Abstract{
$\phantom{ff}$In this section, it is told about how to construct and to solve the n-dimensional twistor equation, and then to investigate its properties. The conclusion of all results of this section is made on the basis \cite[v. 1, pp. 352-357, v. 2, pp. 43-46, p. 463 (eng)]{Penrouz1}.
}

We define the twistor equation as
\begin{equation}
\label{e11.0}
\eta_\Lambda{}_{AB}\nabla_\Psi X^A+\eta_\Psi{}_{AB}\nabla_\Lambda X^A=\frac{2}{n}g_{\Lambda\Psi}\eta^\Phi{}_{AB}\nabla_\Phi X^A\spsd
\end{equation}
We contract it with $\eta^{\Psi}{}^{CB}$
\begin{equation}
\begin{array}{c}
\eta_\Lambda{}_{AB}\nabla^{CB} X^A+\frac{n}{2}\nabla_\Lambda X^C=\frac{2}{n}\eta_\Lambda{}^{CB}\nabla_{AB} X^A\sps\\
\nabla_\Lambda X^C-\eta_\Lambda{}^{CB}\nabla_{AB} X^A+\frac{n}{2}\nabla_\Lambda X^C=\frac{2}{n}\eta_\Lambda{}^{CB}\nabla_{AB} X^A\sps\\
\end{array}
\end{equation}
\begin{equation}
\label{e11.1}
\nabla_\Lambda X^C=\frac{2}{n}\eta_\Lambda{}^{CB}\nabla_{AB} X^A\spsd\\
\end{equation}
The integrability condition of the equation for an arbitrary $X^A$ have the form (the proof is given in Appendix)
\begin{equation}
\label{e11.2}
C_{\Phi\Psi\Lambda\Delta}=0
\end{equation}
which corresponds to a conformally flat space. Let $\Omega$ be an arbitrary scalar field. Consider the conformal rescaling of the metric
\begin{equation}
g_{\Lambda\Psi}\rightarrow\hat g_{\Lambda\Psi}=\Omega g_{\Lambda\Psi}\spsd
\end{equation}
Therefore, we can set
\begin{equation}
\hat \eta_\Lambda{}^{AB}:=\eta_\Lambda{}^{AB}\sps
\hat \eta_\Lambda{}_{AB}:=\Omega \eta_\Lambda{}_{AB}\spsd
\end{equation}
Let's demand the performance of
\begin{equation}
\nabla_\Lambda\eta_\Psi{}^{AB}=0\sps\nabla_\Lambda\eta_\Psi{}_{AB}=0\sps
\hat \nabla_\Lambda\hat\eta_\Psi{}^{AB}=0\sps\hat\nabla_\Lambda\hat\eta_\Psi{}_{AB}=0\spsd
\end{equation}
This will lead to the system
\begin{equation}
\begin{array}{c}
\left\{
\begin{array}{l}
(\hat\nabla_\Lambda -\nabla_\Lambda)\eta_\Psi{}^{AB}=0\sps\\
(\hat\nabla_\Lambda -\nabla_\Lambda)\eta_\Psi{}_{AB}=-(\Omega^{-1}\nabla_\Lambda \Omega)\eta_\Psi{}_{AB}\sps\\
\end{array}
\right.\\ \\
\left\{
\begin{array}{l}
Q_{\Lambda\Psi}{}^\Theta\eta_\Theta{}^{AB}=Q_{\Lambda K}{}^A\eta_\Psi{}^{KB}+Q_{\Lambda K}{}^B\eta_\Psi{}^{AK}\sps\\
Q_{\Lambda\Psi}{}^\Theta\eta_\Theta{}_{AB}=-Q_{\Lambda A}{}^K\eta_\Psi{}_{KB}-Q_{\Lambda B}{}^K\eta_\Psi{}_{AK}+(\Omega^{-1}\nabla_\Lambda \Omega)\eta_\Psi{}_{AB}\sps\\
\end{array}
\right.
\end{array}
\end{equation}
where $Q_{\Lambda\Psi}{}^\Theta$ is \emph{strain tensor}. Then
\begin{equation}
\begin{array}{c}
\left\{
\begin{array}{l}
Q_{\Lambda(\Psi\Omega)}=\frac{2}{N}g_{\Psi\Omega}Q_{\Lambda K}{}^K\sps\\
Q_{\Lambda(\Psi\Omega)}=-\frac{2}{N}g_{\Psi\Omega}Q_{\Lambda K}{}^K+(\Omega^{-1}\nabla_\Lambda \Omega)g_{\Psi\Omega}\sps\\
\end{array}
\right.\\
\Upsilon_\Lambda:=\frac{1}{2}\Omega^{-1}\nabla_\Lambda \Omega\sps Q_{\Lambda(\Psi\Omega)}=g_{\Psi\Omega}\Upsilon_\Lambda\sps
Q_{\Lambda K}{}^K=\frac{N}{2}\Upsilon_\Lambda\sps\\
\tilde Q_{\Lambda\Psi\Omega}:=Q_{\Lambda[\Psi\Omega]}\sps \tilde Q_{\Lambda K}{}^A:=Q_{\Lambda K}{}^A-\frac{1}{N}Q_{\Lambda L}{}^L\delta_K{}^A\spsd
\end{array}
\end{equation}
Therefore,
\begin{equation}
\tilde Q_{\Lambda\Psi\Omega}=\frac{8}{N}A_{\Psi\Omega A}{}^K\tilde Q_{\Lambda K}{}^A\spsd
\end{equation}
If we require the preservation of the twistor equation form
\begin{equation}
\left\{

\right.
\hat X^C:=X^C\sps
\end{equation}
then it would impose the condition
\begin{equation}
Q_{\Lambda K}{}^N=\frac{2}{n}\eta_\Lambda{}^{CN}\eta_\Psi{}_{CB}Q^\Psi{}_K{}^B\sps
\end{equation}
on the strain tensor. Then
\begin{equation}
%
\end{equation}
Whence, we definitively obtain
\begin{equation}
Q_{[\Lambda\Phi\Theta]}=0\sps
Q_{\Lambda[\Phi\Theta]}=\Upsilon_\Phi g_{\Lambda\Theta}-\Upsilon_\Theta g_{\Lambda\Phi}\sps
Q_{\Lambda\Phi\Theta}=\Upsilon_\Lambda g_{\Phi\Theta}+\Upsilon_\Phi g_{\Lambda\Theta}-\Upsilon_\Theta g_{\Lambda\Phi}\spsd
\end{equation}
Try to solve the twistor equation in the flat space ($R_{\Lambda\Psi K}{}^A X^K=\nabla_{\left[\right.\Lambda}\nabla_{\Psi\left.\right]} X^A=0$) under the covariant constancy condition of the connecting operators ($\nabla_\Lambda \eta_\Psi{}^{CB}=0$)
\begin{equation}
\begin{array}{c}
\nabla_\Lambda X^C=\frac{2}{n}\eta_\Lambda{}^{CB}\nabla_{AB} X^A\sps\\
\nabla_\Psi\nabla_\Lambda X^C=\frac{2}{n}\eta_\Lambda{}^{CB}\nabla_{AB} (\nabla_\Psi X^A)\sps\\
\nabla_\Psi\nabla_\Lambda X^C=\frac{2}{n}\eta_\Lambda{}^{CB}\nabla_{AB} (\frac{2}{n}\eta_\Psi{}^{AK}\nabla_{LK} X^L)\sps\\
\nabla_{\left(\right.\Psi}\nabla_{\Lambda\left.\right)} X^C=
\frac{4}{n^2}(\eta_{\left(\right.\Lambda}{}^{CK}\nabla_{\Psi\left.\right)}-\eta_{\left(\right.\Lambda}{}^{CB}\eta_{\Psi\left.\right)}{}_{AB}\nabla^{AK})\nabla_{LK} X^L\sps\\
(1-\frac{2}{n})\nabla_{\left(\right.\Psi}\nabla_{\Lambda\left.\right)} X^C=-\frac{2}{n^2}g_{\Lambda\Psi}\nabla^{CK}\nabla_{LK} X^L=
-\frac{1}{n}g_{\Lambda\Psi}\nabla^\Omega\nabla_\Omega X^C\sps\\
\nabla_\Psi\nabla_\Lambda X^C=0\spsd
\end{array}
\end{equation}
Thus, $\nabla_\Lambda X^C$ is a constant and can be represented as
\begin{equation}
\label{e11.4}
\nabla_\Lambda X^C:=i\eta_\Lambda{}^{CA}\dot Y_A\sps
\end{equation}
where as $\dot Y_A\sps\dot X^C$ denote the constant spinor fields. Integrating this equation, we obtain
\begin{equation}
\label{e11.5}
X^C:=\dot X^C+iR^{CA}\dot Y_A\spsd
\end{equation}
We are interested in the case, where $X^C=0$. Omitting the terms above the spinors, we obtain the following relation
\begin{equation}
\label{e11.3}
X^C=-iR^{CA} Y_A\spsd
\end{equation}
From the geometric point of view, the equation (\ref{e11.4}), rewritten in terms of the operators $\gamma_\Lambda$, is anything other as the derivational equation \cite[eq. (1.2)]{Newfield2} of the normalized Grassmannian,  and (\ref{e11.5}) is the equation \cite[eq. (2.6)]{Newfield2}. Such the normalization is called {\it spinor normalization}. For n = 6, it was constructed in \cite{Andreev7}.

We return to the twistor equation (\ref{e11.0}), but use the covariant constancy operators (\ref{e9.10}) constructed in the previous section. Then for the real representation, the twistor equation (\ref{e11.4}) takes the form
\begin{equation}
\nabla_\myf\Lambda X^\myff C:=i\eta_\myf\Lambda{}^{\myff C\myff A}\dot Y_\myff A=:P_\myf\Lambda{}^\myff C\sps
\end{equation}
and for the complex representation,
\begin{equation}
\begin{array}{c}
\nabla_\Lambda (X_K)^C=i(\eta_K)_\Lambda{}^{C\myff A}\dot Y_\myff A=:(P^*_K)_\Lambda{}^C\sps\\
(P_K)_\Lambda{}_A\nabla_\Psi (X_K)^A+(P_K)_\Psi{}_A\nabla_\Lambda (X_K)^A=\frac{2}{n}g_{\Lambda\Psi}(P_K)^\Phi{}_A\nabla_\Phi (X_K)^A\spsd
\end{array}
\end{equation}
Carrying out the summation on K and defining $x_\Lambda:=\sum\limits_{K=1}^{2N}(P_K)_\Lambda{}_A(X_K)^A$, we obtain the conformal Killing equation
$$
\nabla_\Lambda x_\Psi+\nabla_\Psi x_\Lambda=\frac{2}{n}g_{\Lambda\Psi}\nabla_\Phi x^\Phi\spsd
$$

\section{Spinor formalism for n=6 and n=8}
\Abstract{
$\phantom{ff}$In this section, it is told about how to construct the spinor formalism for the small dimensions. For clarity, we construct various geometric interpretations of some algebraic relations. Presented, how to pass to the structure constants of the octonion algebra with the help of the Cartan triality principle. The conclusion of all results of this section is made on the basis \cite{Andreev1} and the literature to it,  \cite{Buchdahl1}, \cite{Dietmar1}, \cite{Klotz1}, \cite{Stepanovskii1}, \cite{Stepanovskii2}, \cite{Schouten1}, \cite{Scharnhorst1}.
}

\subsection{General isomorphisms}
\Abstract{
$\phantom{ff}$In this subsection, it is told about how to construct the basic isomorphisms of the spinor formalism for n = 6. The conclusion of all results of this subsection is made on the basis \cite{Andreev1}, \cite{Andreev3}.
}

\begin{enumerate}
\item $\mathbb C^6 \cong \Lambda^2\mathbb C^4$.\\
Let $\alpha,\beta, ...=\overline{1,6}$, $a,b,a_1,b_1,...=\overline{1,4}$. Then from (\ref{e6.5}), the identities
\begin{equation}
    \frac{1}{2}\eta^\alpha{}_{aa_1}\eta_\beta{}^{aa_1}=\delta_\alpha{}^\beta\sps
    \eta^\alpha{}_{aa_1}\eta_\alpha{}^{bb_1}=\delta_{aa_1}{}^{bb_1}:=
    2\delta_{\left[\right.a}{}^{\left[\right.b}
    \delta_{a_1\left.\right]}{}^{b_1\left.\right]}\sps
\end{equation}
\begin{equation}
\label{e12.1}
    r^{\alpha}=1/2\cdot \eta^{\alpha}{}_{aa_1}R^{aa_1}\sps
    R^{aa_1}=\eta_{\alpha}{}^{aa_1}r^{\alpha}\sps
\end{equation}
\begin{equation}
    \begin{array}{c}
    g^{\alpha\beta}=1/4\cdot \eta^{\alpha}{}_{aa_1} \eta^{\beta}{}_{bb_1}
    \varepsilon^{aa_1bb_1}\sps
    \varepsilon^{aa_1bb_1}=
    \eta_{\alpha}{}^{aa_1} \eta_{\beta}{}^{bb_1}g^{\alpha\beta}\sps\\[2ex]
    g_{\alpha\beta}=1/4\cdot \eta_{\alpha}{}^{aa_1} \eta_{\beta}{}^{bb_1}
    \varepsilon_{aa_1bb_1}\sps
    \varepsilon_{aa_1bb_1}=
    \eta^{\alpha}{}_{aa_1} \eta^{\beta}{}_{bb_1}g_{\alpha\beta}
    \end{array}
\end{equation}
will follow. This defines the isomorphism between the space $\mathbb C^6$ and the bivector space $\Lambda^2\mathbb C^4$, where $g_{\alpha\beta}$ is the metric tensor of the space $\mathbb C^6$.
\item $SO(6,\mathbb C)\cong SL(4,\mathbb C)/\pm 1$.\\
According to (\ref{e3.15}), the special orthogonal transformation can be represented as
\begin{equation}
S_\alpha{}^\beta\eta_\beta{}^{aa_1}=\eta_\alpha{}^{bb_1} S_b{}^a S_{b_1}{}^{a_1}\spsd
\end{equation}
In this case, $\tilde E_a{}^b$ from (\ref{e6.3a}) is the identity transformation multiplied by the imaginary unit. For non-special transformations, we have the identity
\begin{equation}
S_\alpha{}^\beta\eta_\beta{}^{aa_1}=\eta_\alpha{}_{bb_1} S^{ba} S^{b_1a_1}\spsd
\end{equation}

\item $so(6,\mathbb C)\cong sl(4,\mathbb C)$.\\
Define
\begin{equation}
       A_{\alpha\beta d}{}^c=
       \eta_{\left[\right.\alpha}{}^{ca}
       \eta_{\beta\left.\right]}{}_{da}\spsd
\end{equation}
Then by analogy with (\ref{e8.3}) and (\ref{e10.1}), we obtain
\begin{equation}
\label{e12.2}
    T_{\alpha \beta}= A_{\alpha \beta d}{}^cT_c{}^d \sps
    T_k{}^k=0 \sps T_{\alpha \beta}=-T_{\alpha \beta}\sps
\end{equation}
\begin{equation}
\label{e12.3}

\end{equation}
\end{enumerate}

In addition, the main identity for the 4-vector
\begin{equation}
\label{e12.4}
%
\end{equation}
and the main identity for the 6-vector
\begin{equation}
\label{e12.5}
%
\end{equation}
are executed. The proof of these identities is given in Appendix.

\subsection{On the classification of tensors with the symmetries of the Riemann curvature tensor for n=6. Curvature spinors}
\Abstract{
$\phantom{ff}$In this subsection, it is told about how to associate tensors possessing the symmetries of the curvature tensor to their spinor analogues. This simplifies the classification of such the tensors for n=6. The conclusion of all results of this subsection is made on the basis \cite{Andreev2}.
}

\begin{theorem}
\label{t12.1}
The classification of a bitensor, possessing the properties
\begin{equation}
    R_{\alpha \beta \gamma \delta}=R_{\left[ \alpha \beta \right]
    \left[\gamma \delta] \right.} \sps
    R_{\alpha \beta \gamma \delta}=R_{\gamma \delta \alpha \beta}\sps
    R_{\alpha \beta \gamma \delta}+R_{\alpha \delta \beta \gamma}+
    R_{\alpha \gamma \delta \beta}=0
\end{equation}
and belonging to the tangent bundle $\tau(\mathbb CV_6)$ over the six-dimensional analytic Riemannian space $\mathbb CV_6$, can be reduced to the classification of a tensor $R_a{}^b{}_c{}^d$ of the 4-dimensional complex vector space
\begin{equation}
    R_{\alpha \beta \gamma \delta}= A_{\alpha \beta d}{}^c
    A_{\gamma \delta r}{}^sR_c{}^d{}_s{}^r\spsd
\end{equation}
In addition, the following relations
\begin{equation}
    R_k{}^k{}_s{}^r=R_s{}^r{}_k{}^k=0 \sps  R_c{}^d{}_s{}^r=
    R_s{}^r{}_c{}^d
\end{equation}
are executed. The decomposition
\begin{equation}
    R_c{}^d{}_s{}^r=C_c{}^d{}_s{}^r-P_{cs}{}^{dr}-\frac{1}{40}\cdot
    R(3\delta_s{}^d \delta_c{}^r-2\delta_s{}^r \delta_c{}^d)
\end{equation}
corresponds to the decomposition
\begin{equation}
    R_{\alpha\beta}{}^{\gamma\delta}=C_{\alpha\beta}{}^{\gamma\delta}+
    R_{\left[\alpha \right.}{}^{\left[ \gamma \right.}
    g_{\left.\beta \right]}{}^{\left. \delta \right]}-
    1/10Rg_{\left[\alpha \right.}{}^{\left[ \gamma \right.}
    g_{\left.\beta \right]}{}^{\left. \delta \right]}
\end{equation}
into irreducible components not resulted by an orthogonal transformations. These components will satisfy the following relations
\begin{equation}
    P_{cs}{}^{rd}=-4(R_{\left[c \right.}{}^{\left[r \right.}
    {}_{\left. s \right]}{}^{\left. d\right]}+
    R_k{}^{\left[r \right.}{}_{\left[c \right.}{}^{\left|k\right|}
    \delta_{\left. s\right]}{}^{\left.d\right]})\sps
\end{equation}
\begin{equation}
    C_c{}^d{}_s{}^r=R_{\left(c \right.}{}^{\left(d \right.}{}_
    {\left. s\right)}{}^{\left. r\right)}+
    \frac{1}{40}\cdot R\delta_{\left(s \right.}{}^d\delta_{\left. c\right)}{}^r\sps
    C_c{}^d{}_s{}^r=C_{\left(c \right.}{}^{\left(d \right.}{}_
    {\left. s\right)}{}^{\left.r\right)}\sps
\end{equation}
\begin{equation}
    R=R_\beta{}^\beta=-2\cdot R_k{}^r{}_r{}^k\sps
    P_{kc}{}^{kd}=1/2\cdot R\delta_c{}^d\sps
\end{equation}
\begin{equation}
    R_l{}^d{}_s{}^l=-\frac{1}{8}\cdot R\delta_s{}^d
\end{equation}
the last of which is equivalent of the Bianchi identity.\\
\end{theorem}
\begin{proof}
We have the following equality
\begin{equation}
    R_{\alpha\beta\gamma\delta}=1/16\cdot\eta_{\alpha}{}^{aa_1}
    \eta_{\beta}{}^{bb_1}\eta_{\gamma}{}^{cc_1}
    \eta_{\delta}{}^{dd_1}R_{aa_1bb_1cc_1dd_1}\spsd
\end{equation}
Set
\begin{equation}
    R_c{}^d{}_s{}^r:=\frac{1}{4}R_{ck}{}^{dk}{}_{st}{}^{rt}
    \sps  R_{\beta\gamma}=\frac{1}{4}\eta_{\beta}{}^{cs}\eta_{\gamma}{}_{rd}
    \cdot P_{cs}{}^{rd}\spsd
\end{equation}
From this, the parties
\begin{equation}
\begin{array}{c}
       R_{\alpha \beta \gamma \delta}= A_{\alpha \beta d}{}^c
       A_{\gamma \delta r}{}^sR_c{}^d{}_s{}^r\sps\\
\end{array}%
\end{equation}
will follow. Set
\begin{equation}
       %
\end{equation}
then
\begin{equation}
       R_{\beta\delta}=\frac{1}{4}\eta_\beta{}^{cs}\eta_{\delta rd}P_{cs}{}^{rd}\spsd
\end{equation}
Therefore, the scalar curvature has the form
\begin{equation}
       %
\end{equation}
and the condition
\begin{equation}
       %
\end{equation}
is satisfied. The Bianchi identity can be rewritten as
\begin{equation}
    (A_{\alpha\beta d}{}^cA_{\gamma\delta}{}_r{}^s+
    A_{\alpha\gamma d}{}^cA_{\delta\beta}{}_r{}^s+
    A_{\alpha\delta d}{}^cA_{\beta\gamma}{}_r{}^s)\cdot
    R_c{}^d{}_s{}^r=0\spsd
\end{equation}
Contracting this equation with $A^{\alpha\beta}{}_t{}^lA^{\gamma\delta}{}_m{}^n$, we find that
\begin{equation}
    4R_k{}^l{}_m{}^k\delta_t{}^n+
    4R_r{}^n{}_t{}^r\delta_m{}^l-
    2R_k{}^l{}_t{}^k\delta_m{}^n-
    2R_k{}^n{}_m{}^k\delta_t{}^l-
    2R_k{}^r{}_r{}^k\delta_t{}^n\delta_m{}^l+
    R_r{}^k{}_k{}^r\delta_m{}^n\delta_t{}^l=0\spsd
\end{equation}
The contraction of this equation with $\delta_n{}^t$ leads us to the spinor analog of the Bianchi identity. In this case, all 15 significant equations reserved. Let
\begin{equation}
    C_{\alpha\beta}{}^{\gamma\delta}:=
    A_{\alpha\beta d}{}^cA^{\gamma\delta}{}_r{}^sC_c{}^d{}_s{}^r\sps
\end{equation}
then
\begin{equation}
    C_{\alpha\beta}{}^{\gamma\delta}:=
    R_{\alpha\beta}{}^{\gamma\delta}-
    R_{\left[\alpha \right.}{}^{\left[ \gamma \right.}
    g_{\left.\beta \right]}{}^{\left. \delta \right]}+
    1/10Rg_{\left[\alpha \right.}{}^{\left[ \gamma \right.}
    g_{\left.\beta \right]}{}^{\left. \delta \right]}\sps\\
\end{equation}
\begin{equation}
    R_{\left[\alpha \right.}{}^{\left[ \gamma \right.}
    g_{\left.\beta \right]}{}^{\left. \delta \right]}=
    A_{\alpha\beta d}{}^cA^{\gamma\delta}{}_r{}^s
    \cdot\frac{1}{4}(P_{sc}{}^{dr}-1/2R\delta_s{}^d\delta_c{}^r+
    \frac{1}{4}R\delta_s{}^r\delta_c{}^d)\sps
\end{equation}
\begin{equation}
    g_{\left[\alpha \right.}{}^{\left[ \gamma \right.}
    g_{\left.\beta \right]}{}^{\left. \delta \right]}=
    A_{\alpha\beta d}{}^cA^{\gamma\delta}{}_r{}^s
    \cdot\frac{1}{4}(1/2\delta_s{}^r\delta_c{}^d-2\delta_s{}^d\delta_c{}^r)
\end{equation}
from which we obtain the decomposition into irreducible components not resulted by orthogonal transformations. All calculations are given in Appendix.
\end{proof}

\begin{note}
At n=6, there are the two classification schemes of the Weyl spinor:
\begin{equation}
\begin{array}{cccc}
1.& C_k{}^l{}_m{}^n\phi_l{}^k=\lambda \phi_t{}^n\sps &
2.& C_k{}^l{}_m{}^n\phi^{km}=\lambda \phi^{ln}\spsd
\end{array}
\end{equation}
\end{note}

\begin{corollary}
\label{c12.1}
\begin{enumerate}
    \item
The simplicity conditions of a bivector of the 6-dimensional space $\mathbb C^6$ can be written down as
\begin{equation}
    p_{\left[\alpha\beta\right.}p_{\gamma\beta\left.\right]}=0\spsd
\end{equation}
The coordinates of the bivector can be associated to the traceless complex matrix $4 \times 4$
\begin{equation}
    p_l{}^dp_s{}^l-1/4(p_l{}^kp_k{}^l)\delta_s{}^d=0\spsd
\end{equation}
    \item
A simple bivector of the space $\mathbb C^ 6$, constructed on isotropic vectors ($p^{\alpha \beta} p_{\alpha \beta}=0$), can be associated to a degenerate Rosenfeld null-pair: a covector and a vector of the space $\mathbb C^4$, the contraction of which is zero. In this case, the vector and the covector are determined to within a complex factor.
\end{enumerate}
\end{corollary}
\begin{proof}$ $\\
The proof is similar to the one for Corollary \ref{c10.1}. The only difference lies in the fact that isotropic vectors $r^\alpha$ of the space $\mathbb C^6$ are represented as $r^\alpha\eta_\alpha{}^{ab}=R^{ab}=X^aY^b-X^bY^a$. Such the representation is possible due to the perform of the identities for the isotropic vector $r^\alpha$ ($r^\alpha r_\alpha=0$)
\begin{equation}
24R^{\left[\right.ab}R^{cd\left.\right]}=\varepsilon_{klmn}R^{kl}R^{mn}\varepsilon^{abcd}=4(r^\alpha r_\alpha)\varepsilon^{abcd}=0\sps
\end{equation}
where $\varepsilon_{abcd}$ is a spin-tensor of the form (\ref{e6.5}). Moreover, the spin-tensor is antisymmetric on all 4 indices and satisfies to the parities
\begin{equation}
\begin{array}{c}
\varepsilon_{abcd}\varepsilon^{klmn}=24\delta_{\left[\right.a}{}^k\delta_b{}^l\delta_c{}^m\delta_{n\left.\right]}{}^d\sps
\varepsilon_{abcd}\varepsilon^{almn}=6\delta_{\left[\right.b}{}^l\delta_c{}^m\delta_{n\left.\right]}{}^d\sps
\varepsilon_{abcd}\varepsilon^{abmn}=4\delta_{\left[\right.c}{}^m\delta_{n\left.\right]}{}^d\sps\\
\varepsilon_{abcd}\varepsilon^{abcn}=6\delta_d{}^n\sps
\varepsilon_{abcd}\varepsilon^{abcd}=24\spsd
\end{array}
\end{equation}
Set $x^\alpha\eta_\alpha{}^{ab}:=X^aY^b-Y^aX^b$, $y^\alpha\eta_\alpha{}^{ab}:=Z^aT^b-T^aZ^b$. From the condition $p^{\alpha\beta}p_{\alpha\beta}=0$, the relation $x^\alpha y_\alpha=0$ will follow. This means that $\varepsilon_{abcd}X^aY^bZ^cT^d=0$.
Therefore, the vectors $X^a\sps Y^b\sps Z^c\sps T^d$ are linearly dependent. Set $T^a:=\alpha X^a+\beta Y^a+\gamma Z^a$  and obtain
\begin{equation}
\begin{array}{c}
p_a{}^b:=\frac{1}{2}A_{\alpha\beta}{}_a{}^b p_{\alpha\beta}=\frac{1}{4}\eta_\alpha{}^{kb}\eta_\beta{}^{cd}\varepsilon_{cdka}(x^\alpha y^\beta-y^\alpha x^\beta)=\\
=\frac{1}{4}((X^kY^b-X^bY^k)Z^cT^d-X^cY^d(Z^kT^b-T^kZ^b))\varepsilon_{cdka}=\\
=Z^c(\beta Y^d)X^k\varepsilon_{cdka}Y^b-Z^c(\alpha X^d)Y^k\varepsilon_{cdka}X^b-X^cY^dZ^k\varepsilon_{cdka}(\alpha X^b+\beta Y^b+\gamma Z^b)+\\
+X^cY^d(\gamma Z^k)\varepsilon_{cdka}Z^b=\underbrace{X^cY^dZ^k\varepsilon_{cdka}}_{:=M_a}\underbrace{(-2\alpha X^b-2\beta Y^b)}_{:=N^b}=M_a N^b\spsd
\end{array}
\end{equation}
In this case, $N^b $ and $M_a$ are defined up to transformations
\begin{equation}
    N^b\longmapsto e^\phi N^b\sps
    M_a\longmapsto e^{-\phi} M_a\spsd
\end{equation}
\end{proof}
Note that the pair $(N^b,M_a) $ is \emph{Rosenfeld null-pair} \cite{Rozenfeld1}. In the space $\mathbb C\mathbb P^4=~'\mathbb C^4/'\mathbb C$
(where $'\mathbb C^s=\mathbb C^s/{0}$), $N^b$ defines the point and $M_a$ defines the plane with \emph{incidence condition} $M^b N_a=0$. Therefore, we can construct the space $\mathbb C\mbox{\foreignlanguage{russian}{П}}^4='\mathbb C^*{}^4/'\mathbb C$ which is \emph{dual space} to $\mathbb C\mathbb P^4$.
Then the space $\mathbb C\mathbb P^4\times \mathbb C\mbox{\foreignlanguage{russian}{П}}^4$ is \emph{Rosenfeld null-pair space}. It should be noted that such the spaces
studied for the first time  in \cite{Sintcov1} and  \cite{Kotelnikov1}. \\

Now we substitute $\varepsilon_{ab}{}^{cd}=2\delta_{\left[\right.a}{}^c\delta_{b\left.\right]}{}^d$ into (\ref{e10.8}). Then the differential Bianchi identity will become simpler to
\begin{equation}
    \nabla_{\left[cm\right.}R_{t\left.\right]}{}^k{}_r{}^s=\delta_{\left[m\right.}{}^k\nabla_{c\left|n\right|}R_{t\left.\right]}{}^n{}_r{}^s\spsd
\end{equation}
Contract this equation with $\delta_k{}^c$
\begin{equation}
    \nabla_{c\left(\right.m}R_{t\left.\right)}{}^c{}_r{}^s=0
\end{equation}
then contract with $\delta_s{}^m$
\begin{equation}
    \nabla_{cm}R_t{}^c{}_r{}^m=1/8\nabla_{rt}R
\end{equation}
which is the spinor analogue of the equation
\begin{equation}
    \nabla^\alpha(R_{\alpha\beta}-1/2Rg_{\alpha\beta})=0\spsd
\end{equation}

\subsection{\texorpdfstring{Geometric representation of a twistor in $\mathbb R^6_{(2,4)}$}{Geometric representation of a twistor}}
\Abstract{
$\phantom{ff}$In this subsection, it is told about how to construct a geometric interpretation of isotropic spinors (twistors) on the isotropic cone $\mathbb R^6_{(2,4)}$. This interpretation is similar to the spinor representation in $\mathbb R^4_{(1,3)}$ with that only a difference that the dimension of the flagpole and the flag-plane is incremented. The conclusion of all results of this subsection is made on the basis \cite{Andreev5}, \cite{Rozenfeld3}.
}

Let the metric of the space $\mathbb R^6_{(2,4)}$ have the form
\begin{equation}
    dS^2=dT^2+dV^2-dW^2-dX^2-dY^2-dZ^2
\end{equation}
and let the cross section of the light-cone $K_6$
\begin{equation}
    T^2+V^2-W^2-X^2-Y^2-Z^2=0
\end{equation}
 be set by the plane V+W=1. Let's consider the stereographic projection of this section on the plane (V=0,W=1) with the pole $N(0,\frac{1}{2},\frac{1}{2},0,0,0)$ so that the point P(T,V,W,X,Y,Z) corresponds to $p(t,0,1,x,y,z)$ in the plane (V=0,W=1). Then
\begin{equation}
    T/t=X/x=Y/y=Z/z=-\frac{(V-\frac{1}{2})}{\frac{1}{2}}\spsd
\end{equation}
We make the substitution
\begin{equation}
    \varsigma=ix-y\sps\omega=i(t+z)\sps\eta=i(t-z)
\end{equation}
and obtain
\begin{equation}
    \varsigma =\frac{-iX+Y}{2V-1}\sps\eta=\frac{-i(T+Z)}{2V-1}\sps\omega=\frac{i(Z-T)}{2V-1}\spsd
\end{equation}
Therefore, the metric, induced in the cross-section, has the form
\begin{equation}
    d s^2:=dT^2-dX^2-dY^2-dZ^2=-\frac{ d\varsigma d\bar\varsigma+d\omega d\eta}{(\varsigma\bar\varsigma+\eta\omega)^2}\spsd
\end{equation}
The proof of this fact is given in Appendix. Set
\begin{equation}
    X:=
    \left(

    \right)\spsd
\end{equation}
Then
\begin{equation}
    d s^2=-\frac{det(dX)}{(det(X))^2}\sps\bar X^T+X=0\spsd
\end{equation}
Consider the linear-fractional group L
\begin{equation}
    {\tilde X}=(AX+B)(CX+D)^{-1} \sps
    S:=
    \left(
    %
    \right) \sps
    detS=1\spsd
\end{equation}
The reality condition, imposed on X ($X^*+X=0)$, gives the linear-fractional unitary subgroup  $LU(2,2)$ so that the matrix S satisfies (see Appendix) the relation
\begin{equation}
    S^*\hat E S=\hat E\sps
    \hat E:=\left(
    %
    \right)\spsd
\end{equation}
Further, we define the special basis \cite[v. 2, p. 65, p. 305 (eng)]{Penrouz1}
\begin{equation}
    %
\end{equation}
This relations are remarkable because this shows as the bivector $R^{ab}$ is expressed in terms of its spinor components. Set
\begin{equation}
    %
    \right) \sps
\end{equation}
then
\begin{equation}
    \tilde S^*\tilde E\tilde S=\tilde E\spsd
\end{equation}
The matrixes S form the group isomorphic $SU(2,2)$ so the matrixes $\tilde S$ form the group $SU(2,2)$. A transformation from the group $LU(2,2)$ is called \emph{twistor transformation}. Due to the double covering of the connected identity component of the group $SO(2,4)$ (which is denoted as $SO^+(2,4)$) by the group $SU(2,2)$ and due to the double covering of the conformal group $C^{\uparrow 4}_+(1,3)$ \cite[v. 2, p. 304 (eng)] {Penrouz1} by the group $SO^+(2,4)$, the existence of the isomorphisms
\begin{equation}
    \begin{array}{c}
    SU(2,2)/\{\pm 1 ;\pm i\}\cong
    LU(2,2)\cong C^{\uparrow 4}_+(1,3) \cong
    SO^+(2,4)/\pm 1
    \end{array}
\end{equation}
will imply. This means that the group $LU(2,2)$ exhausts all conformal transformations of the group $\mathbb C^{\uparrow 4}_+(1,3)$. The matrix S is restored up to a factor $\lambda $ such that $\lambda^4=1$ (det(S)=1) from which we obtain the ambiguity. Since we have the equalities
\begin{equation}
    Y=AX+B\ \ \Rightarrow \ \ dX=AdY \sps
    Y=X^{-1}\ \ \Rightarrow \ \ dX=-X^{-1}dXX^{-1}\sps
\end{equation}
where A and B are some constant matrixes, then
\begin{equation}
    \tilde Z^*\ d\tilde X\ \tilde Z= Z^*\ dX\ Z\spsd
\end{equation}
This equation is an invariant under the group LU(2,2). The proof of this fact is considered in Appendix. Other invariant can be obtained with the help of  the identity
\begin{equation}
    Y=AX+B\ \ \Rightarrow \ \
    \frac{\partial}{\partial X}=A^T\frac{\partial}{\partial Y} \sps
    Y=X^{-1}\ \ \Rightarrow \ \ \frac{\partial}{\partial X}
    =-Y^T\frac{\partial}{\partial Y}Y^T
\end{equation}
where A and B are also some constant matrixes. This invariant will have the form
\begin{equation}
    \tilde Z^{-1}\frac{\partial}{\partial \tilde X^T}\tilde Z^{*\ -1}= Z^{-1}\frac{\partial}{\partial X^T} Z^{*\ -1}\spsd
\end{equation}
The proof can be found in Appendix. This means that there is a real vector $\tilde L$ tangent to the hyperboloid obtained in the cross-section of the cone $K_6$ by the plane V + W = 1. This vector is an invariant under transformations of a basis from the group LU(2,2) (i.e. coordinate-independent in the tangent space to this hyperboloid). The vector is uniquely determined by the matrix
\begin{equation}
    \begin{array}{c}
    \hat L:=\frac{1}{\sqrt{2}}
    (Z^{-1}\frac{\partial}{\partial X^T} Z^{*\ -1}-
    \bar Z^{-1}\frac{\partial}{\partial X} Z^{T\ -1})=\\ \\
    =\left(
    \begin{array}{cc}
    0  & 1 \\
    -1 & 0
    \end{array}
    \right)
    (\frac{\partial}{\partial\omega}(-\bar\eta^0\pi^0+\eta^0\bar\pi^0)+
    \frac{\partial}{\partial\eta}(-\bar\eta^1\pi^1+\eta^1\bar\pi^1)+\\ \\
    +\frac{\partial}{\partial\xi}(-\bar\eta^1\pi^0+\eta^0\bar\pi^1)+
    \frac{\partial}{\partial\bar\xi}(\bar\eta^0\pi^1-\eta^1\bar\pi^0))
    \cdot \frac{1}{(det(Z))^2 \sqrt{2}}:=
    \left(
    \begin{array}{cc}
    0  & 1 \\
    -1 & 0
    \end{array}
    \right)\tilde L
    \end{array}
\end{equation}
The norm of this vector in our metric will be such that
\begin{equation}
    \parallel\tilde L\parallel =-\frac{1}{2(det(Y))^2}=-\frac{1}{(V+W)^2}
\end{equation}
 An isotropic vector k is called a vector of \emph{first type unit extension} \cite[v. 1, p. 36, eq. (1.4.16) (eng)]{Penrouz1} in the case when k sets the point belonging to the cross-section of the isotropic cone by the plane V + W = 1. Then $\parallel \tilde L \parallel =- 1 $ and any isotropic vector K collinear with k is defined as
\begin{equation}
    K=(-\parallel\tilde L\parallel)^\frac{1}{2}k\spsd
\end{equation}
However, when V =- W we will obtain a vector with the infinite first type extension. To learn to distinguish between them, it is necessary to set the cross-section of the cone $K_6$ by the plane T+Z=1 and to enter a vector $\tilde{\tilde L}$ with the norm
\begin{equation}
    \parallel\tilde{\tilde L}\parallel =-\frac{1}{(T+Z)^2}
\end{equation}
 in the same way. An isotropic vector k we call a vector of \emph{second type unit extension} in the case when k is sets the point belonging to the cross-section of the isotropic cone by the plane T + Z = 1 and the first type extension  will not be finite. We define \emph{extension} of the vector K as
 \begin{enumerate}
     \item \emph{first type extension} if such the extension is finite;
     \item \emph{second type extension} if the first type extension is infinite.
\end{enumerate}
 Note that the vector $\tilde L $ is not coordinate-independent in the space $\mathbb R^6_{(2,4)}$ although it is an invariant of the tangent space to the hyperboloid obtained in the cross-section of the cone $K_6$ by the plane V+W=1. Now there is a possibility to represent \emph{isotropic twistor} in the space $\mathbb R^6{}_{(2,4)}$ visually. Consider a pair of vectors of equal extension in $\mathbb R^6{}_{(2,4)}$
\begin{equation}
K^\alpha=\eta^\alpha{}_{ab}\ iT^{\left[\right. a}X^{b\left.\right]}\sps
N^\alpha=\eta^\alpha{}_{ab}\ T^{\left[\right. a}Z^{b\left.\right]}\ \Rightarrow\
K^\alpha K_\alpha=0\sps N^\alpha K_\alpha=0\sps N^\alpha N_\alpha=0\spsd
\end{equation}
 We choose a vector $Y^a$ in such a way as to satisfy the conditions
\begin{equation}

\end{equation}
Thus, we obtain the vector basis $X^a, Y^a, Z^a, T^a$ (recall that $Y_a=s_{aa '}Y^{a'}$)
\begin{equation}
%
\end{equation}
Whence,
\begin{equation}
    \varepsilon_{abcd}T^{\left[\right.c}Y^{d\left.\right]}=-2X_{\left[\right.a}Z_{b\left.\right]}\spsd
\end{equation}
Therefore, the vectors
\begin{equation}
    L^\alpha=\eta^\alpha{}_{ab}(-T^{\left[\right.a}Y^{b\left.\right]}+
                                X^{\left[\right.a}Z^{b\left.\right]})\sps
    M^\alpha=\eta^\alpha{}_{ab}(-i)(T^{\left[\right.a}Y^{b\left.\right]}+
                                X^{\left[\right.a}Z^{b\left.\right]})
\end{equation}
satisfy the parities
\begin{equation}
%
\end{equation}
Now we can construct the threevector
\begin{equation}
    P^{\alpha\beta\gamma}=6K^{\left[\right.\alpha}
                           N^\beta L^{\gamma\left.\right]}\spsd
\end{equation}
Knowing $K^\alpha$, we know $T^a$ and $X^a$ up to
\begin{equation}
    X^a\longmapsto\lambda_1 X^a+\mu_1 T^a\sps
    T^a\longmapsto\nu_1 X^a+\xi_1 T^a\sps
    det\left(
    %
    \right)=1\spsd
\end{equation}
And if we know $N^\alpha$ then an arbitrariness in a choice of $T^a$ and $Z^a$ will be such that
\begin{equation}
    Z^a\longmapsto\lambda_2 Z^a+\mu_2 T^a\sps
    T^a\longmapsto\nu_2 Z^a+\xi_2  T^a\sps
    det\left(
    %
    \right)=1\spsd
\end{equation}
Therefor, $\nu_1=\nu_2=0$ and $\xi_2=\xi_1$. For $Y^a$, we obtain
\begin{equation}
    Y^a  \longmapsto  \alpha X^a+\beta Y^a+\gamma Z^a+\delta T^a\spsd
\end{equation}
If we now demand that two such bases are related by a transformation from the group LU(2,2) then we obtain
\begin{equation}
    %
\end{equation}
Thus, the 3-half-plane, spanned by $K^\alpha, N^\alpha, L^\alpha$ is a coordinate-independent in the space $\mathbb R^6_{(2,4)}$. Thus, our design can be presented as follows. The first type extension of the vectors $K^\alpha$ and $N^\alpha $ should be the same. $K^\alpha$ and $N^\alpha$ determine \emph{flagpole}: the set of vectors with
\begin{enumerate}
    \item the first type extension is equal to the first type extension of the vector $K^\alpha$;
    \item the start coinciding with the beginning of the vector $K^\alpha$.
\end{enumerate}
$K^\alpha$, $N^\alpha$, $L^\alpha$ determine the 3-half-plane which we call \emph{flag-plane}. Thus, knowing $K^\alpha$ and $N^\alpha$, we know the twistor $T^a$ up to the phase $\Theta$. In turn, in the 2-plane $(L^\alpha, M^\alpha)$, $2\Theta$ is an angle of the rotation of the flag (3-half-plane $P^{\alpha\beta\gamma}$) around the flagpole $(N^\alpha,K^\alpha)$. Therefore, a rotation of the flag on $2\pi$ will lead to the twistor $-T^a$, and only a rotation on $4\pi$ will return our design to the original state. In addition, collinear twistors can be distinguished from each other using the concept \emph{extension of the vector} for the $K^\alpha$ so that under the transformation $T^a\longmapsto rT^a,\ Y^a\longmapsto r^{-1}Y^a\ (r\in R\backslash \{0\})$, the flagpole is multiplied by r and the flag-plane remains unchanged. Finally, it should be noted that the mentioned geometrical structure is uniquely determined by the twistor $T^a$. In the case of the infinite first type extension of the vector $K^\alpha$, we consider the cone $K_4\subset K_6$ on which the vector $N^\alpha$ lays. But non-zero vectors $K^\alpha$ and $N^\alpha$ have the finite second type extension giving the geometric interpretation of a spinor on the isotropic cone $K_4$.

\subsection{Structure constants of the octonion algebra. Spinor formalism for n = 8}
\Abstract{
$\phantom{ff}$In this subsection, it is told about how on the connecting operators, satisfying the Clifford equation, to construct the structural constants of the octonion algebra. Such the algebra is the special case of the normalized division hypercomplex algebra for n = 8. We consider the axioms of such the algebra. The conclusion of all results of this subsection is made on the basis \cite{Andreev4}.
}

Let the operators $\eta_\alpha{}^{ab}\ (\alpha\ ,\beta\ , ...=\overline{1,6};a\ ,b\ , ...=\overline{1,4})$ be the antisymmetric operators constructed above for the spinor formalism for n = 6. And let the operators $\eta_\Lambda{}^{AB}\ (\Lambda\ ,\Psi\ , ...=\overline{1,8};A\ ,B\ , ...=\overline{1,8})$ for the spinor formalism for n = 8 be constructed according to the scheme (\ref{e6.6}). Then the tensor (\ref{e6.2}), (\ref{e6.5}) will have the symmetries
\begin{equation}
\label{e12.6}
\varepsilon_{AB(CD)}=\frac{1}{2}\varepsilon_{AB}\varepsilon_{CD}\sps
\varepsilon_{A(B|C|D)}=\frac{1}{2}\varepsilon_{AC}\varepsilon_{BD}\sps
\end{equation}
where according to (\ref{e6.3}), $\varepsilon_{AC}=\tilde \varepsilon_{AC}$ is the metric spin-tensor in the spinor space $\mathbb C^8$
\begin{equation}
\label{e12.6a}
\eta_\Lambda{}_{AB}=\frac{1}{4}\eta_\Lambda{}^{CD}\varepsilon_{ABCD}=\eta_\Lambda{}^{CD}\varepsilon_{AD}\varepsilon_{BC}\sps
\eta_\Lambda:=\frac{1}{4}\eta_\Lambda{}^{CD}\varepsilon_{CD}\spsd
\end{equation}
The second identity (\ref{e12.6}) is very important. It leads to the non-degeneracy operator $\varepsilon_{XA}{}^{YB}X^XX_Y$ and, as a result, to the normability of the octonion algebra. In this case, the identity (\ref{e13.5}) splits into three parts: the alternative identity (left and right)  and the flexible identity. As it happens.
\begin{theorem}
\label{t12.2}
For every $r^\Lambda \in \mathbb C^8$, there is the decomposition
\begin{equation}
\label{e12.7}
r^\Lambda=\eta^\Lambda{}_{AB}X^AY^B
\end{equation}
for some $X^A\sps Y^B\in \mathbb C^8$.
\end{theorem}
\begin{proof}
From the Clifford equation  (\ref{e6.1}), the parities
\begin{equation}
R^{AB}R_{CB}=\frac{1}{2}r^\Lambda r_\Lambda\delta_C{}^A\sps
R_{KL}\delta_A{}^B=R_{AR}\varepsilon_{KL}{}^{BR}+R^{BR}\varepsilon_{KL}{}_{AR}
\end{equation}
will follow. Contract this equation with $P^K P^A$ choosing $P^K$ so that $p:=\frac{1}{2}\varepsilon_{AK}P^K P^A\ne 0$ and assuming that $Q_L:=R_{KL}P^K$
\begin{equation}
Q_LP^B=Q_RP^K\varepsilon_{KL}{}^{BR}+pR^B{}_L\spsd
\end{equation}
Then from (\ref{e12.6}), the identities
\begin{equation}
R^{BL}=\frac{1}{p}Q^RP^K(\delta_R{}^L\delta_K{}^B-\varepsilon_K{}^{LB}{}_R)=\frac{1}{p}Q^RP^K\varepsilon^{BL}{}_{KR}\spsd
\end{equation}
will follow. Set
\begin{equation}
X^A:=\frac{1}{\sqrt p}P^A\sps Y^B:=\frac{1}{\sqrt p}Q^B
\end{equation}
that proves this theorem.
\end{proof}

Define
\begin{equation}
P^\Lambda{}_A:=\eta^\Lambda{}_{AB}X^A\sps X^AX_A:=2\sps
\end{equation}
then
\begin{equation}
P^\Lambda{}_AP^\Psi{}_B\varepsilon^{AB}=g^{\Lambda\Psi}\sps P^\Lambda{}_AP^\Psi{}_Bg_{\Lambda\Psi}=\varepsilon_{AB}\spsd
\end{equation}
Thus, the operator $P^\Lambda{}_A$ defines the isomorphism between the spaces $\mathbb C^n\cong \mathbb C^N\ (n=N=8)$
\begin{equation}
r^\Lambda=P^\Lambda{}_BY^B\spsd
\end{equation}
Set
\begin{equation}
\eta_\Lambda{}^{\Psi\Theta}:=\sqrt{2}\eta_\Lambda{}^{AB}P^\Psi{}_AP^\Theta{}_B\sps
\end{equation}
then for these structural constants the alternative octonion identities are executed
\begin{equation}
\label{e12.8}
\begin{array}{c}
\eta_{\Lambda\Theta}{}^\Phi\eta_{(\Psi\Upsilon)}{}^{\Theta}=
\eta_{\Lambda\left(\right.\Psi}{}^\Theta\eta_{|\Theta|\Upsilon\left.\right)}{}^\Phi\sps
\eta_{(\Lambda\Psi)}{}^\Theta\eta_{\Theta\Phi}{}^{\Upsilon}=
\eta_{\left(\right.\Lambda|\Theta|}{}^\Upsilon\eta_{\Psi\left.\right)\Phi}{}^\Theta
\end{array}
\end{equation}
and the central Moufang identity
\begin{equation}
\label{e12.9}
\begin{array}{c}
r^\Phi\eta_{\Phi\Theta}{}^\Omega\eta_{\Lambda\Psi}{}^{\Theta}\eta_{\Omega\Upsilon}{}^\Gamma r^\Upsilon=
r^\Phi\eta_{\Phi\Lambda}{}^\Theta\eta_{\Theta\Omega}{}^{\Gamma}\eta_{\Psi\Upsilon}{}^\Omega r^\Upsilon
\end{array}
\end{equation}
 is executed too. The proof is given in Appendix. Now, in order to enter the  octonion structure constants, we must use the inclusion operator $H_i{}^\Lambda:\ \mathbb R^6\subset\mathbb C^6$. Normalizability of the octonion algebra  will follow from the Clifford equation  (\ref{e6.1}) written out for the structural constants
\begin{equation}
\label{e12.9/1}
(\eta_{\Phi\Theta}{}^\Omega\eta_{\Psi\Xi}{}^\Delta+\eta_{\Psi\Theta}{}^\Omega\eta_{\Phi\Xi}{}^\Delta)g_{\Omega\Delta}=2g_{\Phi\Psi}g_{\Theta\Xi}\spsd
\end{equation}
The operators of the same type (\ref{e10.1})
\begin{equation}
\label{e12.10}
\begin{array}{c}
A_{\Theta\Phi L}{}^X:=\frac{1}{2}\eta_{\left[\right.\Theta}{}^{MX}\eta_{\Phi\left.\right]}{}_{ML}\sps
\tilde A_{\Theta\Phi L}{}^X:=\frac{1}{4}(\eta_{\left[\right.\Theta}{}^{MX}\eta_{\Phi\left.\right]}{}_{ML}+
\eta_{\left[\right.\Theta}{}^{XM}\eta_{\Phi\left.\right]}{}_{LM})
\end{array}
\end{equation}
will satisfy the identities
 \begin{equation}
\label{e12.11}
\begin{array}{cc}
A_{\Lambda\Psi}{}^{AB}A^{\Lambda\Psi}{}_{CD}=\delta_{\left[\right.C}{}^A\delta_{D\left.\right]}{}^B\sps
&\tilde A_{\Lambda\Psi}{}^{AB}\tilde A^{\Lambda\Psi}{}_{CD}=\frac{1}{2}\delta_{\left[\right.C}{}^A\delta_{D\left.\right]}{}^B+
\frac{1}{4}\varepsilon_{\left[\right.C}{}^A{}_{D\left.\right]}{}^B\sps\\
A_{\Lambda\Psi}{}^{AB}A^{\Upsilon\Omega}{}_{AB}=\delta_{\left[\right.\Lambda}{}^\Upsilon\delta_{\Psi\left.\right]}{}^\Omega\sps
&\tilde A_{\Lambda\Psi}{}^{AB}\tilde A^{\Upsilon\Omega}{}_{AB}=\delta_{\left[\right.\Lambda}{}^\Upsilon\delta_{\Psi\left.\right]}{}^\Omega+
\eta^{\left[\right.\Upsilon}\delta_{\left[\right.\Lambda}{}^{\Omega\left.\right]}\eta_{\Psi\left.\right]}\spsd\\
\end{array}
\end{equation}
The proof is given in Appendix.

\subsection{Geometry of the inductive step from n=6 to n=8}
\Abstract{
$\phantom{ff}$In this subsection, it is told about how to construct two quadrics associated to the Cartan triality principle \cite[p. 175 (rus)]{Cartan1}. The construction is carried out in the explicit form justifying the standard scheme of the induction transition from n = 6 to n = 8.  The conclusion of all results of this subsection is made on the basis \cite{Andreev6}, \cite{Cartan1}, \cite[v. 2]{Hodge1}.
}

\subsubsection{Rosenfeld null-pair}
\Abstract{
$\phantom{ff}$In this subsubsection, it is told about how to construct an explicit solution of the twistor equation for n=6.  The conclusion of all results of this subsubsection is made on the basis \cite{Andreev6}, \cite{Rozenfeld1}.
}

Denote by $\mathbb A^{\mathbb C*}$ the spinor 4-dimensional complex vector space. Such the space is dual to the space $\mathbb A^{\mathbb C}$. Then the 8-dimensional complex space $\mathbb T^2$ is formed as the direct sum $\mathbb A^{\mathbb C}\oplus\mathbb A^{\mathbb C*}$. That is, if $X^a$ $(a, b ,...=\overline{1,4})$ are the coordinates of a vector in $A^{\mathbb C}$, and $Y_b$ are the coordinates of a covector in $A^{\mathbb C*}$ then $X^A:=(X^a,Y_b)\sps (A,B,...=\overline{1,8})$ are the coordinates of a vector of $\mathbb T^2$. We will consider the bivector coordinates $r^{ab}$ (\ref{e11.3})
\begin{equation}
\label{e14.1}
    X^a=ir^{ab}Y_b
\end{equation}
as the ones in the complex affine space $\mathbb C \mathbb A_6$. This is a system of 4 linear equations with 6 unknowns.
To determine its rank, we consider \emph{homogeneous equation} $r^{ab} Y_b = 0$ which has nontrivial solutions if and only if the bivector is simple: $r_{kc}r^{ac}=\frac{1}{4}r^{ab}r_{ab}\delta_k{}^d=0$, and therefore can be represented as
\begin{equation}
     r^{ab}_{\mbox{\scriptsize{homogeneous}}}=P^aQ^b-P^bQ^a\sps
\end{equation}
 where $P^a$ and $Q^a$ are defined up to linear combinations of them. From this, it follows that $P^aY_a = 0 \sps Q^ aY_a = 0$. For $X^aY_a = 0$, by $X^a$, $S^a$, $Z^a$, we denote  all solutions, that form a basis. Then our solution takes the form
\begin{equation}
    r^{ab}_{\mbox{\scriptsize{homogeneous}}}=
    \lambda_1 S^{\left[\right.a} X^{b\left.\right]}+
    \lambda_2 X^{\left[\right.a} Z^{b\left.\right]}+
    \lambda_3 S^{\left[\right.a} Z^{b\left.\right]}
\end{equation}
and hence determines a 3-dimensional subspace in the bivector space. From here, \emph{common solution}
\begin{equation}
    r^{ab}=
    r^{ab}_{\mbox{\scriptsize{particular}}}+
    \lambda_1 S^{\left[\right.a} X^{b\left.\right]}+
    \lambda_2 X^{\left[\right.a} Z^{b\left.\right]}+
    \lambda_3 S^{\left[\right.a} Z^{b\left.\right]}
\end{equation}
is obtained, where $r^{ab}_{\mbox{\scriptsize{particular}}}$ is an arbitrary bivector being \emph{particular solution} of (\ref{e14.1}). \\

\subsubsection{\texorpdfstring{Construction of the quadrics $\mathbb CQ_6$ and $\mathbb C \tilde Q_6$}{Construction of the two quadrics}}
\Abstract{
$\phantom{ff}$In this subsubsection, it is told about how to construct a quadric satisfying the Cartan triality principle. The conclusion of all results of this subsubsection is made on the basis \cite{Andreev6}, \cite{Cartan1}, \cite[v. 2]{Hodge1}.
}

The space $\mathbb T^2$ will be a complex space in which the scalar square of a vector is determined by the quadratic form
\begin{equation}
\label{e14.2}
    \varepsilon_{AB}X^AX^B=2X^aY_a
\end{equation}
so that the matrix of the spin-tensor $\varepsilon_{AB}$ has the form
\begin{equation}
    \parallel \varepsilon_{AB}\parallel=\left(
    \begin{array}{cc}
    0 & \delta_a{}^c \\
    \delta^b{}_d  & 0
    \end{array}
    \right)\spsd
\end{equation}
For the fixed $r^{ab}$, the equation (\ref{e14.1}) will define the 4-dimensional subspace in $\mathbb T^2$ which will be the 4-dimensional planar generators of the cone $\varepsilon_{AB} X^AX^B=0$. Thus, in the projective space $\mathbb C \mathbb P_7$, we can consider the quadric $\mathbb CQ_6$ defined by the equation (\ref{e14.2}). 4 basis points of the generator satisfy the condition $\varepsilon_{AB} X^A_IX^B_J=0\sps (I,J,...=\overline{1,4})$. Set
\begin{equation}
    X^A_1:=(X^a,Y_b)\sps
    X^A_2:=(Z^a,T_b)\sps
    X^A_3:=(L^a,N_b)\sps
    X^A_4:=(K^a,M_b)\spsd
\end{equation}
On the basis of the common solution of the equation (\ref{e14.1}), each point of the quadric $\mathbb CQ_6$  can be associated to the 3-dimensional isotropic plane of the space $\mathbb C \mathbb A_6$. The point (t,v,w,x,y,z) of the space $\mathbb C \mathbb A_6$ can be represented by the line  $(\lambda T, \lambda V, \lambda U, $ $\lambda S, \lambda W, \lambda X, \lambda Y, \lambda Z)$ of the space $\mathbb C\mathbb R^8$ having the metric
\begin{equation}
    dL^2=dT^2+dV^2+dU^2+dS^2+dW^2+dX^2+dY^2+dZ^2\spsd
\end{equation}
This line will be a generator of the isotropic cone $\mathbb CK_8$. The intersection of the 7-plane $U-iS=1$ with the cone $\mathbb CK_8 $ has the induced metric
\begin{equation}
    d\tilde L^2=dT^2+dV^2+dW^2+dX^2+dY^2+dZ^2\spsd
\end{equation}
This space has the form of a paraboloid in $\mathbb CK_8$ and is identical to the space $\mathbb C\mathbb R^6$: $U=1+iS=\frac{1}{2}(1-T^2-V^2-W^2-X^2-Y^2-Z^2)$. Every generator of the cone (a set of points belonging to $\mathbb CK_8$ with the constant ratio  T:V:U:S:W:X:Y:Z), not lying on the hyperplane $U=iS$, intersects the paraboloid in the single point. Every generator of the cone, lying on the hyperplane $U=iS$, corresponds to the point belonging to the infinity of the space $\mathbb C\mathbb R^6$. Thus, straight lines  of $\mathbb C\mathbb R^8$, passing through the origin of $\mathbb C\mathbb R^8$, correspond to points of the projective space $\mathbb C \mathbb P_7$. The stereographic projection of this section on the plane (S=0, U=1) with the pole $N(0,0,\frac{1}{2},\frac{i}{2},0,0,0,0)$ maps the point P(T,V,U,S,W,X,Y,Z) of the hyperboloid to the point p(t,v,1,0,w,x,y,z) of the  plane (S=0, U=1)
\begin{equation}
    \begin{array}{c}
    \lambda T=t\sps
    \lambda V=v\sps
    \lambda W=w\sps
    \lambda X=x\sps
    \lambda Y=y\sps
    \lambda Z=z\sps
    \lambda =\frac{1}{U+iS}\sps\\
    r^\alpha r_\alpha=-\frac{U-iS}{U+iS}\sps
    \lambda U=\frac{1}{2}(1-t^2-v^2-w^2-x^2-y^2-z^2)=-\lambda iS+1\spsd\\
    \end{array}
\end{equation}
All generators of the same cone $\mathbb CK_8$ form the quadric $\mathbb C\tilde Q_6 $ in the projective space $\mathbb C \mathbb P_7$
\begin{equation}
    g_{\Lambda\Psi}r^\Lambda r^\Psi=0\spsd
\end{equation}

\subsection{\texorpdfstring{Correspondence  $\mathbb CQ_6\longmapsto \mathbb C\tilde Q_6$}{Correspondence: CQ to C'Q}}
\Abstract{
$\phantom{ff}$In this subsubsection, it is told about how to construct a correspondence between generators of the quadric of different ranks according to the Cartan triality principle.  The conclusion of all results of this subsubsection is made on the basis \cite{Andreev6}, \cite{Cartan1}, \cite[v. 2]{Hodge1}.
}

\begin{enumerate}
\item
The common solution of the equation (\ref{e14.1})
\begin{equation}
    r^{ab}=r^{ab}_{\mbox{\scriptsize{particular}}}+r^{ab}_{\mbox{\scriptsize{homogeneous}}}=
    r^{ab}_{\mbox{\scriptsize{particular}}}+
    \lambda_1 S^{\left[\right.a} X^{b\left.\right]}+
    \lambda_2 X^{\left[\right.a} Z^{b\left.\right]}+
    \lambda_3 S^{\left[\right.a} Z^{b\left.\right]}
\end{equation}
determines the 4-dimensional planar generator of the cone $\mathbb CK_8$. Then for such the generator, the system
\begin{equation}
\label{e14.3}
    \left\{
    \begin{array}{lcl}
    ir^{ab} Y_b & = &  X^a\sps\\
    ir^{ab} T_b & = &  Z^a\sps\\
    ir^{ab} N_b & = &  L^a\sps\\
    ir^{ab} M_b & = &  K^a
    \end{array}
    \right.
\end{equation}
will be determined with the conditions
\begin{equation}
    \begin{array}{c}
     X^a Y_a=0\sps
     Z^a T_a=0\sps
     L^a N_a=0\sps
     K^a M_a=0\sps\\ \\
     X^a T_a=- Z^a Y_a\sps
     X^a N_a=- L^a Y_a\sps
     X^a M_a=- K^a Y_a\sps\\ \\
     Z^a N_a=- L^a T_a\sps
     Z^a M_a=- K^a T_a\sps
     K^a N_a=- L^a M_a\spsd
    \end{array}
\end{equation}
Thus, from the 16 equations with the 6 unknowns $r^{ab}$ only 6 from them  will be significant equations (the 10 communication conditions) that defines the point of $\mathbb C \mathbb A_6$ and hence the point of the quadric $\mathbb C \tilde Q_6$.
\item
If we know only one equation
\begin{equation}
    ir^{ab} Y_b= X^a
\end{equation}
with the condition
\begin{equation}
    X^aY_a=0
\end{equation}
then from the 4 equations, only 3 from them will be significant equations (the 1 communication condition). This means that the point of the quadric $\mathbb CQ_6$  will uniquely define the 3-dimensional planar generator $\mathbb C \mathbb P_3$ belonging to the quadric $\mathbb C \tilde Q_6$.

\begin{figure}\center
\includegraphics[width=0.7\textwidth]{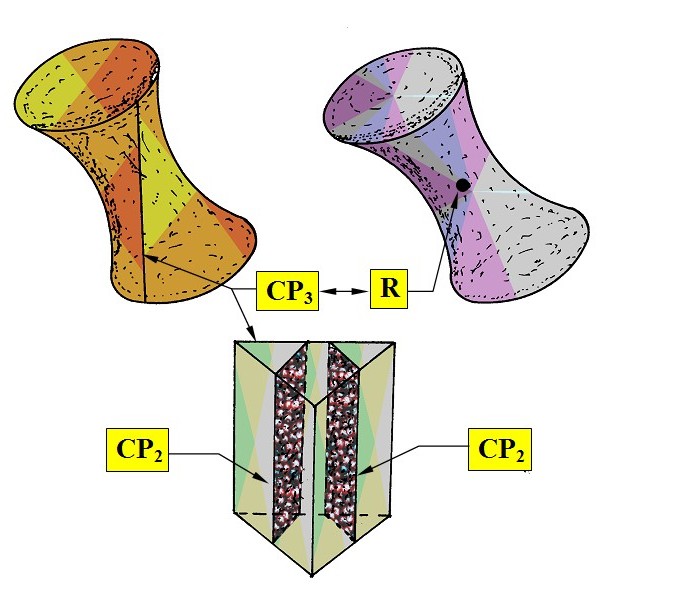}
\caption{Correspondence  $\forall CP_2\subset CP_3 \leftrightarrow R$}
\vspace{1cm}
\includegraphics[width=0.7\textwidth]{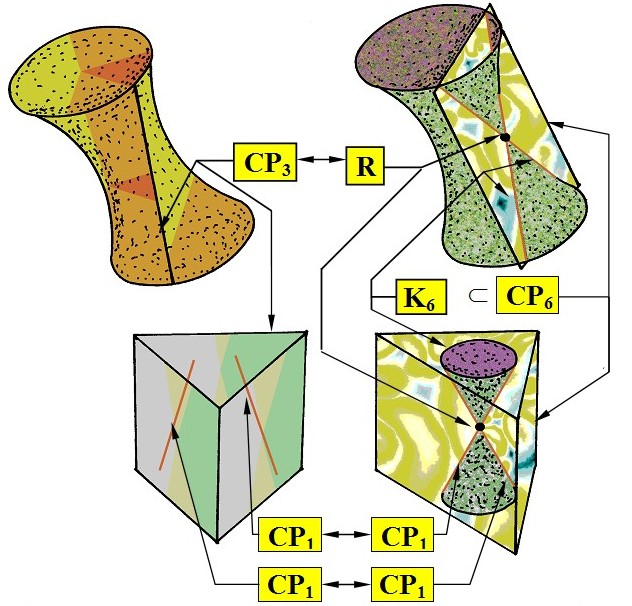}
\caption{Correspondence $CP_3\supset CP_1\leftrightarrow CP_1\subset K_6$}
\end{figure}

\item
If we know the two equations
\begin{equation}
    \left\{
    \begin{array}{lcl}
    ir^{ab}Y_b & = & X^a\sps\\
    ir^{ab}T_b & = & Z^a
    \end{array}
    \right.
\end{equation}
with the conditions
\begin{equation}
    \begin{array}{c}
    X^a Y_a=0\sps
    Z^a T_a=0\sps
    X^a T_a=-Z^a Y_a
    \end{array}
\end{equation}
then from 8 equations, only 5 from them will be significant (the 6 unknowns and the 3 communication conditions). This means that the rectilinear generator  $\mathbb C\mathbb P_1$ of the quadric $\mathbb CQ_6$ will uniquely define the rectilinear generator $\mathbb C \mathbb P_1$ belonging to the quadric $\mathbb C \tilde Q_6$. In this case, the manifold of generators $\mathbb C \mathbb P_1 (\mathbb CQ_6)$, belonging to the same generator $\mathbb C\mathbb P_3(\mathbb CQ_6)$, defines the beam of generators $\mathbb C\mathbb P_1(\mathbb C\tilde Q_6)$ belonging to the quadric $\mathbb C \tilde Q_6$ (this beam is a cone). The center of the beam is determined by the system (\ref{e14.3}).
\item
If we know the three equations
\begin{equation}
    \left\{
    \begin{array}{lcl}
    ir^{ab} Y_b & = &  X^a\sps\\
    ir^{ab} T_b & = &  Z^a\sps\\
    ir^{ab} N_b & = &  L^a
    \end{array}
    \right.
\end{equation}
with the conditions
\begin{equation}
    \begin{array}{c}
     X^a Y_a=0\sps
     Z^a T_a=0\sps
     L^a N_a=0\sps\\
     X^a T_a=- Z^a Y_a\sps
     X^a N_a=- L^a Y_a\sps
     Z^a N_a=- L^a T_a
    \end{array}
\end{equation}
then from the 12 equations only 6 from them will be significant (the 6 unknowns and the 6 communication conditions). This means, that the 2-dimensional generator $\mathbb C\mathbb P_2$ of the quadric $\mathbb CQ_6$ will uniquely define the point of the quadric $\mathbb C \tilde Q_6$. At the same time, the manifold of generators $\mathbb C\mathbb P_2(\mathbb CQ_6)$, belonging to the same generator $\mathbb C\mathbb P_3(\mathbb CQ_6)$, uniquely determines the same point of the quadric $\mathbb C \tilde Q_6$. This point is determined by the system (\ref{e14.3}).
\end{enumerate}

\subsubsection{\texorpdfstring{Geometry of the transition to the connecting operators $\eta_\Lambda{}^{KL}$ for n = 8}{The connection operators for n=8}}
\Abstract{
$\phantom{ff}$In this subsubsection, it is told about how to construct the connecting operators of the spinor formalism for n = 8.  The conclusion of all results of this subsubsection is made on the basis \cite{Andreev6}, \cite{Cartan1}, \cite[v. 2]{Hodge1}.
}

Further we consider the bivector $\hat R^{AB}$ such that its vector components $X_1{}^A, X_2{}^A$ are defined as
\begin{equation}

\end{equation}
This defines the identity chain
\begin{equation}
    %
\end{equation}
Contract the last identity with $Z^m$ that will give
\begin{equation}
    %
\end{equation}
Based on the foregoing, for the quadric $\mathbb CQ_6$, we consider the rectilinear generator defined by the bivector $(T_k X^k\ne 0)$
\begin{equation}
    %
    \right)
    \end{array}
\end{equation}
where $\varepsilon_{AC}$ is the same metric symmetric spin-tensor. At the same time, the equation
\begin{equation}
    R_A{}^C\hat R^{AB}=0
\end{equation}
will be true that means that any spin-tensor $\hat R^{AB}$, representing the generator $\mathbb C \mathbb P_1 (\mathbb CQ_6)$, will contain the same tensor $R^A{}_K$ in its expansion, wherein, the second spin-tensor $P^{KB}$ of the decomposition will be responsible for the position of $\mathbb C \mathbb P_1$ in $\mathbb C \mathbb P_3$. Therefore, there is a reason to assign the bispinor $R^{AB}$ to the point of the quadric $\mathbb C \tilde Q_6$. This point is uniquely determined. In the transition to the space $\mathbb C\mathbb R^8$ on the basis of
\begin{equation}

\end{equation}
For $(r_i)^\Lambda$, to define a generator of the quadric $\mathbb C\tilde Q_6$, it is necessary and sufficient to have the condition
\begin{equation}
    g_{\Lambda\Psi}(r_I)^\Lambda (r_J)^\Psi=0\spsd
\end{equation}
We define some connecting operators $\eta_\Lambda{}^{BC}$ so that to satisfy the conditions
\begin{equation}
     r_\Lambda=\frac{1}{4}\eta_\Lambda{}^{BC}R_{BC}\sps
     r^\Lambda=\frac{1}{4}\eta^\Lambda{}_{BC}R^{BC}\spsd
\end{equation}
Then these operators satisfy the Clifford equation
\begin{equation}
     g_{\Lambda\Psi}\delta_K{}^L=\eta_{\Lambda K}{}^R\eta_\Psi{}^L{}_R+\eta_{\Lambda K}{}^R\eta_\Psi{}^L{}_R\spsd
\end{equation}
Define
\begin{equation}
     \varepsilon_{PQRT}:=\eta^\Lambda{}_{PQ}\eta^\Psi{}_{RT}g_{\Lambda\Psi}\sps
     \varepsilon_{PQRT}=\varepsilon_{RTPQ}
\end{equation}
that will give another metric spin-tensor $\varepsilon_{PQRT}$ with which the help we can raise and lower a pair of indices
\begin{equation}

\end{equation}
The result of the applying of the two metric spin-tensors should be the same. If the matrixes of the tensors $g_{\Lambda \Psi}$ and  $\varepsilon_{KL}$ have the form
\begin{equation}
    %
\end{equation}
then the significant coordinates of the operators $\eta^\Lambda{}_{KL}$ in some basis would be such that
\begin{equation}
    %
\end{equation}
and the significant coordinates of the operators $\eta_\Lambda{}^{KL}$ would be such that
\begin{equation}
    %
\end{equation}
The basis is slightly different from one in which the homogeneous coordinates of $\mathbb C\mathbb R^8$ were defined above. Thus, the bispinor $\hat R^{AB}$ determines the rectilinear generator $\mathbb C\mathbb P_1 (\mathbb CQ_6)$ belonging to the planar generator $\mathbb C\mathbb P_3 (\mathbb CQ_6)$ which determines the point with the coordinates $R^{AB}$ for the quadric $\mathbb C\tilde Q_6$. We define the spin-tensor $\hat{\hat R}^{AB}$
\begin{equation}
    \hat{\hat R}^{AB}:=\hat R^{AK} \hat P_K{}^B\sps
    \hat P_K{}^B:=\left(
    %
    \right)\spsd
\end{equation}
The tensor $\hat{\hat R}^{AB}$ will continue to represent the rectilinear generator $\mathbb C\mathbb P_1 (\mathbb CQ_6)$ belonging to the planar generator $\mathbb C\mathbb P_3 (\mathbb CQ_6)$. We apply the operators $\eta^\Lambda{}_{KL}$ to the spin-tensor $\hat{\hat R}^{AB}$ and obtain
\begin{equation}
    \eta^\Lambda{}_{KL}\hat{\hat R}^{KL}=
    \eta^\Lambda{}_{KL} T_m X^m
    \left(
    %
    \right)=
    \eta^\Lambda{}_{KL}R^{KL}\spsd
\end{equation}
Thus, the operators $\eta^\Lambda{}_{KL}$ take each rectilinear generator $\mathbb C \mathbb P_1 (\mathbb CQ_6)$, belonging to the planar generator $\mathbb C \mathbb P_3 (\mathbb CQ_6)$, to the same generator $\mathbb C \mathbb P_3 (\mathbb CQ_6)$, and this determines the point of the quadric $\mathbb C \tilde Q_6$. In the homogeneous coordinates, the same tensor $R^{KL}$ determines the coordinates of the point R of the space $\mathbb C\mathbb R^8$. \\

Let us find out what is the number of a family which comprises this generator $\mathbb C\mathbb P_3 (\mathbb CQ_6)$. For this purpose, we consider the conditions
\begin{equation}
    \begin{array}{c}
    \varepsilon_{AB}(X_I){}^A(X_J){}^B=0\sps\\
    X^{ABCD}:=\varepsilon^{IJKL}(X_I){}^A(X_J){}^B(X_K){}^C(X_L){}^D\sps
    \end{array}
\end{equation}
where $\varepsilon^{IJKL}$ is a 4-vector antisymmetrical in all indices. Also we consider the 8-vector $e_{ABCDKLMN}$ antisymmetrical in all indices too. If in the condition
\begin{equation}
    \frac{1}{24}e_{ABCDKLMN}X^{ABCD}=\rho\varepsilon_{KR}\varepsilon_{LT}\varepsilon_{MU}
    \varepsilon_{NV}X^{RTUV}\sps \rho^2=1\sps
\end{equation}
$\rho=1$ then we say that the planar generator $\mathbb C\mathbb P_3(\mathbb CQ_6)$ belongs to family I, and if $\rho=-1$ then the planar generator belongs to family II. In our case
\begin{equation}
    (X_I){}^A=( (X_I){}^a, (Y_I)_{b})\sps
\end{equation}
and then
\begin{equation}
    \varepsilon^{IJKL}(X_I){}^1(X_J){}^2(X_K){}^3(X_L){}^4=\rho
    \varepsilon^{IJKL}(X_I){}^1(X_J){}^2(X_K){}^3(X_L){}^4\spsd
\end{equation}
Whence, $\rho=1$. This means that our generators should belong to family I. In addition, there is the tensor $P_\Lambda{}^B:=\eta_\Lambda{}^{AB}X_A$ for $X_A=(1,0,0,0,1,0,0,0)$
\begin{equation}

\end{equation}
it follows that $\rho=1$ and
\begin{equation}
    %
\end{equation}
This shows that the generators $\mathbb C\mathbb P_3 (\mathbb C\tilde Q_6)$ must belong to family I. To obtain family II, we must act by the elementary orthogonal reflection (with determinant equal to -1) on the operator $P_\Lambda{}^K$.

\subsection{\texorpdfstring{Correspondence $\mathbb C\tilde Q_6\longmapsto \mathbb CQ_6$}{Correspondence: C'Q to CQ}}
\Abstract{
$\phantom{ff}$In this subsubsection it is told about how to construct a reverse mapping for the quadrics between generators of different ranks according to the Cartan triality principle. The conclusion of all results of this subsubsection is made on the basis \cite{Andreev6}, \cite{Cartan1}, \cite[v. 2]{Hodge1}.
}

Applying the operators $\eta_\Lambda{}^{KL}$ to $g_{\Lambda\Psi}(r_I){}^\Lambda (r_J){}^\Psi=0$, we obtain
\begin{equation}
      \begin{array}{c}
      (R_I)^{AB}(R_J)_{AB}=0\Leftrightarrow
      ((R_I){}^{AB}-(R_J){}^{AB})((R_K){}_{AB}-(R_L){}_{AB})=0
      \Leftrightarrow\\[2ex]
      ((R_I){}^{ab}-(R_J){}^{ab})((R_K){}_{ab}-(R_L){}_{ab})=0\spsd
      \end{array}
\end{equation}
Here as usual, by I, J, we denote the number of basis points. This defines the system
\begin{equation}
\label{e13.7}
    \left\{
    \begin{array}{lcl}
    i(R_1){}^{ab} Y_b & = &  X^a\sps\\
    i(R_2){}^{ab} Y_b & = &  X^a\sps\\
    i(R_3){}^{ab} Y_b & = &  X^a\sps\\
    i(R_4){}^{ab} Y_b & = &  X^a\sps
    \end{array}
    \right.
    \Leftrightarrow
    \left\{
    \begin{array}{lcl}
    i((R_1){}^{ab}-(R_2){}^{ab}) Y_b & = & 0\sps\\
    i((R_1){}^{ab}-(R_3){}^{ab}) Y_b & = & 0\sps\\
    i((R_3){}^{ab}-(R_4){}^{ab}) Y_b & = & 0\sps\\
    i(R_1){}^{ab} Y_b & = &  X^a\spsd\\
    \end{array}
    \right.
\end{equation}
Next we consider only the right system. It is constructed as follows. A cospinor $Y_a$ which resets 3 different simple bivectors always exists. This statement is reduced to the existence of a covector orthogonal to three linearly independent vectors since any simple bivector is decomposed by the formula $R^{ab}=2P^{\left[\right.a} Q^{b\left.\right]}$. From the fourth equation, the spinor $X^a$ is defined. On the other hand, on the basis that all $(R_I)^{ab}$ have the form
    $R^{ab}_{\mbox{\scriptsize{particular}}}+
    \lambda_1 S^{\left[\right.a} X^{b\left.\right]}+
    \lambda_2 X^{\left[\right.a} Z^{b\left.\right]}+
    \lambda_3 S^{\left[\right.a} Z^{b\left.\right]}$
for fixed $\lambda_1,\lambda_2,\lambda_3$ the equation
\begin{equation}
     ((R_I){}^{ab}-(R_J){}^{ab})((R_K){}_{ab}-(R_L){}_{ab})=0
\end{equation}
is satisfied.
\begin{enumerate}
\item
So, let us know the last equation (\ref{e13.7})
\begin{equation}
     i{R_1}^{ab} Y_b  =   X^a
\end{equation}
then we have the 4 equations, all of which will be significant. Since we have the eight unknowns $(X^a,Y_b)$ for fixed $(R_I){}^{ab}$ then the point of the quadric $\mathbb C \tilde Q_6$ uniquely defines the 3-dimensional planar generator $\mathbb C\mathbb P_3 (\mathbb CQ_6)$.
\item
If we know all the equations of (\ref{e13.7}) with the conditions
\begin{equation}

\end{equation}
that from the 16 equations, only 7 from them will be significant (the 8 unknowns and the 9 communication conditions). Therefore, the generator $\mathbb C \mathbb P_3 (\mathbb C \tilde Q_6)$ will uniquely define the point of the quadric $\mathbb CQ_6$.
\item
If we know the three equations of the system (\ref{e13.7})
\begin{equation}
    \left\{
    %
    \right.
\end{equation}
with the conditions
\begin{equation}
     %
\end{equation}
that from  the 12 equations, only 7 from them will be significant (the 8 unknowns and the 5 communication conditions). This means that the generator $\mathbb C \mathbb P_2 (\mathbb C \tilde Q_6)$ will uniquely define the point of the quadric $\mathbb CQ_6$. In this case, the manifold of the generators $\mathbb C \mathbb P_2 (\mathbb C \tilde Q_6)$, belonging to a single generator $\mathbb C \mathbb P_3 (\mathbb C \tilde Q_6)$, uniquely defines the point of the quadric $\mathbb CQ_6$.
\item
If we know the two equations of the system (\ref{e13.7})
\begin{equation}
    \left\{
    %
    \right.
\end{equation}
with the conditions
\begin{equation}
     ((R_1){}^{ab}-(R_2^{ab})((R_1){}_{ab}-(R_2){}_{ab})=0\sps
      (R_1){}^{ab}(R_2){}_{ab}=0\sps
\end{equation}
then from  the 8 equations, only 6 from them will be significant (the 8 unknowns and the 2 communication conditions). This means that the rectilinear generator  $\mathbb C\mathbb P_1$ of the quadric $\mathbb C\tilde Q_6$ will uniquely define the rectilinear generator $\mathbb C \mathbb P_1$ belonging to the quadric $\mathbb CQ_6$. In this case, the manifold of generators $\mathbb C \mathbb P_1 (\mathbb C\tilde Q_6)$, belonging to the same generator $\mathbb C\mathbb P_3(\mathbb C\tilde Q_6)$, defines the beam of generators $\mathbb C\mathbb P_1(\mathbb CQ_6)$ belonging to the quadric $\mathbb CQ_6$ (this beam is a cone). The center of the beam is determined by the system (\ref{e13.7}).
\end{enumerate}

\subsubsection{On the Majorana spinors.}
\Abstract{
$\phantom{ff}$In this subsubsection, it is told about the Majorana condition in terms of an involution in the spinor space.  The conclusion of all results of this subsubsection is made on the basis \cite{Zee1}, \cite{Penrouz1}.
}

As an illustration, we consider the Infeld-van der Waerden symbols \cite[v. 1, p. 123 (eng)]{Penrouz1}. Let the Clifford equation, given on the vector space $\mathbb C^4$ in a special basis, have the form
\begin{equation}
\eta_\Lambda\sigma_\Psi+\eta_\Psi\sigma_\Lambda=2\delta_{\Lambda\Psi}\sps (\Lambda,\Psi,...=\overline{0,3})\spsd
\end{equation}
If we specify the real inclusion $H_i{}^\Lambda :\mathbb R^4_{(1,3)}\subset \mathbb C^4$ then the involution $S_A{}^{A'}:=\varepsilon^{A'B'}S_{AB'}$
\begin{equation}
S_\Lambda{}^{\Psi '}\bar \eta_{\Psi '}{}^{A'B'}=\eta_{\Lambda}{}^{BA} S_A{}^{A'}S_B{}^{B'}\sps
(A,A',...=\overline{1,2})\spsd
\end{equation}
is defined according to Corollary \ref{c6.3}, Case II). The Infeld-van der Waerden symbols are determined as follows
\begin{equation}
\begin{array}{c}
g_i{}^{AA'}:=H_i{}^\Lambda \eta_\Lambda{}^{AB}S_B{}^{A'}\sps \\
\overline{g_i{}^{AD'}}=\bar g_i{}^{A'D}:=\bar H_i{}^{\Lambda '} \bar\eta_{\Lambda '}{}^{A'B'}\bar S_{B'}{}^D=
\bar H_i{}^{\Lambda '} \bar S_{\Lambda '}{}^\Lambda\eta_\Lambda{}^{DC}S_C{}^{A'}=
H_i{}^\Lambda\eta_\Lambda{}^{DC}S_C{}^{A'}=g_i{}^{DA'}\sps\\
\bar g_i{}^{A'D}=g_i{}^{DA'}\sps
(i,j,...=\overline{0,3})\spsd
\end{array}
\end{equation}
By this equation, the involution $S_B{}^{A '}$ is introduced in the definition of the symbols $g_i{}^ {AA'}$. The presence of the metric antisymmetric spin-tensor $\varepsilon_{AB}$ allows to carry out such the operation with which we can raise and lower the single index only. We now construct the Dirac operators
\begin{equation}
\gamma_i=
\left(
\begin{array}{cc}
 0           & (g^T)_i{}_{AA'} \\
 g_i{}^{BB'} & 0
\end{array}
\right)\spsd
\end{equation}
In a special basis, they will have the form
\begin{equation}
\begin{array}{c}
\gamma_0=
\left(
\begin{array}{rrrr}
 0 & 0 & 1 & 0\\
 0 & 0 & 0 & 1\\
 1 & 0 & 0 & 0\\
 0 & 1 & 0 & 0
\end{array}
\right)\sps
\gamma_1=
\left(
\begin{array}{rrrr}
 0 & 0 & 0 &-1\\
 0 & 0 &-1 & 0\\
 0 & 1 & 0 & 0\\
 1 & 0 & 0 & 0
\end{array}
\right)\sps\\
\gamma_2=
\left(
\begin{array}{rrrr}
 0 & 0 & 0 &-i\\
 0 & 0 & i & 0\\
 0 &-i & 0 & 0\\
 i & 0 & 0 & 0
\end{array}
\right)\sps
\gamma_3=
\left(
\begin{array}{rrrr}
 0 & 0 &-1 & 0\\
 0 & 0 & 0 & 1\\
 1 & 0 & 0 & 0\\
 0 &-1 & 0 & 0
\end{array}
\right)\sps\\
\gamma_i\bar\gamma_j+\gamma_j\bar\gamma_i=2g_{ij}\sps\\
\gamma_i\bar\gamma_2+\gamma_2\bar\gamma_i=0\sps
\gamma_2\bar\gamma_2=-1\sps\\
\gamma_i\bar\gamma_2+\gamma_2\bar\gamma_i=0\sps i\ne 2\sps\\
\bar\gamma_2\gamma_i\bar\gamma_2+\bar\gamma_2\gamma_2\bar\gamma_i=0\sps\\
\bar\gamma_2\gamma_i\bar\gamma_2=\bar\gamma_i\sps\\
\end{array}
\end{equation}
Accordingly, the involution $S_A{}^{A'}$ will generate the involution $S$ for the complex Dirac operators $\bar\gamma_i = \bar S\gamma_i\bar S$, and in this special basis, it will be executed $||S||=||\gamma^2||$. Thus, we can define the Majorana spinor as
\begin{equation}
 \psi^i=\gamma^2 \bar\psi^i
\end{equation}
which corresponds to the presentation \cite[p. 98, Appendix E (eng)]{Zee1}.\\

In the transition to the space $\mathbb R^6_{(2,4)}$, the Majorana condition will have the form ($A,A',...=\overline{1,4}$, $a,a',...=\overline{1,2}$) ($\eta_\Lambda{}^{AB}=-\eta_\Lambda{}^{BA}$)
\begin{equation}
\begin{array}{c}
S_\Lambda{}^{\Lambda '}\bar\eta_{\Lambda '}{}^{A'B'}=\eta_\Lambda{}_{AB}\underbrace{(i\tilde S^{AA'})}_{:=S^{AA'}}\underbrace{(i\tilde S^{BB'})}_{:=S^{BB'}}\sps S^{AA'}=-\bar S^{A'A}\sps\\
\bar X^{A'}=(\bar X^{a'},\bar Y_{b'})=(\bar S^{a'}{}_aX^a,\bar S_{b'}{}^bY_b)=
\left(
\begin{array}{cc}
0                & \bar S^{a'}{}_a            \\
\bar S_{b'}{}^b  & 0 \\
\end{array}
\right)
\left(
\begin{array}{c}
Y_b\\X^a
\end{array}
\right)=
\bar S^{A'A}X_A\spsd
\end{array}
\end{equation}
However, in such the spinor space according to (\ref{e6.8}), there is no metric tensors, except $S_{AA '}$, capable to raise and to lower single indexes. The Majorana condition for the space $\mathbb R^6$ will be the same. In this case, we use the systems ($a,a',...= \overline {1,4}$)
\begin{equation}
\label{e14.4}
\left\{
\begin{array}{c}
iR^{ab}Y_b=X^a\sps\\
iR^{ab}N_b=L^a\sps\\
\end{array}
\right.
\left\{
\begin{array}{c}
-iR_{ab}Z^b=T_a\sps\\
-iR_{ab}K^b=M_a\spsd
\end{array}
\right.
\end{equation}
If our space would be a complex $\mathbb C\mathbb R^6$ then this system would establish a one-to-one correspondence between the respective rectilinear generators of the quadrics. The representation freedom would be dictated by the homogeneous solution of the system which would be isotropic. But, because of the real inclusion $\mathbb R^6 \subset \mathbb C\mathbb R^6$, the isotropic vectors are represented only by the zero-vector.
\begin{equation}
\begin{array}{cc}
R^{ab}=\frac{-i}{Y^k Y_k}(X_dY_c \varepsilon^{abcd}+2X^{\left[\right. a}Y^{b\left.\right]})\sps &
R^{ab}=\frac{-i}{N^k N_k}(L_dN_c \varepsilon^{abcd}+2L^{\left[\right. a}N^{b\left.\right]})\sps\\
Z^b:=Y^b=S^{bb'}\bar Y_{b'}\sps T_a:=X_a=S_{aa'}\bar X^{a'}\sps &
K^b:=N^b=S^{bb'}\bar N_{b'}\sps M_a:=L_a=S_{aa'}\bar L^{a'}\sps \\
X^a Y_a=0\sps &
L^a N_a=0\spsd
\end{array}
\end{equation}
Therefore, one can establish the one-to-one correspondence between a point of the quadric $\mathbb C\tilde Q_6$ and a rectilinear generator $\mathbb C\mathbb P_1(\mathbb CQ_6)$. Since on a rectilinear generator of the quadrics $\mathbb CQ_6$ (the left system (\ref{e14.4})), we construct the conjugate rectilinear generator $\mathbb C\mathbb P_1(\mathbb CQ_6)$ (the right system (\ref{e14.4})) then this will define the 3-dimensional planar generator $\mathbb C\mathbb P_3(\mathbb CQ_6)$ which will uniquely identify the point to the coordinates $R^{ab}$ on the quadric $\mathbb C\tilde Q_6$. This is the geometric meaning of the Majorana condition for the inclusion $\mathbb R_6\subset \mathbb C \mathbb R_6$.
\begin{equation}
X^a X_a+Y^a Y_a(-\frac{1}{4} R^{ab}R_{ab})=0\sps
L^a L_a+N^a N_a(-\frac{1}{4} R^{ab}R_{ab})=0\spsd
\end{equation}
(Obviously, $R^{ab} R_{ab}>0\sps X^aX_a= S_{aa'}X^a\bar X^{a'}>0\sps Y^aY_a= S_{aa'}Y^a\bar Y^{a'} >0$ for nonzero vectors.) The condition can be rewritten as follows
\begin{equation}
(X^a,Y_b)(X_a,\frac{-X^d X_d}{Y^c Y_c} Y^b)=0\sps
(L^a,N_b)(L_a,\frac{-L^d L_d}{N^c N_c} N^b)=0\spsd
\end{equation}
 Incidentally, $R^{ab}$ is a null vector of the isotropic cone of the space $R^8_{(1,7)}$. The corresponding quadric cut out by a section of the cone.
\begin{example}
Thus, let $Y^kY_k=2 \sps X^aY_a=0$ then in the special basis
\begin{equation}
\begin{array}{c}
iR^{12}=-i\ \overline{R^{34}}=X^{\left[\right. 1}Y^{2\left.\right]}+\overline{X^{\left[\right. 3} Y^{4\left.\right]}}\sps
iR^{13}=i\ \overline{R^{24}}=X^{\left[\right. 1}Y^{3\left.\right]}-\overline{X^{\left[\right. 2} Y^{4\left.\right]}}\sps\\
iR^{14}=-i\ \overline{R^{23}}=X^{\left[\right. 1}Y^{4\left.\right]}+\overline{X^{\left[\right. 2} Y^{3\left.\right]}}\spsd
\end{array}
\end{equation}
\end{example}

\subsubsection{Theorem on two quadrics.}
\Abstract{
$\phantom{ff}$In this subsubsection, it is told about how to interpret the Cartan triality principle on the basis of previous results and to bring them together. The conclusion of all results of this subsubsection is made on the basis \cite{Andreev6}, \cite{Cartan1}, \cite[v. 2, p. 258 (rus)]{Hodge1}.
}

Thus, the theorem is proved.
\begin{theorem}
\label{theorem9}(the triality principle for the two B-cylinders).\\
In the projective space $\mathbb C \mathbb P_7$, there are two quadrics (two B - cylinders) with the following main properties
\begin{enumerate}
    \item The planar generator $\mathbb C \mathbb P_3$ of a one quadric will define one-to-one the point R on the other quadric.
    \item The planar generator $\mathbb C \mathbb P_2$ of a one quadric will uniquely define the point R on the other quadric. But the point R of the second quadric can be associated to the manifold of planar generators $\mathbb C \mathbb P_2$ belonging to the same planar generator $\mathbb C \mathbb P_3$ of the first quadric.
    \item The rectilinear generator $\mathbb C \mathbb P_1$ of a one quadric will define one-to-one the rectilinear generator $\mathbb C \mathbb P_1$ of the other quadric. And all the rectilinear generators belonging to the same planar generator $\mathbb C \mathbb P_3$ of the first quadric define the beam centered at R belonging to the second quadric.
\end{enumerate}
\end{theorem}

This theorem is actually the generalization of the Klein correspondence. Prove this.
\begin{proof}$ $\\
On the quadric $\mathbb CQ_6$, we consider only those generators which have the form $X^A=(0,Y_b)$. The manifold of such generators is diffeomorphic to $\mathbb C \mathbb P_3$. In this case, each generator can be associated to the point of the quadric $\mathbb CQ_4 \subset \mathbb C \tilde Q_6$. Then $R^{ab} Y_b = 0$. Until the end of the proof, we set $\bf{A,B,A',B',...}$$=\overline{1,2}$. In addition, we consider the spinor representation of a twistor according to \cite[eq. (6.1.24) and (6.2.18) (eng)]{Penrouz1}.
\begin{equation}

\end{equation}
Therefore, the equation $R^{ab}Y_b=0$ can be rewritten as a system of two equations
\begin{equation}
     \left\{
     %
     \right.
\end{equation}
and only one of them will be significant
\begin{equation}
     ir^{{\mbox{\scriptsize\bf AA}}'}
      \bar\pi_{{\mbox{\scriptsize\bf A}}'}=
      \omega^{\mbox{\scriptsize\bf A}}\spsd
\end{equation}
Here we use the operation of conjugation and the metric spin-tensor $\bar\varepsilon_{\mbox{\scriptsize\bf A'B'}}, \varepsilon_{\mbox{\scriptsize\bf AB}}$ with the help of which spinor indices rise and lover. This spinors pass each other by the conjugation. This defines the system
\begin{equation}
     %
     \right.
      Y_b=( \pi_{\mbox{\scriptsize\bf B}},
      \bar\omega^{{\mbox{\scriptsize\bf B}}'})\sps
      T_b=( \eta_{\mbox{\scriptsize\bf B}},
      \bar\xi^{{\mbox{\scriptsize\bf B}}'})\spsd
     \end{array}
\end{equation}
This system coincides with the system \cite[eq. (6.2.14)]{Penrouz1} which, in turn, leads to the Klein correspondence.
\end{proof}
     It should be noted in conclusion that from this theorem, the Cartan triality principle implies:
     There are the 3 diffeomorphic manifolds:
\begin{enumerate}
      \item the manifold of all points of the quadric;
      \item the manifold of I-family maximal planar generators;
      \item the manifold of II-family maximal planar generators.
\end{enumerate}
This is true, because the two constructed quadrics can be identified, for example, by means of the spin-tensor $P_\Lambda{}^L$. The manifold of points of the quadric is diffeomorphic to the maximal planar generator manifold. In addition, since the Cartan triality principle is performed then the operators $\eta^i{}_{KL}$ for the inclusion $\mathbb R^8 \subset \mathbb C^8$ define the octave algebra. This assertion is based on the results given in the monography \cite[v. 2, pp. 460-462]{Penrouz1} which deals with the structure constants of this algebra.

\subsection{Explicit construction of the Lie spin-operators for n=4.}
\Abstract{
$\phantom{ff}$In this subsection, it is told about how possible in a certain basis to get a representation of the operators $P$ with which the help we can construct the spinor analog of the Lie operators.  The conclusion of all results of this subsection is made on the basis \cite{Bilyalov1}, \cite[v. 2, pp. 101-103 (eng)]{Penrouz1}.
}

Let on the space $\mathbb R^8_{(4,4)}$, the metric
\begin{equation}
dS^2=-dU^2+dS^2+dV^2-dW^2+dX^2-dY^2-dT^2+dZ^2
\end{equation}
be given then in some basis, the connecting operators $\eta_\Lambda{}^{KL}$ have the form
\begin{equation}

\end{equation}
and the significant coordinates of the operators $\eta^\Lambda{}_{KL}$ will be such that
\begin{equation}
    %
\end{equation}
Therefore, in the same basis,
\begin{equation}
    %
\end{equation}
Hence, Algorithm \ref{a6.1} will take the form
\newpage
\begin{algorithm}$ $\\
\label{a14.1}
1. On the first step, as the operators (\ref{e3.12}) and (\ref{e3.12/1}), it is used the operators
\begin{equation}
%
\right)}_{:=\eta_{1_\pm}{}_\myf\Lambda{}^{AB}}\sps\\
\alpha_1=\frac{1}{2}(\eta_3+\eta_4)\sps
\beta_1=\frac{1}{2}(\eta_3-\eta_4)\sps
\gamma_1=\frac{1}{2}(\eta_2-\eta_1)\sps
\delta_1=\frac{1}{2}(\eta_2+\eta_1)\spsd
\end{array}
\end{equation}
Here, $\varepsilon^{\myff a\myff b}$ is the antisymmetric metric spin-tensor for the connecting operators $\eta_\alpha{}_{\myff p}{}^{\myff l}$ ($a,b,...,\myff a,\myff b,...=\overline{1,2},\ \myf\Lambda,...,\myff A,\myff B,... =\overline{1,8},\ A,B,...=\overline{1,4}$).\\
2. On the second step, as the operators (\ref{e3.12}) and (\ref{e3.12/1}), it is used the operators
\begin{equation}

\end{equation}
($\myf\Lambda,...=\overline{1,8},\ \myff A,\myff B,...=\overline{1,4},\ A,B,...=\overline{1,2}$).\\
3. For the tangent space, the operators $m_\Lambda{}^\myf\Lambda$ have the form
\begin{equation}
%
\end{equation}
Then the operators
\begin{equation}
\underbrace{\left(
%
\right)}_{:=(\eta_{1,2,3,4}){}_\Lambda{}^{AB}}\\
\end{equation}
will be the connecting operators for the complex representation (n=4). Thus, it is constructed the transition from the real representation of the connecting operators for n = 8 to their complex counterparts for n = 4.
\end{algorithm}

Thus, the operators $2((\tilde M^*_K)^T)_\myff A{}^A$, according to (\ref{e3.13/1}) and Algorithm \ref{a6.1}, have the form
\begin{equation}
\label{e14.6}
\left(
%
\right)\spsd
\end{equation}
In this case, $p_1,p_2$ are taking one of the two values $\pm 1$ providing the 4 different variants of $\tilde M^*_K$. It remains to apply these operators and the operators $m_\Lambda{}^\myf\Lambda$, $\bar m_{\Lambda '}{}^\myf\Lambda$ to the operators $P_\myf\Lambda{}^\myff A$
\begin{equation}
\label{e14.5}
%
\right)\spsd
\end{array}
\end{equation}
\begin{note}
$(P^*_K)_\Lambda{}^A\ne(P_K)_\Lambda{}^A=g_{\Lambda\Psi}(P_K)^\Lambda{}_B\varepsilon^{AB}$ where $\varepsilon^{AB}$ is the antisymmetric metric spinor.
\end{note}
It is obvious that for the 4 variants $K=\overline{1,4}$, the two will be significantly only: $(P^*_2)_\Lambda{}^A$, $(P^*_3)_\Lambda{}^A$ and
$(\bar P^*_1)_{\Lambda '}{}^A$, $(\bar P^*_4)_{\Lambda '}{}^A$ ($\Lambda,...=\overline{1,4}$). Under the condition of the ordinary and covariant differentiation constancy of such the operators, the connection coefficients $\Gamma_{\myf\Lambda\Psi}{}^\Phi$ decay into the 8 significant components
\begin{equation}
\Gamma_{\myf\Lambda\Psi}{}^\Phi\rightarrow
\left\{
\begin{array}{l}
(\Gamma_K)_{\Lambda A}{}^B:=\Gamma_{\Lambda\Psi}{}^\Phi (P_K)^\Psi{}_A(P^*_K)_\Phi{}^B\sps\\
(\Gamma_K)_{\Lambda A'}{}^{B'}:=\Gamma_{\Lambda\Psi}{}^\Phi (P_K)^\Psi{}_{A'}(P^*_K)_\Phi{}^{B'}\sps\\
(\bar \Gamma_K)_{\Lambda 'A'}{}^{B'}:=\bar\Gamma_{\Lambda '\Psi '}{}^{\Phi '} (\bar P_K)^{\Psi '}{}_{A'}(\bar P^*_K)_{\Phi '}{}^{B'}\sps\\
(\bar \Gamma_K)_{\Lambda 'A}{}^B:=\bar\Gamma_{\Lambda '\Psi '}{}^{\Phi '} (\bar P_K)^{\Psi '}{}_A(\bar P^*_K)_{\Phi '}{}^B\spsd
\end{array}
\right.
\end{equation}
In turn, the coefficients $\Gamma_{\myf\Lambda\Psi}{}^\Phi$ are defined by the equation $\nabla_\myf\Lambda m_\Psi{}^\myf\Psi=0$. Thus, Theorems \ref{theorem9.1}, \ref{theorem9.2} remain valid in this case. In the case of a real inclusion, with the help of the operators $H_i{}^\Lambda$, under the condition of the covariant and ordinary differentiation constancy of the operators $(P^*_2)_i{}^A$, $(P^*_3)_i{}^A$ and $(P^*_1)_i{}^A$, $(P^*_4)_i{}^A$, the connection coefficients $\Gamma_{\myf\Lambda i}{}^j$ decay into the 4 significant components
\begin{equation}
\Gamma_{\myf\Lambda i}{}^j\rightarrow
\left\{
\begin{array}{l}
(\Gamma_K)_{kA}{}^B:=\Gamma_{ki}{}^j (P_K)^i{}_A(P^*_K)_j{}^B\sps\\
(\bar \Gamma_K)_{kA'}{}^{B'}:=\Gamma_{ki}{}^j (\bar P_K)^i{}_{A'}(\bar P^*_K)_j{}^{B'}\spsd\\
\end{array}
\right.
\end{equation}
However, the real condition restricts the choice of the spinor $X^\myff A$ in the determination of the operators $P_\myf\Lambda{}^\myff A$
\begin{equation}
X^\myff A=\tilde S_\myff B{}^\myff A X^\myff B\sps
X^\myff A=\tilde {\tilde S}^{\myff B\myff A} X_\myff B\sps
\end{equation}
where $\tilde S_\myff B{}^\myff A$ ($S^{\myff B\myff A}$) is a real representation of the complex involution $\tilde S_B{}^{A'}$ ($\tilde S^{BA'}$) which is the image of the involution $S_\Lambda{}^{\Psi '}$ in the tangent space by Corollary \ref{c6.3}. We need to add that there is the non-special orthogonal transformation
\begin{equation}
(-\eta_\Lambda\eta^\Psi+\delta_\Lambda{}^\Psi)\eta_\Psi{}^{AB}=\eta_\Lambda{}^{BA}
\end{equation}
performing the transposition of the connecting operators (\ref{e12.6a}).

\begin{example}
Consider the inclusion $\mathbb R^4_{(1,3)}\subset \mathbb C^4$ with
\begin{equation}
\bar H_i{}^{\Lambda '}=\left(
\begin{array}{cccc}
1 & 0 & 0 & 0\\
0 & 1 & 0 & 0\\
0 & 0 & 1 & 0\\
0 & 0 & 0 &-i\\
\end{array}
\right)\sps
H^i{}_\Lambda =\left(
\begin{array}{cccc}
1 & 0 & 0 & 0\\
0 & 1 & 0 & 0\\
0 & 0 & 1 & 0\\
0 & 0 & 0 &-i\\
\end{array}
\right)\sps
S_\Lambda{}^{\Lambda '}=\left(
\begin{array}{cccc}
1 & 0 & 0 & 0\\
0 & 1 & 0 & 0\\
0 & 0 & 1 & 0\\
0 & 0 & 0 &-1\\
\end{array}
\right)\spsd
\end{equation}
For the values of the complex representation operators from Algorithm \ref{a14.1}, the metric takes the form
\begin{equation}
g_{\Lambda\Psi}=\left(
\begin{array}{cccc}
1 & 0 & 0 & 0\\
0 &-1 & 0 & 0\\
0 & 0 &-1 & 0\\
0 & 0 & 0 & 1\\
\end{array}
\right)\sps
g_{ij}=\left(
\begin{array}{cccc}
1 & 0 & 0 & 0\\
0 &-1 & 0 & 0\\
0 & 0 &-1 & 0\\
0 & 0 & 0 &-1\\
\end{array}
\right)\spsd
\end{equation}
Then the definition (\ref{e14.5}) will have the form
\begin{equation}
\begin{array}{c}
(P^*_K)_i{}^A:=
\frac{1+i}{4}\left(
\begin{array}{cc}
 0          &  p_1-p_2   \\
 0          &  1-p_1p_2  \\
 p_1-p_2    &  0         \\
 i(1-p_1p_2)&  0         \\
\end{array}
\right)\sps
(P_K)^i{}_A:=
\frac{1-i}{4}\left(
\begin{array}{cc}
  0          &  p_1-p_2   \\
  0          &  1-p_1p_2  \\
  p_1-p_2    &  0         \\
-i(1-p_1p_2) &  0         \\
\end{array}
\right)\spsd
\end{array}
\end{equation}
Since $x^i:=(P_2)^i{}_A(X_2){}^A+(P_3)^i{}_A(X_3){}^A$ then from the equations (\ref{e3.13/1}), (\ref{e12.6}) and Corollary (\ref{c6.2}) imply that $(X_K){}^A:=x^i(P^*_K)_i{}^A$, $(X^*_K){}_A:=x^i(P_K)_i{}_A$; then from the equations (\ref{e3.13/1}), (\ref{e12.6}), Corollary (\ref{c6.2}) and the condition $(M_K)^i{}^A(M_K)_i{}_A=0,\ (M^*_K)^i{}^A(M^*_K)_i{}_A=0$ (see (\ref{e14.6})) imply that $(X^*_2){}_C=(P_2)^i{}_C(P_3)_i{}_B(X_3){}^B$. Therefore, we can construct the one-to-one mapping ($a,b,...=\overline{1,4}$)
\begin{equation}
x^i\cdot\underbrace{((P^*_2)_i{}^A,(P^*_3)_i{}^B)}_{:=P_i{}^a}=\underbrace{((X_2){}^A,(X_3){}^B)}_{:=X^a}\sps
x^i\cdot\underbrace{((P_2)_i{}_A,(P_3)_i{}_B)}_{:=P_i{}_a}=\underbrace{((X^*_2){}_A,(X^*_3){}_B)}_{:=X_a}\spsd
\end{equation}
Then on the spin-pair $X^a=((X_2){}^A,(X_3){}^B)$, the metric and the involution of the form
\begin{equation}
\begin{array}{c}
\varepsilon_{ab}=
\left(
\begin{array}{cc}
                      0 & (P_2)^i{}_A(P_3)_i{}_B \\
 (P_3)^i{}_C(P_2)_i{}_D & 0 \\
\end{array}
\right)=
\left(
\begin{array}{cccc}
 0 & 0 &-i & 0\\
 0 & 0 & 0 & i\\
-i & 0 & 0 & 0\\
 0 & i & 0 & 0\\
\end{array}
\right)\sps\\
S_a{}^{b'}=
\left(
\begin{array}{cc}
 (P_2)^i{}_A(\bar P^*_2)_i{}^{C'} & (P_2)^i{}_A(\bar P^*_3)_i{}^{D'} \\
 (P_3)^i{}_B(\bar P^*_2)_i{}^{C'} & (P_3)^i{}_A(\bar P^*_3)_i{}^{D'} \\
\end{array}
\right)=
\left(
\begin{array}{cccc}
 0 & 0 & i & 0\\
 0 &-i & 0 & 0\\
 i & 0 & 0 & 0\\
 0 & 0 & 0 &-i\\
\end{array}
\right)
\end{array}
\end{equation}
are defined. Therefor, it is possible to construct \emph{Lie pair-spin operators}
\begin{equation}
L_x(Y)^a=x^i\nabla_iY^a-P^j{}_bY^bP_i{}^a\nabla_j x^i\sps
L_X(Y)^a=X^c\nabla_cY^a-Y^c\nabla_c X^a\spsd
\end{equation}
Note, that the expansion $x^i:=(P_2)^i{}_A(X_2){}^A+(P_3)^i{}_A(X_3){}^A$ is an one of the vector into the two isotropic vectors.
And for the isotropic vectors, \emph{Lie spinor operators} were constructed in \cite[v. 2, pp. 101-103 (eng)]{Penrouz1}.

The restricted Lorentz transformation converts the vector $x^i$ to the vector $\tilde x^i$. This induces the spinor transformation: $(X_K)^A\longmapsto (\tilde X_K)^A$. The spinor transformations act on the operators $(P^*_K)_i{}^A,\ (P_K)^i{}_A$, more precisely, on the controlling spinor $X^\myff A$ appearing in the definition (\ref{e9.10}) of such the operators. In turn, this induces the spin-pair transformation of the spin-pair $X^a=((X_2){}^A,(X_3){}^B)\longmapsto \tilde X^a$.
\end{example}

\section{Acknowledgments}
$\phantom{ff}$This article consists of the answers to frequently asked questions. Therefore, the author is grateful to all those who has truly formulated these questions that has given the chance to find the answer on them. Special thanks to K. Scharnhorst (for the copies of articles, in particular), without who this publication was not.

\newpage

\section{Appendix}
\subsection{Proof of the metric tensor and involution relations}
$\phantom{ff}$We need to prove the identity (\ref{e6.3}).\\
Case I)., the metric tensor $\varepsilon_{AB}=-\varepsilon_{BA}$.
\begin{equation}

\end{equation}
Case I)., the metric tensor $\tilde\varepsilon_{AB}=\tilde\varepsilon_{BA}$.
\begin{equation}
%
\end{equation}

\newpage
\subsection{Proof of the connecting operator relations}
$\phantom{ff}$We need to prove the identity (\ref{e6.4})
\begin{equation}
\eta_\Lambda{}^{AB}\eta_\Psi{}_{AD}\eta^\Omega{}^{CD}\eta^\Theta{}_{CB}=
\frac{N}{4}(g_{\Lambda\Psi}g^{\Omega\Theta}+\delta_\Lambda{}^\Theta\delta_\Psi{}^\Omega-\delta_\Psi{}^\Theta\delta_\Lambda{}^\Omega)\spsd
\end{equation}
\begin{enumerate}
\item
First step.
\begin{equation}
%
\end{equation}
\end{proof}
\end{enumerate}

\subsection{Proof of the infinitesimal transformation relations}
$\phantom{ff}$We need to prove the identity (\ref{e8.2}).\\
\begin{enumerate}
\item Case 1).
\begin{equation}
%
\end{equation}
\end{enumerate}
Substitute in this equation
\begin{equation}
T_C{}^A=\frac{1}{2}T^{\Theta\Phi}\eta_\Phi{}^{AB}\eta_\Theta{}_{CB}
\end{equation}
and obtain
\begin{equation}
%
\end{equation}

\subsection{Proof of the Lie bracket identity}
$\phantom{ff}$We need to prove the identity (\ref{e9.7}) (see (\ref{e9.11})).\\
\begin{equation}
%
\end{equation}

\subsection{Proof of the curvature tensor relations}
$\phantom{ff}$We need to prove the identity (\ref{e10.2}).\\
The curvature tensor in the spinor bundle is given by
\begin{equation}
(\mathcal R_K)_{\Lambda\Psi C}{}^A=2(\partial_{\left[\right.\Lambda}(\Gamma_K)_{\Psi\left.\right] C}{}^A+(\Gamma_K)_{\left[\right.\Lambda |L|}{}^A(\Gamma_K)_{\Psi\left.\right] C}{}^L)\spsd
\end{equation}
Then $\partial_\Lambda(\varepsilon_K)_{AB}{}^{CD}=0$ under the condition (\ref{e9.9}) means
\begin{equation}
\partial_{\left[\right.\Lambda}(\Gamma_K)_{\Psi\left.\right] A}{}^C=
\underbrace{-\frac{2}{n-4}((\varepsilon_K)_{AB}{}^{CD}-\frac{n-2}{N}\delta_A{}^C\delta_B{}^D)}_{:=(\tilde\varepsilon_K)_{AB}{}^{CD}}\partial_{\left[\right.\Lambda}(\Gamma_K)_{\Psi\left.\right] D}{}^B\spsd
\end{equation}
In turn, $\nabla_\Lambda(\varepsilon_K)_{AB}{}^{CD}=\partial_\Lambda(\varepsilon_K)_{AB}{}^{CD}=0$ under the condition (\ref{e9.9}) means
\begin{equation}

\end{equation}
Whence,
\begin{equation}
(\mathcal R_K)_{\Lambda\Psi C}{}^M=(\mathcal R_K)_{\Lambda\Psi L}{}^D\tilde\varepsilon_{CD}{}^{ML}\sps
\end{equation}

\subsection{Proof of the auxiliary identities for the theorem on the spin analogs of the curvature tensor}
$\phantom{ff}$We need to prove the identity (\ref{e10.3}).\\
\begin{equation}
%
\end{equation}
whence the demanded follows.
\end{proof}

We need to prove the identity (\ref{e10.4}).\\
\begin{equation}
%
\right|\\
=(-1+\frac{4}{n-4})
\eta_{\left[\right. \Lambda_1}{}^{A_1A_2}\eta_{\Lambda_2\left.\right]}{}_{A_1A_3}
\eta_{\left[\right. \Lambda_3}{}^{A_4A_3}\eta_{\Lambda_4\left.\right]}{}_{A_4A_5}
\eta_{\left[\right. \Lambda_5}{}^{A_6A_5}\eta_{\Lambda_6\left.\right]}{}_{A_6A_2}+\\
+(\frac{8}{n-4}-4)\frac{N}{2}
  g_{\left[\right. \Lambda_3 | \left[\right. \Lambda_2}
  g_{\Lambda_1 \left.\right] | \left[\right. \Lambda_5}
  g_{\Lambda_6 \left.\right] | \Lambda_4 \left]\right.}\sps
\end{array}
\end{equation}
whence the demanded follows.
\end{proof}
We need to prove the identity (\ref{e10.5}).\\
\begin{equation}

\end{equation}
whence the demanded follows.
\end{proof}
\begin{note} It is obvious that these calculations are not applicable for n $<$8, since $\frac{8}{n-4}-4$ will not make the sense in this case.
\end{note}

\subsection{Proof of the twistor equation relations}
$\phantom{ff}$We need to prove  (\ref{e11.2}).\\
\begin{equation}
%
\end{equation}
Thus, the integrability conditions take the form
\begin{equation}
%
\end{equation}
We assume that this condition is satisfied for any $X^K$. We use the Bianchi identity  $2R^{\Delta[\Phi\Psi]\Lambda}=R^{\Psi\Phi\Delta\Lambda}$:
\begin{equation}
%
\end{equation}
Finally,
\begin{equation}
R^{\Phi\Psi\Lambda\Delta}-\frac{4}{n-2}R^{\left[\Lambda \right.|\left[ \Phi \right.}g^{\left.\Psi \right]|\left. \Delta \right]}+
\frac{2}{(n-1)(n-2)}Rg^{\left[\Delta \right.|\left[ \Psi \right.}g^{\left.\Phi \right]|\left. \Lambda \right]}=C^{\Phi\Psi\Lambda\Delta}=0\spsd
\end{equation}

\newpage
\subsection{Proof of the identities for n = 6 and n=8}
\subsubsection{Proof of the fundamental identities for n = 6}
$\phantom{ff}$We need to prove  (\ref{e12.3}).\\
\begin{equation}
       A_{\alpha\beta d}{}^c=
       \eta_{\left[\right.\alpha}{}^{ca}
       \eta_{\beta\left.\right]}{}_{da}\sps
\end{equation}
\begin{equation}

\end{equation}

\newpage
\subsubsection{Proof of the 4-vector identities}
$\phantom{ff}$We need to prove  (\ref{e12.4}).\\
\begin{equation}
       %
\end{equation}

\subsubsection{Proof of the 6-vector identities}
$\phantom{ff}$We need to prove  (\ref{e12.5}).\\
\begin{equation}
       %
\end{equation}
Whence,
\begin{equation}
       \tilde e=\frac{1}{8}i\spsd
\end{equation}

\newpage
\subsubsection{Proof of the Bianchi identity}
$\phantom{ff}$The Bianchi identity has the form
\begin{equation}
       %
\end{equation}
Let's contract it with $A^{\alpha\beta}{}_t{}^lA^{\gamma\delta}{}_m{}^n$
\begin{equation}
       %
\end{equation}
that finishes the proof.

\subsubsection{Proof of the identities related to the Weyl tensor}
$\phantom{ff}$Since,
\begin{equation}
    %
\end{equation}
then it will be true the following relation
\begin{equation}
       %
\end{equation}
Contract it with $A^{\alpha\beta}{}_k{}^lA_{\gamma\delta t}{}^n$
\begin{equation}
       %
\end{equation}
that finishes the proof.

\subsubsection{\texorpdfstring{Proof of the relations associated with the metric induced in the cross section of the cone $K_6$}{Proof of the relations associated with the metric induced in the cross section of the cone}}
$\phantom{ff}$Let the cone section of $K_6$
\begin{equation}
     T^2+V^2-W^2-X^2-Y^2-Z^2=0
\end{equation}
by the plane V+W=1 be set. We make a stereographic projection of the resulting hyperboloid onto the plane (V = 0, W = 1).
\begin{equation}
%
\end{equation}
We make the substitution
\begin{equation}
%
\end{equation}
Therefore, there is a reason to set
\begin{equation}
    X:=
    \left(
    %
    \right)\spsd
\end{equation}
Then
\begin{equation}
    d s^2=-\frac{det(dX)}{(det(X))^2}\sps \bar X^T+X=0\spsd
\end{equation}

\subsubsection{Proof of the first invariant relations}
$\phantom{ff}$Consider the fractional linear group L
\begin{equation}
    {\tilde X}=(AX+B)(CX+D)^{-1} \sps
    S:=
    \left(
    %
    \right) \sps
    detS=1\spsd
\end{equation}
Let there be two consecutive transformations
\begin{equation}
    \tilde X=(AX+B)(CX+D)^{-1} \sps
    \tilde{\tilde X}=(\tilde A\tilde X+\tilde B)(\tilde C\tilde X+\tilde D)^{-1}\sps
\end{equation}
then
\begin{equation}
%
\end{equation}
For the unitary fractional-linear transformations, we have
\begin{equation}
%
\end{equation}
Therefore, the identities
\begin{equation}
%
\end{equation}
are true. The proof of the second is that
\begin{equation}
%
\end{equation}
Then we will obtain the identity chain
\begin{equation}
%
\end{equation}
Multiply both parts by $CdX$
\begin{equation}
%
\end{equation}
Thus, the first invariant turns out.

\subsubsection{Proof of the second invariant relations}
$\phantom{ff}$Let
\begin{equation}
    \hat X=AX+B\sps
\end{equation}
or in the componentwise record
\begin{equation}
    %
\end{equation}
or in the abbreviated form
\begin{equation}
%
\end{equation}
Now let
\begin{equation}
     \hat X=X^{-1}\sps
\end{equation}
or in the componentwise record
\begin{equation}
%
\end{equation}
or in the abbreviated form
\begin{equation}
%
\end{equation}
It is obvious that $\tilde D $ can always be chosen so that $det(\tilde{\tilde {\tilde C}}) \ne 0 $. So
\begin{equation}
%
\end{equation}
that will give the second invariant.

\subsubsection{Proof of the octave alternative identity}
$\phantom{ff}$Let us prove the identity (\ref{e12.8}). If $P_{\Psi A}=\eta_{\Psi XA}X^X$ then
\begin{equation}
%
\end{equation}

\newpage
\subsubsection{Proof of the central Moufang identity}
$\phantom{ff}$Let us prove the identity (\ref{e12.9}). Note that $r^\Phi\eta_{\Phi\Theta}{}^\Omega$ and $\eta_{\Omega\Upsilon}{}^\Gamma r^\Upsilon$ are orthogonal transformations.
 \begin{equation}
%
\end{equation}
\end{enumerate}
This is direct proof. The simpler proof based on the formula (\ref{e8.1}, case 2).

\newpage
\subsubsection{Proof of the A-operator identities for n = 8}
$\phantom{ff}$Let us prove the identity (\ref{e12.11}).
\begin{enumerate}
\item First.\\
\begin{equation}
%
\end{equation}
\item Third.\\
On the basis of the identity (\ref{e6.4})
\begin{equation}
\eta_\Lambda{}^{AB}\eta_\Psi{}_{AD}\eta^\Omega{}^{CD}\eta^\Theta{}_{CB}=
2(g_{\Lambda\Psi}g^{\Omega\Theta}+\delta_\Lambda{}^\Theta\delta_\Psi{}^\Omega-\delta_\Psi{}^\Theta\delta_\Lambda{}^\Omega)\sps
\end{equation}
we obtain
\begin{equation}
%
\end{equation}

}
\newpage
\References{
\bibitem{Andreev1}
К. В. Андреев [K.V. Andreev]: Спинорный формализм и геометрия шестимерных римановых пространств [Spinorny\u\i\ formalizm i geometriya shestimernykh rimanovykh prostranstv]. Кандидатская диссертация [Kandidatskaya dissertatsiya], Уфа [Ufa], 1997, [in Russian: Spinor formalism and the geometry of six-dimensional Riemannian spaces. Ph. D. Thesis], \href{http://arxiv.org/abs/1204.0194}{arXiv:1204.0194v1}.
\bibitem{Andreev2}
К. В. Андреев [K.V. Andreev]: О структуре тензора кривизны 6-мерных римановых пространств [O strukture tenzora krivizny 6-mernykh rimanovykh prostranstv].  Вестик Башкирского унверситета [Vestnik Bashkirskogo Universiteta], 2(1996)44-47, [in Russian: On the structure of the curvature tensor of 6-dimensional Riemannian spaces].
\bibitem{Andreev3}
К. В. Андреев [K.V. Andreev]: О спинорном формализме при n=6 [O spinornom formalizme pri n=6]. Изв. Высш. Учебн. Завед., Матем. [Izv. Vyssh. Uchebn. Zaved., Mat.], 1(2001)11-23 (The article is freely available online at the Mathnet site: \url{http://mi.mathnet.ru/eng/ivm838}, the two figures of the article are missing in the electronic version, however). English translation: K. V. Andreyev: On spinor formalism for n = 6. Russian Mathematics (Iz. VUZ), 45:1(2001)9-20.]
\bibitem{Andreev4}
К. В. Андреев [K.V. Andreev]: О структурных константах алгебры октав. Уравнение Клиффорда [O strukturnykh konstantakh algebry oktav. Uravnenie Klifforda]. Изв. Высш. Учебн. Завед., Матем. [Izv. Vyssh. Uchebn. Zaved. Mat.],  3(2001)3-6. English translation: K. V. Andreyev: Structure constants of the algebra of octaves. The Clifford equation. Russian Mathematics (Iz. VUZ), 45:3(2001)1-4 (Mathnet URL: \url{http://mi.mathnet.ru/ivm856}).
\bibitem{Andreev5}
К. В. Андреев [K.V. Andreev]: О твисторах 6-мерного пространства [O tvistorakh 6-mernogo prostranstva]. ВИНИТИ-2469-B-98 [VINITI-2469-V-98], август [Aug] 1998, 23 с [23pp. Paper deponed on Aug 3, 1998 at VINITI (Moscow), ref. \No 2469-V 98], [in Russian: On twistors of 6-dimensional space].
\bibitem{Andreev6}
К. В. Андреев [K.V. Andreev]: Принцип тройственности для двух квадрик [Printsip tro\u\i stvennosti dlya dvykh kvadrik]. ВИНИТИ-2470-B-98 [VINITI-2470-V-98], август [Aug] 1998. 19 с. [19pp. Paper deponed on Aug 3, 1998 at VINITI (Moscow), ref. \No 2470-V 98.], [in Russian: Triality principle for two quadrics].
\bibitem{Andreev7}
К. В. Андреев [K.V. Andreev]: О внутренних геометриях многообразия плоских образующих 6-мерной квадрики [O vnutrennikh geometriyakh mnogoobraziya ploskikh obrazuyushchikh 6-merno\u\i\ kvadriki], Изв. Высш. Учебн. Завед., Матем. [Izv. Vyssh. Uchebn. Zaved. Mat.], 6(1998)3-8. English translation: K. V. Andreyev: On intrinsic geometries of the manifold of plane generators of a 6-dimensional quadric, Izv. Vyssh. Uchebn. Zaved. Mat., 42:6(1998)1-6 (Mathnet URL: \url{http://mi.mathnet.ru/ivm436}).
\bibitem{Baez1}
John C. Baez: The octonions, Bull. Amer. Math. Soc. (N.S.) 39(2002)145–205, \href{http://arxiv.org/abs/math/0105155}{arXiv:math.RA/0105155v4}. Russian translation: Баэз Джон С. [Ba\`ez Dzhon S]: Октонионы [Oktoniony]. Гиперкомплексные числа в геометрии и физике [Giperkomleksnye chisla v geometrii i fizike], 5:1(vol. 3)(2006)120-177.
\bibitem{Bilyalov1}
Р.Ф. Билялов, Б.С. Никитин: Спиноры в произвольных реперах. Ковариантная производная и производная Ли спиноров [Spinory v proizvol'nykh reperakh. Kovariantnaya proizvodnaya Li spinorov]. Изв. Высш. Учебн. Завед., Матем. [Izv. Vyssh. Uchebn. Zaved. Mat.], 6(1998)9-19. English translation: R. F. Bilyalov, B. S. Nikitin. Spinors in arbitrary frames. The covariant derivative and the Lie derivative of spinors. Izv. Vyssh. Uchebn. Zaved. Mat., 42:6(1998)7-15 (Mathnet URL: \url{http://mi.mathnet.ru/ivm440}).
\bibitem{Berger1}
Berger, Marcel: Geometry. II. Translated from the French by M. Cole and S. Levy. Universitext. Springer-Verlag, Berlin, 1987. Berger, Marcel: Geometry. I. Translated from the French by M. Cole and S. Levy. Universitext. Springer-Verlag, Berlin, 1987. Russian translation: М. Бергер [M. Berger]. Геометрия [Geometriya], т. 1, 2 [v. 1, 2]. Мир [Mir], Москва [Moskva],  1984. Russian translation by Ю.Н. Сударев [Yu.N. Sudarev], А.В. Пажитнов [A.V. Pazhitnov] and С.В. Чмутов [S.V. Chmutov] under edition И.Х. Сабитов [I.Kh. Sabitov].
\bibitem{Buchdahl1}
H.A. Buchdahl: On the calculus of for-spinors. Proceedings of the Royal Society of London, Series A, Mathematical and Physical Sciences, 303(1968)355-378 (\href{http://dx.doi.org/10.1098/rspa.1968.0055}{DOI:10.1098/rspa.1968.0055})
\bibitem{Dietmar1}
Dietmar Salamon: Spin Geometry and Seiberg-Witten invariants. Warwick, 1996.
\bibitem{Klotz1}
F. S. Klotz: Twistors and the conformal group. Journal of Mathematical Phisics, 15(1974)2242-2247, (\href{http://dx.doi.org/10.1063/1.1666606}{DOI:10.1063/1.1666606}).
\bibitem{Kotelnikov1}
А.П. Котельников [A.P. Kotel'nikov]: Винтовое счисление и некоторые приложения его к геометрии и механике [Vintovoe schislenie i nekotorye prilozheniya ego k geometrii i mekhanike]. Типо-литография Императорского Университета [Tipo-litografiya Imperatorskogo Universiteta], Казань [Kazan'], 1895. [in Russian: Screw Calculus and Some of Its Applications to Geometry and Mechanics]. Second reprint: А.П. Котельников [A.P. Kotel'nikov]: Винтовое счисление и некоторые приложения его к геометрии и механике [Vintovoe schislenie i nekotorye prilozheniya ego k geometrii i mekhanike]. КомКнига [Komkniga], Москва [Moskva], 2006.
\bibitem{Newfield1}
Э.Г. Нейфельд [\`E.G. Ne\u\i fel'd]: Об инволюциях в комплексных пространствах [Ob involyutsiyakh v kompleksnykh prostranstvakh]. Тр. Геом. Семин. [Tr. Geom. Semin.], Казанский университет [Kazanski\u\i\ universitet] (Выпуск [Vypusk]) 19(1989)71-82 (Mathnet URL: \url{http://mi.mathnet.ru/eng/kutgs98}). [in Russian: Involutions in complex spaces].
\bibitem{Newfield2}
Э.Г. Нейфельд [\`E.G. Ne\u\i fel'd]: Нормализация комплексных грассманианов и квадрик [Normalizatsiya kompleksnykh grassmanianov i kvadrik]. Тр. Геом. Семин. [Tr. Geom. Semin.], Казанский университет [Kazanski\u\i\ universitet] (Выпуск [Vypusk]) 20(1990)58-69 (Mathnet URL: \url{http://mi.mathnet.ru/kutgs82}). [in Russian: Normalization of complex Grassmannians and quadrics].
\bibitem{Norden1}
А.П. Норден [A.P. Norden]: О комплексном представлении тензоров пространства Лоренца [O kompleksnom predstavlenii tenzorov prostranstva Lorentsa]. Изв. Высш. Учебн. Завед., Матем. [Izv. Vyssh. Uchebn. Zaved., Mat.] (1959) No. 1 (8), 156-164 (Mathnet URL: \url{http://mi.mathnet.ru/eng/ivm2415}). [in Russian: On a complex representation of the tensors of Lorentz space].
\bibitem{Postnicov1}
М.М. Постников. [M.M. Postnikov] Группы и алгебры Ли [Gruppy i algebry Li]. Наука [Nauka], Москва [Moskva], 1986. [in Russian: M. Postnikov: Lie Groups and Lie Algebras. Lectures in Geometry, Semester 5. Mir, Moscow, 1986; URSS Publishing, Moscow, 1994]. The main ideas of the hypercomplex number constraction on the base of the Bott periodicity are given in the lectures 13-16.
\bibitem{Rozenfeld1}
Б.А. Розенфельд [B.A. Rozenfel'd]: Проективная геометрия как метрическая геометрия [Proektivnaya geometriya kak metricheskaya geometriya]. Труды семинара по векторному и тензорному анализу с их приложениями к геометрии, механике и физике [Trudy seminara po vektornomu i tenzornomu analizu s ikh prilozheniyami k geometrii, mekhanike, fizike], Москва [Moskva], 8(1950)328-354, [in Russian: Projective geometry as the metric geometry. Proceedings of the Seminar on Vector and tensor analysis with their applications.]
\bibitem{Rozenfeld2}
Б.А. Розенфельд [B.A. Rozenfel'd]: Неевклидовы геометрии [Neevklidovy geometrii]. ГИТТО [GITTO], Москва [Moskva], 1955. [in Russian: Non-Euclidean Geometries]. The Cartan triality principle is given on the page 534.
\bibitem{Zee1}
A. Zee: Quantum field theory in a nutshell. Princeton university press, 2003. Russian translation: Зи Энтони [Zi \`Entoni]. Квантовая теория в двух словах [Kvantovaya teoriya v dvukh slovakh]. РХД [RKHD], Москва [Moskva], Ижевск [Izhevsk], 2009. Russian translation by В.Г. Войткевич [V.G. Vo\u\i tkevich] and Ю.В. Колесниченко [Yu.V. Kolesnichenko] under edition И.В. Полюбина [I.V. Polyubina].
\newpage
\bibitem{Penrouz1}
R. Penrose, W. Rindler: Spinors and Space-Time. Vol. 1: Two-Spinor Calculus and Relativistic Fields. Vol. 2: Spinor and Twistor Methods in Space-Time Geometry. Cambridge Monogr. Math. Phys. Cambridge University Press, Cambridge, Vol. 1: 1984, Vol. 2: 1986. Russian translation: Р. Пенроуз [R. Penrouz], В. Риндлер [V. Rindler]: Спиноры и пространство-время [Spinory i prostranstvo-vremya]. Мир [Mir], Москва [Moskva], т. 1 [Tom 1]: 1987, т. 2 [Tom 2]: 1988. Russian translation by Е.М. Серебрянный [E.M. Serebryanny\u\i] and З.А. Штейнгард [Z.A. Shеу\u\i ngard] under edition Д.М. Гальцов [D.M. Gal'tsov].
\bibitem{Sintcov1}
Д.М.Синцов [D.M. Sintsov]: Теория коннексов в пространстве в связи с теорией дифференциальных уравнений в частных производных первого порядка. [Teoriya konneksov v prostranstve v svyazi s teorie\u\i\ differentsial'nykh uravneni\u\i\ v chastnykh proizvodnykh pervogo poryadka]. Типо-литография Императорского Университета [Tipo-litografiya Imperatorskogo Universiteta], Казань [Kazan'], 1894. [in Russian: Theory of connexes in space in relation to the theory of first order partial differential equations].
\bibitem{Stepanovskii1}
Ю.П. Степановский [Yu. P. Stepanpovski\u\i]: Спиноры 6-мерного пространства и их применение к описанию поляризованных частиц со спином 1/2 [Spinory 6-mernogo prostranstva i ikh primenenie k opisaniyu polyarizovannykh chastits so spinom 1/2]. Проблемы Ядерной Физики и Космических Лучей [Problemy Yaderno\u\i\ Fiziki i Kosmicheskikh Luche\u\i], Харьков [Khar'kov], 4(1976) 9-21, [in Russian: Spinors of a sixdimensional space and their application to the description of polarized particles with spin 1/2].
\bibitem{Stepanovskii2}
Ю.П. Степановський [Yu. P. Stepanpovski\u\i]. Алгебра матриць Дiрака у шестiмiрному виглядi [Algebra matrits Diraka u shestimirnomu vyglyadi]. Украiнський Фiзичний Журнал [Ukrains'ky\u\i\ Fizychny\u\i\ Zhurnal]. Kиiв [Kyiv], т. XI, 8(1968) 813-824, in Ukraine: Yu. P. Stepanovskii: Algebra of Dirac matrices in six-dimensional form. Ukrainian Journal of Physics, Kiev, v.XI, 8(1968) 813-824.
\bibitem{Schouten1}
J.A. Schouten, J. Haantjes: Konforme Feldtheorie II; $R_6$ und Spinuraum [Conformal field theory II and spin space]. Annali della R.Scuola Normale Superiore di Piza, Scienze Fisiche e Matematiche, 2. Series, 4(1935) 175-189.
\bibitem{Husemoller1}
Dale Husemoller: Fibre bundles. McGraw-Hill Book Company, 1966. Russian translation: Д. Хьюзмоллер [D. Kh'yuzmoller]. Расслоенные пространства [Rassloennye prostranstva]. Мир [Mir], Москва [Moskva], 1970. Russian translation by В.А. Исаковских [V.A. Isakovskikx] under edition М.М. Постников [M.M. Postnikov].
\bibitem{Scharnhorst1}
K. Scharnhorst  and J.-W. van Holten: Nonlinear Bogolyubov-Valatin transformations: 2 modes. Annals of Physics (New York), 326(2011)2868-2933 [\href{http://arxiv.org/abs/1002.2737}{arXiv:1002.2737v3}, NIKHEF preprint NIKHEF/2010-005] (\href{http://dx.doi.org/10.1016/j.aop.2011.05.001}{DOI: 10.1016/j.aop.2011.05.001}
\bibitem{Cartan1}
\'{E}. Cartan: Le\c{c}ons sur la Th\'{e}orie des Spineurs, 2 Vols.. Vol.\ I: Les Spineurs de l'Espace a Trois Dimensions. Actual. Sci. Ind., Vol.\ 643, Expos\'{e}s G\'{e}om., Vol.\ 9. Vol.\ II: Les Spineurs de l'Espace a n $>$ 3 dimensions. Les Spineurs en G\'{e}om\'{e}trie Riemanienne. Actual. Sci. Ind., Vol.\ 701, Expos\'{e}s G\'{e}om., Vol.\ 11. Hermann, Paris, 1938. English translation: The Theory of Spinors. Hermann, Paris, 1966. Reprinted: Dover Publications, Inc., New York, 1981. Russian translation: Э. Картан [\`E Kartan]: Теория спиноров [Teoriya spinorov]. Платон [Platon], Москва [Moskva], 1997. Russian translation by П. А. Широков [P.A. Shirokov].
\bibitem{Hodge1}
W.V.D. Hodge, D. Pedoe: Methods of Algebraic Geometry, Vol. 2. Cambridge University Press, Cambridge, 1952. Russian edition: В.Д. Ходж [V. D. Khodzh], Д. Пидо [D. Pido]: Методы алгебраической геометрии [Metody algebraichesko\u\i\ geometrii]. Т. 2 [Tom 2]. ИЛ [IL], Москва [Moskva], 1954. Russian translation by А.И. Узков [A.I. Uzkov].
\bibitem{Albert1}
A.A. Albert: Quadratic Forms Permitting Composition. Ann. of Math. 43(1942)161-177.
\newpage
\bibitem{Rozenfeld3}
Хуа Ло-гэн [Khua Lo-g\`en], Б.А. Розенфельд [B.A. Rozenfel'd]: Геометрия прямоугольных матриц и ее применение к вещественной проективной и неевклидовой геометрии.
[Geometriya pryamougol'nykh matrits i ee primenenie k veshchestvenno\u\i\ proektivno\u\i\ i neevklidovo\u\i\ geometrii]. Изв. Высш. Учебн. Завед., Матем. [Izv. Vyssh. Uchebn. Zaved., Matem.] (1957) No. 1, 233-247 (Mathnet URL: \url{http://mi.mathnet.ru/ivm3038}). English translation: Hua Loo-geng (Hua Loo-keng), B.A. Rozenfel'd: The geometry of rectangular matrices and its application to real-projective and non-euclidean geometry. Chin. Math. 8(1966)726-737.
\bibitem{Andreev0}
К. В. Андреев [K.V. Andreev]: О спинорном формализме при четной размерности базового пространства [O spinornom formalizme pri chetno\u\i\ razmernosti bazovogo prostranstva]. ВИНИТИ-298-B-11 [VINITI-298-V-11], июнь [Jun] 2011, 138 с. [138pp. Paper deponed on Jun 16, 2011 at VINITI (Moscow), ref. \No 298-V 11],
[in Russian: On the spinor formalism for the base space of even dimension].
}
  
\newpage
\renewcommand{\proofname}{Доказательство.}
\renewcommand{\refname}{Литература}
\def\ptctitle{Содержание}
\def\plttitle{Список таблиц}
\def\plftitle{Список рисунков}
\renewcommand{\figurename}{Рис.}
\renewcommand{\tablename}{Таблица}
\newtheorem{theoremr}{Теорема}[section]
\newtheorem{lemmar}{Лемма}[section]
\newtheorem{noter}{Замечание}[section]
\newtheorem{corollaryr}{Следствие}[section]
\newtheorem{definitionr}{Определение}[section]
\newtheorem{algorithmr}{Алгоритм}[section]

\def\References#1{{\normalsize\baselineskip=12pt\begin{thebibliography}{99}{#1}\end{thebibliography}}}
\newcounter{dividepage}
\setcounter{dividepage}{-\thepage}
\addtocounter{dividepage}{1}
\newcounter{myfootpage}
\newcounter{mypage}
\setcounter{mypage}{\thepage}
\addtocounter{mypage}{1}
\newcounter{pagemy}

\makeatletter
\def\addcontentsline#1#2#3{
  \begingroup
    \let\label\@gobble
    \ifx\@currentHref\@empty
      \Hy@Warning{%
        No destination for bookmark of \string\addcontentsline,%
        \MessageBreak destination is added%
      }%
      \phantomsection
    \fi
    \expandafter\ifx\csname toclevel@#2\endcsname\relax
      \begingroup
        \def\Hy@tempa{#1}%
        \ifx\Hy@tempa\Hy@bookmarkstype
          \Hy@WarningNoLine{%
            bookmark level for unknown #2 defaults to 0%
          }%
        \else
          \Hy@Info{bookmark level for unknown #2 defaults to 0}%
        \fi
      \endgroup
      \expandafter\gdef\csname toclevel@#2\endcsname{0}%
    \fi
    \edef\Hy@toclevel{\csname toclevel@#2\endcsname}%
    \Hy@writebookmark{\csname the#2\endcsname}%
      {#3}%
      {\@currentHref}%
      {\Hy@toclevel}%
      {#1}%
    \ifHy@verbose
      \begingroup
        \def\Hy@tempa{#3}%
        \@onelevel@sanitize\Hy@tempa
        \let\temp@online\on@line
        \let\on@line\@empty
        \Hy@Info{%
          bookmark\temp@online:\MessageBreak
          thecounter {\csname the#2\endcsname}\MessageBreak
          text {\Hy@tempa}\MessageBreak
          reference {\@currentHref}\MessageBreak
          toclevel {\Hy@toclevel}\MessageBreak
          type {#1}%
        }%
      \endgroup
    \fi
    \setcounter{pagemy}{\themypage}
    \addtocounter{pagemy}{\thedividepage}
    \addtocontents{#1}{\protect\contentsline{#2}{#3}{\thepage(\thepagemy)}{\@currentHref}}
  \endgroup
}
\renewcommand{\@oddhead}{\setcounter{myfootpage}{\thepage}\addtocounter{myfootpage}{\thedividepage}}
\renewcommand{\@evenhead}{\setcounter{myfootpage}{\thepage}\addtocounter{myfootpage}{\thedividepage}}
\renewcommand{\@oddfoot}{\addtocounter{mypage}{1}\thepage\hbox to 170mm{\quad\hrulefill\quad \themyfootpage}}
\renewcommand{\@evenfoot}{\addtocounter{mypage}{1}\thepage\hbox to 170mm{\quad\hrulefill\quad \themyfootpage}}
\makeatother
 \topmargin=-35mm
 \hsize=170mm
 \textwidth=170mm
 \linewidth=170mm
 \textheight=260mm
 \oddsidemargin=0mm
 \evensidemargin=0mm
\def\Abstract#1{{\normalsize\baselineskip=18pt\emph{\begin{quotation}\noindent\hcorrection{#1}\end{quotation}}}}
\def\References#1{{\normalsize\baselineskip=22pt\begin{thebibliography}{99}{#1}\end{thebibliography}}}
\def\UDK#1{\rightline{УДК {#1}}}
\addtocontents{toc}{\eject}
\label{originb}
\thispagestyle{empty}
\begin{center}
МИНИСТЕРСТВО ОБРАЗОВАНИЯ И НАУКИ\\
РОССИЙСКОЙ ФЕДЕРАЦИИ\\
ГОСУДАРСТВЕННОЕ ОБРАЗОВАТЕЛЬНОЕ УЧРЕЖДЕНИЕ\\
ВЫСШЕГО ПРОФЕССИОНАЛЬНОГО ОБРАЗОВАНИЯ\\
БАШКИРСКИЙ ГОСУДАРСТВЕННЫЙ УНИВЕРСИТЕТ\\
\end{center}
\vspace{7cm}

\UDK{514.744.2}
\vspace{1cm}
\Author{К.В. Андреев}
\vspace{1cm}
\begin{center}
\Title{О спинорном формализме при четной размерности базового пространства.}

\vspace{6cm}
УФА-2011
\end{center}
\newpage
 
\renewcommand{\baselinestretch}{1.5}
\foreignlanguage{russian}{
\part{Русская редакция}
\parttoc
\partlot
\partlof
\newpage
\section{Введение}
В статье\footnote{Статья приведена с исправлениями и оригинальной нумерацией страниц.} рассматриваются ответы на следующие вопросы при произвольном четном $n\ge~4$:
\begin{enumerate}
\setlength{\itemsep}{1.mm}
\item Что такое алгебра Клиффорда и как ее построить?
\item Что такое вещественная и комплексная реализация?
\item Что такое инволюция и как она помогает при переходе к вещественным вложениям?
\item Как построить вещественную и комплексную реализацию образующих (связующих операторов) алгебры Клиффорда?
\item Как инволюция действует на образующие (связующие операторы) алгебры Клиффорда?
\item Как привести комплексную ортогональную матрицу к блочно-диагональному виду?
\item Как построить основные изоморфизмы (в т.ч. двулистные накрытия) и иные соотношения в явном виде с помощью связующих операторов?
\item Как построить частные решения уравнения Клиффорда для связующих операторов при четном $n\ge 4$?
\item Как по связующим операторам построить структурные константы алгебр (без деления) гиперкомплексных чисел (седенионов) для $n\ mod\ 8=0$?
\item Как ввести и согласовать связности в касательном и спинорном расслоении?
\item Как построить аналоги операторов Ли и каковы условия их построения?
\item Как построить спиноры кривизны?
\item Какова связь между твисторным уравнением, деривационным уравнением нормализованного грассманиана и конформным уравнением Киллинга?
\item Почему отличается спинорный формализм при $n\le 8$ от $n>8$ и как его построить для малых размерностей?
\item Как построить геометрическое представление спинора для $\mathbb R^6_{(2,4)}$?
\item Как построить обобщение принципа тройственности Картана с соответствием Клейна и какова геометрия такого обобщения при $n=8$?
\item Как построить спинорные аналоги операторов Ли при $n=4$ в явном виде?
\end{enumerate}
\newpage

\section{Алгебра Клиффорда}
\Abstract{
\indent В этом параграфе рассказывается о том, как построить вещественную 2n-мерную алгебру Клиффорда с произвольным функционалом G согласно
\cite[лекция 13, с. 258-299]{Postnikovr1}. В случае, когда функционал G в подходящем базисе имеет диагональный вид с одинаковым количеством
$\ll+\gg$ и $\ll-\gg$ на главной диагонали, образующие такой алгебры Клиффорда порождают образующие комплексной n-мерной алгебры Клиффорда, которая
изоморфна алгебре комплексных матриц $\mathbb C(2^{\frac{n}{2}})$. Вывод всех результатов этого параграфа сделан на основе \cite{Postnikovr1}.
}

Пусть $\mathcal V$ - векторное пространство над $\mathbb R$. Тогда $\mathcal V$  будет модулем над $\mathbb R$: 1). a(x+y)=ax+ay,
2). (a+b)x=ax+bx,  3).~(ab)x=a(bx),  4). 1$\cdot$ x=x, где x,y $\in\ \mathcal V$, a,b $\in\ \mathbb R$.
Рассмотрим в алгебре $T_0(\mathcal V)=T_0{}^0(\mathcal V)\oplus ...\oplus T_0{}^q(\mathcal V)\oplus ...$ идеал
\begin{equation}
J(G):=\{T\otimes (x\otimes x - \frac{1}{2}G(x));\ T\in T_0(\mathcal V), x\in T_0{}^1(\mathcal V)\}
\end{equation}
и определим алгебру Клиффорда как $CL(G^{2n}):=T_0(\mathcal V)/J(G)$. Построим представление такой алгебры в алгебру многочленов
\begin{equation}
\begin{array}{ll}
\alpha :\mathcal V\longmapsto \mathcal A\sps & \tilde\alpha : T_0(\mathcal V)\longmapsto \mathcal A\sps \\
\alpha(\mathcal V)=B^\myf\Lambda x_\myf\Lambda\sps &
\tilde\alpha(T_0(\mathcal V))=A+B^\myf\Lambda x_\myf\Lambda+C^{\myf\Lambda\myf\Psi}x_\myf\Lambda x_\myf\Psi +...\ \spsd\\
\end{array}
\end{equation}
Здесь и далее $\myf\Lambda\sps\myf\Psi\sps ... =\overline{1,2n}$. Пусть теперь выполнено $\hat\alpha (J(G))=0$ для некоторого отображения
$\hat\alpha :CL(G)\longmapsto C\mathcal A$. Это будет означать, что
\begin{equation}
\label{re2.1}
x_\myf\Lambda x_\myf\Psi + x_\myf\Psi x_\myf\Lambda=G(x_\myf\Lambda , x_\myf\Psi)\spsd
\end{equation}
Поэтому форма $G(x_\myf\Lambda , x_\myf\Psi)$ в подходящем базисе имеет диагональный вид. Пусть для такого базиса
выполняются соотношения
\begin{equation}
x_\myf\Lambda x_\myf\Psi + x_\myf\Psi x_\myf\Lambda =0\sps (\myf\Lambda\ne \myf\Psi);\ \ x_\myf\Lambda{}^2=\pm 1\sps
\end{equation}
что определит отображение
\begin{equation}
\hat\alpha (Cl(G^{2n}))=A+B^\myf\Lambda x_\myf\Lambda+C^{[\myf\Lambda\myf\Psi ]}x_\myf\Lambda x_\myf\Psi +...
\end{equation}
В свою очередь, это означает, что алгебра Клиффорда конечна и $dim\ CL(G^{2n})=2^{2n}\sps dim\ \mathcal V =2n$.
Построим теперь комплексную алгебру Клиффорда $CL(g^n)$. Для этого рассмотрим вещественную алгебру Клиффорда
$CL(G^{2n}_{(n,n)})$. Будем считать, что $(x_{2\Lambda})^2=1\sps (x_{2\Lambda-1})^2=-1$. Зададим
\begin{equation}
\sqrt 2z_\Lambda=x_{2\Lambda}+ix_{2\Lambda -1}\sps \sqrt 2\bar z_\Lambda=x_{2\Lambda}-ix_{2\Lambda -1}\spsd
\end{equation}
Здесь и далее $\Lambda\sps\Psi\sps ... =\overline{1,n}$. Тогда
\begin{equation}

\end{equation}
Поэтому факторизация комплексной алгебры $T_0(\mathcal V^\mathbb C)= T_0(\mathcal V^\mathbb C)(z)$ по идеалу
$J(g)=\{T\otimes (z\otimes z-\frac{1}{2}g(z));\ \ T\in T_0(\mathcal V^\mathbb C)\sps z \in T_0{}^1(\mathcal V^\mathbb C)\}$
определит комплексную алгебру $CL(g^n)\cong T_0(\mathcal V^\mathbb C)/J(g)$.
\begin{theoremr}
\label{rtheorem2.1}
Для алгебр Клиффорда существует разложение
\begin{equation}
%
\end{equation}
Это даст возможность определить базис в $\mathcal V^{2n+2}$ как
\begin{equation}
X_\myf\Lambda=x_\myf\Lambda x_{2n+1}x_{2n+2}\sps
X_{2n+1}=x_{2n+2}\sps X_{2n+2}=x_{2n+1}
\end{equation}
с условиями
\begin{equation}
%
\end{equation}
Тогда для любого $a\in CL(G_{(n+1,n+1)}^{2n+2})$ выполнено разложение
\begin{equation}
\label{re2.7}
%
\end{equation}
Проведем замену базиса
\begin{equation}
%
\end{equation}
Тогда разложение (\ref{re2.7}) примет вид
\begin{equation}
a=(-a_1+a_4)y_{2n+1}y_{2n+2}+(a_2+a_3)y_{2n+1}+(a_2-a_3)y_{2n+2}+(a_1+a_4)y_{2n+2}y_{2n+1}\spsd
\end{equation}
Следовательно, любой элемент $a\in CL(G_{(1,1)}^2)$ разлагается как
\begin{equation}
%
\end{equation}
а для любых двух элементов $a,b\in CL(G_{(n+1,n+1)}^{2n+2})$ выполнено
\begin{equation}
%
\end{equation}
В то же время для любых двух элементов $a,b\in CL(G_{(1,1)}^2)$ верно тождество
\begin{equation}
%
\end{equation}
Что даст возможность поставить в соответствие образующим алгебры Клиффорда $CL(G_{(1,1)}^{2})$ их матричное представление
\begin{equation}
x_{2n+1}=
\left(
\begin{array}{rr}
0 & 1\\
1 & 0
\end{array}%
\right)\spsd
\end{equation}
Это означает, что для любых двух элементов $a,b\in CL(G_{(n+1,n+1)}^{2n+2})$ в аналогичном матричном представлении
$c_1,c_2,c_3,c_4\in CL(G_{(n,n)}^{2n})$. Аналогично доказываются и два других представления.
\end{proof}
\begin{theoremr}
\label{rtheorem2.2}
Комплексная алгебра Клиффорда при четном n изоморфна алгебре комплексных матриц вида
\begin{equation}
CL(g)\cong \mathbb C(2^{n/2})\spsd
\end{equation}
\end{theoremr}
\begin{proof}
Доопределим, согласуя с доказательством теоремы \ref{rtheorem2.1},
\begin{equation}
%
\end{equation}
Поставим в соответствие образующим $CL(G_{(2,2)}^{4})$ их матричное представление
\begin{equation}
\label{e2.4}
%
\end{equation}
Тогда для образующих $CL(g^2)$ в базисе, подобном базису примера \ref{rex2.2}\footnote{C точностью до множителя $x_\myf\Lambda\longmapsto \frac{1}{\sqrt{2}}x_\myf\Lambda$, $z_\Lambda\longmapsto \frac{1}{\sqrt{2}}z_\Lambda$.}, получим
\begin{equation}
\label{e2.5}
\sqrt{2}z_{n+1}=
\left(
%
\right)\sps
\end{equation}
что и докажет нашу теорему. В общем виде данное соответствие будет построено ниже.
\end{proof}

Таким образом вещественная алгебра Клиффорда $CL(G_{(n,n)}^{2n})$ изоморфна вещественной алгебре матриц размерности n
\begin{equation}
CL(G^{2n}_{(n,n)})\cong \mathbb R(2^n)\spsd
\end{equation}
Поэтому существует отображение
\begin{equation}
\label{re2.6}
\gamma:\ \ \mathbb R(2^n)\longmapsto (\mathcal V^\mathbb R)^{2n}\spsd
\end{equation}
На основании (\ref{re2.1}) операторы $\gamma_\myf\Lambda$ должны необходимо удовлетворять уравнению
\begin{equation}
\label{re2.8}
\gamma_\myf\Lambda\gamma_\myf\Psi+\gamma_\myf\Psi\gamma_\myf\Lambda=G_{\myf\Lambda\myf\Psi}\sps
\end{equation}
которое и называется уравнением Клиффорда. В то же время комплексная алгебра Клиффорда представляется матричной
алгеброй вида $\mathbb C(2^{n/2})$. Умножение в комплексной алгебре Клиффорда делегирует матричное умножение в
$\mathbb C(2^{n/2})$. Поэтому можно построить отображение
\begin{equation}
\label{re2.9}
\gamma:\ \ \mathbb C(2^{n/2})\longmapsto (\mathcal V^\mathbb C)^n\spsd
\end{equation}
Из (\ref{re2.2}) следует, что операторы $\gamma_\Lambda$ должны необходимо удовлетворять комплексному уравнению Клиффорда
\begin{equation}
\label{re2.10}
\gamma_\Lambda\gamma_\Psi+\gamma_\Psi\gamma_\Lambda=g_{\Lambda\Psi}\spsd
\end{equation}

\section{Комплексная и вещественные реализации}
\Abstract{
\indent В этом параграфе рассказывается о том, как по вещественной реализации 2n-мерного псевдориманова многообразия со слоями касательного расслоения,
изоморфными  $\mathbb R^{2n}_{(n,n)}$, локально построить его n-мерное комплексное представление. Для этого строится соответствующая
комплексная репараметризация атласа в окрестности некоторой точки, что индуцирует в касательном расслоении операторы Нейфельда \cite{Neifeldr1},
связанные с аффинором Нордена \cite{Nordenr1}. Соответственно, в спинорном расслоении связующими операторами $\gamma_{\tilde\Lambda}$
индуцируются с точностью до знака аналоги операторов Нейфельда. Это даст возможность перейти к комплексной  реализации связующих операторов
$\gamma_\Lambda$. Вывод всех результатов этого параграфа сделан на основе \cite{Neifeldr1}, \cite{Nordenr1}.
}

Под комплексным аналитическим римановым пространством $\mathbb CV_n$ в дальнейшем будем понимать аналитическое комплексное многообразие,
снабженное аналитической квадратичной метрикой, т.е. метрикой, определенной с помощью симметрического невырожденного тензора
$g_{\Lambda\Psi}$ (здесь $\Lambda\sps\Psi\sps...=\overline{1,n}$; $\myf\Lambda\sps\myf\Psi\sps ...=\overline{1,2n}$),
координаты которого - аналитические функции координат точки. Этому тензору соответствует комплексная риманова связность без кручения,
коэффициенты которой определяются символами Кристофеля и поэтому являются аналитическими функциями. Касательное расслоение этого многообразия
$\tau^\mathbb C(\mathbb CV_n)$ имеет слои $\tau_x^\mathbb C\cong \mathbb C\mathbb R^n$, то есть слои, изоморфные n-мерному комплексному евклидовому
пространству, метрика которого определяется значением метрического тензора в данной точке. Вещественная реализация $V_{2n}$ нашего
$\mathbb CV_n$ имеет касательное расслоение $\tau^\mathbb R(V_{2n})$ со слоями, изоморфными $\mathbb R^{2n}_{(n,n)}$. Пусть на $V_{2n}$
задан атлас $(U;x^\myf\Lambda)$. Рассмотрим репараметризацию этого атласа $(U;w^\Lambda)$ такую, что $w^\Lambda=\frac{1}{\sqrt{2}}(u^\Lambda(x^\myf\Lambda)+
iv^\Lambda(x^\myf\Lambda))$, которая локально разрешима как $x^\myf\Lambda=x^\myf\Lambda(u^\Lambda,v^\Lambda)$. Положим
\begin{equation}
\label{re1.1}
 m^\Lambda{}_\myf\Lambda:=\frac{1}{\sqrt{2}}(
 \frac{\partial u^\Lambda}{\partial x^\myf\Lambda}+i
 \frac{\partial v^\Lambda}{\partial x^\myf\Lambda})=:\frac{\partial w^\Lambda}{\partial x^\myf\Lambda}\sps
 m_\Lambda{}^\myf\Lambda:=\frac{1}{\sqrt{2}}(
 \frac{\partial x^\myf\Lambda}{\partial u^\Lambda}-i
 \frac{\partial x^\myf\Lambda}{\partial v^\Lambda})=:\frac{\partial x^\myf\Lambda}{\partial w^\Lambda}\spsd
\end{equation}
 Тогда
\begin{equation}
 \Delta_\myf\Lambda{}^\myf\Psi:=m^\Lambda{}_\myf\Lambda m_\Lambda{}^\myf\Psi=\frac{1}{2}((
 \underbrace{
 \frac{\partial u^\Lambda}{\partial x^\myf\Lambda}\ \frac{\partial x^\myf\Psi}{\partial u^\Lambda}+
 \frac{\partial v^\Lambda}{\partial x^\myf\Lambda}\ \frac{\partial x^\myf\Psi}{\partial v^\Lambda}}
 _{=\delta_\myf\Lambda{}^\myf\Psi})+
 i(\underbrace{
   \frac{\partial v^\Lambda}{\partial x^\myf\Lambda}\ \frac{\partial x^\myf\Psi}{\partial u^\Lambda}-
   \frac{\partial u^\Lambda}{\partial x^\myf\Lambda}\ \frac{\partial x^\myf\Psi}{\partial v^\Lambda}}
   _{=:f_\myf\Lambda{}^\myf\Psi}))\sps
\end{equation}
что определит комплексную структуру f на $V_{2n}$, где $\Delta_\myf\Lambda{}^\myf\Psi$ - аффинор Нордена. Определим $\bar m^{\Psi '}{}_\myf\Psi:=\overline{m^\Psi{}_\myf\Psi}$ и положим
\begin{equation}
 0=:m^\Lambda{}_\myf\Lambda \bar m_{\Psi '}{}^\myf\Lambda=
 \frac{1}{2}((
 \underbrace{
 \frac{\partial u^\Lambda}{\partial x^\myf\Lambda}\ \frac{\partial x^\myf\Lambda}{\partial u^{\Psi '}}-
 \frac{\partial v^\Lambda}{\partial x^\myf\Lambda}\ \frac{\partial x^\myf\Lambda}{\partial v^{\Psi '}}}
 _{\equiv 0})+
 i\ (\frac{\partial u^\Lambda}{\partial x^\myf\Lambda}\ \frac{\partial x^\myf\Lambda}{\partial v^{\Psi '}}+
    \frac{\partial v^\Lambda}{\partial x^\myf\Lambda}\ \frac{\partial x^\myf\Lambda}{\partial u^{\Psi '}}
   ))\sps
\end{equation}
\begin{equation}
 \delta_\Psi{}^\Lambda=:m^\Lambda{}_\myf\Lambda m_\Psi{}^\myf\Lambda=
 \frac{1}{2}((
 \underbrace{
 \frac{\partial u^\Lambda}{\partial x^\myf\Lambda}\ \frac{\partial x^\myf\Lambda}{\partial u^\Psi}+
 \frac{\partial v^\Lambda}{\partial x^\myf\Lambda}\ \frac{\partial x^\myf\Lambda}{\partial v^\Psi}}
 _{=2\delta_\Psi{}^\Lambda})-
 i\ (\frac{\partial u^\Lambda}{\partial x^\myf\Lambda}\ \frac{\partial x^\myf\Lambda}{\partial v^\Psi}-
    \frac{\partial v^\Lambda}{\partial x^\myf\Lambda}\ \frac{\partial x^\myf\Lambda}{\partial u^\Psi}
   ))\spsd
\end{equation}
 Откуда $\frac{\partial u^\Lambda}{\partial x^\myf\Lambda}\ \frac{\partial x^\myf\Lambda}{\partial v^\Psi}\equiv 0$
 (поскольку $v^{\Psi '}=\delta_\Psi{}^{\Psi '} v^\Psi$ и $u^{\Psi '}=\delta_\Psi{}^{\Psi '} u^\Psi$, а тензор $\delta_\Psi{}^{\Psi '}$
 имеет единичную матрицу). Следовательно
\begin{equation}
 \frac{\partial u^\Lambda}{\partial x^\myf\Lambda} f_\myf\Psi{}^\myf\Lambda=
 \frac{\partial u^\Lambda}{\partial x^\myf\Lambda}\
 (\frac{\partial v^\Psi}{\partial x^\myf\Psi}\ \frac{\partial x^\myf\Lambda}{\partial u^\Psi}-
  \frac{\partial u^\Psi}{\partial x^\myf\Psi}\ \frac{\partial x^\myf\Lambda}{\partial v^\Psi})=
 \frac{\partial v^\Lambda}{\partial x^\myf\Psi}
\end{equation}
 есть условия Коши-Римана, если в данном базисе комплексная структура имеет канонический вид
\begin{equation}
f_\myf\Lambda{}^\myf\Psi=\left(
\begin{array}{cc}
  0 & E \\
 -E & 0
\end{array}\right)\sps
\end{equation}
где E - единичная матрица размера $n\times n$. Легко проверить,  что  $f^2=-E$. Если теперь потребовать, чтобы карты $\mathbb CV_n$ были
согласованы голоморфными преобразованиями, то таким образом можно отождествить два многообразия: $\mathbb CV_n$ и $V_{2n}$. При этом можно
построить изоморфизм между слоями касательных расслоений: $\tau^\mathbb C(\mathbb CV_n)$ и $\tau^\mathbb C(V_{2n})$, так же как и между слоями их
вещественных реализаций: $\tau^\mathbb R(\mathbb CV_n)$ и $\tau^\mathbb R(V_{2n})$.\\
\indent Пусть в том же базисе, в котором комплексная структура имеет канонический вид, наша метрика такая
\begin{equation}
 G_{\myf\Lambda\myf\Psi}=
 \left(
 \begin{array}{cc}
  E &  0\\
  0 & -E
 \end{array}
 \right)\sps
\end{equation}
где E - единичная матрица размера $n\times n$. За соответствие $V_{2n}\longleftrightarrow \mathbb CV_{n}$ отвечают специальные операторы
$m_\Lambda{}^\myf\Lambda\sps m^\Lambda{}_\myf\Lambda$, которые по определению должны удовлетворять следующей системе уравнений
\begin{equation}
\label{re1.2}
 \left\{
 \begin{array}{l}
  m_\Lambda{}^\myf\Lambda m^\Psi{}_\myf\Lambda=\delta_\Lambda{}^\Psi\sps
  \\ \\
  m_\Lambda{}^\myf\Lambda \bar m^{\Psi '}{}_\myf\Lambda=0\sps
  \\ \\
  m_\Lambda{}^\myf\Lambda m^\Lambda{}_\myf\Psi=\frac{1}{2}(\delta_\myf\Psi{}^\myf\Lambda+if_\myf\Psi{}^\myf\Lambda):=
  \triangle_\myf\Psi{}^\myf\Lambda\spsd
  \\
 \end{array}
 \right.
\end{equation}
 Тогда из среднего уравнения будет следовать
\begin{equation}
 m^\Psi{}_\myf\Omega=m^\Lambda{}_\myf\Omega (m_\Lambda{}^\myf\Lambda m^\Psi{}_\myf\Lambda)=(m^\Lambda{}_\myf\Omega m_\Lambda{}^\myf\Lambda)
 m^\Psi{}_\myf\Lambda=\frac{1}{2}m^\Psi{}_\myf\Omega+\frac{i}{2} f_\myf\Omega{}^\myf\Lambda m^\Psi{}_\myf\Lambda\spsd
\end{equation}
 Откуда
\begin{equation}
\label{re1.3}
 m^\Psi{}_\myf\Omega=i\ f_\myf\Omega{}^\myf\Lambda m^\Psi{}_\myf\Lambda
\end{equation}
 опять есть  условия Коши-Римана. Поэтому в присутствии метрики $G_{\myf\Lambda\myf\Psi}$ на $\mathbb R^{2n}_{(n,n)}$
 в комплексной реализации индуцируется метрика
\begin{equation}
\label{re1.4}

\end{equation}
В стандартной комплексной реализации (где вместо пространства $\mathbb R^{2n}_{(n,n)}$ берется пространство $\mathbb R^{2n}$) подразумевается,
что $g_{\Lambda '\Psi ' }=\bar g_{\Lambda ' \Psi '}=0$, и это приводит к эрмитовости метрики $g_{\Lambda '\Psi }$. Однако, в нашем случае
следует воспользоваться условиями
\begin{equation}
\label{re1.5}
 g_{\Lambda '\Psi  }=\bar g_{\Lambda \Psi '}=0\spsd
\end{equation}
 Это приведет к некоторым ограничениям на метрический тензор
\begin{equation}
%
\end{equation}
 Поэтому наша псевдоевклидова метрика с канонической комплексной структурой удовлетворяет уравнению
\begin{equation}
\label{re1.6}
 f_\myf\Omega{}^\myf\Lambda G_{\myf\Lambda\myf\Psi}f_\myf\Theta{}^\myf\Psi=-G_{\myf\Omega\myf\Theta}\sps
\end{equation}
$$
\left(
%
\right)\spsd
$$
 Кроме того, будет выполнено
\begin{equation}
\label{re1.7}
 G_{\myf\Lambda\myf\Psi}=
 m^\Lambda{}_\myf\Lambda m^\Psi{}_\myf\Psi g_{\Lambda\Psi}+
 \bar m^{\Lambda '}{}_\myf\Lambda \bar m^{\Psi '}{}_\myf\Psi \bar g_{\Lambda '\Psi '}\spsd
\end{equation}

\vspace{4mm}
\begin{example} При n=1 для $\mathbb R^2_{(1,1)}$
\begin{equation}
%
\end{equation}
\end{example}

Теперь операторы (\ref{re1.2}) можно применить к операторам (\ref{re2.6}). Для этого рассмотрим разложение при n=2 (N=1)
\begin{equation}
\label{re1.8}
 f_\myf\Psi{}^\myf\Lambda\gamma_\myf\Lambda=\tilde F\gamma_\myf\Psi \tilde {\tilde F}\sps \tilde F{}^4=\tilde {\tilde F}{}^4=E\spsd
\end{equation}
Доказательство этой формулы будет дано в следующем параграфе (\ref{re3.7}). Из (\ref{re1.8}) будет следовать
\begin{equation}
%
\end{equation}
Это означает, что можно положить при n=2 (N=1)
\begin{equation}
\label{re1.9}
 %
 \right.  \\
 \end{array}
\end{equation}
Здесь $\myf A\sps\myf B\sps\myf C\sps\myf D, ... =\overline{1,4}$, $\hat A,\hat B,\hat C,\hat D, ... =\overline{1,2}$. Заметим, что
можно ввести операцию сопряжения (не комплексного!) такую, что  операторы $\tilde m_{\pm}\ (\tilde{\tilde m}_{\pm})$ и
$\tilde m_{\mp}\ (\tilde{\tilde m}_{\mp})$ относительно такой операции будут сопряжены друг другу. Следовательно, в качестве операторов
$\gamma_\Lambda$ можно взять либо $\gamma_+{}_\Lambda$, либо $\gamma_-{}_\Lambda$
\begin{equation}
\label{re1.10}
\gamma_\pm{}_\Lambda{}_{\hat A}{}^{\hat B}:=
m{}_\Lambda{}^\myf\Lambda\gamma_\myf\Lambda{}_\myf C{}^\myf D
\tilde m_\pm{}_{\hat A}{}^\myf C \tilde {\tilde m}_\pm{}^{\hat B}{}_\myf D\spsd
\end{equation}

\vspace{4mm}

\begin{example}
\label{rex2.2}
При n=2 для $\mathbb R^4_{(2,2)}$
\begin{equation}
\begin{array}{c}
 m_\Lambda{}^\myf\Lambda=\frac{1}{\sqrt{2}}
\left(
\begin{array}{cccc}
  1 & i & 0 & 0 \\
  0 & 0 & 1 & i
\end{array}
\right)\sps
 (m^T)_\myf\Lambda{}^\Lambda=\frac{1}{{\sqrt{2}}}
\left(
\begin{array}{cc}
  1 & 0 \\
 -i & 0 \\
  0 & 1 \\
  0 &-i
\end{array}
\right)\sps
 g_{\Lambda\Psi}=
\left(
\begin{array}{cc}
  1 &  0 \\
  0 &  1
\end{array}
\right)\sps\\
 G_{\myf\Lambda\myf\Psi}=
\left(
\begin{array}{cccc}
  1 &  0 & 0 &  0 \\
  0 & -1 & 0 &  0 \\
  0 &  0 & 1 &  0 \\
  0 &  0 & 0 & -1 \\
\end{array}
\right)\sps
 f_\myf\Lambda{}^\myf\Psi=
\left(
\begin{array}{cccc}
  0 & 1 & 0 &  0 \\
 -1 & 0 & 0 &  0 \\
  0 & 0 & 0 &  1 \\
  0 & 0 &-1 &  0 \\
\end{array}\right)\spsd
\end{array}
\end{equation}
$\gamma_\myf\Lambda$ определены согласно (\ref{re2.8}), $\gamma_\Lambda$ определены согласно (\ref{re2.10}).
\begin{equation}
\begin{array}{c}
 m_{\hat A}{}^\myf B:=\tilde m_+{}_{\hat A}{}^\myf B=\tilde {\tilde m}_+{}_{\hat A}{}^\myf B=
 \frac{1}{{\sqrt{2}}}\left(
\begin{array}{cccc}
  1 & i & 0 & 0 \\
  0 & 0 & 1 & 1
\end{array}
\right)\sps\\
 (m^T)_\myf B{}^{\hat A}=(\tilde m_+^T){}_\myf B{}^{\hat A}=(\tilde {\tilde m}_+^T){}_\myf B{}^{\hat A}=
 \frac{1}{{\sqrt{2}}}\left(
\begin{array}{cc}
  1 & 0 \\
 -i & 0 \\
  0 & 1 \\
  0 & 1
\end{array}
\right)\sps\\
\tilde m_-{}_{\hat A}{}^\myf B=\tilde {\tilde m}_-{}_{\hat A}{}^\myf B=
 \frac{1}{{\sqrt{2}}}\left(
\begin{array}{cccc}
  1 &-i & 0 & 0 \\
  0 & 0 & 1 &-1
\end{array}
\right)\sps\\
 (\tilde m_-^T){}_\myf B{}^{\hat A}=(\tilde {\tilde m}_-^T){}_\myf B{}^{\hat A}=
 \frac{1}{{\sqrt{2}}}\left(
\begin{array}{cc}
  1 & 0 \\
  i & 0 \\
  0 & 1 \\
  0 &-1
\end{array}
\right)\sps\\
F^\myf A{}_\myf B:=\tilde F^\myf A{}_\myf B=\tilde {\tilde F}^\myf A{}_\myf B=
\left(
\begin{array}{cccc}
  0 & 1 & 0 &  0 \\
 -1 & 0 & 0 &  0 \\
  0 & 0 & 0 &  1 \\
  0 & 0 & 1 &  0 \\
\end{array}\right)\spsd
\end{array}
\end{equation}
Кстати, это еще раз говорит о том, что оператор $F$ является матричным корнем 4 степени из единицы. Квадрат $F^2$ соответствует в пространстве
$\mathbb R^4_{(2,2)}$ тривиальному преобразованию, умноженному на -1.
\end{example}

В случае произвольного n алгоритм понижения спинорной размерности следующий.
\begin{algorithmr}
\label{ra2.1}
Комплексная структура представима в виде произведения элементарных преобразований размерности $4 \times 4$. Каждое I-ое элементарное
преобразование порождает свой оператор $(m_I)_\Lambda{}^{\tilde\Lambda}$, который отвечает за переход от вещественной реализации
подпространства $\mathbb (R_I)^4_{(2,2)}\subset\mathbb R^{2n}_{(2n,2n)}$ к комплексной реализации $\mathbb (C_I)^2\subset R^{2n-4}_{(2n-2,2n-2)}\oplus\mathbb (C_I)^2$
и, следовательно, понижает размерность пространства на 2. Соответственно, в спинорном пространстве генерируются операторы
$\tilde{\tilde m}_{I_\pm} ,\ \tilde m_{I_\pm}$, способные понизить размерность спинорного пространства в 2 раза. Очевидно, что требуется
$n/2$ таких шагов $(I=\overline{1,\frac{n}{2}})$. Это понизит спинорную размерность в $2^{\frac{n}{2}}=2N$ раз и приведет к комплексному
матричному представлению размерности $2N \times 2N$. При этом операторы $\tilde{\tilde m}_{I_\pm} ,\ \tilde m_{I_\pm}$ будут удовлетворять
следующим соотношениям
\begin{equation}
\label{re1.11}
 \begin{array}{c}
 \left\{
 \begin{array}{l}
  \tilde m_{I_\pm}{}_{\hat A}{}^\myf B \tilde m_{I_\pm}{}^{\hat C}{}_\myf B=\delta_{\hat A}{}^{\hat C}\sps
  \\ \\
  \tilde m_{I_\pm}{}_{\hat A}{}^\myf B \tilde m_{I_\mp}{}^{\hat C}{}_\myf B=0\sps
  \\ \\
  \tilde m_{I_\pm}{}_{\hat A}{}^\myf B \tilde m_{I_\pm}{}^{\hat A}{}_\myf C=\tilde\triangle_{I_\pm}{}_\myf C{}^\myf B\sps
  \\
 \end{array}
 \right.
 \left\{
 \begin{array}{l}
  \tilde{\tilde m}_{I_\pm}{}_{\hat A}{}^\myf B \tilde{\tilde m}_{I_\pm}{}^{\hat C}{}_\myf B=\delta_{\hat A}{}^{\hat C}\sps
  \\ \\
  \tilde{\tilde m}_{I_\pm}{}_{\hat A}{}^\myf B \tilde{\tilde m}_{I_\mp}{}^{\hat C}{}_\myf B=0\sps
  \\ \\
  \tilde{\tilde m}_{I_\pm}{}_{\hat A}{}^\myf B \tilde{\tilde m}_{I_\pm}{}^{\hat A}{}_\myf C=\tilde{\tilde\triangle}_{I_\pm}{}_\myf C{}^\myf B\spsd
  \\
 \end{array}
 \right.  \\
 \end{array}
\end{equation}
Здесь для I шага $\myf A\sps\myf B\sps\myf C\sps\myf D, ... =\overline{1,\frac{(2N)^2}{2^{I-1}}}$, $\hat A,\hat B,\hat C,\hat D, ... =\overline{1,\frac{(2N)^2}{2^I}}$.
Тогда можно построить операторы ($\tilde m_{J_{z_J}}:=\tilde m_{J_\pm}{}_{\hat A}{}^\myf B$,
$\tilde{\tilde m}^*_{J_{z_J}}:=\tilde{\tilde m}_{J_\pm}{}^{\hat A}{}_\myf B$)
\begin{equation}
\label{re1.12}
\begin{array}{l}
\tilde M_K:=\tilde m_{\frac{n}{2}_{z_{\frac{n}{2}}}}\tilde m_{{\frac{n}{2}-1}_{z_{\frac{n}{2}-1}}}\cdot ...\cdot\tilde m_{2_{z_2}}\tilde m_{1_{z_1}}\sps
\tilde{\tilde M}_K:=\tilde{\tilde m}^*_{\frac{n}{2}_{z_{\frac{n}{2}}}}\tilde{\tilde m}^*_{{\frac{n}{2}-1}_{z_{\frac{n}{2}-1}}}\cdot ...\cdot\tilde{\tilde m}^*_{2_{z_2}}\tilde{\tilde m}^*_{1_{z_1}}\sps\\
\end{array}
\end{equation}
где $z_J\ (J=\overline{1,\frac{n}{2}})$  равно 0 при выборе знака $\ll-\gg$ или 1 при выборе знака $\ll+\gg$, тогда $K=\sum\limits_{J=1}^{\frac{n}{2}}z_J\cdot 2^{J-1}+1$.
С помощью этих операторов можно определить
\begin{equation}
\label{re1.13}
(\gamma_K)_\Lambda:=\tilde M_K (m_\Lambda{}^\myf\Lambda\gamma_\myf\Lambda)\tilde{\tilde M}_K\spsd
\end{equation}
\end{algorithmr}
Соответствующий пример будет разобран ниже (алгоритм \ref{ra6.1}).

\section{Вещественные вложения. Инволюции}
\Abstract{
\indent В этом параграфе рассказывается о том, как в комплексную реализацию $\mathbb CV_n$ 2n-мерного псевдориманова пространства локально вложить
действительное  n-мерное (псевдо-)риманово пространство. Для этого рассматривается вещественная поверхность в $\mathbb CV_n$ с вещественной
параметризацией. Это индуцирует в касательном расслоении оператор вложения $H_i{}^\Lambda$ \cite{Neifeldr1}, с помощью которого
можно получить слой вещественного касательного расслоения, снабженный (псевдо-)евклидовой метрикой. Сигнатура такой метрики будет
существенно зависеть от вида оператора вложения. Вывод всех результатов этого параграфа сделан на основе \cite{Neifeldr1}.
}

Вещественное (псевдо-)риманово пространство $V_n$ будем  рассматривать как поверхность вещественной размерности n  в пространстве
$\mathbb CV_n$, т.е. определять с помощью параметрического уравнения
\begin{equation}
\label{re5.1}
    w^\Lambda=w^\Lambda(u^i)\sps (\Lambda, \Psi ,... ,i,j,g,h=\overline{1,n})\sps
\end{equation}
где $w^\Lambda$ - комплексные координаты точки $x$ базы, а $u^i$ - параметры: локальные координаты точки пространства $V_n$.  Частные производные
$(\partial_i w^\Lambda =:H_i{}^\Lambda)$ определяют локальное вложение вещественного касательного пространства поверхности (\ref{re5.1}) в комплексное
касательное пространство $\tau_x^\mathbb C$ следующим образом
\begin{equation}
    H:\tau_x^\mathbb R\longmapsto\tau_x^\mathbb C\sps
\end{equation}
\begin{equation}
\begin{array}{c}
    w^\Lambda=w^\Lambda(u^i(t))\sps r^\Lambda:=\frac{dw^\Lambda}{dt}=
    H_i{}^\Lambda\frac{du^i}{dt}=:H_i{}^\Lambda r^i\\[2ex]
    \frac{du^i}{dt}\in\tau_x^\mathbb R\longmapsto \frac{dw^\Lambda}{dt}\in\tau_x^\mathbb C\sps
\end{array}
\end{equation}
где дифференцирование ведется вдоль вещественной кривой $\gamma (t)$ поверхности (\ref{re5.1}). Так как матрица
$\parallel H_i{}^\Lambda \parallel$ есть невырожденная якобиева матрица, то существует оператор $H^i{}_\Lambda$ такой, что
\begin{equation}
\label{re5.2}
    \left\{
\begin{array}{c}
    H^i{}_\Lambda H_i{}^\Psi=\delta_\Lambda{}^\Psi\sps\\
    H^i{}_\Lambda H_j{}^\Lambda=\delta_j{}^i\spsd
\end{array}
    \right.
\end{equation}
Отсюда следует, что оператор $H_i{}^\Lambda$ определяет в комплексном пространстве инволюцию
\begin{equation}
\label{re5.3}
    S_\Lambda {}^{\Psi '}=H^i{}_\Lambda\bar  H_i{}^{\Psi '}\sps
\end{equation}
где координаты $\bar H_i{}^{\Psi '}$ комплексно сопряжены координатам $H_i{}^\Psi$. Поэтому
\begin{equation}
    r^i=H^i{}_\Lambda r^\Lambda=\overline{H^i{}_\Lambda r^\Lambda}
    \ \ \Rightarrow \ \ S_\Lambda{}^{\Psi '}r^\Lambda=\bar r^{\Psi '}\spsd
\end{equation}
Это есть необходимое и достаточное условие того, что вектор $r^\Lambda\in \tau^\mathbb C_x$ будет вещественным. При этом
\begin{equation}
    S_\Lambda{}^{\Psi '}\bar S_{\Psi '}{}^\Phi=\delta_\Lambda{}^\Phi\spsd
\end{equation}
Метрику $V_n$ (вещественного (псевдо-)риманова пространства) определим условием
\begin{equation}
\label{re5.4}
    g_{\Lambda\Psi}r^\Lambda r^\Psi=\overline{g_{\Lambda\Psi}r^\Lambda r^\Psi}\sps
    \forall \bar r^{\Psi '}=S_\Lambda{}^{\Psi '} r^\Lambda\spsd
\end{equation}
Это означает, что вещественный тензор пространства $V_n$ определяется как тензор, самосопряженный относительно указанной эрмитовой инволюции
\begin{equation}
    g_{\Lambda\Psi}=S_\Lambda{}^{\Phi '}S_\Psi{}^{\Theta '}\bar g_{\Phi '\Theta '}\spsd
\end{equation}
Поэтому тензор
\begin{equation}
\label{re5.5}
    g_{ij}:=H_i{}^\Lambda H_j{}^\Psi g_{\Lambda\Psi}=\overline{H_i{}^\Lambda H_j{}^\Psi g_{\Lambda\Psi}}
\end{equation}
будет метрическим тензором $V_n\subset \mathbb CV_n$. Вид метрики $g_{ij}$  существенно зависит от структуры оператора $H_i{}^\Lambda$ и,
следовательно, от тензора инволюции  $S_\Lambda{}^{\Psi '}$.
\begin{example}
Пусть имеется комплексно-евклидово пространство $\mathbb C\mathbb R^4$. Определить вложение действительного пространства в комплексное можно
тремя существенно различными способами
\begin{equation}
\begin{array}{c}
\begin{array}{rrr}
H_i{}^\Lambda=
\left(
\begin{array}{cccc}
 1 & 0 & 0 & 0 \\
 0 & 1 & 0 & 0 \\
 0 & 0 & 1 & 0 \\
 0 & 0 & 0 & 1
\end{array}
\right)\sps &
H_i{}^\Lambda=
\left(
\begin{array}{cccc}
 1 & 0 & 0 & 0 \\
 0 & 1 & 0 & 0 \\
 0 & 0 & i & 0 \\
 0 & 0 & 0 & i
\end{array}
\right)\sps &
H_i{}^\Lambda=
\left(
\begin{array}{cccc}
 1 & 0 & 0 & 0 \\
 0 & i & 0 & 0 \\
 0 & 0 & i & 0 \\
 0 & 0 & 0 & i
\end{array}
\right)\spsd
\end{array}\\
G_{\Lambda\Psi}=
\left(
\begin{array}{cccc}
 1 & 0 & 0 & 0 \\
 0 & 1 & 0 & 0 \\
 0 & 0 & 1 & 0 \\
 0 & 0 & 0 & 1
\end{array}
\right)\sps\\
\begin{array}{rrr}
g_{ij}=
\left(
\begin{array}{cccc}
 1 & 0 & 0 & 0 \\
 0 & 1 & 0 & 0 \\
 0 & 0 & 1 & 0 \\
 0 & 0 & 0 & 1
\end{array}
\right)\sps &
g_{ij}=
\left(
\begin{array}{cccc}
 1 & 0 & 0 & 0 \\
 0 & 1 & 0 & 0 \\
 0 & 0 &-1 & 0 \\
 0 & 0 & 0 &-1
\end{array}
\right)\sps &
g_{ij}=
\left(
\begin{array}{cccc}
 1 & 0 & 0 & 0 \\
 0 &-1 & 0 & 0 \\
 0 & 0 &-1 & 0 \\
 0 & 0 & 0 &-1
\end{array}
\right)\spsd
\end{array}
\end{array}
\end{equation}
\end{example}

\section{Элементарные преобразования ортогональной группы}
\Abstract{
\indent В этом параграфе рассказывается о том, как привести псевдоортогональное преобразование к блочно-диагональному виду, а затем
перейти к комплексной реализации такого преобразования. Вывод всех результатов этого параграфа сделан на основе \cite{Bergerr1},
\cite{Rosenfeldr2}.
}

Рассмотрим ортогональные  преобразования в пространстве $\mathbb R^{2n}_{(n,n)}$, которые задаются соотношением
\begin{equation}
\label{re4.2}
G_{\myf\Lambda\myf\Psi}S_\myf\Omega{}^\myf\Lambda S_\myf\Gamma{}^\myf\Psi=G_{\myf\Omega\myf\Gamma}\spsd
\end{equation}
При этом,  базис выберем так, что $\parallel G_{\myf\Lambda\myf\Psi}\parallel$ будет иметь в нем диагональный вид.
\begin{theoremr}
\label{rtheorem4.1}
Конформные преобразования пространства $\mathbb R^{2n-2}_{(n-1,n-1)}$ образуют группу $O(n,n)$, состоящую из:\\
\begin{tabular}{lll}
1. вращений из $O(n-1,n-1)$, & 3. трансляций,  & 5.  суперпозиции  1-4.\\
2. дилатаций,                & 4. инверсий,
\end{tabular}\\
Тогда любое преобразование из $O(n,n)$ представимо в виде произведения элементарных преобразований
\begin{equation}
\label{re4.1}
S_{\myf\Lambda_1}{}^{\myf\Lambda_{J+1}}=\pm\prod_{I=1}^J(\pm (r_I)_{\myf\Lambda_I}(r_I)^{\myf\Lambda_{I+1}}-
\delta_{\myf\Lambda_I}{}^{\myf\Lambda_{I+1}})\sps (r_I)_{\myf\Lambda_I}(r_I)^{\myf\Lambda_I}=\pm 2\spsd
\end{equation}
\end{theoremr}
\begin{proof}
Элементарные преобразования из (\ref{re4.1}) действительно ортогональны
\begin{equation}
(\pm r_\myf\Lambda r^\myf\Psi-\delta_\myf\Lambda{}^\myf\Psi)
(\pm r_\myf\Psi r^\myf\Theta-\delta_\myf\Psi{}^\myf\Theta)=
(\pm 2)r_\myf\Lambda r^\myf\Theta\mp r_\myf\Lambda r^\myf\Theta\mp r_\myf\Lambda r^\myf\Theta+\delta_\myf\Lambda{}^\myf\Theta=
\delta_\myf\Lambda{}^\myf\Theta\spsd
\end{equation}
Обычные вращения из группы $O(2)$ представимы в виде (\ref{re4.1}), при этом, $r^\myf\Lambda=\sqrt{2}(\cos\frac{\alpha}{2}\sps\sin\frac{\alpha}{2}$),
и имеется два несвязных класса вращений
\begin{equation}
\begin{array}{rr}
a).\left(
\begin{array}{rr}
 \cos\alpha & \sin\alpha\\
 \sin\alpha &-\cos\alpha
\end{array}
\right) \sps
b).\left(
\begin{array}{rr}
 \cos\alpha & \sin\alpha\\
-\sin\alpha & \cos\alpha
\end{array}
\right) \spsd
\end{array}
\end{equation}
Однако преобразование b). получается суперпозицией двух преобразований из a).
\begin{enumerate}
\item
Преобразования из группы $O(1,1)$ представляют собой буст одного из 4 видов
\begin{equation}
\begin{array}{cccc}
a).&\left(
\begin{array}{rr}
 \ch\theta & \sh\theta\\
 \sh\theta & \ch\theta
\end{array}
\right) \sps &
b).&\left(
\begin{array}{rr}
 \ch\theta &-\sh\theta\\
 \sh\theta &-\ch\theta
\end{array}
\right) \sps
\\
c).&\left(
\begin{array}{rr}
-\ch\theta & \sh\theta\\
-\sh\theta & \ch\theta
\end{array}
\right) \sps&
d).&\left(
\begin{array}{rr}
-\ch\theta &-\sh\theta\\
-\sh\theta &-\ch\theta
\end{array}
\right) \spsd
\end{array}
\end{equation}
Преобразование b). представимо в виде (\ref{re4.1}), где $r^\myf\Lambda=\sqrt{2}(\ch\frac{\theta}{2}\sps\sh\frac{\theta}{2})$.
Преобразование a). есть суперпозиция двух преобразований вида b). Преобразования c). и d). отличаются лишь знаком от
преобразований b). и a). соответственно. Этим описываются 4 несвязные компоненты группы $O(1,1)$.
\item
Пусть одномерная дилатация в $\mathbb R^1$ имеет вид: $\tilde x=\lambda x$. Рассмотрим световой конус в
$\mathbb R^3_{(1,2)}$, заданный как $T^2-Z^2-X^2=0$. Рассечем его плоскостью $T+Z=1$, а затем выполним стереографическую проекцию сечения
на прямую $T=1$, $Z=0$, что индуцирует одномерное пространство $\mathbb R^1$ с единственной координатой $x=\frac{X}{T-Z}$.
Тогда $\tilde T+\tilde Z=\lambda(T+Z)$, $\tilde T-\tilde Z=\lambda^{-1}(T-Z)$, $\tilde X=X$.
При $\lambda>0$ данное преобразование будет бустом вида a).; при $\lambda<0$ данное преобразование будет бустом вида d).
\item
Рассмотрим трансляцию $\tilde x = x+a$ пространства $\mathbb R^1$. Это приведет к соотношениям:
$\tilde T-\tilde Z=T-Z$, $\tilde T+\tilde Z=T+Z+2aX+a^2(T-Z)$, $\tilde X=X+a(T-Z)$. Неподвижным вектором данного преобразования будет
изотропный вектор вида $(b,0,b)$. Никаким псевдоортогональным преобразованием невозможно перевести изотропный вектор в неизотропный,
поскольку такие преобразования сохраняют величину $T^2-X^2-Z^2$ неизменной. Это означает, что невозможно подействовать такими преобразованиями
на базис так, чтобы изотропный вектор был направлен вдоль неизотропной координатной оси. Однако, это не мешает разложить дилатацию на
элементарные преобразования: вращение, буст и еще одно такое же вращение. Для этого в качестве вращения возьмем элементарное преобразование
\begin{equation}
\left(
\begin{array}{rrr}
 1 &          0 &          0 \\
 0 & \cos\alpha &-\sin\alpha \\
 0 & \sin\alpha & \cos\alpha
\end{array}
\right)
\end{equation}
с $\tg\alpha =\frac{a}{2}$, а в качестве буста вида a). - элементарное преобразование
\begin{equation}
\left(
\begin{array}{rrr}
 \ch\theta & \sh\theta & 0\\
 \sh\theta & \ch\theta & 0\\
         0 &         0 & 1
\end{array}
\right)
\end{equation}
с $\ch\theta =\frac{a^2}{2}+1$. Тогда указанная композиция будет иметь вид
\begin{equation}
\begin{array}{c}
\left(
\begin{array}{ccc}
 1 &          0 &          0 \\
 0 & \frac{1}{\sqrt{1+\frac{a^2}{4}}} &-\frac{\frac{a}{2}}{\sqrt{1+\frac{a^2}{4}}} \\
 0 & \frac{\frac{a}{2}}{\sqrt{1+\frac{a^2}{4}}} & \frac{1}{\sqrt{1+\frac{a^2}{4}}}
\end{array}
\right)
\left(
\begin{array}{ccc}
 1+\frac{a^2}{2}         & a\sqrt{1+\frac{a^2}{4}} & 0\\
 a\sqrt{1+\frac{a^2}{4}} & 1+\frac{a^2}{2}         & 0\\
                       0 &                       0 & 1
\end{array}
\right)\cdot\\
\cdot\left(
\begin{array}{ccc}
 1 &          0 &          0 \\
 0 & \frac{1}{\sqrt{1+\frac{a^2}{4}}} &-\frac{\frac{a}{2}}{\sqrt{1+\frac{a^2}{4}}} \\
 0 & \frac{\frac{a}{2}}{\sqrt{1+\frac{a^2}{4}}} & \frac{1}{\sqrt{1+\frac{a^2}{4}}}
\end{array}
\right)=
\left(
\begin{array}{ccc}
 1+\frac{a^2}{2} & a &  -\frac{a^2}{2} \\
               a & 1 &              -a \\
   \frac{a^2}{2} & a & 1-\frac{a^2}{2}
\end{array}
\right)\spsd
\end{array}
\end{equation}
\item
Инверсия $\tilde x =\frac{1}{x}$ в $\mathbb R^1$  индуцирует вращение $\tilde T=T$,  $\tilde Z=-Z$, $\tilde X=X$.
\item
Суперпозиции всех одномерных преобразований вида 1-4 вместе с вращениями $O(n-1,n-1)$ представляют группу конформных преобразований
пространства $\mathbb R^{2n-2}_{(n-1,n-1)}$, которая будет изоморфна группе $O(n,n)$.
\end{enumerate}
\end{proof}

\begin{example}
Рассмотрим группу псевдоортогональных преобразований $O(1,2)$ пространства $\mathbb R^3_{(1,2)}$. Как известно, собственные значения любого
такого преобразования являются корнями многочлена третьей степени. Одно из таких значений обязательно должно быть вещественным числом.
Поскольку псевдоортогональные преобразования удовлетворяют соотношению (\ref{re4.2}), то квадрат такого вещественного числа есть единица.
Поэтому любое преобразование из группы $O(1,2)$ обладает неподвижной осью. Однако, не всегда эту ось можно совместить с координатной осью,
используя только псевдоортогональные преобразования базиса.
\begin{enumerate}
\item
Рассмотрим композицию вращения и буста для пространства $\mathbb R^3_{(1,2)}$
\begin{equation}
\left(

\right) \spsd
\end{equation}
Пусть эта композиция вектор $(a,b,c)$ оставит неподвижным
\begin{equation}
\left\{
%
\end{equation}
Проведем преобразование базиса
\begin{equation}
\left(
%
\right)\spsd
\end{equation}
Поскольку теперь $|r|^2=b^2(1/\sh^2\frac{\theta}{2}-\tg^2\frac{\alpha}{2})$, положим $b:=\pm\frac{1}{\sqrt{|1/\sh^2\frac{\theta}{2}-\tg^2\frac{\alpha}{2}}|}$.
Тогда существует преобразование базиса
\begin{equation}
\left(
%
\end{equation}
В случае a). неподвижна координатная ось (1,0,0), а в случае b). - ось (0,0,1).
Если $1/\sh\frac{\theta}{2}=\pm\tg\frac{\alpha}{2}$, то никаким псевдоортогональным вращением базиса невозможно упростить первоначальную суперпозицию.
\item
Рассмотрим композицию двух бустов для пространства $\mathbb R^3_{(1,2)}$
\begin{equation}
\left(
%
\right) \spsd
\end{equation}
Пусть эта композиция оставит вектор $(a,b,c)$  неподвижным
\begin{equation}
\left\{
%
\end{equation}
Проведем преобразование базиса
\begin{equation}
\left(
%
\right)\spsd
\end{equation}
Поскольку теперь $|r|^2=a^2(1/\ch^2\frac{\theta}{2}-\th^2\frac{\psi}{2})$, положим $a:=\pm\frac{1}{\sqrt{|1/\ch^2\frac{\theta}{2}-\th^2\frac{\psi}{2}}|}$.
Тогда существует преобразование базиса
\begin{equation}
\left(
%
\end{equation}
В случае a). неподвижна координатная ось (1,0,0), а в случае b). - ось (0,0,1).
Если $1/\ch^2\frac{\theta}{2}=\pm\th^2\frac{\psi}{2}$, то никаким псевдоортогональным вращением базиса невозможно упростить первоначальную суперпозицию.
\end{enumerate}
\end{example}

\begin{example}
Рассмотрим суперпозицию элементарных преобразований из группы $O(1,2)$
\begin{equation}
%
\end{equation}
Решением этой системы уравнений будет
\begin{equation}
%
\end{equation}
Опять же исключительным случаем является $\ch^2\tilde\psi\sin^2\tilde\alpha\geq 1$. Поэтому, не всегда возможно переставить
в суперпозиции элементарные преобразования, в отличие от ортогональной группы $O(3)$.
\end{example}

\begin{example}
Рассмотрим теперь преобразования из группы $O(2,2)$. Любое преобразование из этой группы  согласно теореме
Картана-Дьедонне \cite[т. 2, с. 33]{Bergerr1} можно представить с помощью четырех преобразований вида (\ref{re4.1})
\begin{equation}
\pm r_\myf\Lambda r^{\myf\Psi}-\delta_\myf\Lambda{}^\myf\Psi\sps r_\myf\Lambda r^\myf\Lambda=\pm 2\spsd
\end{equation}
Однако, всегда можно подобрать разложение таким образом, чтобы два из таких преобразований имели диагональную матрицу. Элементы главной
диагонали такой матрицы $I_\myf\Lambda{}^\myf\Psi$ по модулю равны 1. Поэтому (\ref{re4.1}) может быть переписано как
\begin{equation}

\end{equation}
Существует 3 существенно различных взаимоисключающих варианта.
\begin{enumerate}
\item
В этом случае псевдоортогональное преобразование может иметь блочно-диагональный вид
\begin{equation}
%
\end{equation}
\item
В этом случае псевдоортогональное преобразование может иметь блочно-диагональный вид
\begin{equation}
%
$$
\item
В этом случае псевдоортогональное преобразование не будет иметь блочно-диагонального вида, но может быть представлено как
\begin{equation}
%
\end{equation}
\end{enumerate}
\end{example}
Согласно все той же теореме Картана-Дьедонне любое преобразование из группы $O(2n,2n)$ в некотором базисе будет иметь
блочно-диагональный вид, состоящий из блоков $O(2,2)$. Это означает, что выполнено (\ref{re4.1})
\begin{equation}
\label{re4.3}
%
\end{equation}
Пусть теперь в пространстве $\mathbb R^{2n}_{(n,n)}$ задана каноническая комплексная структура, тогда в разложении (\ref{re4.1}) можно выделить
$J/2$ сомножителей так, чтобы было выполнено
\begin{equation}
%
\end{equation}
Этим для любого псевдо-ортогонального преобразования $S_\myf\Lambda{}^\myf\Psi$ определятся соотношения
\begin{equation}
\label{re4.4}
%
\end{equation}
Таким образом определится комплексная симметрия
\begin{equation}
%
\end{equation}
Кроме того,
\begin{equation}
r_\Lambda r^\Lambda=m_\Lambda{}^\myf\Psi r_\myf\Psi m^\Lambda{}_\myf\Phi r^\myf\Phi=
\frac{1}{2}(\delta_\myf\Phi{}^\myf\Psi+if_\myf\Phi{}^\myf\Psi)r_\myf\Psi r^\myf\Phi=
\frac{1}{2}r_\myf\Psi r^\myf\Psi=1\sps
\end{equation}
что окончательно подтверждает существование разложения
\begin{equation}
\label{re4.5}
%
\end{equation}
Однако, следует отметить, что  не всегда  $(r_I)_{\Lambda}(r_J)^{\Lambda}=0\sps (I\ne J)$ ввиду существования изотропных векторов,
что и было показано в предыдущем примере. Что касается блочной структуры преобразований из группы $O(n,\mathbb C)$, то размерность блоков тоже
будет  равна четырем.

\begin{example}
Пусть имеется трехмерное псевдоевклидово пространство $\mathbb R^3_{(1,2)}$. Рассмотрим преобразование
\begin{equation}
S_i{}^j=
\left(
%
\right)\spsd
\end{equation}
Определим оператор вложения как
\begin{equation}
H^i{}_\Lambda=
\left(
%
\right)
\spsd
\end{equation}
Это определит комплексификацию нашего преобразования
\begin{equation}
S_\Lambda{}^\Psi=H^i{}_\Lambda S_i{}^j H_j{}^\Psi=
\left(
%
\right)
\end{equation}
c комплексно-евклидовой метрикой. Чтобы перейти к действительной реализации, нам потребуются операторы следующего вида
\begin{equation}
(m^T)_\myf\Lambda{}^\Lambda=
\frac{1}{\sqrt{2}}\left(
%
\right)\spsd
\end{equation}
Соответственно, действительная реализация нашего преобразования будет иметь вид
\begin{equation}
S_\myf\Lambda{}^\myf\Psi=m^\Lambda{}_\myf\Lambda S_\Lambda{}^\Psi m_\Psi{}^\myf\Psi+
\bar m^{\Lambda '}{}_\myf\Lambda S_{\Lambda '}{}^{\Psi '}\bar m_{\Psi '}{}^\myf\Psi=
\left(
%
\right)\spsd
\end{equation}
Видно, что такая реализация имеет блочно-диагональную структуру, однако комплексная реализация такой структуры не имеет.
\end{example}

Наряду с преобразованиями (\ref{re4.4}) имеет смысл рассматривать псевдоортогональные преобразования
\begin{equation}
\label{re4.6}
%
\end{equation}
Комплексная реализация таких преобразований будет являться инволюциями на пространстве $\mathbb C^n$,
ответственными за некоторое вещественное вложение. Поэтому в разложении (\ref{re4.1}) можно выделить $J/2$ сомножителей так,
чтобы было выполнено
\begin{equation}
%
\end{equation}
Таким образом собственные значения преобразования $S_\Lambda{}^{\Psi '}$ суть $\pm 1$. Это автоматически означает, что
такое преобразование ортогонально диагонализируемо в комплексной области, а вещественная реализация в некотором базисе имеет
блочно-диагональный вид
\begin{equation}
\left(
%
\right)
\end{equation}
и тоже диагонализируема. За более подробным изложением следует обратиться к \cite[c. 79-126]{Rosenfeldr2}, \cite[т. 2]{Bergerr1}.

\section{Уравнение Клиффорда. Двулистное накрытие}
\Abstract{
\indent В этом параграфе рассказывается о том, как перейти к редуцированным спинорам Картана. Это даст возможность построить взаимно однозначное отображение между
векторами пространства $\mathbb C^n$ и бивекторами спинорного пространства $\mathbb C^N$, что приведет к построению двулистного накрытия
$Spin(n,\mathbb C)/\{\pm 1\}\cong SO(n,\mathbb C)$. Вывод всех результатов этого параграфа сделан на основе \cite{Postnikovr1},
\cite{Penroser1}.
}

Пусть имеется некоторое псевдориманово пространство $V_{2n}$, где n-число четное. Оно является вещественной реализацией некоторого комплексного
пространства $\mathbb CV_n$. Пусть в каждой точке $V_{2n}$ задан набор невырожденных операторов $\gamma_\myf\Lambda$.
$\myf\Lambda,\ \myf\Psi,\ ...=\overline{1,2n}$. При этом будем считать, что каждый оператор
$\gamma_\myf\Lambda:=m_\myf\Lambda{}^\Lambda\gamma_\Lambda+\bar m_\myf\Lambda{}^{\Lambda '}\gamma_{\Lambda '}$
может быть представлен вещественной матрицей $(2N)^2 \times (2N)^2$, где $N=2^{n/2-1}$, и удовлетворяют условию (\ref{re2.1})
\begin{equation}
\label{re3.0}
 \gamma_\myf\Lambda\gamma_\myf\Psi+\gamma_\myf\Psi\gamma_\myf\Lambda=G_{\myf\Lambda\myf\Psi}\sps
\end{equation}
где $G_{\myf\Lambda\myf\Psi}=G_{\myf\Psi\myf\Lambda}$ - невырожденный метрический тензор вида (\ref{re1.4})-(\ref{re1.5}), заданный в
вещественной реализации касательного расслоения к $V_{2n}$. Назовем комплексной реализацией операторов $\gamma_\myf\Lambda$ операторы (\ref{re1.13})
(если не важно, какой из K операторов будет выбран, то номер K будем опускать)
\begin{equation}
\label{re3.1}
\gamma_\Lambda:=m_\Lambda{}^\myf\Lambda (\tilde M\gamma_\myf\Lambda \tilde {\tilde M})\spsd
\end{equation}
Они будут удовлетворять уравнению
\begin{equation}
\label{re3.2}
 \gamma_\Lambda\gamma_\Psi+\gamma_\Psi\gamma_\Lambda=g_{\Lambda\Psi}\sps
 \bar \gamma_{\Lambda '}\bar \gamma_{\Psi '}+\bar \gamma_{\Psi '}\bar \gamma_{\Lambda '}=g_{\Lambda '\Psi '}\spsd
\end{equation}
Пусть $r^\Lambda$ - вектор комплексного касательного расслоения $\tau^\mathbb C$. В качестве базы можно рассматривать как комплексную
$\mathbb CV_n$, так и вещественную реализацию $V_{2n}$. Слой этого расслоения изоморфен пространству $\mathbb C^n$, вещественной реализацией
которого будет псевдоевклидово пространство $R_{(n,n)}^{2n}$, где n - индекс псевдоевклидовой метрики (количество $"+"$ на главной диагонали).
Таким образом,  вектору из пространства $\mathbb C^n$($\mathbb R^{2n}_{(n,n)})$ можно однозначно сопоставить некоторый оператор из пространства
$\mathbb C^{2N}$ по правилу
\begin{equation}
\label{re3.3}
 R:=r^\Phi\gamma_\Phi\spsd
\end{equation}
Поскольку из уравнения Клиффорда следует, что
\begin{equation}
tr(\gamma_\Lambda\gamma_\Psi)=N g_{\Lambda\Psi}\sps
\end{equation}
то любому такому оператору R можно однозначно сопоставить вектор пространства $\mathbb C^n$($R^{2n}_{(n,n)}$), взяв след от обеих частей
соответствия (\ref{re3.3}) с $\gamma^\Lambda:=g^{\Lambda\Psi}\gamma_\Psi$
\begin{equation}
 g^{\Lambda\Psi} tr(\gamma_\Psi R)=r^\Phi g^{\Lambda\Psi}tr(\gamma_\Psi\gamma_\Phi)\sps
\end{equation}
и тогда
\begin{equation}
 \frac{1}{N} tr(\gamma^\Psi R)=r^\Psi \spsd
\end{equation}
Таким образом, получено взаимно однозначное соответствие между векторами $r^\Lambda$ пространства $\mathbb C^n$($\mathbb R^{2n}_{(n,n)})$ и
операторами R (специального вида, конечно же), действующими на пространстве $\mathbb C^{2N}$. Используя уравнение Клиффорда (\ref{re3.2}),
можно написать тождество
\begin{equation}
\label{re3.4}
 R\gamma_\Psi R=
 r^\Lambda\gamma_\Lambda \gamma_\Psi \gamma_\Phi r^\Phi=
 r^\Lambda (g_{\Lambda\Psi}-\gamma_\Psi\gamma_\Lambda)
 \gamma_\Phi r^\Phi= (r_\Psi r^\Phi-\frac{1}{2}(r_\Lambda r^\Lambda)\delta_\Psi{}^\Phi)
 \gamma_\Phi\spsd
\end{equation}
Пусть $r_\Lambda r^\Lambda:=\pm 2$. Известно, что любое (псевдо-)ортогональное преобразование $S_\Psi{}^\Phi$ может быть представлено как
конечное произведение элементарных преобразований вида $S_I:=(r_I){}_\Psi (r_I)^\Phi-\frac{1}{2}((r_I)_\Lambda (r_I)^\Lambda)\delta_\Psi{}^\Phi$
(I пробегает конечные значения от 1 до какого-нибудь конечного J, например). Доказательству этого факта посвящен предыдущий параграф. Определим
\begin{equation}
 R_{I}:=(r_{I}){}^\Lambda\gamma_\Lambda\sps
 S=\prod\limits_{I=1}^{J} S_I=S_1S_2\ldots S_J\sps
\end{equation}
\begin{equation}
 \tilde S=\prod\limits_{I=1}^J R_I=R_1 R_2 \ldots R_J \sps
 \tilde{\tilde S}:=\prod\limits_{I=J}^1 R_I = R_J\ldots R_2 R_1\spsd
\end{equation}
Здесь $\tilde S$ есть правое произведение операторов $R_I$, а $\tilde{\tilde S}$ есть левое произведение операторов $R_I$. Поэтому матрицы
операторов $\tilde S$ и $\tilde{\tilde S}$ в общем случае различны и никак друг с другом не связаны. Это позволит переписать соответствие
(\ref{re3.4}) для ортогонального преобразования $S_\Lambda{}^\Psi$ как
\begin{equation}
\label{re3.5}
 S_\Lambda{}^\Psi\gamma_\Psi=\tilde S\gamma_\Lambda\tilde{\tilde S}\sps  \tilde{\tilde S}=\tilde S^{-1}\spsd
\end{equation}
Уравнение (\ref{re3.4}) справедливо и для вещественной реализации, что даст
\begin{equation}
\label{re3.6}
 S_\myf\Lambda{}^\myf\Psi\gamma_\myf\Psi=\tilde S\gamma_\myf\Lambda\tilde{\tilde S}\sps
 f_\myf\Theta{}^\myf\Lambda S_\myf\Lambda{}^\myf\Psi f_\myf\Psi{}^\myf\Phi=-S_\myf\Theta{}^\myf\Phi\spsd
\end{equation}
Соответственно, уравнение (\ref{re3.6}) для инволюции перепишется как
\begin{equation}
\label{re3.7}
 S_\myf\Lambda{}^\myf\Psi\gamma_\myf\Psi=\tilde S\gamma_\myf\Lambda\tilde{\tilde S}\sps
 f_\myf\Theta{}^\myf\Lambda S_\myf\Lambda{}^\myf\Psi f_\myf\Psi{}^\myf\Phi=S_\myf\Theta{}^\myf\Phi\sps
 S_\myf\Lambda{}^\myf\Psi S_\myf\Psi{}^\myf\Phi = \delta_\myf\Lambda{}^\myf\Phi\spsd
\end{equation}
Тогда для комплексной реализации получим тождество
\begin{equation}
\label{re3.8}
 S_\Lambda{}^{\Psi '}\bar \gamma_{\Psi '}=\bar{\tilde S}\gamma_\Lambda \tilde {\tilde S}\sps
 S_\Lambda{}^{\Psi '}\bar S_{\Psi '}{}^\Phi =\delta_\Lambda{}^\Phi\sps \tilde S\bar{\tilde S} = \pm E\sps
 \tilde {\tilde S}\bar{\tilde {\tilde S}} = \pm E\spsd
\end{equation}
Теперь возможно комплексифицировать уже и вещественную реализацию $\mathbb R^{2n}_{(n,n)}$, рассматривая вектор $r^\myf\Lambda$ в том же базисе,
но с комплексными коэффициентами. При этом $\gamma_\myf\Lambda$ останутся прежними. В этом случае $S_\myf\Lambda{}^\myf\Psi$ из (\ref{re3.6})
будет уже комплексным преобразованием. В частности, в качестве такого преобразования можно рассмотреть преобразование $if_\myf\Psi{}^\myf\Phi$
ввиду (\ref{re1.6}). Это означает, что комплексная структура разложима как
\begin{equation}
\label{re3.9}
 if_\myf\Lambda{}^\myf\Psi\gamma_\myf\Psi=\tilde I\tilde F\gamma_\myf\Lambda \tilde {\tilde F}\tilde {\tilde I}\spsd
\end{equation}
Для дальнейшего исследования этого соответствия необходимо изучить структуру операторов $\gamma_\myf\Lambda$.
Не ограничивая общности, можно рассматривать базис, в котором метрический тензор $G_{\myf\Lambda\myf\Psi}$  на главной диагонали
имеет $\pm 1$, a остальные компоненты равны 0
\begin{equation}
\label{re3.10}
 \gamma_\myf\Lambda\gamma_\myf\Psi=-\gamma_\myf\Psi\gamma_\myf\Lambda\sps \myf\Lambda\neq\myf\Psi\spsd
\end{equation}
Построим оператор
\begin{equation}
 \gamma_0:=\prod_{\myf\Lambda =1}^{2n} \gamma_\myf\Lambda\spsd
\end{equation}
Тогда из (\ref{re3.10}) будет следовать
\begin{equation}
\label{re3.11}
 \gamma_0\gamma_\myf\Lambda=-\gamma_\myf\Lambda\gamma_0 \sps
\end{equation}
$$
 (\gamma_0-\lambda E)\gamma_\myf\Lambda=-\gamma_\myf\Lambda(\gamma_0+\lambda E)\spsd
$$
Поэтому
\begin{equation}
 det(\gamma_0-\lambda E)=0=det(\gamma_0+\lambda E)
\end{equation}
будет уравнением на собственные значения оператора $\gamma_0$. Видно, что указанные значения парны: $\pm\lambda$. Поэтому
положительным и отрицательным собственным значениям будут соответствовать 2 различных собственных подпространства. Тогда
существует базис  в котором  возможно блочное представление оператора $\gamma_0$ (блоки имеют размер $N^2\times N^2$)
\begin{equation}
\gamma_0=\left(
\begin{array}{cc}
\zeta_0   & 0 \\
 0         & \xi_0
\end{array}\right)\sps
\end{equation}
где $\zeta_0$ отвечают за положительные собственные значения, а  $\xi_0$ -  за отрицательные. Определим
\begin{equation}
\gamma_\myf\Lambda=\left(
\begin{array}{cc}
\zeta_\myf\Lambda   & \sigma_\myf\Lambda \\
\eta_\myf\Lambda & \xi_\myf\Lambda
\end{array}\right)\spsd
\end{equation}
С учетом этого определения условие (\ref{re3.10}) приведет к соотношениям
\begin{equation}
\zeta_0\zeta_\myf\Lambda=-\zeta_\myf\Lambda\zeta_0\sps \xi_0\xi_\myf\Lambda=-\xi_\myf\Lambda\xi_0\spsd
\end{equation}
Это означает, что $\zeta_0$ и $\xi_0$ должны иметь парные положительные и отрицательные значения (даже если $\zeta_\myf\Lambda$ и
$\xi_\myf\Lambda$ вырождены, но не равны 0). Поэтому $\zeta_\myf\Lambda=\xi_\myf\Lambda\equiv 0$. Таким образом
\begin{equation}
\gamma_\myf\Lambda=\left(
\begin{array}{cc}
0              & \sigma_\myf\Lambda \\
\eta_\myf\Lambda  & 0
\end{array}\right)\sps
\end{equation}
и уравнение (\ref{re3.0}) равносильно системе
\begin{equation}
\label{re3.12} \left\{
\begin{array}{c}
 \eta_\myf\Lambda\sigma_\myf\Psi+\eta_\myf\Psi\sigma_\myf\Lambda=G_{\myf\Lambda\myf\Psi}\sps\\
 \sigma_\myf\Lambda\eta_\myf\Psi+\sigma_\myf\Psi\eta_\myf\Lambda=G_{\myf\Lambda\myf\Psi}\spsd
\end{array}\right.
\end{equation}
Отметим, что если перейти к другому базису в пространстве $\mathbb R^{(2N)^2}$ с помощью невырожденного оператора S из общей линейной
группы $\tilde{\gamma}_\myf\Lambda:=S\gamma_\myf\Lambda S^{-1}$, то уравнение Клиффорда (\ref{re3.11}), вообще говоря, не сохранит
свой вид.

Вернемся теперь к комплексной структуре.  Ее квадрат равен $-E$ - эта матрица представляет некоторое ортогональное преобразование.
Согласно (\ref{re3.12}) при n=2, N=1  такое преобразование представимо по  формуле (\ref{re3.6}) как
\begin{equation}
{\tilde F}^2=\tilde{\tilde F}^2=
\left(
\begin{array}{rr}
-E & 0 \\
 0 & E
\end{array}
\right)\spsd
\end{equation}
Если комплексная структура $f_\myf\Lambda{}^\myf\Psi$ имеет в данном базисе канонический вид, то в разложении (\ref{re3.9})
$F:=\tilde F=\tilde {\tilde F}$, $I:=\tilde I=\tilde {\tilde I}$ тоже имеет блочный вид (из-за блочной структуры $\gamma_\myf\Lambda$ и
соответствия (\ref{re3.4}), написанного для вещественной реализации)
\begin{equation}
F=
\left(
\begin{array}{rr}
\hat F & 0 \\
0 & \hat {\hat F}^T
\end{array}
\right)\sps
\hat F^2=-E\sps \hat {\hat F}^2=E\sps
I=
\left(
\begin{array}{rr}
iE & 0 \\
 0 & E
\end{array}
\right)\spsd
\end{equation}
Что касается преобразования I, то оно определяется из соотношения
\begin{equation}
\begin{array}{c}
i\gamma_\Lambda=\frac{1}{2}(i(\gamma_\Lambda)+(-i)(-\gamma_\Lambda))=\\
=\underbrace{\frac{1}{2}((1+i)E+(1-i){\tilde F}^2)}_{:=\tilde I}\gamma_\Lambda\underbrace{\frac{1}{2}((1+i)E+(1-i){\tilde{\tilde F}}^2)}_{:=\tilde{\tilde I}}\spsd
\end{array}
\end{equation}
Ввиду того, что количество элементарных преобразований, составляющих комплексную структуру $f_\myf\Lambda{}^\myf\Psi$ всегда
кратно двум (поскольку размерность вещественной реализации 2n при четном n кратна 4), то блочная структура будет именно
по главной диагонали. Следовательно, для блочных составляющих $\gamma_\myf\Lambda$ будет верно
\begin{equation}
f_\myf\Lambda{}^\myf\Psi\sigma_\myf\Psi=\hat F \sigma_\myf\Lambda\hat{\hat F}^T\sps
f_\myf\Lambda{}^\myf\Psi\eta_\myf\Psi=\hat {\hat F}^T\eta_\myf\Lambda\hat F\sps
\end{equation}
что определит положительную реализацию (\ref{re1.9}) для верхнего знака и отрицательную реализацию (\ref{re1.9}) для нижнего знака.
Если теперь положить ($\myf A,\myf B,\myf C,\myf D, ... =\overline{1,4}$, $\hat A,\hat B,\hat C,\hat D, ... =\overline{1,2}$,
$\tilde{\tilde A},\tilde{\tilde B},\tilde{\tilde C},\tilde{\tilde D}, ... =\overline{1,2}$, $A,B,C,D ... =\overline{1,1}$)
\begin{equation}
\tilde m_{\pm}{}_{\hat A}{}^{\tilde B}:=
\left(

\end{equation}
Чтобы построить пошаговые операторы (\ref{re3.12/1}) для произвольного четного n, воспользуемся алгоритмом \ref{ra2.1}
($\myf A,\myf B,\myf C,\myf D, ... =\overline{1,\frac{(2N)^2}{2^{I-1}}}$, $\hat A,\hat B,\hat C,\hat D, ... =\overline{1,\frac{(2N)^2}{2^I}}$,
$\tilde{\tilde A},\tilde{\tilde B},\tilde{\tilde C},\tilde{\tilde D}, ... =\overline{1,\frac{2N^2}{2^{I-1}}}$, $A,B,C,D ... =\overline{1,\frac{2N^2}{2^I}}$,
$N=2^{n/2-1}$, $I=\overline{1,\frac{n}{2}}$)
\begin{equation}
\label{re3.12/2}
 %
 \right.  \\
 \end{array}
\end{equation}
При этом
\begin{equation}
(\tilde\triangle_{I_\pm}){}_{\tilde{\tilde C}}{}^{\tilde{\tilde A}}:=
\frac{1}{2}(\delta_{\tilde{\tilde C}}{}^{\tilde{\tilde A}}\pm\hat I_{\tilde{\tilde C}}{}^{\tilde{\tilde K}}\hat F_{\tilde{\tilde K}}{}^{\tilde{\tilde A}})\sps
(\tilde{\tilde\triangle}_{I_\pm}){}_{\tilde{\tilde C}}{}^{\tilde{\tilde A}}:=
\frac{1}{2}(\delta_{\tilde{\tilde C}}{}^{\tilde{\tilde A}}\pm\hat{\hat I}_{\tilde{\tilde C}}{}^{\tilde{\tilde K}}\hat{\hat F}_{\tilde{\tilde K}}{}^{\tilde{\tilde A}})\spsd
\end{equation}
Теперь возможно определить операторы
\begin{equation}
\label{re3.13/1}
\begin{array}{cc}
\tilde M_K:=\tilde m_{\frac{n}{2}_{z_{\frac{n}{2}}}}\tilde m_{{\frac{n}{2}-1}_{z_{\frac{n}{2}-1}}}\cdot ...\cdot\tilde m_{2_{z_2}}\tilde m_{1_{z_1}}\sps &
\tilde{\tilde M}_K:=\tilde{\tilde m}_{\frac{n}{2}_{z_{\frac{n}{2}}}}\tilde{\tilde m}_{{\frac{n}{2}-1}_{z_{\frac{n}{2}-1}}}\cdot ...\cdot\tilde{\tilde m}_{2_{z_2}}\tilde{\tilde m}_{1_{z_1}}\sps\\
\end{array}
\end{equation}
где $z_J\ (J=\overline{1,\frac{n}{2}})$  равно 0 при выборе знака $\ll-\gg$ или 1 при выборе знака $\ll+\gg$, тогда $K=\sum\limits_{J=1}^{\frac{n}{2}}z_J\cdot 2^{J-1}+1$. Если не важно, какой из K операторов будет выбран, то номер K будем опускать. Используя эти операторы, можно определить ($\tilde M:=\tilde M_C{}^\myff A$, $\tilde M^*:=\tilde M^{*C}{}_\myff A$, $\tilde{\tilde M}:=\tilde{\tilde M}_C{}^\myff A$, $\tilde{\tilde M}^*:=\tilde{\tilde M}^{*C}{}_\myff A$)
\begin{equation}
\label{re3.13/2}
\sigma_\Lambda:=\tilde M (m_\Lambda{}^\myf\Lambda\sigma_\myf\Lambda)\tilde{\tilde M}^T\sps
\eta_\Lambda:=\tilde{\tilde M}^* (m_\Lambda{}^\myf\Lambda\eta_\myf\Lambda)(\tilde M^*){}^T\spsd
\end{equation}
Для дальнейших выкладок необходимо наложить условие
\begin{equation}
tr(\eta_\myf\Psi \sigma_\myf\Phi)=
\frac{N}{2}g_{\myf\Psi\myf\Phi}\spsd
\end{equation}
Переходя в (\ref{re3.12}) к комплексной реализации с помощью операторов (\ref{re3.13/1}), можно получить
\begin{equation}
\label{re3.13}
\left\{
\begin{array}{c}
 \eta_\Lambda\sigma_\Psi+\eta_\Psi\sigma_\Lambda=g_{\Lambda\Psi}\sps\\
 \sigma_\Lambda\eta_\Psi+\sigma_\Psi\eta_\Lambda=g_{\Lambda\Psi}\sps\\
tr(\eta_\Psi \sigma_\Lambda)=\frac{N}{2}g_{\Psi\Lambda}\spsd
\end{array}\right.
\end{equation}
Это приведет согласно (\ref{re3.3}) к
\begin{equation}
\label{re3.14}
 R_\eta:=r^\Lambda\eta_\Lambda\sps
 R_\sigma:=r^\Lambda\sigma_\Lambda\sps
 \frac{2}{N}tr((R_\sigma)\eta_\Lambda)=r_\Lambda\sps
 \frac{2}{N}tr((R_\eta)\sigma_\Lambda)=r_\Lambda\sps
\end{equation}
\begin{equation}
 R=\left(
 \begin{array}{cc}
   0       & R_\sigma \\
   R_\eta  & 0
 \end{array}\right)\sps
\end{equation}
 где тензоры $R_\eta,\ R_\sigma$ имеют размерность уже $N \times N$. Следовательно, (\ref{re3.4}) разобьется на пару уравнений
\begin{equation}
 R_\sigma\eta_\Psi R_\sigma=
 (r_\Psi r^\Phi-\frac{1}{2}(r_\Omega r^\Omega)\delta_\Psi{}^\Phi) \sigma_\Phi\sps
 R_\eta\sigma_\Psi R_\eta=
 (r_\Psi r^\Phi-\frac{1}{2}(r_\Omega r^\Omega)\delta_\Psi{}^\Phi) \eta_\Phi\spsd
\end{equation}
Поскольку
\begin{equation}
R_\sigma R_\eta=\frac{1}{2}(r_\Omega r^\Omega)E\sps
\end{equation}
то есть возможность определить формализацию операторов $\eta_\Lambda\ , \sigma_\Lambda$ следующим образом
\begin{equation}
\eta_\Lambda:=\eta_\Lambda{}^{AB}\sps
\sigma_\Lambda:=\sigma_\Lambda{}_{AB}=\eta_\Lambda{}_{BA}\sps
(A,B,... =\overline{1,N})\spsd
\end{equation}
Тогда (\ref{re3.5}) можно переписать как
\begin{equation}
\label{re3.15}
 I).\  S_\Psi{}^\Lambda\eta_\Lambda={\tilde S}^T\eta_\Psi\tilde{\tilde S}\sps
 II).\ S_\Psi{}^\Lambda\eta_\Lambda=\tilde{\tilde S}\sigma_\Psi\tilde S^T\sps
\end{equation}
но уже здесь операторы $\tilde{\tilde S}$ и $\tilde S$ будут иметь размерность $N \times N$. Однако, теперь разложение этих операторов
для собственного $S_\Lambda{}^\Psi\ (\det\parallel S_\Lambda{}^\Psi\parallel =1)$ (случай I) будет иметь вид
\begin{equation}
 \tilde S^T=\prod_{I=1}^{J}
 (R_\eta)_{2I-1}(R_\sigma)_{2I}={(R_\eta)}_1{(R_\sigma)}_2{(R_\eta)}_3{(R_\sigma)}_4 \ldots {(R_\eta)}_{2J-1}{(R_\sigma)}_{2J}\sps
\end{equation}
\begin{equation}
 \tilde {\tilde S}=\prod_{I=J}^{1} (R_\sigma)_{2I}(R_\eta)_{2I-1}=
 {(R_\sigma)}_{2J}{(R_\eta)}_{2J-1}\ldots {(R_\sigma)}_4{(R_\eta)}_3{(R_\sigma)}_2{(R_\eta)}_1 \sps
\end{equation}
 а для несобственного преобразования $S_\Lambda{}^\Psi \ (\det\parallel S_\Lambda{}^\Psi\parallel =-1)$ (случай II) соответственно
\begin{equation}
 \tilde S^T=(R_\eta)_{1}\prod_{I=1}^{J} (R_\sigma)_{2I}(R_\eta)_{2I+1}={(R_\eta)}_1{(R_\sigma)}_2{(R_\eta)}_3
 \ldots {(R_\sigma)}_{2J}{(R_\eta)}_{2J+1}\sps
\end{equation}
\begin{equation}
 \tilde {\tilde S}=(\prod_{I=J}^{1} (R_\eta)_{2I+1}(R_\sigma)_{2I})(R_\eta)_1=
 {(R_\eta)}_{2J+1}{(R_\sigma)}_{2J}\ldots {(R_\eta)}_3{(R_\sigma)}_2{(R_\eta)}_1 \spsd
\end{equation}
Таким образом, уравнение (\ref{re3.15}) определяет алгебраическую реализацию двулистного накрытия $Spin(n,\mathbb C)/\{\pm 1\}\cong SO(n,\mathbb C)$.
Соответственно, для инволюции  (\ref{re3.8})  определит
\begin{equation}
\label{re3.16}
\begin{array}{llll}
 I).&\  S_\Psi{}^{\Lambda '}\eta_{\Lambda '}=\tilde S{}^T\eta_\Psi\tilde{\tilde S}\sps &
 II).&\ S_\Psi{}^{\Lambda '}\eta_{\Lambda '}=\tilde{\tilde S}^T\sigma_\Psi\tilde S\sps\\
 &\tilde S \bar{\tilde S}=\pm E\sps \tilde{\tilde S}\bar{\tilde{\tilde S}}=\pm E\spsd &
 &\tilde S =\pm \bar{\tilde S}\sps \tilde{\tilde S}=\pm \bar{\tilde{\tilde S}}\spsd
\end{array}
\end{equation}
Нижняя строка соотношений наследуется вещественной реализацией инволюции как в касательном, так и в спинорном пространстве.
А вещественная реализация инволюции в касательном пространстве есть ортогональное преобразование, квадрат которого
имеет единичную матрицу. Соответственно, в вещественной реализации спинорного пространства наследуется это свойство,
и квадрат спинорных инволюций (случай I) в вещественной реализации с точностью до знака тоже является тождественным преобразованием.
Осталось перейти к комплексной реализации по правилу  (\ref{re1.10}) .

\vspace{12mm}
\begin{example} 2n=2, N=1.
\label{rex3.1}

Если $g_{\myf\Lambda\myf\Psi}=\left(

\end{equation}
 Таким образом, тождество $tr(\eta_\myf\Lambda\sigma_\myf\Psi)=\frac{1}{2}g_{\myf\Lambda\myf\Psi}$ неверно. Тем не менее можно воспользоваться
(\ref{re3.5}) и (\ref{re3.6}). Положим
\begin{equation}
%
\end{equation}
Тогда для любого вектора $x^\myf\Lambda=(T,X)$ соответствия $X=x^\myf\Lambda\gamma_\myf\Lambda\sps
(S_1)_\myf\Lambda{}^\myf\Psi x_\myf\Psi \longleftrightarrow R_1 X R_1$ примут вид
\begin{equation}
 \left(%
\end{equation}
Тогда для любого вектора $x^\myf\Lambda=(T,X)$ соответствия $X=x^\myf\Lambda\gamma_\myf\Lambda\sps (S_2)_\myf\Lambda{}^\myf\Psi x_\myf\Psi
\longleftrightarrow R_2 X R_2$ примут вид
\begin{equation}
\left(%
\right)\spsd
\end{equation}
 Теперь возможно построить собственное преобразование $S:=S_1S_2$, для которого $S_\myf\Lambda{}^\myf\Psi x_\myf\Psi
 \longleftrightarrow R_1 R_2X R_2 R_1$ и $\tilde S=R_1 R_2\sps \tilde{\tilde S}=R_2 R_1$. Явная запись соответствия будет иметь вид
\begin{equation}
 \left(%
 \right)\spsd
\end{equation}
Очевидно, что SO(2) исчерпывется\ такими\  преобразованиями\ \ $S_\myf\Lambda{}^\myf\Psi$, тогда $\left(%
\right)\cong U(1) $. При этом $\tilde{\tilde S}=\overline{\tilde S}$ (комплексносопряжены). Этим
доказано, что $SO(2)\cong U(1)$.
\end{example}

\vspace{12mm}

\begin{example} 2n=2, N=1.
\label{rex3.2}

Если $g_{\myf\Lambda\myf\Psi}=\left(%
\end{equation}
Таким образом тождество $tr(\eta_\myf\Lambda\sigma_\myf\Psi)=\frac{1}{2}g_{ij}$ неверно. Явная запись соответствия
$S_\myf\Lambda{}^\myf\Psi  x_\myf\Psi \longleftrightarrow \tilde S X \tilde {\tilde S}$ будет иметь вид
\begin{equation}
 \left(%
 \right)\sps
\end{equation}
 что определит соответствие $SO^+(1,1)\cong R$ (аддитивная), где $SO^+(1,1)$ - связная компонента единицы с $S_1{}^1>0$.
\end{example}

\vspace{12mm}

\begin{example} 2n=4, N=2.
\label{rex3.3}
Если $g_{ij}=\left(%
\right)$, то можно построить
\begin{equation}
%
\right)$.
Легко проверить, что $tr(\eta_i\sigma_j)=tr(\sigma_i\eta_j)=g_{ij}$. Вычисления в этом случае более громоздки, но
аналогичны случаю n=2 (и частично будут проведены ниже). Просто отметим тот факт, что здесь действует накрытие
$SO^+(2,2)\cong SL(2)\times SL(2)/{\pm 1}$. Однако теперь хотелось бы знать, как перейти от случая n=4 к случаю n=2.
\end{example}

\vspace{12mm}

\begin{example} $n=2 \subset n=4,\ N=2.$

\begin{equation}
%
\right)\sps\\
 x^\myf\Lambda=(T,Y,Z,X)\sps x^\Lambda= m^\Lambda{}_\myf\Lambda x^\myf\Lambda =\frac{1}{2}(T-Y+i(Z-X),-i(T+Y)+Z+X)\sps
\end{array}
\end{equation}
тогда любое собственное ортогональное преобразование $S_\myf\Lambda{}^\myf\Psi \in SO^+(2,2)$$ (\det\parallel S_\Lambda{}^\Psi\parallel =1$
и определитель левого верхнего минора  $2\times 2$  - положителен) представимо в виде конечного произведения элементарных преобразований.
Поскольку $fSf= -S$ из-за наличия комплексной структуры, поэтому такие преобразования будут двух видов
\begin{equation}

\end{equation}
 Комплексную реализацию $S_\myf\Lambda{}^\myf\Psi$ можно определить как $S_\Lambda{}^\Psi=m_\Lambda{}^\myf\Lambda S_\myf\Lambda{}^\myf\Psi m^\Psi{}_\myf\Psi$
\begin{equation}
 (S_1)_\Lambda{}^\Psi=\left(
 %
\right)\spsd
\end{equation}
Поэтому $S=S_1 S_2$ описывает группу $SO(2,\mathbb C)$ полностью. Пусть теперь
\begin{equation}
 %
\end{equation}
Построим $(S_1)_\myf\Lambda{}^\myf\Theta=((r_1)_\myf\Lambda (r_1)^\myf\Psi -\delta_\myf\Lambda{}^\myf\Psi)
((r_2)_\myf\Psi (r_2)^\myf\Theta +\delta_\myf\Psi{}^\myf\Theta)$. При  этом $(r_1)_\myf\Lambda (r_1)^\myf\Lambda=
-(r_2)_\myf\Lambda (r_2)^\myf\Lambda=2$. Пусть $\eta_\myf\Lambda$ и $\sigma_\myf\Lambda$ такие же как в предыдущем примере, то
\begin{equation}
%
\right)$ представляет аддитивную группу R.  Если в соответствующих уравнениях сделать замену $\eta$ на  $\sigma$, то
\begin{equation}
%
\end{equation}
\end{example}

\vspace{12mm}

\begin{example} $n=2 \subset 2n=4,\ N=2.$

Пусть тензор инволюции в касательном слое для вещественной реализации $V_{4}$ представляется тензором (слева вложение для метрики индекса 2,
справа - для индекса 1)
\begin{equation}
%
\end{equation}
 и выполнены все тождества предыдущих примеров. Тогда для комплексной реализации $\mathbb CV_{2}$ он будет иметь вид
 $S_\Lambda{}^{\Psi '}=m_\Lambda{}^\myf\Lambda S_\myf\Lambda{}^\myf\Psi\bar m ^{\Lambda '}{}_\myf\Psi$
\begin{equation}
%
\end{equation}
С другой стороны,  у инволюции существует образ (буковка T означает транспонирование) и в некотором базисе
$S_\myf\Lambda{}^\myf\Psi \eta_\myf\Psi=\tilde S^T\eta_\myf\Psi\tilde S$
\begin{equation}
%
\end{equation}
В первом случае получается (индекс 2)
\begin{equation}
%
\right)\sps
\end{equation}
 и мы опять приходим к изоморфизму $SO(2)\cong U(1)$
\begin{equation}
%
\end{equation}
Во втором случае (индекс 1)
\begin{equation}
%
\right)\sps
\end{equation}
 и мы опять приходим к изоморфизму $SO^+(1,1)\cong R$ (аддитивная)
\begin{equation}
%
\end{equation}
\end{example}
Следующий пример достоин отдельного параграфа.

\section{Кватернионы}
\Abstract{
\indent В этом параграфе рассказывается о том, как связана алгебра кватернионов с клиффордовой алгеброй и о том, как построить в явном виде двулистные
накрытия, соответствующие этому случаю. Вывод всех результатов этого параграфа сделан на основе \cite{Postnikovr1}.
}
\newpage
Известно, что алгебру Клиффорда $CL(g^n)$ можно представить как
\begin{equation}
%
\end{equation}
что и называется $Z_2$ градуировкой. При этом $CL^0(g^n)$ есть подалгебра $CL(g^n)$.  Если рассматривать n=4 с 8-мерной алгеброй Клиффорда
$CL^0(g^n)$, то базис указанной алгебры будет иметь вид
\begin{equation}
 %
\end{equation}
Из уравнения Клиффорда (\ref{re2.2}) будет следовать
\begin{equation}
 e_\Lambda e_\Psi=1\sps\Lambda = \Psi \sps
 e_\Lambda e_\Psi=-e_\Psi e_\Lambda \sps \Lambda \neq \Psi
\end{equation}
в некотором базисе
\begin{equation}
 %
\end{equation}
который будет представлять собой базис, подобный базису алгебры октав, но с другой нормировкой. Для этого необходимо рассмотреть вложение $\mathbb R^4\subset\mathbb C^4$. Затем можно найти представление указанного базиса с помощью построенных выше операторов $\gamma_i=H_i{}^\Lambda\gamma_\Lambda$.
Но такой базис, в свою очередь, получается прямой суммой двух кватернионных базисов, за что отвечает изоморфизм
$CL^0(g^n_{(n,0)})\cong \mathbb H\oplus \mathbb H$. Для того, чтобы построить соответствующее  представление указанного базиса, необходимо
потребовать для частей операторов $\gamma_i$ выполнение условия $tr(\eta_i\sigma_j)=\frac{N}{2}g_{ij}$.  И тогда, действительно, возможно
ограничиться кватернионным  базисом (как раз имеющим 2 генератора i и j (k=ij) и кватернионную единицу I)
\begin{equation}

\end{equation}
Пусть теперь на пространстве $\mathbb R^4\subset \mathbb C^4$ заданы генерирующие элементарные ортогональные преобразования векторы
\begin{equation}
%
\end{equation}
При этом, как и прежде, преобразования $(r_I)^i (r_I)_j-\delta_j{}^i $  $(I=\overline{1,6})$ генерируют все вращения из группы O(4).
Соответственно, на пространстве $\mathbb C^2$ ($N=2^{2/2-1}=2$) будут индуцированы их образы
\begin{equation}
%
\end{equation}
Чтобы упростить выкладки, ограничимся случаем, когда $\phi_1 =\phi_2 =\phi_4 =\frac{\pi}{2}$. Тогда
\begin{equation}
%
\end{equation}
Этим построено двулистное накрытие $SO(3)\cong SU(2)/\{\pm 1\}$ и поэтому $Spin(3)\cong SU(2)$
\begin{equation}
 {S_1}_i{}^j \eta_j= A_1{}^T\eta_i A_1^*\spsd
\end{equation}
Если теперь положить $\phi_3 =\phi_5 =\phi_6 =\frac{\pi}{2}$ и провести аналогичные выкладки, то получим
\begin{equation}
%
 \right) =:A_2{}^{\sharp} \spsd
\end{array}
\end{equation}
Поэтому
\begin{equation}
 {S_2}_i{}^j \eta_j= A_2{}^T\eta_i A_2^{\sharp}\spsd
\end{equation}
Преобразования $A_1,A_2$ - представители группы SU(2), которую полностью и описывают. Собирая два уравнения вместе, получим, что
\begin{equation}
 S_i{}^j \eta_j:= (S_1S_2)_i{}^j \eta_j=
 A_1{}^TA_2{}^T\eta_j A_2^\sharp A_1^*:=\tilde S {}^T\eta_j \tilde{\tilde S}\spsd
\end{equation}
Вообще говоря, зная преобразование $A_1{}^TA_2{}^T$, ничего нельзя сказать про $A_2^\sharp A_1^*$. Поэтому преобразования $\tilde S{}^T$ и
$\tilde{\tilde S}$ суть различны. Это означает, что существует двулистное накрытие $SO(4)\cong SU(2) \times SU(2)/\{\pm 1\}$. И поэтому
$Spin(4)\cong SU(2) \times SU(2)$.

\section{Частные решения уравнения Клиффорда}
\Abstract{
\indent В этом параграфе рассказывается о том, как построить частные решения уравнения Клиффорда, которые позволят однозначно построить двулистное
накрытие $Spin(n,\mathbb C)/\{\pm 1\}\cong SO(n,\mathbb C)$. Такие решения дадут возможность в дальнейшем однозначно продолжить риманову связность
без кручения в спинорное расслоение. Вывод всех результатов этого параграфа сделан на основе предыдущих выкладок.
}

Для того, чтобы найти некоторые частные решения уравнения Клиффорда (\ref{re3.13}), перепишем его как
\begin{equation}
\label{re6.1}
\eta_\Lambda{}^{AB}\eta_\Psi{}_{CB}+\eta_\Psi{}^{AB}\eta_\Lambda{}_{CB}=g_{\Lambda\Psi}\delta_C{}^A\spsd
\end{equation}
\begin{example}
\label{rex6.1}
n=4, N=2.\\
В этом случае воспользуемся изоморфизмом $\mathbb C^4\cong \mathbb C(2)$, который строится с помощью связующих операторов
$\eta^\Lambda{}_{AB}$ (\ref{re3.14})
\begin{equation}
r^\Lambda=\frac{1}{2}\eta^\Lambda{}_{AB} R^{AB}\spsd
\end{equation}
Пусть метрика в $\mathbb C^4$ имеет вид
\begin{equation}
g_{\Lambda\Psi}=
\left(

\right)\spsd
\end{equation}
Очевидно, что любой тензор $R^{AB}$ разлагается на 3 симметричных и 1 кососимметричную части
\begin{equation}
%
\end{equation}
где $(\varepsilon_I)^{ab}$ берутся из предыдущего примера, там как раз n=4, N=2. Это один из возможных вариантов (4 симметричные и 2
кососимметрические части). Если теперь домножить справа все полученные $(\varepsilon_I)^{AB}$ ($I =\overline{1,2,3,4,\tilde 1,\tilde 2}$) на
\begin{equation}
\left(
%
\right)\sps
\end{equation}
то возможно построить новые тензоры (6 кососимметричных частей)
\begin{equation}
%
\right)\sps
\end{array}
\end{equation}
что даст возможность определить
\begin{equation}
R^{AB}=-R^{BA}=R_{\tilde 1}(\varepsilon_{\tilde 1})^{AB}+R_{\tilde 2}(\varepsilon_{\tilde 2})^{AB}+
R_{\tilde 3}(\varepsilon_{\tilde 3})^{AB}+R_{\tilde 4}(\varepsilon_{\tilde 4}^{AB})+
R_{\tilde 5}(\varepsilon_{\tilde 5})^{AB}+R_{\tilde 6}(\varepsilon_{\tilde 6}^{AB})\spsd
\end{equation}
\end{example}

Для того, чтобы получить подобную конструкцию для любого четного n, дадим следующие определения.
\begin{definitionr}
\label{rd6.1}
Пусть в пространстве $\mathbb C^n$ $(n\ge 4)$ определено n ортогональных векторов вида $(\eta_I){}^\Lambda$
\begin{equation}
(\eta_I){}^\Lambda (\eta_J){}_\Lambda=2\delta_{IJ}\sps \Lambda ,\ \Psi , \ ... = \overline{1,n}\sps
\end{equation}
которым в спинорном пространстве поставим в соответствие спин-тензоры
\begin{equation}
(\varepsilon_I)_{AB}:=(\eta_I)^\Lambda \eta_\Lambda{}_{AB}\spsd
\end{equation}
Тогда среди n векторов  $(\eta_I){}^\Lambda$ найдутся $\frac{n}{2}+q$ векторов таких, что
\begin{equation}
\eta^\Psi{}_{(AB)}=\frac{1}{2}\sum_{Q=1}^{\frac{n}{2}+q}(\eta_Q){}^\Psi(\varepsilon_Q)_{AB}\sps
(\varepsilon_Q)_{AB}=(\varepsilon_Q)_{BA}
\end{equation}
и $\frac{n}{2}+\tilde q$ векторов таких, что
\begin{equation}
\eta^\Psi{}_{[AB]}=\frac{1}{2}\sum_{\tilde Q =1}^{\frac{n}{2}+\tilde q}(\eta_{\tilde Q}){}^\Psi(\varepsilon_{\tilde Q}){}_{AB}\sps
(\varepsilon_{\tilde Q}){}_{AB}=-(\varepsilon_{\tilde Q}){}_{BA}\sps q+\tilde q=0 \spsd
\end{equation}
Кроме того, потребуем, чтобы вещественная реализация таких операторов удовлетворяла аналогичным соотношениям.
\end{definitionr}

\begin{corollaryr}
Из соотношения
\begin{equation}
\label{re6.3/1}
\tilde T_A{}^B(\varepsilon_I)_{BC}+\tilde{\tilde T}_C{}^B(\varepsilon_I)_{AB}=0\sps
\end{equation}
выполненного для всех $I=\overline{1,n}$, следует, что $\tilde T_A{}^B=\tilde{\tilde T}_A{}^B=0$.
\end{corollaryr}
\begin{proof}
Воспользуемся тождеством
\begin{equation}

\end{equation}
Следовательно, $\tilde{\tilde T}_A{}^D=\tilde T_A{}^D=0$.
\end{proof}

Таким образом, на образующие тензоры $(\varepsilon_I)_{AB}$ уравнением Клиффорда (\ref{re6.1}) накладывается условие
\begin{equation}
\label{re6.2}
%
\end{equation}

\begin{definitionr}
На пространстве бивекторов определен 4-валентный тензор, с помощью которого можно поднимать и опускать парные индексы
\begin{equation}
\label{re6.5}
%
\end{equation}
\end{definitionr}

\begin{noter}
Следует отметить, что для действительных вложений условие $(\eta_I){}^i (\eta_J){}_i=2\delta_{IJ}$ не всегда выполнимо. Его следует
модифицировать до $(\eta_I){}^i (\eta_J){}_i=2g_{IJ}$, где $g_{IJ}=0\sps I\ne J$ и $ g_{II}=\pm 1$, при этом знак выбирается в зависимости от
индекса метрики и номера вектора. Это буквально означает, что метрика в некотором базисе имеет диагональный вид с $\pm 1$ на главной диагонали.
Тогда уравнение (\ref{re6.2}) примет вид
\begin{equation}
\label{re6.2a}
\tag{\ref{re6.2}$'$}
%
\end{equation}
В случае комплексного пространства всегда можно ограничиться только $+1$ с евклидовой метрикой $\delta_{IJ}$. Этого будет достаточно
для доказательства основных тождеств. К действительным же пространствам можно перейти, используя соответствующий оператор вложения.
\end{noter}
Поэтому возможно два существенно различных способа в зависимости от кратности n
\begin{equation}
%
\end{equation}
определения либо пары метрических тензоров $\varepsilon_{AB}=-\varepsilon_{BA}$, $\tilde\varepsilon_{AB}=\tilde\varepsilon_{BA}$
в случае I)., либо пары инволюций $E_A{}^B$, $\tilde E_A$ в случае II). (доказательство в приложении)
\begin{equation}
\label{re6.3}
%
\end{equation}

\begin{noter}
Для действительных вложений тензоры из (\ref{re6.3}) домножаются на  $g:=(\prod\limits_{P=1}^{(n/2+q)}\sqrt{g_{PP}})$, где  в качестве
корня $\sqrt{g_{\tilde P\tilde P}}$ при $g_{\tilde P\tilde P}=-1$ выбирается любой из двух вариантов $\pm i$.
\end{noter}

Покажем, что такие операторы существуют. Для n=4 возможно представление, описанное в примере \ref{rex6.1}. Пусть построено некоторое
представление $\eta_\alpha{}^{ab}\ (\alpha\ ,\beta\ , ...=\overline{1,n-2};a\ ,b\ , ...=\overline{1,2^{(n-2)/2-1}})$. Операторы
$\eta_\Lambda{}^{AB}\ (\Lambda\ ,\Psi\ , ...=\overline{1,n};A\ ,B\ , ...=\overline{1,2^{n/2-1}})$ строятся следующим образом.
Пусть тензор $g_{\alpha\beta}$ содержит только +1 на главной диагонали, тогда
\begin{equation}
\label{re6.6}
\begin{array}{lcclcc}
\eta_\alpha{}^{AB}&=&
\left(
\begin{array}{cc}
   \eta_\alpha{}^{ab} & 0                    \\
                    0 & -(\eta^T)_\alpha{}_{cd}
\end{array}
\right)\sps &
\eta^\alpha{}_{AB}&=&
\left(
\begin{array}{cc}
  \eta^\alpha{}_{pk} & 0                    \\
                   0 & -(\eta^T)^\alpha{}^{lm}
\end{array}
\right)\sps\\ \\
\eta_{n-1}{}^{AB}&=&
\frac{1}{\sqrt{2}}\left(
\begin{array}{cc}
             0 & i\delta^a{}_d\\
-i\delta_c{}^b & 0
\end{array}
\right)\sps &
\eta^{n-1}{}_{AB}&=&
\frac{1}{\sqrt{2}}\left(
\begin{array}{cc}
             0 & -i\delta_p{}^m\\
 i\delta^l{}_k & 0
\end{array}
\right)\sps\\ \\
\eta_{n}{}^{AB}&=&
\frac{1}{\sqrt{2}}\left(
\begin{array}{cc}
              0 & \delta^a{}_d\\
 \delta_c{}^b & 0
\end{array}
\right)\sps &
\eta^{n}{}_{AB}&=&
\frac{1}{\sqrt{2}}\left(
\begin{array}{cc}
              0 & \delta_p{}^m\\
 \delta^l{}_k & 0
\end{array}
\right)\sps
\end{array}
\end{equation}
При общей схеме построения (\ref{re6.6}) уравнение
\begin{equation}
\tilde T_A{}^B(\varepsilon_I)_{BC}+\tilde{\tilde T}_C{}^B(\varepsilon_I)_{AB}=0
\end{equation}
для n=4 приводит к $\tilde T_A{}^B=\tilde{\tilde T}_A{}^B=0$, что проверяется непосредственно. Пусть данное уравнение выполнено для
некоторого четного n-2. Тогда для n уравнение распадется на части такого вида:
\begin{enumerate}
\item
\begin{equation}
\tilde T_a{}^b(\varepsilon_I)_{bc}+\tilde{\tilde T}_c{}^b(\varepsilon_I)_{ab}=0\sps
\tilde T^a{}_b(\varepsilon_I)^{bc}+\tilde{\tilde T}^c{}_b(\varepsilon_I)^{ab}=0\sps I=\overline{1,n-2}\sps
\end{equation}
что немедленно дает $\tilde T_a{}^b=\tilde{\tilde T}_a{}^b=0\sps \tilde T^a{}_b=\tilde{\tilde T}^a{}_b=0$.
\item
\begin{equation}
\begin{array}{c}
\tilde T_{ab}(\varepsilon_{n+1})^b{}_c+\tilde{\tilde T}_{cb}(\varepsilon_{n+1})_a{}^b=\tilde T_{ac}-\tilde{\tilde T}_{ca}=0\sps\\
\tilde T^{ab}(\varepsilon_{n+1})_b{}^c+\tilde{\tilde T}^{cb}(\varepsilon_{n+1})^a{}_b=-\tilde T^{ac}+\tilde{\tilde T}^{ca}=0\spsd
\end{array}
\end{equation}
\item
\begin{equation}
\begin{array}{c}
\tilde T_{ab}(\varepsilon_{n+2})^b{}_c+\tilde{\tilde T}_{cb}(\varepsilon_{n+2})_a{}^b=\tilde T_{ac}+\tilde{\tilde T}_{ca}=0\sps\\
\tilde T^{ab}(\varepsilon_{n+2})_b{}^c+\tilde{\tilde T}^{cb}(\varepsilon_{n+2})^a{}_b=\tilde T^{ac}+\tilde{\tilde T}^{ca}=0\sps
\end{array}
\end{equation}
что немедленно дает $\tilde T_{ab}=\tilde{\tilde T}_{ab}=0\sps \tilde T^{ab}=\tilde{\tilde T}^{ab}=0$. Таким образом опять $\tilde T_A{}^B=\tilde{\tilde T}_A{}^B=0$.
\end{enumerate}
Пусть теперь имеются вещественные связующие операторы $\eta_\myf\Lambda{}^{\myff A\myff B}$. При этом существует набор ортогональных
векторов такой, что часть $(\varepsilon_Q)^{\myff A\myff B}$ симметрична, а оставшаяся часть $(\varepsilon_{\tilde Q})^{\myff A\myff B}$
кососимметрична. Покажем, что возможно перейти к комплексной реализации таких операторов, сохраняющих указанное свойство и, следовательно,
удовлетворяющих определению \ref{rd6.1}.
\begin{algorithmr}$ $\\
\label{ra6.1}
1.  На первом шаге используем в качестве операторов (\ref{re3.12}) и (\ref{re3.12/1}) операторы
\begin{equation}

\end{equation}
В качестве операторов $(\tilde m_{1_-})^A{}_{\myff A}$ можно взять оператор
$
\left(
%
\right)\spsd
$
Если теперь воспользоваться операторами $m_\Lambda{}^\myf\Lambda$, переводящими подпространство $\mathbb R^4_{(2,2)}$ в подпространство
$\mathbb C^2$, а пространство $\mathbb R^{2n-4}_{(n-2,n-2)}$ оставляющее без изменений, то можно построить связующие операторы
$\eta_\Lambda{}^{ab}$, соответствующие пространству $\mathbb C^2\oplus \mathbb R^{2n-4}_{(n-2,n-2)}$, при этом размерность спинорного
пространства будет понижена в 2 раза. \\
\newpage
2. На втором шаге воспользуемся операторами
\begin{equation}
%
\right)}_{:=(\tilde m^T_2){}_{\myff B}{}^B}\spsd
\end{array}
\end{equation}
Если теперь воспользоваться операторами $m_\Lambda{}^\myf\Lambda$, переводящими подпространство $\mathbb R^4_{(2,2)}$ в подпространство
$\mathbb C^2$, а пространство $\mathbb C^2\oplus \mathbb R^{2n-8}_{(n-4,n-4)}$ оставляющее без изменений, то можно построить связующие операторы
$\eta_\Lambda{}^{ab}$, соответствующие пространству $\mathbb C^4\oplus \mathbb R^{2n-8}_{(n-4,n-4)}$, при этом размерность спинорного
пространства будет понижена в 2 раза.\\
3. Поступая аналогично, можно получить пространство $\mathbb C^{n-2}\oplus \mathbb R^{4}_{(2,2)}$ и соответствующие связующие операторы.
Очевидно, что на каждом шаге у связующих операторов модифицируются только компоненты, отвечающие за переход от действительной реализации
подпространства $\mathbb R^4_{(2,2)}$ к его комплексной реализации $\mathbb C^2$.\\
4. Заключительный шаг осуществляется с помощью операторов из примера \ref{rex2.2}.\\
\end{algorithmr}

\begin{corollaryr}
\label{rc6.1}
Связующие операторы из определения \ref{rd6.1} удовлетворяют тождеству (доказательство в приложении):
\begin{equation}
\label{re6.4}
\eta_\Lambda{}^{AB}\eta_\Psi{}_{AD}\eta^\Omega{}^{CD}\eta^\Theta{}_{CB}=
\frac{N}{4}(g_{\Lambda\Psi}g^{\Omega\Theta}+\delta_\Lambda{}^\Theta\delta_\Psi{}^\Omega-\delta_\Psi{}^\Theta\delta_\Lambda{}^\Omega)\spsd
\end{equation}
\end{corollaryr}

\begin{corollaryr}
\label{rc6.2}
Для ортогональных преобразований (\ref{re3.15}) существует два варианта представлений для каждого случая:
\begin{enumerate}
\renewcommand{\theenumi}{\Roman{enumi})}
\item Собственные вращения.
\begin{equation}
S_\Lambda{}^\Psi\eta_\Psi{}^{AB}=\eta_\Lambda{}^{CD}\tilde S_C{}^A\tilde{\tilde S}_D{}^B\spsd
\end{equation}
\begin{enumerate}[1.]
\item n mod 4 =2\\
\begin{equation}
\tilde{\tilde S}_A{}^B=-E_A{}^C\tilde S_C{}^DE_D{}^B\sps
\tilde{\tilde S}_A{}^B=-\tilde E_A{}^C\tilde S_C{}^D\tilde E_D{}^B\spsd
\end{equation}
\item n mod 4 =0\\
\begin{equation}
\begin{array}{c}
\tilde{\tilde S}_A{}^B\tilde{\tilde S}_C{}^D\varepsilon_{BD}=\varepsilon_{AC}\sps
\tilde S_A{}^B\tilde S_C{}^D\varepsilon_{BD}=\varepsilon_{AC}\sps\\
\tilde{\tilde S}_A{}^B\tilde{\tilde S}_C{}^D\tilde\varepsilon_{BD}=\tilde\varepsilon_{AC}\sps
\tilde S_A{}^B\tilde S_C{}^D\tilde\varepsilon_{BD}=\tilde\varepsilon_{AC}\spsd
\end{array}
\end{equation}
\end{enumerate}
\item Несобственные вращения.
\begin{equation}
S_\Lambda{}^\Psi\eta_\Psi{}^{BA}=\eta_\Lambda{}_{CD}\tilde S^{CA}\tilde{\tilde S}^{DB}\spsd
\end{equation}
\begin{enumerate}[1.]
\item n mod 4 =2\\
\begin{equation}
\tilde{\tilde S}^{CA}=-E_K{}^A\tilde S^{KL}E_L{}^C\sps
\tilde{\tilde S}^{CA}=\tilde E_K{}^A\tilde S^{KL}\tilde E_L{}^C\spsd
\end{equation}
\item n mod 4 =0\\
\begin{equation}
\begin{array}{c}
\tilde{\tilde S}^{AB}\tilde{\tilde S}^{CD}\varepsilon_{BD}=\varepsilon^{AC}\sps
\tilde S^{AB}\tilde S^{CD}\varepsilon_{BD}=\varepsilon^{AC}\sps\\
\tilde{\tilde S}^{AB}\tilde{\tilde S}^{CD}\tilde\varepsilon_{BD}=\tilde\varepsilon^{AC}\sps
\tilde S^{AB}\tilde S^{CD}\tilde\varepsilon_{BD}=\tilde\varepsilon^{AC}\spsd
\end{array}
\end{equation}
\end{enumerate}
\end{enumerate}
\end{corollaryr}

\begin{corollaryr}
\label{rc6.3}
Для инволюций (\ref{re3.16}) существует два варианта разложения:
\begin{enumerate}
\renewcommand{\theenumi}{\Roman{enumi})}
\item Аналог собственных вращений.
\begin{equation}
S_\Lambda{}^{\Psi '}\bar\eta_{\Psi '}{}^{A'B'}=\eta_\Lambda{}^{CD}\tilde S_C{}^{A'}\tilde{\tilde S}_D{}^{B '}\spsd
\end{equation}
\begin{equation}
\tilde{\tilde S}_A{}^{B'}\bar{\tilde{\tilde S}}_{B'}{}^D=\pm\delta_A{}^D\sps
\tilde S_A{}^{B'}\bar{\tilde S}_{B'}{}^D=\pm\delta_A{}^D\spsd
\end{equation}
\item Аналог несобственных вращений.
\begin{equation}
S_\Lambda{}^{\Psi'}\bar\eta_{\Psi'}{}^{B'A'}=\eta_\Lambda{}_{CD}\tilde S^{CA'}\tilde{\tilde S}^{DB'}\spsd
\end{equation}
\begin{equation}
\tilde{\tilde S}^{AB'}=\pm\bar{\tilde{\tilde S}}{}^{B'A}\sps
\tilde S{}^{AB'}=\pm\bar{\tilde S}{}^{B'A}\spsd
\end{equation}
\end{enumerate}
\end{corollaryr}

Будем считать, что операторы $\eta_\Lambda{}^{AB}(\sigma_\Lambda{}_{AB})$ строятся либо по схеме (\ref{re6.6})($\alpha=\overline{1,n-2}$)
\begin{equation}
\label{re6.7}
\begin{array}{lc}
a). &
\begin{array}{c}
\eta_\Lambda{}^{AB}=
\left(
\begin{array}{cc}
 \eta_\alpha{}^{ab}                            &  \frac{1}{2}(i\eta_{n-1}+\eta_n)\delta^a{}_d  \\
\frac{1}{2}(-i\eta_{n-1}+\eta_n)\delta_c{}^b   &                     -(\eta^T)_\alpha{}_{cd}
\end{array}
\right)\sps\\
\sigma_\Lambda{}_{AB}=
\left(
\begin{array}{cc}
           (\eta^T)_\alpha{}_{ab}              &   \frac{1}{2}(i\eta_{n-1}+\eta_n)\delta_a{}^d \\
 \frac{1}{2}(-i\eta_{n-1}+\eta_n)\delta^c{}_b  &                            -\eta_\alpha{}^{cd}
\end{array}
\right)\spsd
\end{array}
\end{array}
\end{equation}
либо по схеме
\begin{equation}
\label{re6.7a}
\tag{\ref{re6.7}$'$}
\begin{array}{lc}
b). &
\begin{array}{c}
\eta_\Lambda{}^{AB}=
\left(
\begin{array}{cc}
 \eta_\alpha{}^{ab}                            &  \frac{1}{2}(i\eta_{n-1}+\eta_n)\delta^a{}_d  \\
\frac{1}{2}(i\eta_{n-1}-\eta_n)\delta_c{}^b    &                     -(\eta^T)_\alpha{}_{cd}
\end{array}
\right)\sps\\
\sigma_\Lambda{}_{AB}=
\left(
\begin{array}{cc}
           (\eta^T)_\alpha{}_{ab}              &   \frac{1}{2}(i\eta_{n-1}+\eta_n)\delta_a{}^d \\
 \frac{1}{2}(i\eta_{n-1}-\eta_n)\delta^c{}_b   &                            -\eta_\alpha{}^{cd}
\end{array}
\right)\spsd
\end{array}
\end{array}
\end{equation}
Это означает, что из предыдущей размерности наследуются спиноры сопряжения (\ref{re3.16})
\begin{equation}
\begin{array}{lllll}
I). & a). &
S_A{}^{A'}:=
\left(
\begin{array}{cc}
  0       & iS_{aa'}\\
-iS^{bb'} &  0
\end{array}
\right)\sps &
b). &
S_A{}^{A'}:=
\left(
\begin{array}{cc}
  0       & iS_{aa'}\\
 iS^{bb'} &  0
\end{array}
\right)\sps\\
II). & a). &
S_{AA'}:=
\left(
\begin{array}{cc}
  0          & iS_a{}^{a'}\\
-iS^b{}_{b'} &  0
\end{array}
\right)\sps &
b). &
S_{AA'}:=
\left(
\begin{array}{cc}
  0          & iS_a{}^{a'}\\
 iS^b{}_{b'} &  0
\end{array}
\right)\sps
\end{array}
\end{equation}
которые определяются с точностью до знака.

\begin{example}
В таблице \ref{rt6.1} приведены различные варианты действительных вложений для n=2  (см. примеры \ref{rex3.1} и \ref{rex3.2}).
\end{example}

\begin{example}
В таблице \ref{rt6.2} приведены  различные варианты действительных вложений для n=4  (см. пример \ref{rex3.3}).
\end{example}

\begin{example}
В таблице \ref{rt6.3} приведены различные варианты действительных вложений для n=6.
В случае, когда $\tilde\eta_\Lambda{}^{AC}:=\eta_\Lambda{}^{AB}\varepsilon_B{}^C$ кососимметричен (см. пример \ref{rex6.2}), должны быть выполнены соотношения:
\begin{equation}

\end{equation}
Напомним, что $\eta_i:=H_i{}^\Lambda\eta_\Lambda$ согласно (\ref{re5.2}).
\end{example}
\renewcommand{\multirowsetup}{\centering}
\renewcommand{\baselinestretch}{1.2}
\setcounter{mypage}{175}
\begin{table}[!hp]
\caption{Вид матрицы тензора S для соответствующих действительных вложений
         2-мерных пространств в комплексное для $r^i=(T,Z)$.}
\label{rt6.1}
\begin{center}
%
\end{center}
\caption{Вид матрицы тензора S для соответствующих действительных вложений
         4-мерных пространств в комплексное для $r^i=(T,X,Y,Z)$.}
\label{rt6.2}
\begin{center}
%
\end{center}
\end{table}
\setcounter{mypage}{176}
\begin{table}[!hp]
\caption{Вид матрицы тензора S для соответствующих действительных вложений
         6-мерных пространств в комплексное для $r^i=(T,X, V,W,Y,Z)$.}
\label{rt6.3}
\begin{center}
%
\end{center}
\end{table}
\renewcommand{\baselinestretch}{1.5}

\setcounter{mypage}{177}
\section{О структурных константах алгебр без деления. Седенионы}
\Abstract{
\indent В этом параграфе рассказывается о том, как по связующим операторам, удовлетворяющим уравнению Клиффорда, построить структурные константы
алгебры седенионов. Такие алгебры строятся индуктивным переходом для n mod 8=0, исходя из периодичности Ботта. Рассматривается аксиоматика
таких алгебр. Вывод всех результатов этого параграфа сделан на основе \cite{Baezr1}, \cite{Husemollerr1}.
}

Воспользуемся тем фактом, что для n mod 8=0 существуют такие связующие операторы $\eta_\Lambda{}^{AB}$, индуктивное построение которых
не является удвоением алгебр, а связано с периодичностью Ботта. Согласно определению \ref{rd6.1} это означает, что
\begin{equation}
\eta_\Lambda{}^{AB}+\eta_\Lambda{}^{BA}=\eta_\Lambda\varepsilon^{AB}\sps \eta_\Lambda:=(\eta_{n})_\Lambda\sps
\end{equation}
где $\varepsilon^{AB}$ - симметрический метрический спин-тензор на спинорном пространстве. Это даст возможность положить
\begin{equation}
\begin{array}{cccc}
I).  & P_\Lambda{}^A:=\eta_\Lambda{}^{BA}X_B\sps P_\Lambda{}^AP_\Psi{}_A=g_{\Lambda\Psi}\sps &
II). & P_\Lambda{}^A:=\eta_\Lambda{}^{AB}X_B\sps P_\Lambda{}^AP_\Psi{}_A=g_{\Lambda\Psi}
\end{array}
\end{equation}
для некоторого $X_B\ (X^AX_A=2)$. Тогда можно определить некоторые структурные константы следующим образом
\begin{equation}
\begin{array}{cccc}
I).  & \eta_{\Lambda\Psi}{}^\Theta:=\sqrt{2}\eta_\Lambda{}^{AB}P_\Psi{}_AP^\Theta{}_B\sps
II). & \eta_{\Lambda\Psi}{}^\Theta:=\sqrt{2}\eta_\Lambda{}^{BA}P_\Psi{}_AP^\Theta{}_B\spsd
\end{array}
\end{equation}
Оба варианта равноправны, поэтому в дальнейшем будем рассматривать I). вариант. Будут ли они определять некоторую групповую алгебру? Ответ совсем
неоднозначен, и вот почему.
\begin{enumerate}
\item
По любым двум элементам $(r_1)^\Lambda\sps(r_2)^\Psi$ можно всегда единственным образом построить их произведение
\begin{equation}
\label{re13.1}
(r_1,r_2)^\Theta:=(r_1)^\Lambda(r_2)^\Psi\eta_{\Lambda\Psi}{}^\Theta\spsd
\end{equation}
Поэтому первая аксиома умножения выполнена.
\item Существует единичный элемент по умножению $e:=\frac{1}{\sqrt{2}}\eta^\Lambda$.
\begin{proof}
\begin{equation}
\label{re13.2}
\begin{array}{c}
\frac{1}{\sqrt{2}}\eta^\Lambda\eta_{\Lambda\Psi}{}^\Theta=\varepsilon^{AB}P_\Psi{}_AP^\Theta{}_B=\delta_\Psi{}^\Theta\sps\\
\frac{1}{\sqrt{2}}\eta^\Psi\eta_{\Lambda\Psi}{}^\Theta=\eta_\Lambda{}^{AB}\varepsilon_{XA}X^XP^\Theta{}_B=P_\Lambda{}^BP^\Theta{}_B=\delta_\Lambda{}^\Theta\spsd
\end{array}
\end{equation}
\end{proof}
Поэтому вторая аксиома умножения выполнена.
\item Для любого неизотропного элемента существует обратный элемент по умножению
\begin{equation}
\label{re13.3}
r^{-1}:=\frac{1}{<r,r>}(2<r,e>e-r)\sps <r_1,r_2>:=g_{\Lambda\Psi}(r_1)^\Lambda(r_2)^\Psi\spsd
\end{equation}
\begin{proof}
\begin{equation}
\begin{array}{c}
(r^{-1})^\Psi=\frac{1}{r^\Phi r_\Phi}r^\Omega(\eta_\Omega\eta^\Psi-\delta_\Omega{}^\Psi)\sps\\
r^\Lambda \frac{1}{r^\Phi r_\Phi}r^\Omega(\eta_\Omega\eta^\Psi-\delta_\Omega{}^\Psi)\eta_{\Lambda\Psi}{}^\Theta=
\frac{\sqrt{2}}{r^\Phi r_\Phi}R^{AB}r^\Omega(\eta_\Omega\eta^\Psi-\delta_\Omega{}^\Psi)\eta_\Psi{}_{XA}X^XP^\Theta{}_B=\\
=\frac{\sqrt{2}}{r^\Phi r_\Phi}R^{AB}R_{AX}X^XP^\Theta{}_B=\frac{1}{\sqrt{2}}X^BP^\Theta{}_B=\frac{1}{\sqrt{2}}\eta^\Theta\sps
\\
\frac{1}{r^\Phi r_\Phi}r^\Omega(\eta_\Omega\eta^\Lambda-\delta_\Omega{}^\Lambda)r^\Psi \eta_{\Lambda\Psi}{}^\Theta=
\frac{\sqrt{2}}{r^\Phi r_\Phi}R^{BA}R_{XA}X^XP^\Theta{}_B=\frac{1}{\sqrt{2}}\eta^\Theta\spsd
\end{array}
\end{equation}
Эти тождества, верные для произвольного неизотропного r, равносильны соотношению
\begin{equation}
\eta_{\Lambda\Psi}{}^\Omega+\eta_\Lambda{}^\Omega{}_\Psi=\biggl(\frac{1}{\sqrt{2}}\eta_\Lambda\biggl)\delta_{\Psi}{}^\Omega\spsd
\end{equation}
\end{proof}
Поэтому третья аксиома умножения выполнена частично. Для вложения $\mathbb R^n\subset \mathbb C^n$ $(n\ mod\ 8=0)$ эта аксиома выполнена для
любого ненулевого элемента. Соответственно, с помощью операторов вложения $H_i{}^\Lambda$ определены структурные константы $\eta_{ij}{}^k$,
которые, однако, не для всех спиноров $X^A$  будут действительными.
\item Коммутативный закон не выполняется. Однако существует тождество
\begin{equation}
\label{re13.4}

\end{equation}
\end{proof}
Частный случай этого тождества
\begin{equation}
\label{re13.4a}
\tag{\ref{re13.4}$'$}
r^2=-<r,r>e+2<r,e>r
\end{equation}
приводит к (\ref{re13.3}) и означает выполнение степенной ассоциативности.  Поэтому четвертая аксиома умножения не выполнена. Но ей можно найти замену.
\item Ассоциативный закон не выполняется. Однако существует альтернативно-эластичное тождество
\begin{equation}
\label{re13.5}
%
\end{equation}
\end{proof}
Поэтому пятая аксиома умножения не выполнена. Но ей можно найти замену.
\end{enumerate}
Согласно общей теории такая алгебра имеет делители 0 при $n>8$. Действительно,
\begin{equation}
%
\end{equation}
Произведение $P^\Omega{}_BP_\Omega{}^C$ есть вырожденный оператор. Если принять, что известен вектор $r_2$, спинор $X^A$, то неизвестными будут
координаты вектора $r_1$. Но, поскольку $P^\Omega{}_BP_\Omega{}^C$ вырожден, указанная система должна иметь нетривиальное решение при некоторых
значениях вектора $r_2$. Однако, существует такой базис алгебры, для которого
\begin{equation}
\label{re13.6}
%
\end{equation}
Наиболее интересными для изучения представляются алгебры, для которых можно построить такой базис (\ref{re13.6}), элементы которого
не являются делителями нуля и удовлетворяют тождествам альтернативности (такую алгебру назовем алгеброй седенионов)
\begin{equation}
(r_i(r_i,r_j))=((r_i,r_i),r_j)\sps ((r_j,r_i),r_i)=(r_j,(r_i,r_i))\spsd
\end{equation}
Конечно, не для всех алгебр такой базис существует. Поэтому гиперкомплексные алгебры будут классифицироваться по максимальному количеству базисных элементов, не являющихся делителями нуля. Индуктивное построение метрической групповой альтернативно-эластичной алгебры более высокой размерности выглядит следующим образом.
\begin{algorithmr}
\begin{enumerate}
\item Пусть нам известны кососимметричные связующие операторы $\eta_\alpha{}^{ab}$ для n mod 8 =6. Тогда связующие операторы $\eta_\Lambda{}^{AB}$
для n mod 8 =0 строятся по схеме
\begin{equation}
\label{re8.7}

\end{equation}
что обеспечит наличие симметрического метрического спинора $\tilde\varepsilon_{AB}\equiv\varepsilon_{AB}$ для n mod 8 = 0.

\item Пусть нам известны связующие операторы $\eta_\alpha{}^{ab}$ для n mod 8 =0. Тогда связующие операторы $\eta_\Lambda{}^{AB}$
для n mod 8 =2 строятся по схеме
\begin{equation}
\label{re8.8}
%
\end{equation}

\item Пусть нам известны связующие операторы $\eta_\alpha{}^{ab}$ для n mod 8 =2. Тогда связующие операторы $\eta_\Lambda{}^{AB}$
для n mod 8 =4 строятся по схеме
\begin{equation}
\label{re8.9}
%
\end{equation}
обеспечивая для n mod 8 = 4 наличие метрического кососимметрического спинора
\begin{equation}
\varepsilon_{AB}:=(\varepsilon_1)_{AC}(\varepsilon_2)^{DK}(\varepsilon_3)_{BK}=
\left(
%
\right)\sps
\end{equation}
где $(\varepsilon_1)_{ab}$ - метрический симметричный спинор для n mod 8=0.

\item Пусть нам известны связующие операторы $\eta_\alpha{}^{ab}$ для n mod 8 =4. Тогда связующие операторы $\eta_\Lambda{}^{AB}$
для n mod 8 =6 строятся по схеме
\begin{equation}
\label{re8.10}
%
\end{equation}
Кроме того, дополнительно проведем преобразование связующих операторов по схеме
\begin{equation}
\label{re8.10a}
\tag{\ref{re8.10}$'$}
\tilde\eta_\Lambda{}^{AC}:=\eta_\Lambda{}^{AB}\varepsilon_B{}^C\sps
\varepsilon_B{}^C:=
\left(
%
\right)\sps
\end{equation}
где $\varepsilon_{ab}$ $(n/2+q=3)$ для n mod 8 = 4 будет метрическим кососимметрическим спинором. Тогда для следующей реализации n mod 8 = 6
операторы $\tilde\eta_\Lambda{}^{AC}=-\tilde\eta_\Lambda{}^{CA}$ будут полностью кососимметричны $(n/2+q)=0$. Легко проверить выполнение
уравнения Клиффорда (\ref{re6.1}) для таких операторов. Если дальнейшее построение для n mod 8 = 0 вести на основании уже таких операторов,
тогда будет существовать симметричный метрический спинор $\varepsilon_{AB}\equiv(\varepsilon_1)_{AB}$, $n/2+q=1$.
Для n mod 8 = 2 уже будет существовать инволюция при $n/2+q=2$. И, снова, для n mod 8 = 4 получаем $n/2+q=3$. Круг замкнулся.
Начало же положено в примерах \ref{rex6.1}-\ref{rex6.2}.

\item Пусть имеется алгебра над $\mathbb R^n$ со структурными константами, сгенерированными от связующих операторов $\eta_\Lambda{}^{AB}$ с метрическим спинором
$\varepsilon^{XZ}$ и оператором вложения $H_i{}^\Lambda$. Будем считать, что метрический тензор $g_{\Lambda\Psi}$ на главной диагонали содержит одни $\ll+\gg$.
Тогда можно построить кососимметричные связующие операторы для пространства $\mathbb C^{n+6}$
\begin{equation}

\end{equation}
\item Переход к связующим операторам пространства $\mathbb C^{n+8}$ осуществляется так:
\begin{equation}
%
\right)\spsd
$
После этого, можно перейти к связующим операторам пространства $\mathbb R^{n+8}\subset\mathbb C^{n+8}$ с помощью соответствующего оператора вложения.
А уже такие операторы генерируют структурные константы алгебры размерности $n+8$.
\end{enumerate}
\end{algorithmr}

\setcounter{mypage}{182}
\section{Инфинитизимальные преобразования}
\Abstract{
\indent В этом параграфе рассказывается о том, как по связующим операторам, удовлетворяющим уравнению Клиффорда, построить инфинитизимальные
преобразования. Вывод всех результатов этого параграфа сделан на основе \cite[т. 1, с. 224-226]{Penroser1}.
}

Согласно следствию \ref{rc6.2} для собственных ортогональных преобразований верно разложение
\begin{equation}
\label{re8.1}
\begin{array}{ll}
1). & S_\Lambda{}^\Psi\eta_\Psi{}^{AB}=-\eta_\Lambda{}^{CD}S_C{}^A E_D{}^K S_K{}^M E_M{}^B\sps \\
2). & S_\Lambda{}^\Psi\eta_\Psi{}^{AB}=\eta_\Lambda{}^{CD}S_C{}^A (\eta_I)^\Omega S_\Omega{}^\Phi\eta_\Phi{}^{KB}\varepsilon_{KL}S_X{}^L \varepsilon^{MX}(\varepsilon_I)_{MD}\spsd
\end{array}
\end{equation}
Пусть имеется $S_\Lambda{}^\Psi (\lambda)$ - однопараметрическое семейство ортогональных преобразований
\begin{equation}
S_\Lambda{}^\Psi (\lambda) S_\Omega{}^\Phi (\lambda) g_{\Psi\Phi}=g_{\Lambda\Omega}\sps
S_\Lambda{}^\Psi (0):=\delta_\Lambda{}^\Psi\sps
\end{equation}
что индуцирует однопараметрические преобразования в спинорном пространстве
\begin{equation}
S_C{}^A(\lambda)\sps S_C{}^A(0)=\delta_C{}^A\spsd
\end{equation}
Соответствующие инфинитизимальные преобразования определятся как
\begin{equation}
\begin{array}{ll}
T_\Lambda{}^\Psi=\left.\left[\frac{d}{d\lambda} S_\Lambda{}^\Psi(\lambda)\right]\right|_{\lambda =0}\sps &
\begin{array}{lll}
&& T_{\Lambda\Psi}=-T_{\Psi\Lambda}\sps
\end{array}\\
T_C{}^A=\left.\left[\frac{d}{d\lambda}S_C{}^A(\lambda)\right]\right|_{\lambda =0}\sps &
\begin{array}{ll}
1).& T_C{}^A=-E_C{}^B T_B{}^D E_D{}^A\sps\\
2).& T_C{}^A=-T^A{}_C\spsd\\
\end{array}
\end{array}
\end{equation}
Дифференцируя в нуле (\ref{re8.1}), получим тождество (выкладки в приложении)
\begin{equation}
\label{re8.2}
T_\Lambda{}^\Psi\eta_\Psi{}^{AB}=\eta_\Lambda{}^{CB}T_C{}^A-\eta_\Lambda{}^{AD}(\varepsilon_I){}^{KB}T_K{}^M(\varepsilon_I){}_{MD}+
\eta_\Lambda{}^{AD}(\varepsilon_I){}_{MD}(\eta_I)^\Omega T_\Omega{}^\Phi\eta_\Phi{}^{MB}\spsd
\end{equation}
Частным решением этого уравнения будет
\begin{equation}
\label{re8.3}
T_A{}^C=\frac{1}{2}T^{\Theta\Phi}\eta_\Phi{}^{AB}\eta_\Theta{}_{CB}\sps
\end{equation}
что проверяется прямой подстановкой. Однородное же решение уравнения
\begin{equation}
\eta_\Lambda{}^{CB}T_C{}^A-\eta_\Lambda{}^{AD}(\varepsilon_I){}^{KB}T_K{}^M(\varepsilon_I){}_{MD}=0
\end{equation}
перепишется как  (\ref{re6.3/1}), что означает тривиальность такого решения.

\section{Комплексная и вещественная связности}
\Abstract{
\indent В этом параграфе рассказывается о том, как по частным решениям уравнения Клиффорда построить продолжение римановой связности
без кручения в спинорное расслоение. Рассматривается два различных варианта. Это даст возможность построить спинорные аналоги
операторов Ли. Вывод всех результатов этого параграфа сделан на основе \cite{Bilyalovr1}, \cite[т. 1, с. 282-285]{Penroser1}.
}

Для вещественной реализации касательного расслоения $\tau^\mathbb R (V_{2n})$ уравнения геодезических примут вид
\begin{equation}
\frac{d^2x^\myf\Omega}{ds^2}+\Gamma_{\myf\Phi\myf\Psi}{}^\myf\Omega\ \frac{dx^\myf\Phi}{ds}\frac{dx^\myf\Psi}{ds}=0\spsd
\end{equation}
Для параметризации атласа $w^\Theta=\frac{1}{\sqrt{2}}(u^\Theta(x^\myf\Theta)+iv^\Theta(x^\myf\Theta))$
\begin{equation}

\end{equation}
Домножим обе части на $\frac{dw^\Lambda}{dx^\myf\Omega}=m^\Lambda{}_\myf\Omega$. Учитывая тождества (\ref{re1.2}) для операторов
$m^\Lambda{}_\myf\Omega$ и $m^{\Lambda '}{}_\myf\Omega$, получим
\begin{equation}
%
\end{equation}
Для комплексной реализации касательного расслоения $\tau^\mathbb C (V_{2n})$ уравнения геодезических примут вид
\begin{equation}
\frac{d^2w^\Lambda}{ds^2}+(\Gamma_{\myf\Phi\Theta}{}^\Lambda\ \frac{dw^\Theta}{ds}+
\Gamma_{\myf\Phi\Theta '}{}^\Lambda\ \frac{dw^{\Theta '}}{ds})\ \frac{dx^\myf\Phi}{ds}=0\spsd
\end{equation}
Сравнивая эти два уравнения и определение (\ref{re1.2}) операторов $m^\Lambda{}_\myf\Omega$, $m_\Lambda{}^\myf\Omega$, получим
\begin{equation}
\left\{
%
\right.
\end{equation}
принимая во внимание тождество
\begin{equation}
\frac{\partial^2x^\myf\Omega}{\partial w^\Theta ds}=\frac{\partial^2x^\myf\Omega}{\partial w^\Theta  \partial x^\myf\Phi}\frac{dx^\myf\Phi}{ds}=
\frac{dx^\myf\Phi}{ds}\partial_\myf\Phi\frac{\partial x^\myf\Omega}{\partial w^\Theta}=\frac{dx^\myf\Phi}{ds}\partial_\myf\Phi m_\Theta{}^\myf\Omega\spsd
\end{equation}
Другая запись этой системы такая
\begin{equation}
%
\end{equation}
Сложим эти два уравнения
\begin{equation}
\partial_\myf\Phi m^\Lambda{}_\myf\Lambda-\Gamma_{\myf\Phi\myf\Lambda}{}^\myf\Omega m^\Lambda{}_\myf\Omega+
\Gamma_{\myf\Phi\Xi}{}^\Lambda m^\Xi{}_\myf\Lambda+\Gamma_{\myf\Phi\Xi '}{}^\Lambda \bar m^{\Xi '}{}_\myf\Lambda=0\spsd
\end{equation}
Свернем его с $\bar m_{\Psi '}{}^\myf\Lambda$
\begin{equation}
%
\end{equation}
Поэтому связность в касательном расслоении необходимо доопределить системой
\begin{equation}
\left\{
%
\end{equation}
Домножим это уравнение на $\frac{i}{2}\bar m_{\Psi '}{}^\myf\Omega m^\Lambda{}_\myf\Lambda$
\begin{equation}
\frac{i}{2}\bar m_{\Psi '}{}^\myf\Omega m^\Lambda{}_\myf\Lambda\partial_\myf\Phi f_\myf\Omega{}^\myf\Lambda=
\bar m_{\Psi '}{}^\myf\Omega m^\Lambda{}_\myf\Lambda\Gamma_{\myf\Phi\myf\Omega}{}^\myf\Lambda \spsd
\end{equation}
Отсюда
\begin{equation}
\Gamma_{\myf\Phi\Psi '}{}^\Lambda=0\spsd
\end{equation}
Поэтому можно записать
\begin{equation}
\nabla_\myf\Phi m^\Lambda{}_\myf\Lambda:=\partial_\myf\Phi m^\Lambda{}_\myf\Lambda-
\Gamma_{\myf\Phi\myf\Lambda}{}^\myf\Omega m^\Lambda{}_\myf\Omega+\Gamma_{\myf\Phi\Psi}{}^\Lambda m^\Psi{}_\myf\Lambda=0\spsd
\end{equation}
Таким образом, если известны коэффициенты связности в одной из реализации (вещественной или комплексной), то по этому уравнению
однозначно восстанавливаются  коэффициенты связности в другой реализации (комплексной или вещественной соответственно).
\newpage
\noindent Если теперь в вещественной реализации касательного расслоения задана связность, согласованная с метрикой
\begin{equation}
\label{re9.0}
\nabla_\myf\Lambda G_{\myf\Psi\myf\Omega}=0\sps
\end{equation}
то в спинорное расслоение ее можно продолжить условием
\begin{equation}
\label{re9.2}
\nabla_\myf\Lambda\eta_\myf\Psi{}^{\myff A \myff B}=0\spsd
\end{equation}
Пусть теперь I). n mod 4 =0, n>2, II). n mod 4 =2, n>4, тогда
\begin{equation}

\end{equation}
Таким образом, получаем, что
\begin{equation}
\label{re9.1}
%
\end{equation}
Такой вид связности возможен при нормировке спинорного базиса согласно \cite[т. 1, с. 284]{Penroser1}. Однако, можно ослабить условия (\ref{re9.0}) и (\ref{re9.2}), вспомнив, что мы рассматриваем связность в касательном расслоении. Следовательно, необходимо туда и вернуться, получая соотношение
\begin{equation}
\label{re9.3}
\eta^\myf\Psi{}_{\myff A\myff B}\nabla_\myf\Lambda\eta_\myf\Omega{}^{\myff A\myff B}=0\sps
\end{equation}
которое расписывается как
\begin{equation}
\label{re9.4}
\eta^\myf\Psi{}_{\myff A\myff B}\partial_\myf\Lambda \eta_\myf\Omega{}^{\myff A\myff B}-
\Gamma_{\myf\Lambda\myf\Omega}{}^\myf\Theta\eta^\myf\Psi{}_{\myff A\myff B}\eta_\myf\Theta{}^{\myff A\myff B}+
\Gamma_{\myf\Lambda\myff D}{}^{\myff A}\eta^\myf\Psi{}_{\myff A\myff B}\eta_\myf\Omega{}^{\myff D\myff B}+
\Gamma_{\myf\Lambda\myff D}{}^{\myff B}\eta^\myf\Psi{}_{\myff A\myff B}\eta_\myf\Omega{}^{\myff A\myff D}=0\spsd
\end{equation}
 Поэтому для того, чтобы однозначно продолжить построенную связность в спинорное расслоение, необходимо потребовать
\begin{equation}
\begin{array}{c}
\Gamma_{\myf\Lambda \myff L}{}^{\myff M}:=\frac{1}{N^2}\Gamma_{\myf\Lambda \myff K}{}^{\myff A}(
\eta_\myf\Psi{}_{\myff L\myff B}\eta_\myf\Omega{}^{\myff M\myff B} \eta^\myf\Psi{}_{\myff A\myff C}\eta^\myf\Omega{}^{\myff K\myff C}
-\frac{(2n-2)}{4}\delta_{\myff L}{}^{\myff M}\delta_{\myff K}{}^{\myff A})\sps\\ \\
\Gamma_{\myf\Lambda \myff K}{}^{\myff A}:=-\frac{2}{2n-4}\Gamma_{\myf\Lambda \myff C}{}^{\myff D}(\varepsilon_{\myff D\myff K}{}^{\myff C\myff A}-
\frac{2n-2}{2N^2}\delta_{\myff K}{}^{\myff A}\delta_{\myff D}{}^{\myff C})\spsd
\end{array}
\end{equation}
Действительно,
\begin{equation}
\begin{array}{c}
\eta_\myf\Lambda{}^{\myff A\myff B}\eta_\myf\Psi{}_{\myff K\myff B} \cdot (-\frac{2}{2n-4})(\varepsilon_{\myff D\myff A}{}^{\myff C\myff K}-
\frac{2n-2}{2N^2}\delta_{\myff A}{}^{\myff K}\delta_{\myff D}{}^{\myff C})=\eta_{\left[\right.\myf\Lambda}{}^{\myff B\myff C}\eta_{\myf\Psi\left.\right]}{}_{\myff B\myff D}-\\
-\frac{1}{2n-4}(\frac{2n}{2}-\frac{2n-2}{2N^2}\cdot 2N^2)g_{\myf\Lambda\myf\Psi}\delta_{\myff D}{}^{\myff C}=
\eta_\myf\Lambda{}^{\myff B\myff C}\eta_\myf\Psi{}_{\myff B\myff D}\spsd
\end{array}
\end{equation}

\begin{noter}
\label{rn9.1}
Следует отметить, что при n=4 все выводы остаются верными, а последнее тождество примет вид
\begin{equation}
\begin{array}{c}
\eta_\myf\Lambda{}^{\myff A\myff C}\eta_\myf\Psi{}_{\myff K\myff C}=(\eta_\myf\Lambda{}^{\myff C\myff A}+\eta_\myf\Lambda\varepsilon^{\myff A\myff C})\eta_\myf\Psi{}_{\myff K\myff C}=
\eta_\myf\Lambda{}^{\myff C\myff A}(\eta_\myf\Psi{}_{\myff C\myff K}+\eta_\myf\Psi\varepsilon_{\myff K\myff C})+\eta_\myf\Lambda\eta_\myf\Psi{}_{\myff K}{}^{\myff A}=\\
=\eta_\myf\Lambda{}^{\myff C\myff A}\eta_\myf\Psi{}_{\myff C\myff K}-\eta_\myf\Psi\eta_\myf\Lambda{}_{\myff K}{}^{\myff A}+\eta_\myf\Lambda\eta_\myf\Psi{}_{\myff K}{}^{\myff A}\spsd
\end{array}
\end{equation}
Таким образом, свертка (\ref{re9.4}) c $\eta_\myf\Psi\eta^\myf\Omega$ определит выражение $\Gamma_{\myf\Lambda\myff A}{}^{\myff A}$, после подстановки
которого в (\ref{re9.4}) однозначно определятся коэффициенты $\Gamma_{\myf\Lambda\myff A}{}^{\myff D}$. Для этого достаточно свернуть (\ref{re9.4}) с
$\eta_\myf\Psi{}_{\myff K\myff L}\eta^\myf\Omega{}^{\myff M\myff L}$. Это обусловлено тождеством
$\eta_\Psi{}_{\myff A\myff B}\eta^\Psi{}_{\myff C\myff D}=\varepsilon_{\myff A\myff B\myff C\myff D}=\varepsilon_{\myff A\myff C}\varepsilon_{\myff B\myff D}$,
где $\varepsilon_{\myff B\myff D}$ - как и прежде, кососимметрический метрический спинор (см. пример \ref{rex6.1}, который легко адаптируется для вложения $\mathbb R^4\subset\mathbb C^4$).
Но в таком случае условие (\ref{re9.3}) равносильно (\ref{re9.2}).
\end{noter}

Если выполнить симметризацию выражения (\ref{re9.4}) по $\Omega$  и опущенному $\Psi$, то из аналога (\ref{re6.5}) получим
\begin{equation}
\begin{array}{c}
\underbrace{\frac{N^2}{2}(\partial_\myf\Lambda G_{\myf\Psi\myf\Omega}-
2\Gamma_{\myf\Lambda(\myf\Omega\myf\Psi)})}_{=\frac{N^2}{2}\nabla_\Lambda G_{\myf\Psi\myf\Omega}}+
\underbrace{(\frac{1}{2N^2}\eta_\myf\Psi{}_{\myff A\myff B}\eta_\myf\Omega{}_{\myff C\myff D}\partial_\Lambda\varepsilon^{\myff A\myff B\myff C\myff D}+
\Gamma_{\myf\Lambda\myff B}{}^{\myff B}G_{\myf\Psi\myf\Omega})}_
{=\frac{1}{2N^2}\eta_{\myf\Psi}{}_{\myff A\myff B}\eta_\myf\Omega{}_{\myff C\myff D}\nabla_\Lambda \varepsilon^{\myff A\myff B\myff C\myff D}}=0\spsd
\end{array}
\end{equation}
Если метрический тензор ковариантно постоянен, тогда
\begin{equation}
\eta_\myf\Psi{}_{\myff A\myff B} \eta_\myf\Omega{}_{\myff C\myff D}\nabla_\myf\Lambda\varepsilon^{\myff A\myff B\myff C\myff D}=0\spsd
\end{equation}
Соответственно, комплексная связность определится соотношениями
\begin{equation}
\nabla_\Lambda:=m_\Lambda{}^\myf\Lambda\nabla_\myf\Lambda\sps
\bar\nabla_{\Lambda '}:=\bar m_{\Lambda '}{}^\myf\Lambda\nabla_\myf\Lambda\sps
\end{equation}
что приведет в отсутствии кручения к связности
\begin{equation}
\nabla_\Lambda g_{\Psi\Phi}=0\sps  \bar\nabla_{\Lambda '} \bar g_{\Psi '\Phi '}=0\sps
\end{equation}
которая продолжается в комплексное спинорное расслоение условиями (см. (\ref{re3.13/1}))
\begin{equation}
\label{re9.8}

\end{equation}
При этом следует учесть, что верхние соотношения
\begin{equation}
%
\end{equation}
определяют различные коэффициенты связности $(\tilde\Gamma_K)_\Lambda{}_A{}^C$ и $(\tilde{\tilde\Gamma}_K)_\Lambda{}_A{}^C$,
а нижние соотношение расписывается как
\begin{equation}
%
\end{equation}
Чтобы однозначно продолжить связность, можно положить
\begin{equation}
\label{re9.9}
%
\end{equation}
Поэтому, из ковариантного постоянства метрического тензора
\begin{equation}
%
\end{equation}
будет следовать
\begin{equation}
\label{re9.5}
\sum\limits_{K=1}^{2N}(\eta_K)_\Psi{}_{AB} (\eta_K)_\Omega{}_{CD}\nabla_\Lambda(\varepsilon_K)^{ABCD}=0\spsd
\end{equation}
Теперь обратимся к вещественным вложениям. Уравнение
\begin{equation}
\nabla_\Lambda H_i{}^\Psi=0
\end{equation}
распишется как
\begin{equation}
\partial_\Lambda H_i{}^\Psi -\Gamma_{\Lambda i}{}^j H_j{}^\Psi+\Gamma_{\Lambda\Phi}{}^\Psi H_i{}^\Phi=0\spsd
\end{equation}
Соответственно в спинорное расслоение вещественная связность продолжается условием
\begin{equation}
\sum\limits_{I=1}^{2N}(\eta_K)^i{}_{AB}\nabla_k(\eta_K)_j{}^{AB}=0\spsd
\end{equation}
Это означает, что на вещественном пространстве связность, согласованная с метрикой будет иметь вид
\begin{equation}
\nabla_i g_{ik}=0\sps
\end{equation}
где
\begin{equation}
\nabla_i:=H_i{}^\Lambda\nabla_\Lambda\spsd
\end{equation}

\begin{example}
n=4. Индекс метрики равен 1.\\
Пусть
\begin{equation}
g^i{}_{AB'}:=(\eta_1)^i{}_{AB}\bar S_{B'}{}^B\sps g_i{}^{AB'}:=(\eta_1)_i{}^{AB}S_B{}^{B'}\sps \overline{g_j{}^{A'B}}=\bar g_j{}^{B'A}=g_j{}^{AB'}\spsd
\end{equation}
Тогда частный случай условий согласования связностей может иметь вид
\begin{equation}
g^i{}_{AB'}\nabla_k g_j{}^{AB'}+\bar g^i{}_{B'A}\nabla_k \bar g_j{}^{B'A}=0\spsd
\end{equation}
Это условие расписывается как
\begin{equation}
\begin{array}{c}
g^i{}_{AB'}(\partial_k g_j{}^{AB'}-\Gamma_{kj}{}^lg_l{}^{AB'}+\tilde{\tilde\Gamma}_{kC}{}^Ag_j{}^{CB'}+\tilde\Gamma_{kC'}{}^{B'}g_j{}^{AC'}+
\partial_k g_j{}^{AB'}-\Gamma_{kj}{}^lg_l{}^{AB'}+\\+\overline{\tilde{\tilde\Gamma}}_{kC'}{}^{B'}g_j{}^{AC'}+\overline{\tilde\Gamma}_{kC}{}^Ag_j{}^{CB'})=0\spsd
\end{array}
\end{equation}
Поэтому это соотношение ввиду замечания \ref{rn9.1} перепишется так
\begin{equation}
\partial_k g_j{}^{AB'}-\Gamma_{kj}{}^lg_l{}^{AB'}+\frac{1}{2}(\tilde{\tilde\Gamma}_{kC}{}^A+\overline{\tilde\Gamma}_{kC}{}^A)g_j{}^{CB'}+
\frac{1}{2}\overline{(\tilde{\tilde\Gamma}_{kC}{}^B+\overline{\tilde\Gamma}_{kC}{}^B)}g_j{}^{AC'}=0\spsd
\end{equation}
Определим $\Gamma_{kC}{}^A:=\frac{1}{2}(\tilde{\tilde\Gamma}_{kC}{}^A+\overline{\tilde\Gamma}_{kC}{}^A)$, что приведет к связности \cite[т. 1, с. 284]{Penroser1}
\begin{equation}
\nabla_k g_j{}^{AB'}=0\spsd
\end{equation}
\end{example}

Этим доказана теорема.
\begin{theoremr}
Рассмотрим псевдориманово многообразие $V_{2n}$ $(n\ge 4)$. Пусть в касательном расслоении со слоями, изоморфными $R^{2n}_{(n,n)}$, введена
некоторая связность без кручения. Тогда условие согласования связностей в спинорном и касательном расслоении может иметь вид
\begin{equation}
\eta^\myf\Psi{}_{\myff A\myff B}\nabla_\myf\Lambda \eta_\myf\Omega{}^{\myff A\myff B}=0\spsd
\end{equation}
Чтобы продолжение было однозначным, потребуем
\begin{equation}

\end{equation}
Для продолжения  связности на комплексную реализацию $\mathbb CV_{n}$ со слоями, изоморфными $\mathbb C^n$, необходимо, чтобы
\begin{equation}
\nabla_\myf\Lambda m_\myf\Psi{}^\Psi=0\sps\nabla_\myf\Lambda \bar m_\myf\Psi{}^{\Psi '}=0\sps
\nabla_\Lambda:= m^\myf\Lambda{}_\Lambda\nabla_\myf\Lambda\sps
\bar\nabla_{\Lambda '}:= \bar m^\myf\Lambda{}_{\Lambda '}\nabla_\myf\Lambda\spsd
\end{equation}
Тогда условие согласования связностей примет вид
\begin{equation}
%
\end{equation}
Чтобы продолжение было однозначным, потребуем
\begin{equation}
%
\end{equation}
В случае вложения действительного пространства в комплексное необходимо выполнение соотношений
\begin{equation}
\nabla_\Lambda H_i{}^\Psi=0\sps
\bar\nabla_{\Lambda '}\bar H_i{}^{\Psi '}=0\sps \nabla_i:=H_i{}^\Lambda \nabla_\Lambda\spsd
\end{equation}
Тогда условие согласования связностей примет вид
\begin{equation}
\sum\limits_{I=1}^{2N}(\eta_K)^i{}_{AB}\nabla_k(\eta_K)_j{}^{AB}=0\spsd
\end{equation}
\end{theoremr}

\begin{corollaryr}
В случае, если в касательном расслоении многообразия $V_{2n}$ задана связность без кручения, согласованная с метрикой
\begin{equation}
\nabla_\myf\Lambda G_{\myf\Psi\myf\Omega}=0\sps
\end{equation}
то в касательном расслоении комплексной реализации $\mathbb CV_{n}$  индуцируется связность
\begin{equation}
\nabla_\Lambda g_{\Omega\Psi}=0\sps \bar\nabla_{\Lambda '}\bar g_{\Omega '\Psi '}=0\spsd
\end{equation}
Для вложения $V_n\subset\mathbb CV_{n}$ индуцируется связность
\begin{equation}
\nabla_i g_{ik}=0\spsd
\end{equation}
\end{corollaryr}

Построим операторы $P^\myf\Psi{}_{\myff A}:=\eta^\myf\Psi{}_{\myff B\myff A}X^{\myff B}$ и $P_\myf\Psi{}^{\myff A}:=\eta_\myf\Psi{}^{\myff B\myff A}Y_{\myff B}$
такие, что $X^{\myff A}Y_{\myff A}=2$ и $P_\myf\Lambda{}^{\myff A}P_\myf\Psi{}_{\myff A}=G_{\myf\Lambda\myf\Psi}$. Это всегда можно сделать при $n\ge 8$.
Тогда условие согласования связностей для вещественной реализации может иметь и такой вид
\begin{equation}
\nabla_\myf\Lambda P_\myf\Psi{}^{\myff A}=0\sps
\end{equation}
которое распишется как
\begin{equation}
\partial_\myf\Lambda P_\myf\Psi{}^{\myff A}-\Gamma_{\myf\Lambda\myf\Psi}{}^\myf\Phi P_\myf\Phi{}^{\myff A}+\Gamma_\myf\Lambda{}_{\myff C}{}^{\myff A}P_\myf\Psi{}^{\myff C}=0\spsd
\end{equation}
Поэтому можно однозначно определить связность в спинорном расслоении как
\begin{equation}
\label{re9.6}
\Gamma_\myf\Lambda{}_{\myff C}{}^{\myff A}:=P^\myf\Psi{}_{\myff C}\partial_\myf\Lambda P_\myf\Psi{}^{\myff A}+\Gamma_{\myf\Lambda\myf\Psi}{}^\myf\Phi P^\myf\Psi{}_{\myff C}P_\myf\Phi{}^{\myff A}\spsd
\end{equation}
Для комплексной реализации определены операторы
\begin{equation}
\label{re9.10}
\begin{array}{c}
(P_K)^\Psi{}_A:=m_\myf\Psi{}^\Psi P^\myf\Psi{}_{\myff A}(\tilde M_K)_A{}^{\myff A}\sps
(P_K)^{\Psi '}{}_A:=\bar m_\myf\Psi{}^{\Psi '} P^\myf\Psi{}_{\myff A}(\tilde M_K)_A{}^{\myff A}\sps\\
(P^*_K)_\Psi{}^A:=m^\myf\Psi{}_\Psi P_\myf\Psi{}^{\myff A}(\tilde M^*_K)^A{}_{\myff A}\sps
(P^*_K)_{\Psi '}{}^A:=\bar m^\myf\Psi{}_{\Psi '} P_\myf\Psi{}^{\myff A}(\tilde M^*_K)^A{}_{\myff A}\spsd\\
\end{array}
\end{equation}

\begin{theoremr}
\label{rtheorem9.1}
Рассмотрим псевдориманово многообразие $V_{2n}$ $(n\ge 4)$. Пусть в касательном расслоении со слоями, изоморфными $R^{2n}_{(n,n)}$, введена
некоторая связность без кручения. Тогда условие согласования связностей в спинорном и касательном расслоении может иметь вид
\begin{equation}
\nabla_\myf\Lambda P_\myf\Psi{}^{\myff A}=0\spsd
\end{equation}
Чтобы продолжение было однозначным, потребуем
\begin{equation}
\Gamma_\myf\Lambda{}_{\myff C}{}^{\myff A}:=-P^\myf\Psi{}_{\myff C}\partial_\myf\Lambda P_\myf\Psi{}^{\myff A}+\Gamma_{\myf\Lambda\myf\Psi}{}^\myf\Phi P^\myf\Psi{}_{\myff C}P_\myf\Phi{}^{\myff A}\spsd
\end{equation}
Для продолжения  связности на комплексную реализацию $\mathbb CV_{n}$ со слоями, изоморфными $\mathbb C^n$, необходимо, чтобы
\begin{equation}
\nabla_\myf\Lambda m_\myf\Psi{}^\Psi=0\sps\nabla_\myf\Lambda \bar m_\myf\Psi{}^{\Psi '}=0\sps
\nabla_\Lambda:= m^\myf\Lambda{}_\Lambda\nabla_\myf\Lambda\sps
\bar\nabla_{\Lambda '}:= \bar m^\myf\Lambda{}_{\Lambda '}\nabla_\myf\Lambda\spsd
\end{equation}
Тогда условие согласования связностей примет вид
\begin{equation}

\end{equation}
В случае вложения действительного пространства в комплексное необходимо выполнение соотношений
\begin{equation}
\nabla_\Lambda H_i{}^\Psi=0\sps
\bar\nabla_{\Lambda '}\bar H_i{}^{\Psi '}=0\sps \nabla_i:=H_i{}^\Lambda \nabla_\Lambda\spsd
\end{equation}
Тогда условие согласования связностей примет вид
\begin{equation}
\nabla_i (P_K)^j{}_A=0\sps\nabla_i (P_K)_j{}^A=0\spsd\\
\end{equation}
\end{theoremr}

Теперь появилась возможность построить операторы
\begin{equation}
%
\end{equation}
Эти операторы будут инвариантами связности только в том случае, когда
\begin{equation}
%
\end{equation}
что обеспечит инвариантность операторов. Поэтому будут выполнены тождества
\begin{equation}
\label{re9.7}
%
\end{equation}
При n=4 подобные связности рассматривались в \cite{Bilyalovr1}. Однако, там в качестве $(P_K)_\Phi{}^\Psi{}_C{}^A$ рассматривалось
выражение вида $(\eta_K)^\Psi{}_{DC} (\eta_K)_\Phi{}^{DA}$, что вызывало кручение таких операторов.

\begin{theoremr}
\label{rtheorem9.2}
Операторы
\begin{equation}
%
\end{equation}
будут являться инвариантами комплексной связности из теоремы \ref{rtheorem9.1} при условии постоянства операторов
\begin{equation}
%
\end{equation}
Такие операторы будут удовлетворять тождествам
\begin{equation}
%
\end{equation}
и являться естественным продолжением операторов $L_x(y)^\Psi=x^\Omega\partial_\Omega y^\Psi-y^\Omega\partial_\Omega x^\Psi$,
$\bar L_{\bar x}(\bar y)^{\bar\Psi}=\bar x^{\Omega '}\bar \partial_{\Omega '}\bar y^{\Psi '}-\bar y^{\Omega '}\bar\partial_{\Omega '}\bar x^{\Psi '}$
в спинорное расслоение
\begin{equation}
%
\end{equation}
\end{theoremr}
\newpage

\section{О классификации тензоров с симметриями тензора кривизны. Спиноры кривизны}
\Abstract{
\indent В этом параграфе рассказывается о том, как сопоставить тензорам, обладающим симметриями тензора кривизны, их спинорные аналоги. Это упрощает
проблему классификации таких тензоров для малых размерностей. Вывод всех результатов этого параграфа сделан на основе \cite[т. 1, с. 285-303]{Penroser1}.
}

Ограничим себя случаем, когда $(\Gamma_K)_{\Lambda A}{}^B=(\tilde\Gamma_K)_{\Lambda A}{}^B=(\tilde{\tilde\Gamma}_K)_{\Lambda A}{}^B$. Тогда тензоры кривизны можно вычислить по формулам
\begin{equation}
\nabla_{\left[\right.\Lambda}\nabla_{\Psi\left.\right]}r^\Omega=R_{\Lambda\Psi\Phi}{}^\Omega r^\Phi\sps
\nabla_{\left[\right.\Lambda}\nabla_{\Psi\left.\right]}X^A=\sum\limits_{K=1}^{2N}(\mathcal R_K)_{\Lambda\Psi C}{}^A X^C\spsd
\end{equation}
Для этого рассмотрим согласование комплексных связностей, индуцируемое формулами (\ref{re9.8}), (\ref{re9.9}) при ковариантном постоянстве
метрического тензора.  Потребуем, чтобы
\begin{equation}
\nabla_\Lambda (\varepsilon_K)_{AB}{}^{CD}=\partial_\Lambda (\varepsilon_K)_{AB}{}^{CD}=0\spsd
\end{equation}
Следовательно,
\begin{equation}
\sum\limits_{K=1}^{2N}(\nabla_\Theta(\eta_K)^\Psi{}_{AB})(\nabla_\Lambda(\eta_K)_\Omega{}^{AB})=0\spsd
\end{equation}
Поэтому условия интегрируемости (\ref{re9.8}) примут вид
\begin{equation}
\begin{array}{c}
\sum\limits_{K=1}^{2N}(\eta_K)^\Theta{}_{AB}\nabla_{\left[\right.\Lambda}\nabla_{\Psi\left.\right]}(\eta_K)_\Omega{}^{AB}=0\sps\\
N^2R_{\Lambda\Psi\Omega}{}^\Theta=\sum\limits_{K=1}^{2N}
((\mathcal R_K)_{\Lambda\Psi C}{}^A(\eta_K)^\Theta{}_{AB}(\eta_K)_\Omega{}^{CB}+
(\mathcal R_K)_{\Lambda\Psi C}{}^B(\eta_K)^\Theta{}_{AB}(\eta_K)_\Omega{}^{AC})\spsd
\end{array}
\end{equation}
Определим
\begin{equation}
\label{re10.1}
(A_K)_{\Theta\Phi L}{}^X:=\frac{1}{2}(\eta_K)_{\left[\right.\Theta}{}^{MX}(\eta_K)_{\Phi\left.\right]}{}_{ML}\spsd
\end{equation}
Тогда из (\ref{re6.4}) будет следовать
\begin{equation}
(A_K)_{\Phi\Theta L}{}^X (A_K)_{\Omega\Gamma X}{}^L=\frac{N}{8}g_{\Phi\left[\right.\Gamma}g_{\Omega\left.\right]\Theta}\spsd
\end{equation}
Поэтому можно положить
\begin{equation}
\label{re10.2}
\begin{array}{c}
(R_K)_{\Lambda\Psi C}{}^N:=-R_{\Lambda\Psi \Theta \Phi}(A_K)^{\Theta\Phi}{}_C{}^N\sps
R_{\Lambda\Psi \Theta \Phi}=\frac{8}{N}(A_K)_{\Theta\Phi}{}_C{}^N (R_K)_{\Lambda\Psi N}{}^C\sps\\[2ex]
(R_K)_{\Lambda\Psi C}{}^M=-\frac{2}{n-4}(R_K)_{\Lambda\Psi L}{}^D\varepsilon_{CD}{}^{ML}\spsd
\end{array}
\end{equation}
Очевидно, что $(R_K)_{\Lambda\Psi C}{}^N=(\mathcal R_K)_{\Lambda\Psi C}{}^N$, например, если условие согласования связностей имеет вид
\begin{equation}
\label{re10.2/1}
\nabla_\Lambda(\eta_K)_\Psi{}^{AB}=0\spsd
\end{equation}
Для дальнейших выкладок следует воспользоваться тождествами (номер K будем опускать, поскольку все остальные
выкладки не зависят от этого номера)
\begin{equation}
\label{re10.3}
\eta_{\left[\right. \Lambda_1}{}^{A_1A_2}\eta_{\Lambda_2\left.\right]}{}_{A_1A_3}
\eta_{\left[\right. \Lambda_3}{}^{A_4A_3}\eta_{\Lambda_4\left.\right]}{}_{A_4A_2}=
\frac{N}{2} g_{\Lambda_1 \left[\right. \Lambda_4}
            g_{\Lambda_3 \left.\right] \Lambda_2}\spsd
\end{equation}
\begin{equation}
\label{re10.4}
\eta_{\left[\right. \Lambda_1}{}^{A_1A_2}\eta_{\Lambda_2\left.\right]}{}_{A_1A_3}
\eta_{\left[\right. \Lambda_3}{}^{A_4A_3}\eta_{\Lambda_4\left.\right]}{}_{A_4A_5}
\eta_{\left[\right. \Lambda_5}{}^{A_6A_5}\eta_{\Lambda_6\left.\right]}{}_{A_6A_2}=
N g_{\left[\right. \Lambda_3 | \left[\right. \Lambda_2}
  g_{\Lambda_1 \left.\right] | \left[\right. \Lambda_6}
  g_{\Lambda_5 \left.\right] | \Lambda_4 \left]\right.}\spsd
\end{equation}
\begin{equation}
\label{re10.5}
\begin{array}{c}
\eta_{\left[\right. \Lambda_1}{}^{A_1A_2}\eta_{\Lambda_2\left.\right]}{}_{A_1A_3}
\eta_{\left[\right. \Lambda_3}{}^{A_4A_3}\eta_{\Lambda_4\left.\right]}{}_{A_1A_5}
\eta_{\left[\right. \Lambda_5}{}^{A_6A_5}\eta_{\Lambda_6\left.\right]}{}_{A_6A_7}
\eta_{\left[\right. \Lambda_7}{}^{A_8A_7}\eta_{\Lambda_8\left.\right]}{}_{A_8A_2}=\\
=\frac{N}{4}
(g_{\Lambda_1\left[\right. \Lambda_8} g_{\Lambda_7 \left.\right]\Lambda_2}g_{\Lambda_3\left[\right. \Lambda_6} g_{\Lambda_5 \left.\right]\Lambda_4}-
g_{\Lambda_1\left[\right. \Lambda_6} g_{\Lambda_5 \left.\right]\Lambda_2}g_{\Lambda_4\left[\right. \Lambda_8} g_{\Lambda_7 \left.\right]\Lambda_3}+\\
+g_{\Lambda_1\left[\right. \Lambda_4} g_{\Lambda_3 \left.\right]\Lambda_2}g_{\Lambda_5\left[\right. \Lambda_8} g_{\Lambda_7 \left.\right]\Lambda_6})+
N(
g_{\left[\right. \Lambda_3 | \left[\right. \Lambda_2}
g_{\Lambda_1 \left.\right]   \left[\right. \Lambda_8}
g_{\Lambda_7 \left.\right]   \left[\right. \Lambda_6}
g_{\Lambda_5 \left.\right] | \Lambda_4 \left.\right]}+\\
+g_{\left[\right. \Lambda_5 | \left[\right. \Lambda_4}
g_{\Lambda_3 \left.\right]   \left[\right. \Lambda_7}
g_{\Lambda_8 \left.\right]   \left[\right. \Lambda_2}
g_{\Lambda_1 \left.\right] | \Lambda_6 \left.\right]}+
g_{\left[\right. \Lambda_3 | \left[\right. \Lambda_7}
g_{\Lambda_8 \left.\right]   \left[\right. \Lambda_5}
g_{\Lambda_6 \left.\right]   \left[\right. \Lambda_1}
g_{\Lambda_2 \left.\right] | \Lambda_4 \left.\right]})\spsd
\end{array}
\end{equation}
Доказательство в приложении.

\begin{theoremr}
Классификацию битензора, обладающего свойствами ($\Lambda ,\Psi ,...=\overline{1,n}$, $A,B ,...=\overline{1,N}$)
\begin{equation}
    R_{\Lambda \Psi \Phi \Theta}=R_{\left[ \Lambda \Psi\right]
    \left[\Phi \Theta] \right.} \sps
    R_{\Lambda \Psi  \Phi \Theta}=R_{\Phi \Theta \Lambda \Psi}
\end{equation}
и принадлежащего касательному расслоению $\tau(\mathbb CV_n)$ над аналитическим псевдоримановым пространством
$\mathbb CV_n$, можно свести к классификации тензора $R_A{}^B{}_C{}^D$ N-мерного комплексного спинорного пространства такого, что
\begin{equation}
    R_C{}^D{}_S{}^R:=R_{\Lambda \Psi \Phi \Theta}A^{\Lambda \Psi}{}_C{}^D
    A^{\Phi \Theta}{}_S{}^R\sps A_{\Lambda \Psi C}{}^D:=\frac{1}{2}\eta_{\left[\Lambda\right.}{}^{AD}\eta_{\left.\Psi\right]}{}_{AC}\spsd
\end{equation}
Кроме того, выполнены соотношения
\begin{equation}
    R_K{}^K{}_S{}^R=R_S{}^R{}_K{}^K=0 \sps  R_C{}^D{}_S{}^R=
    R_S{}^R{}_C{}^D\spsd
\end{equation}
Разложение
\begin{equation}
\begin{array}{c}
    R_C{}^K{}_M{}^A=C_C{}^K{}_M{}^A-\frac{4}{N(n-2)}R_G{}^D{}_N{}^P\varepsilon^{ABKL}\varepsilon^{GN}{}_{ML}\varepsilon_{DPCB}-\\
    -\big(\frac{1}{2(n-1)(n-2)}-\frac{1}{4(n-2)}\big)R\varepsilon^{ABKL}\varepsilon_{CBML}-
    \big(\frac{1}{4(n-2)}-\frac{n}{8(n-1)(n-2)}\big)R\delta_C{}^A\delta_M{}^K
\end{array}
\end{equation}
соответствует разложению тензора $R_{\Lambda \Psi}{}^{\Phi \Theta}$
\begin{equation}
    R_{\Lambda \Psi}{}^{\Phi \Theta}=C_{\Lambda \Psi}{}^{\Phi \Theta}+
    \frac{4}{n-2} R_{\left[\Lambda \right.}{}^{\left[ \Phi \right.}
    g_{\left.\Psi \right]}{}^{\left. \Theta \right]}-
    \frac{2}{(n-1)(n-2)}Rg_{\left[\Lambda \right.}{}^{\left[ \Phi \right.}
    g_{\left.\Psi \right]}{}^{\left. \Theta \right]}
\end{equation}
на неприводимые ортогональными  преобразованиями компоненты. При этом тождество Бианки
\begin{equation}
    R_{\Lambda \Psi  \Phi \Theta}+R_{\Lambda \Theta \Psi \Phi}+
    R_{\Lambda \Phi \Theta \Psi}=0
\end{equation}
будет выглядеть как
\begin{equation}
    R_L{}^D{}_S{}^L=-\frac{1}{8}\cdot R\delta_S{}^D\spsd
\end{equation}
\end{theoremr}

\begin{proof}$ $\\
\begin{enumerate}
\item Первый шаг. Тождество Бианки $R_{[\Lambda \Psi \Phi] \Theta}=0$.
\begin{equation}

\end{equation}
\item Второй шаг. Тензор Риччи $R_{\Lambda \Psi}$.\\
Воспользуемся двумя вспомогательными тождествами.
\begin{equation}
%
\end{equation}
Это определит тождество
\begin{equation}
\eta_{\left[\right.\Gamma}{}_{|CD|}\eta_{\Omega\left.\right]}{}^{CL}R_{\Lambda\Psi L}{}^D=
-\eta_{\left[\right.\Gamma}{}^{CN}\eta_{\Omega\left.\right] MN}R_{\Lambda\Psi C}{}^M\spsd
\end{equation}
Поэтому
\begin{equation}
%
\end{equation}
\item Третий шаг. Скалярная кривизна R.\\
\begin{equation}
%
\end{equation}
\end{enumerate}
Теперь остается собрать все результаты вместе и получить аналог разложения на неприводимые ортогональными преобразованиями компоненты.
\end{proof}

\begin{corollaryr}$ $
\label{rc10.1}
\begin{enumerate}
    \item
    Условия простоты бивектора n-мерного пространства $\mathbb C^n$ записываются в виде
\begin{equation}
\label{re10.6}
    p^{\left[\Lambda\Psi\right.}p^{\Phi\Omega\left.\right]}=0\spsd
\end{equation}
    Координатам такого бивектора можно сопоставить бесследовую комплексную матрицу такую, что
\begin{equation}
\label{re10.7}
    p_L{}^Dp_S{}^L-\frac{1}{N}(p_L{}^Kp_K{}^L)\delta_S{}^D=0\spsd
\end{equation}
    \item
    Простому  бивектору  пространства $\mathbb C^n$ с условием $p^{\Lambda\Psi}p_{\Lambda\Psi}=0$
    можно поставить в соответствие нильпотентный оператор с индексом 2: $p_L{}^Dp_S{}^L=0$.\\
\end{enumerate}
\end{corollaryr}
\begin{proof}$ $\\
    1). Бивектор прост тогда и только тогда, когда имеет место разложение
\begin{equation}
    p^{\Lambda\Psi}=x^\Lambda y^\Psi-y^\Lambda x^\Psi\spsd
\end{equation}
    Поэтому условия (\ref{re10.6}) выполнены автоматически. Обратно, если выполнены условия (\ref{re10.6}), то
    их можно расписать как
\begin{equation}
    p^{\Lambda\Psi}p^{\Phi\Omega}-
    p^{\Lambda\Phi}p^{\Psi\Omega}+
    p^{\Psi\Phi}p^{\Lambda\Omega}=0\spsd
\end{equation}
    Свернем это уравнение с такими ненулевыми ковекторами $t_\Omega$  и $z_\Phi$, что $p^{\Phi\Omega}z_\Phi t_\Omega \ne 0$
\begin{equation}
    p^{\Lambda\Psi}=\frac{1}{p^{\Theta\Xi}z_\Theta t_\Xi}
    (p^{\Lambda\Phi}z_\Phi p^{\Psi\Omega}t_\Omega-
     p^{\Psi\Phi}z_\Phi p^{\Lambda\Omega}t_\Omega)\spsd
\end{equation}
    Положим
\begin{equation}
    x^\Lambda:=\frac{1}{p^{\Theta\Xi}z_\Theta t_\Xi}p^{\Lambda\Phi}z_\Phi\sps
    y^\Psi:=\frac{1}{p^{\Theta\Xi}z_\Theta t_\Xi}p^{\Psi\Omega}t_\Omega\sps
\end{equation}
    откуда и будет следовать условие простоты бивектора. Поскольку тензор $R_{\Lambda\Psi\Phi\Omega}=p_{\Lambda\Psi}p_{\Phi\Omega}$
    удовлетворяет условиям теоремы классификации, то формула (\ref{re10.7}) есть прямое следствие тождества Бианки.\\

    2). В условиях первого пункта добавится соотношение $p^{\Lambda\Psi}p_{\Lambda\Psi}=0$, которое примет вид $p_L{}^Kp_K{}^L=0$,
    откуда и следует существование нильпотентного оператора с индексом 2.
\end{proof}
Построим аналог дифференциального тождества Бианки при условии ковариантного постоянства связующих операторов
\begin{equation}
\nabla_{\left[\right.\Lambda}R_{\Psi\Phi\left.\right]\Theta\Omega}=0\spsd
\end{equation}
Для этого свернем его с выражением $\eta^\Lambda{}_{AB}A^{\Psi\Phi}{}_C{}^D A^{\Theta\Omega}{}_K{}^L$
\begin{equation}
\begin{array}{c}
\eta^\Lambda{}_{AB}A^{\Psi\Phi}{}_C{}^D(\eta_{\Lambda}{}^{MN} A_{\Psi\Phi}{}_P{}^Q-\eta_{\Psi}{}^{MN} \eta_{\Lambda}{}^{XQ}\eta_\Phi{}_{XP})\nabla_{MN}R_Q{}^P{}_K{}^L=\\
=-\frac{N^2}{16}\nabla_{AB}R_C{}^D{}_K{}^L-\frac{N}{8}(\varepsilon_{YCXP}\varepsilon_{AB}{}^{XQ}\nabla^{YD}-
\varepsilon^{YD}{}_{XP}\varepsilon_{AB}{}^{XQ}\nabla_{YC})R_Q{}^P{}_K{}^L=0\spsd
\end{array}
\end{equation}
Это приведет к дифференциальному спинорному тождеству Бианки
\begin{equation}
\label{re10.8}
\nabla_{AB}R_C{}^D{}_K{}^L=\frac{4}{N}(\varepsilon^{YD}{}_{XP}\varepsilon_{AB}{}^{XQ}\nabla_{YC}-\frac{1}{2}\delta_C{}^D\varepsilon_{AB}{}^{XQ}\nabla_{XP})R_Q{}^P{}_K{}^L\spsd
\end{equation}

\section{Твисторное уравнение}
\Abstract{
\indent В этом параграфе рассказывается о том, как построить и решить n-мерное твисторное уравнение, а затем исследовать его свойства. Вывод всех результатов
этого параграфа сделан на основе \cite[т. 1, с. 419-426, т. 2, с. 56-60, с. 545]{Penroser1}.
}

Определим твисторное уравнение как
\begin{equation}
\label{re11.0}
\eta_\Lambda{}_{AB}\nabla_\Psi X^A+\eta_\Psi{}_{AB}\nabla_\Lambda X^A=\frac{2}{n}g_{\Lambda\Psi}\eta^\Phi{}_{AB}\nabla_\Phi X^A\spsd
\end{equation}
Свернем его с $\eta^{\Psi}{}^{CB}$
\begin{equation}
\begin{array}{c}
\eta_\Lambda{}_{AB}\nabla^{CB} X^A+\frac{n}{2}\nabla_\Lambda X^C=\frac{2}{n}\eta^\Lambda{}^{CB}\nabla_{AB} X^A\sps\\
\nabla_\Lambda X^C-\eta_\Lambda{}^{CB}\nabla_{AB} X^A+\frac{n}{2}\nabla_\Lambda X^C=\frac{2}{n}\eta_\Lambda{}^{CB}\nabla_{AB} X^A\sps\\
\end{array}
\end{equation}
\begin{equation}
\label{re11.1}
\nabla_\Lambda X^C=\frac{2}{n}\eta_\Lambda{}^{CB}\nabla_{AB} X^A\spsd\\
\end{equation}
Условия интегрируемости данного уравнения для произвольного $X^A$ имеют вид (доказательство в приложении)
\begin{equation}
\label{re11.2}
C_{\Phi\Psi\Lambda\Delta}=0\sps
\end{equation}
что соответствует конформно-плоскому пространству. Пусть $\Omega$ - произвольное скалярное поле. Рассмотрим конформное изменение масштаба
метрики
\begin{equation}
g_{\Lambda\Psi}\rightarrow\hat g_{\Lambda\Psi}=\Omega g_{\Lambda\Psi}\spsd
\end{equation}
Поэтому можно положить
\begin{equation}
\hat \eta_\Lambda{}^{AB}:=\eta_\Lambda{}^{AB}\sps
\hat \eta_\Lambda{}_{AB}:=\Omega \eta_\Lambda{}_{AB}\spsd
\end{equation}
Потребуем выполнения
\begin{equation}
\nabla_\Lambda\eta_\Psi{}^{AB}=0\sps\nabla_\Lambda\eta_\Psi{}_{AB}=0\sps
\hat \nabla_\Lambda\hat\eta_\Psi{}^{AB}=0\sps\hat\nabla_\Lambda\hat\eta_\Psi{}_{AB}=0\spsd
\end{equation}
Это приведет к системе
\begin{equation}
\begin{array}{c}
\left\{
\begin{array}{l}
(\hat\nabla_\Lambda -\nabla_\Lambda)\eta_\Psi{}^{AB}=0\sps\\
(\hat\nabla_\Lambda -\nabla_\Lambda)\eta_\Psi{}_{AB}=-(\Omega^{-1}\nabla_\Lambda \Omega)\eta_\Psi{}_{AB}\sps\\
\end{array}
\right.\\ \\
\left\{
\begin{array}{l}
Q_{\Lambda\Psi}{}^\Theta\eta_\Theta{}^{AB}=Q_{\Lambda K}{}^A\eta_\Psi{}^{KB}+Q_{\Lambda K}{}^B\eta_\Psi{}^{AK}\sps\\
Q_{\Lambda\Psi}{}^\Theta\eta_\Theta{}_{AB}=-Q_{\Lambda A}{}^K\eta_\Psi{}_{KB}-Q_{\Lambda B}{}^K\eta_\Psi{}_{AK}+(\Omega^{-1}\nabla_\Lambda \Omega)\eta_\Psi{}_{AB}\sps\\
\end{array}
\right.
\end{array}
\end{equation}
где $Q_{\Lambda\Psi}{}^\Theta$ - тензор деформации. Тогда
\begin{equation}
\begin{array}{c}
\left\{
\begin{array}{l}
Q_{\Lambda(\Psi\Omega)}=\frac{2}{N}g_{\Psi\Omega}Q_{\Lambda K}{}^K\sps\\
Q_{\Lambda(\Psi\Omega)}=-\frac{2}{N}g_{\Psi\Omega}Q_{\Lambda K}{}^K+(\Omega^{-1}\nabla_\Lambda \Omega)g_{\Psi\Omega}\sps\\
\end{array}
\right.\\
\Upsilon_\Lambda:=\frac{1}{2}\Omega^{-1}\nabla_\Lambda \Omega\sps Q_{\Lambda(\Psi\Omega)}=g_{\Psi\Omega}\Upsilon_\Lambda\sps
Q_{\Lambda K}{}^K=\frac{N}{2}\Upsilon_\Lambda\sps\\
\tilde Q_{\Lambda\Psi\Omega}:=Q_{\Lambda[\Psi\Omega]}\sps \tilde Q_{\Lambda K}{}^A:=Q_{\Lambda K}{}^A-\frac{1}{N}Q_{\Lambda L}{}^L\delta_K{}^A\spsd
\end{array}
\end{equation}
Поэтому
\begin{equation}
\tilde Q_{\Lambda\Psi\Omega}=\frac{8}{N}A_{\Psi\Omega A}{}^K\tilde Q_{\Lambda K}{}^A\spsd
\end{equation}
Если потребовать сохранения вида твисторного уравнения
\begin{equation}
\left\{

\right.
\hat X^C:=X^C\sps
\end{equation}
то это наложит на тензор деформации условие
\begin{equation}
Q_{\Lambda K}{}^N=\frac{2}{n}\eta_\Lambda{}^{CN}\eta_\Psi{}_{CB}Q^\Psi{}_K{}^B\spsd
\end{equation}
Тогда
\begin{equation}
%
\end{equation}
Откуда окончательно имеем
\begin{equation}
Q_{[\Lambda\Phi\Theta]}=0\sps
Q_{\Lambda[\Phi\Theta]}=\Upsilon_\Phi g_{\Lambda\Theta}-\Upsilon_\Theta g_{\Lambda\Phi}\sps
Q_{\Lambda\Phi\Theta}=\Upsilon_\Lambda g_{\Phi\Theta}+\Upsilon_\Phi g_{\Lambda\Theta}-\Upsilon_\Theta g_{\Lambda\Phi}\spsd
\end{equation}
Попробуем решить твисторное уравнение в плоском пространстве ($R_{\Lambda\Psi K}{}^A X^K=\nabla_{\left[\right.\Lambda}\nabla_{\Psi\left.\right]} X^A=0$)
при условии ковариантного постоянства связующих операторов ($\nabla_\Lambda \eta_\Psi{}^{CB}=0$)
\begin{equation}
\begin{array}{c}
\nabla_\Lambda X^C=\frac{2}{n}\eta_\Lambda{}^{CB}\nabla_{AB} X^A\sps\\
\nabla_\Psi\nabla_\Lambda X^C=\frac{2}{n}\eta_\Lambda{}^{CB}\nabla_{AB} (\nabla_\Psi X^A)\sps\\
\nabla_\Psi\nabla_\Lambda X^C=\frac{2}{n}\eta_\Lambda{}^{CB}\nabla_{AB} (\frac{2}{n}\eta_\Psi{}^{AK}\nabla_{LK} X^L)\sps\\
\nabla_{\left(\right.\Psi}\nabla_{\Lambda\left.\right)} X^C=
\frac{4}{n^2}(\eta_{\left(\right.\Lambda}{}^{CK}\nabla_{\Psi\left.\right)}-\eta_{\left(\right.\Lambda}{}^{CB}\eta_{\Psi\left.\right)}{}_{AB}\nabla^{AK})\nabla_{LK} X^L\sps\\
(1-\frac{2}{n})\nabla_{\left(\right.\Psi}\nabla_{\Lambda\left.\right)} X^C=-\frac{2}{n^2}g_{\Lambda\Psi}\nabla^{CK}\nabla_{LK} X^L=
-\frac{1}{n}g_{\Lambda\Psi}\nabla^\Omega\nabla_\Omega X^C\sps\\
\nabla_\Psi\nabla_\Lambda X^C=0\spsd
\end{array}
\end{equation}
Таким образом $\nabla_\Lambda X^C$ является константой и представимо как
\begin{equation}
\label{re11.4}
\nabla_\Lambda X^C:=i\eta_\Lambda{}^{CA}\dot Y_A\sps
\end{equation}
где через $\dot Y_A\sps\dot X^C$ будем обозначать постоянные спинорные поля. Проинтегрировав это уравнение, получим
\begin{equation}
\label{re11.5}
X^C:=\dot X^C+iR^{CA}\dot Y_A\spsd
\end{equation}
Нас будет интересовать случай, когда $X^C=0$. Опуская точки над спинорами, получаем следующее соотношение
\begin{equation}
\label{re11.3}
X^C=-iR^{CA} Y_A\spsd
\end{equation}
С геометрической точки зрения уравнение  (\ref{re11.4}), переписанное в терминах операторов $\gamma_\Lambda$, есть ничто иное как деривационное
уравнение нормализованного грассманиана \cite[уравнение (1.2)]{Neifeldr2}, а (\ref{re11.5}) есть уравнение \cite[уравнение (2.6)]{Neifeldr2}.
Подобная нормализация называется спинорной нормализацией. При n=6 она построена в \cite{Andreevr7}.

Вернемся к твисторному уравнению (\ref{re11.0}), но воспользуемся ковариантно постоянными операторами (\ref{re9.10}), построенными в предыдущем параграфе.
Тогда твисторное уравнение (\ref{re11.4}) для действительной реализации примет вид
\begin{equation}
\nabla_\myf\Lambda X^\myff C:=i\eta_\myf\Lambda{}^{\myff C\myff A}\dot Y_\myff A=:P_\myf\Lambda{}^\myff C\sps
\end{equation}
а для комплексной реализации
\begin{equation}
\begin{array}{c}
\nabla_\Lambda (X_K)^C=i\eta_\Lambda{}^{C\myff A}\dot Y_\myff A=(P^*_K)_\Lambda{}^C\sps\\
(P_K)_\Lambda{}_A\nabla_\Psi (X_K)^A+(P_K)_\Psi{}_A\nabla_\Lambda (X_K)^A=\frac{2}{n}g_{\Lambda\Psi}(P_K)^\Phi{}_A\nabla_\Phi (X_K)^A\spsd
\end{array}
\end{equation}
Выполним суммирование по K и, определяя $x_\Lambda:=\sum\limits_{K=1}^{2N}(P_K)_\Lambda{}_A(X_K)^A$, получим конформное
уравнение Киллинга
\begin{equation}
\nabla_\Lambda x_\Psi+\nabla_\Psi x_\Lambda=\frac{2}{n}g_{\Lambda\Psi}\nabla_\Phi x^\Phi\spsd
\end{equation}

\section{Спинорный формализм для n=6 и n=8}
\Abstract{
\indent В этом параграфе рассказывается о том, как построить спинорный формализм при малых размерностях. Для наглядности строятся различные геометрические
интерпретации некоторых алгебраических соотношений. Показывается, как на основании принципа тройственности Картана перейти к структурным константам
алгебры октав. Вывод всех результатов этого параграфа сделан на основе \cite{Andreevr1} и литературе к ней, \cite{Buchdahlr1}, \cite{Dietmar1},
\cite{Klotzr1}, \cite{Stepanovskiir1}, \cite{Stepanovskiir2}, \cite{Schoutenr1}, \cite{Scharnhorstr1}.
}
\subsection{Основные изоморфизмы}
\Abstract{
\indent В этом параграфе рассказывается о том, как построить основные изоморфизмы спинорного формализма при n=6.  Вывод всех результатов этого
параграфа сделан на основе \cite{Andreevr1}, \cite{Andreevr3}.
}

\begin{enumerate}
\item $\mathbb C^6 \cong \Lambda^2\mathbb C^4$.\\
Пусть $\alpha,\beta, ...=\overline{1,6}$, $a,b,a_1,b_1,...=\overline{1,4}$. Тогда из (\ref{re6.5}) будут следовать тождества
\begin{equation}
    \frac{1}{2}\eta^\alpha{}_{aa_1}\eta_\beta{}^{aa_1}=\delta_\alpha{}^\beta\sps
    \eta^\alpha{}_{aa_1}\eta_\alpha{}^{bb_1}=\delta_{aa_1}{}^{bb_1}:=
    2\delta_{\left[\right.a}{}^{\left[\right.b}
    \delta_{a_1\left.\right]}{}^{b_1\left.\right]}\sps
\end{equation}
\begin{equation}
\label{re12.1}
    r^{\alpha}=1/2\cdot \eta^{\alpha}{}_{aa_1}R^{aa_1}\sps
    R^{aa_1}=\eta_{\alpha}{}^{aa_1}r^{\alpha}\sps
\end{equation}
\begin{equation}
    \begin{array}{c}
    g^{\alpha\beta}=1/4\cdot \eta^{\alpha}{}_{aa_1} \eta^{\beta}{}_{bb_1}
    \varepsilon^{aa_1bb_1}\sps
    \varepsilon^{aa_1bb_1}=
    \eta_{\alpha}{}^{aa_1} \eta_{\beta}{}^{bb_1}g^{\alpha\beta}\sps\\[2ex]
    g_{\alpha\beta}=1/4\cdot \eta_{\alpha}{}^{aa_1} \eta_{\beta}{}^{bb_1}
    \varepsilon_{aa_1bb_1}\sps
    \varepsilon_{aa_1bb_1}=
    \eta^{\alpha}{}_{aa_1} \eta^{\beta}{}_{bb_1}g_{\alpha\beta}\sps
    \end{array}
\end{equation}
определяющие изоморфизм между пространством $\mathbb C^6$ и пространством бивекторов $\Lambda^2\mathbb C^4$, а
$g_{\alpha\beta}$ будет метрическим тензором пространства $\mathbb C^6$.

\item $SO(6,\mathbb C)\cong SL(4,\mathbb C)/\pm 1$.\\
Согласно (\ref{re3.15}) собственные ортогональные преобразования представимы как
\begin{equation}
S_\alpha{}^\beta\eta_\beta{}^{aa_1}=\eta_\alpha{}^{bb_1} S_b{}^a S_{b_1}{}^{a_1}\spsd
\end{equation}
При этом $\tilde E_a{}^b$ из (\ref{re6.3a}) будет тождественным преобразованием, домноженным на мнимую единицу.
Для несобственных преобразований справедливо тождество
\begin{equation}
S_\alpha{}^\beta\eta_\beta{}^{aa_1}=\eta_\alpha{}_{bb_1} S^{ba} S^{b_1a_1}\spsd
\end{equation}

\item $so(6,\mathbb C)\cong sl(4,\mathbb C)$.\\
Определим
\begin{equation}
       A_{\alpha\beta d}{}^c=
       \eta_{\left[\right.\alpha}{}^{ca}
       \eta_{\beta\left.\right]}{}_{da}\spsd
\end{equation}
Тогда по аналогии с (\ref{re8.3}) и (\ref{re10.1}) получим
\begin{equation}
\label{re12.2}
    T_{\alpha \beta}= A_{\alpha \beta d}{}^cT_c{}^d \sps
    T_k{}^k=0 \sps T_{\alpha \beta}=-T_{\alpha \beta}\sps
\end{equation}
\begin{equation}
\label{re12.3}

\end{equation}
\end{enumerate}
Кроме того, интересны тождества для 4-вектора
\begin{equation}
\label{re12.4}
%
\end{equation}
Доказательство этих тождеств вынесено в приложение.

\newpage
\subsection{О классификации тензоров с симметриями тензора кривизны для n=6. Спиноры кривизны}
\Abstract{
\indent В этом параграфе рассказывается о том, как сопоставить тензорам, обладающим симметриями тензора кривизны, их спинорные аналоги. Это упрощает
проблему классификации таких тензоров для малых размерностей. Вывод всех результатов этого параграфа сделан на основе \cite{Andreevr2}.
}

\begin{theoremr}
\label{rt12.1}
    Классификацию битензора, обладающего свойствами
\begin{equation}
    R_{\alpha \beta \gamma \delta}=R_{\left[ \alpha \beta \right]
    \left[\gamma \delta] \right.} \sps
    R_{\alpha \beta \gamma \delta}=R_{\gamma \delta \alpha \beta}\sps
    R_{\alpha \beta \gamma \delta}+R_{\alpha \delta \beta \gamma}+
    R_{\alpha \gamma \delta \beta}=0
\end{equation}
    и принадлежащего касательному расслоению $\tau(\mathbb CV_6)$ над шестимерным аналитическим псевдоримановым пространством
    $\mathbb CV_6$, можно свести к классификации тензора $R_a{}^b{}_c{}^d$ 4-мерного комплексного векторного пространства такого, что
\begin{equation}
    R_{\alpha \beta \gamma \delta}= A_{\alpha \beta d}{}^c
    A_{\gamma \delta r}{}^sR_c{}^d{}_s{}^r\spsd
\end{equation}
    Кроме того, выполнены следующие соотношения
\begin{equation}
    R_k{}^k{}_s{}^r=R_s{}^r{}_k{}^k=0 \sps  R_c{}^d{}_s{}^r=
    R_s{}^r{}_c{}^d\spsd
\end{equation}
    Разложение
\begin{equation}
    R_c{}^d{}_s{}^r=C_c{}^d{}_s{}^r-P_{cs}{}^{dr}-\frac{1}{40}\cdot
    R(3\delta_s{}^d \delta_c{}^r-2\delta_s{}^r \delta_c{}^d)
\end{equation}
    соответствует разложению тензора
\begin{equation}
    R_{\alpha\beta}{}^{\gamma\delta}=C_{\alpha\beta}{}^{\gamma\delta}+
    R_{\left[\alpha \right.}{}^{\left[ \gamma \right.}
    g_{\left.\beta \right]}{}^{\left. \delta \right]}-
    1/10Rg_{\left[\alpha \right.}{}^{\left[ \gamma \right.}
    g_{\left.\beta \right]}{}^{\left. \delta \right]}
\end{equation}
    на неприводимые ортогональными  преобразованиями компоненты, которые будут удовлетворять следующим соотношениям
\begin{equation}
    P_{cs}{}^{rd}=-4(R_{\left[c \right.}{}^{\left[r \right.}
    {}_{\left. s \right]}{}^{\left. d\right]}+
    R_k{}^{\left[r \right.}{}_{\left[c \right.}{}^{\left|k\right|}
    \delta_{\left. s\right]}{}^{\left.d\right]})\sps
\end{equation}
\begin{equation}
    C_c{}^d{}_s{}^r=R_{\left(c \right.}{}^{\left(d \right.}{}_
    {\left. s\right)}{}^{\left. r\right)}+
    \frac{1}{40}\cdot R\delta_{\left(s \right.}{}^d\delta_{\left. c\right)}{}^r\sps
    C_c{}^d{}_s{}^r=C_{\left(c \right.}{}^{\left(d \right.}{}_
    {\left. s\right)}{}^{\left.r\right)}\sps
\end{equation}
\begin{equation}
    R=R_\beta{}^\beta=-2\cdot R_k{}^r{}_r{}^k\sps
    P_{kc}{}^{kd}=1/2\cdot R\delta_c{}^d\sps
\end{equation}
\begin{equation}
    R_l{}^d{}_s{}^l=-\frac{1}{8}\cdot R\delta_s{}^d\sps
\end{equation}
    последнее из которых является эквивалентом тождества Бианки.\\
\end{theoremr}
\begin{proof}
     Верно следующее равенство
\begin{equation}
    R_{\alpha\beta\gamma\delta}=1/16\cdot\eta_{\alpha}{}^{aa_1}
    \eta_{\beta}{}^{bb_1}\eta_{\gamma}{}^{cc_1}
    \eta_{\delta}{}^{dd_1}R_{aa_1bb_1cc_1dd_1}\spsd
\end{equation}
    Положим
\begin{equation}
    R_c{}^d{}_s{}^r:=\frac{1}{4}R_{ck}{}^{dk}{}_{st}{}^{rt}
    \sps  R_{\beta\gamma}=\frac{1}{4}\eta_{\beta}{}^{cs}\eta_{\gamma}{}_{rd}
    \cdot P_{cs}{}^{rd}\spsd
\end{equation}
    Из этого следует, что
\begin{equation}
    R_{\alpha \beta \gamma \delta}= A_{\alpha \beta d}{}^c
    A_{\gamma \delta r}{}^sR_c{}^d{}_s{}^r\sps
\end{equation}
\begin{equation}
    \begin{array}{c}
       R_{\beta\delta}=R_{\alpha\beta}{}^{\alpha}{}_\delta=
       A_{\alpha\beta d}{}^cA^\alpha{}_{\delta r}{}^s
       R_c{}^d{}_s{}^r=
       (\eta_\beta{}^{cs}\eta_{\delta rd}+
       \eta_\beta{}^{ck}\eta_{\delta kr}\delta_d{}^s)
       R_c{}^d{}_s{}^r=\\
       =\frac{1}{4}
       \eta_\beta{}^{cs}\eta_{\delta rd}\cdot 4
       (R_{\left[\right.c}{}^{\left[\right.d}
       {}_{s\left.\right]}{}^{r\left.\right]}-
       R_{\left[\right.c}{}^k
       {}_{|k|}{}^{\left[\right.r}
       \delta_{s\left.\right]}{}^{d\left.\right]})\spsd
    \end{array}
\end{equation}
     Положим
\begin{equation}
       \begin{array}{c}
       P_{cs}{}^{rd}:=
       -4(R_{\left[\right.c}{}^{\left[\right.r}
       {}_{s\left.\right]}{}^{d\left.\right]}-
       R_{\left[\right.c}{}^k
       {}_{|k|}{}^{\left[\right.r}
       \delta_{s\left.\right]}{}^{d\left.\right]})
       \end{array}\sps
\end{equation}
       тогда
\begin{equation}
       R_{\beta\delta}=\frac{1}{4}\eta_\beta{}^{cs}\eta_{\delta rd}P_{cs}{}^{rd}\spsd
\end{equation}
       Поэтому скалярная кривизна имеет вид
\begin{equation}
       \begin{array}{c}
       R=R_\beta{}^\beta=\frac{1}{4}
       \eta_\beta{}^{cc_1}\eta^\beta{}_{aa_1}P_{cc_1}{}^{aa_1}=
       \frac{1}{4}\varepsilon_{aa_1}{}^{cc_1}P_{cc_1}{}^{aa_1}=
       \frac{1}{2}P_{aa_1}{}^{aa_1}=-2R_k{}^r{}_r{}^k\sps
       \end{array}
\end{equation}
       и выполнено условие
\begin{equation}
       \begin{array}{c}
       P_{ks}{}^{kd}=
       -4(R_{\left[\right.k}{}^{\left[\right.k}
       {}_{s\left.\right]}{}^{r\left.\right]}+
       R_{\left[\right.r}{}^k
       {}_{|k|}{}^{\left[\right.r}
       \delta_{s\left.\right]}{}^{d\left.\right]})=\\
       =-4(-\frac{1}{2}R_k{}^d{}_s{}^k+
       \frac{1}{4}(R_k{}^r{}_r{}^k\delta_s{}^d+
       4R_k{}^d{}_s{}^k-2R_k{}^d{}_s{}^k))=
       -R_k{}^r{}_r{}^k\delta_s{}^d=\frac{1}{2}R\delta_s{}^d\spsd
       \end{array}
\end{equation}
       Тождество Бианки  можно переписать следующим образом
\begin{equation}
    (A_{\alpha\beta d}{}^cA_{\gamma\delta}{}_r{}^s+
    A_{\alpha\gamma d}{}^cA_{\delta\beta}{}_r{}^s+
    A_{\alpha\delta d}{}^cA_{\beta\gamma}{}_r{}^s)\cdot
    R_c{}^d{}_s{}^r=0\spsd
\end{equation}
    Свернув это уравнение с $A^{\alpha\beta}{}_t{}^lA^{\gamma\delta}{}_m{}^n$, получим, что
\begin{equation}
    4R_k{}^l{}_m{}^k\delta_t{}^n+
    4R_r{}^n{}_t{}^r\delta_m{}^l-
    2R_k{}^l{}_t{}^k\delta_m{}^n-
    2R_k{}^n{}_m{}^k\delta_t{}^l-
    2R_k{}^r{}_r{}^k\delta_t{}^n\delta_m{}^l+
    R_r{}^k{}_k{}^r\delta_m{}^n\delta_t{}^l=0\spsd
\end{equation}
    Свертка этого уравнения c $\delta_n{}^t$ и приведет нас к спинорному аналогу тождества Бианки. При этом все 15 существенных  уравнений
    сохранены. Пусть
\begin{equation}
    C_{\alpha\beta}{}^{\gamma\delta}:=
    A_{\alpha\beta d}{}^cA^{\gamma\delta}{}_r{}^sC_c{}^d{}_s{}^r\sps
\end{equation}
    тогда
\begin{equation}
    C_{\alpha\beta}{}^{\gamma\delta}:=
    R_{\alpha\beta}{}^{\gamma\delta}-
    R_{\left[\alpha \right.}{}^{\left[ \gamma \right.}
    g_{\left.\beta \right]}{}^{\left. \delta \right]}+
    1/10Rg_{\left[\alpha \right.}{}^{\left[ \gamma \right.}
    g_{\left.\beta \right]}{}^{\left. \delta \right]}\sps
\end{equation}
\begin{equation}
    R_{\left[\alpha \right.}{}^{\left[ \gamma \right.}
    g_{\left.\beta \right]}{}^{\left. \delta \right]}=
    A_{\alpha\beta d}{}^cA^{\gamma\delta}{}_r{}^s
    \cdot\frac{1}{4}(P_{sc}{}^{dr}-1/2R\delta_s{}^d\delta_c{}^r+
    \frac{1}{4}R\delta_s{}^r\delta_c{}^d)\sps
\end{equation}
\begin{equation}
    g_{\left[\alpha \right.}{}^{\left[ \gamma \right.}
    g_{\left.\beta \right]}{}^{\left. \delta \right]}=
    A_{\alpha\beta d}{}^cA^{\gamma\delta}{}_r{}^s
    \cdot\frac{1}{4}(1/2\delta_s{}^r\delta_c{}^d-2\delta_s{}^d\delta_c{}^r)\sps
\end{equation}
    откуда получим разложение на неприводимые ортогональными преобразованиями компоненты. Все выкладки приведены в приложении.
\end{proof}

\begin{noter}
При n=6 возможно два варианта классификационных схем спинора Вейля:
\begin{equation}
\begin{array}{cccc}
1.& C_k{}^l{}_m{}^n\phi_l{}^k=\lambda \phi_t{}^n\sps &
2.& C_k{}^l{}_m{}^n\phi^{km}=\lambda \phi^{ln}\spsd
\end{array}
\end{equation}
\end{noter}

\begin{corollaryr}
\label{rc12.1}
\begin{enumerate}
    \item
    Условия простоты бивектора 6-мерного пространства $\mathbb C^6$ записываются как
\begin{equation}
    p_{\left[\alpha\beta\right.}p_{\gamma\beta\left.\right]}=0\spsd
\end{equation}
    Координатам такого бивектора можно сопоставить бесследовую комплексную матрицу $4\times 4$  такую, что выполнено условие
\begin{equation}
    p_l{}^dp_s{}^l-1/4(p_l{}^kp_k{}^l)\delta_s{}^d=0\spsd
\end{equation}
\item
    Простому  бивектору пространства $\mathbb C^6$, построенному на изотропных векторах ($p^{\alpha\beta}p_{\alpha\beta}=0$),
    можно сопоставить вырожденную  нуль-пару Розенфельда: ковектор и вектор пространства $\mathbb C^4$, свертка которых есть нуль.
    При этом указанные вектор и ковектор определятся с точностью до комплексного множителя.\\
\end{enumerate}
\end{corollaryr}
\begin{proof} Доказательство аналогично доказательству следствия \ref{rc10.1}. Единственное отличие заключается в том, что изотропные векторы $r^\alpha$
пространства $\mathbb C^6$ представимы в виде $r^\alpha\eta_\alpha{}^{ab}=R^{ab}=X^aY^b-X^bY^a$. Такое представление возможно из-за выполнения
тождества для изотропного $r^\alpha$ ($r^\alpha r_\alpha=0$)
\begin{equation}
24R^{\left[\right.ab}R^{cd\left.\right]}=\varepsilon_{klmn}R^{kl}R^{mn}\varepsilon^{abcd}=4(r^\alpha r_\alpha)\varepsilon^{abcd}=0\sps
\end{equation}
где $\varepsilon_{abcd}$ - тензор вида (\ref{re6.5}). При этом такой тензор кососимметричен по всем 4 индексам и удовлетворяет соотношениям
\begin{equation}
\begin{array}{c}
\varepsilon_{abcd}\varepsilon^{klmn}=24\delta_{\left[\right.a}{}^k\delta_b{}^l\delta_c{}^m\delta_{n\left.\right]}{}^d\sps
\varepsilon_{abcd}\varepsilon^{almn}=6\delta_{\left[\right.b}{}^l\delta_c{}^m\delta_{n\left.\right]}{}^d\sps
\varepsilon_{abcd}\varepsilon^{abmn}=2\delta_{\left[\right.c}{}^m\delta_{n\left.\right]}{}^d\sps\\
\varepsilon_{abcd}\varepsilon^{abcn}=\delta_d{}^n\sps
\varepsilon_{abcd}\varepsilon^{abcd}=24\spsd
\end{array}
\end{equation}
Положим $x^\alpha\eta_\alpha{}^{ab}:=X^aY^b-Y^aX^b$, $y^\alpha\eta_\alpha{}^{ab}:=Z^aT^b-T^aZ^b$. Из условия $p^{\alpha\beta}p_{\alpha\beta}=0$
будет следовать $x^\alpha y_\alpha=0$. Это означает, что $\varepsilon_{abcd}X^aY^bZ^cT^d=0$. Поэтому векторы $X^a\sps Y^b\sps Z^c\sps T^d$
линейно зависимы. Положим $T^a:=\alpha X^a+\beta Y^a+\gamma Z^a$ и получим
\begin{equation}
\begin{array}{c}
p_a{}^b:=\frac{1}{2}A_{\alpha\beta}{}_a{}^b p_{\alpha\beta}=\frac{1}{4}\eta_\alpha{}^{kb}\eta_\beta{}^{cd}\varepsilon_{cdka}(x^\alpha y^\beta-y^\alpha x^\beta)=\\
=\frac{1}{4}((X^kY^b-X^bY^k)Z^cT^d-X^cY^d(Z^kT^b-T^kZ^b))\varepsilon_{cdka}=\\
=Z^c(\beta Y^d)X^k\varepsilon_{cdka}Y^b-Z^c(\alpha X^d)Y^k\varepsilon_{cdka}X^b-X^cY^dZ^k\varepsilon_{cdka}(\alpha X^b+\beta Y^b+\gamma Z^b)+\\
+X^cY^d(\gamma Z^k)\varepsilon_{cdka}Z^b=\underbrace{X^cY^dZ^k\varepsilon_{cdka}}_{:=M_a}\underbrace{(-2\alpha X^b-2\beta Y^b)}_{:=N^b}=M_a N^b\spsd
\end{array}
\end{equation}
При этом $N^b$ и $M_a$ определены с точностью до преобразования
\begin{equation}
    N^b\longmapsto e^\phi N^b\sps
    M_a\longmapsto e^{-\phi} M_a\spsd
\end{equation}
\end{proof}
Отметим, что пара $(N^b,M_a)$ будет является нуль-парой Розенфельда \cite{Rosenfeldr1}. В пространстве $\mathbb CP^4='\mathbb C^4/'\mathbb C$
(где $'\mathbb C^s=\mathbb C^s/{0}$) $N^b$ определит точку, а $M_a$ - плоскость с условием инцидентности $M^b N_a=0$.
Поэтому можно построить пространство $\mathbb C\mbox{П}^4='\mathbb C^*{}^4/'\mathbb C$, двойственное пространству  $\mathbb CP^4$.
Тогда пространство $\mathbb CP^4\times \mathbb C\mbox{П}^4$ будет пространством нуль-пар Розенфельда. Следует отметить, что такие пространства
изучались впервые  Синцовым \cite{Sintcovr1} и Котельниковым \cite{Kotelnikovr1}.\\

Подставим теперь $\varepsilon_{ab}{}^{cd}=2\delta_{\left[\right.a}{}^c\delta_{b\left.\right]}{}^d$ в (\ref{re10.8}). Тогда
дифференциальные тождества Бианки упростятся до
\begin{equation}
    \nabla_{\left[cm\right.}R_{t\left.\right]}{}^k{}_r{}^s=\delta_{\left[m\right.}{}^k\nabla_{c\left|n\right|}R_{t\left.\right]}{}^n{}_r{}^s\spsd
\end{equation}
Свернем это уравнение с $\delta_k{}^c$
\begin{equation}
    \nabla_{c\left(\right.m}R_{t\left.\right)}{}^c{}_r{}^s=0\sps
\end{equation}
свертка которого с $\delta_s{}^m$ даст
\begin{equation}
    \nabla_{cm}R_t{}^c{}_r{}^m=1/8\nabla_{rt}R\sps
\end{equation}
что является спинорным аналогом
\begin{equation}
    \nabla^\alpha(R_{\alpha\beta}-1/2Rg_{\alpha\beta})=0\spsd
\end{equation}

\subsection{\texorpdfstring{Геометрическое представление твистора в $\mathbb R^6_{(2,4)}$}{Геометрическое представление твистора}}
\Abstract{
\indent В этом параграфе рассказывается о том, как построить геометрическую интерпретацию изотропного спинора (твистора) на изотропном конусе $\mathbb R^6_{(2,4)}$.
Это представление аналогично представлению спинора в $\mathbb R^4_{(1,3)}$ с той лишь разницей, что размерность флага и полотнища увеличиваются
на единицу. Вывод всех результатов этого параграфа сделан на основе \cite{Andreevr5}.
}

    Пусть метрика пространства $\mathbb R^6_{(2,4)}$ имеет вид
\begin{equation}
    dS^2=dT^2+dV^2-dW^2-dX^2-dY^2-dZ^2\sps
\end{equation}
    и пусть задано сечение светового конуса $K_6$
\begin{equation}
    T^2+V^2-W^2-X^2-Y^2-Z^2=0
\end{equation}
    плоскостью V+W=1. Рассмотрим стереографическую проекцию этого сечения    на    плоскость (V=0,W=1) с полюсом
    $N(0,\frac{1}{2},\frac{1}{2},0,0,0)$ так,  что  точке  P(T,V,W,X,Y,Z) соответствует $p(t,0,1,x,y,z)$ на плоскости V=0. Тогда
\begin{equation}
    T/t=X/x=Y/y=Z/z=-\frac{(V-\frac{1}{2})}{\frac{1}{2}}\spsd
\end{equation}
    Сделаем замену
\begin{equation}
    \varsigma=ix-y\sps\omega=i(t+z)\sps\eta=i(t-z)
\end{equation}
    и получим
\begin{equation}
    \varsigma =\frac{-iX+Y}{2V-1}\sps\eta=\frac{-i(T+Z)}{2V-1}\sps\omega=\frac{i(Z-T)}{2V-1}\spsd
\end{equation}
    Поэтому индуцированная в сечении метрика имеет вид
\begin{equation}
    d s^2:=dT^2-dX^2-dY^2-dZ^2=-\frac{ d\varsigma d\bar\varsigma+d\omega d\eta}{(\varsigma\bar\varsigma+\eta\omega)^2}\spsd
\end{equation}
    Доказательство этого факта вынесено в приложение. Положим
\begin{equation}
    X:=
    \left(

    \right)\spsd
\end{equation}
    Тогда
\begin{equation}
    d s^2=-\frac{det(dX)}{(det(X))^2}\sps\bar X^T+X=0\spsd
\end{equation}
    Рассмотрим группу дробно-линейных преобразований L
\begin{equation}
    {\tilde X}=(AX+B)(CX+D)^{-1} \sps
    S:=
    \left(
    %
    \right) \sps
    detS=1\spsd
\end{equation}
    Условие действительности, накладываемое на X ($X^*+X=0)$, даст подгруппу унитарных дробно-линейных преобразований
    $LU(2,2)$ так, что матрица S удовлетворяет условию (см. приложение)
\begin{equation}
    S^*\hat E S=\hat E\sps
    \hat E:=\left(
    %
\end{equation}
    Эта формула примечательна тем, что в ней показано выражение бивектора $R^{ab}$ через его спинорные компоненты. Положим
\begin{equation}
    %
    \right) \sps
\end{equation}
    тогда
\begin{equation}
    \tilde S^*\tilde E\tilde S=\tilde E\spsd
\end{equation}
    Матрицы S образуют группу, изоморфную $SU(2,2)$, следовательно  матрицы $\tilde S$ образуют саму группу  $SU(2,2)$. Назовем преобразования с
    матрицей из группы $LU(2,2)$ твисторными преобразованиями.  Ввиду двулистности накрытия связной компоненты единицы группы $SO(2,4)$
    (которая обозначается через $SO^+(2,4)$) группой $SU(2,2)$ и  двулистности накрытия группы конформных преобразований
    $\mathbb C^{\uparrow 4}_+(1,3)$ \cite[т.2, с.359]{Penroser1} группой $SO^+(2,4)$ следует существование цепочки изоморфизмов
\begin{equation}
    \begin{array}{c}
    SU(2,2)/\{\pm 1 ;\pm i\}\cong
    LU(2,2)\cong \mathbb C^{\uparrow 4}_+(1,3) \cong
    SO^e(2,4)/\pm 1\spsd
    \end{array}
\end{equation}
    Это означает, что группа $LU(2,2)$ исчерпывает все  конформные преобразования из группы $\mathbb C^{\uparrow 4}_+(1,3)$.
    При этом матрица S  восстанавливается с точностью до множителя $\lambda$ такого, что $\lambda^4=1$ (det(S)=1), откуда и
    появляется указанная неоднозначность.  На основании того, что верны тождества
\begin{equation}
    Y=AX+B\ \ \Rightarrow \ \ dX=AdY \sps
    Y=X^{-1}\ \ \Rightarrow \ \ dX=-X^{-1}dXX^{-1}\sps
\end{equation}
    где A и B - некоторые постоянные матрицы, имеем
\begin{equation}
    \tilde Z^*\ d\tilde X\ \tilde Z= Z^*\ dX\ Z\spsd
\end{equation}
    Это уравнение инвариантно относительно преобразований   из группы LU(2,2). Доказательство этого факта рассмотрено в приложении.
    Другой инвариант можно получить,  рассматривая тождества
\begin{equation}
    Y=AX+B\ \ \Rightarrow \ \
    \frac{\partial}{\partial X}=A^T\frac{\partial}{\partial Y} \sps
    Y=X^{-1}\ \ \Rightarrow \ \ \frac{\partial}{\partial X}
    =-Y^T\frac{\partial}{\partial Y}Y^T\sps
\end{equation}
    где A и B - тоже некоторые постоянные матрицы.  Он будет иметь вид
\begin{equation}
    \tilde Z^{-1}\frac{\partial}{\partial \tilde X^T}\tilde Z^{*\ -1}= Z^{-1}\frac{\partial}{\partial X^T} Z^{*\ -1}\spsd
\end{equation}
    Доказательство можно найти в приложении.  Это означает, что существует действительный касательный  вектор $\tilde L$ к гиперболоиду,
    полученному сечением конуса  $K_6$ плоскостью V+W=1. Этот вектор инвариантен относительно преобразований базиса из группы LU(2,2) (т.е.
    координатно-независимый на касательном  пространстве к данному гиперболоиду) и однозначно определенный матрицей
\begin{equation}
    \begin{array}{c}
    \hat L:=\frac{1}{\sqrt{2}}
    (Z^{-1}\frac{\partial}{\partial X^T} Z^{*\ -1}-
    \bar Z^{-1}\frac{\partial}{\partial X} Z^{T\ -1})=\\ \\
    =\left(
    \begin{array}{cc}
    0  & 1 \\
    -1 & 0
    \end{array}
    \right)
    (\frac{\partial}{\partial\omega}(-\bar\eta^0\pi^0+\eta^0\bar\pi^0)+
    \frac{\partial}{\partial\eta}(-\bar\eta^1\pi^1+\eta^1\bar\pi^1)+\\ \\
    +\frac{\partial}{\partial\xi}(-\bar\eta^1\pi^0+\eta^0\bar\pi^1)+
    \frac{\partial}{\partial\bar\xi}(\bar\eta^0\pi^1-\eta^1\bar\pi^0))
    \cdot \frac{1}{(det(Z))^2 \sqrt{2}}:=
    \left(
    \begin{array}{cc}
    0  & 1 \\
    -1 & 0
    \end{array}
    \right)\tilde L
    \end{array}
\end{equation}
    Норма этого вектора в нашей метрике  будет такой
\begin{equation}
    \parallel\tilde L\parallel =-\frac{1}{2(det(Y))^2}=-\frac{1}{(V+W)^2}
\end{equation}
    Назовем изотропный вектор k вектором, имеющим единичную протяженность первого типа \cite[т. 1, с. 57]{Penroser1} в том
    случае, когда k будет задавать точку на изотропном конусе, принадлежащую сечению плоскостью V+W=1. Тогда $\parallel\tilde L\parallel=-1$
    и любой изотропной вектор K,   коллинеарный k, определится как
\begin{equation}
    K=(-\parallel\tilde L\parallel)^\frac{1}{2}k\spsd
\end{equation}
    Однако, при V=-W получаются векторы с бесконечной протяженностью  первого типа. Чтобы научиться их различать можно задавать сечение
    $K_6$  плоскостью T+Z=1 и ввести подобным  образом некоторый вектор $\tilde{\tilde L}$ с нормой
\begin{equation}
    \parallel\tilde{\tilde L}\parallel =-\frac{1}{(T+Z)^2}\spsd
\end{equation}
    Назовем изотропный вектор k вектором, имеющим единичную  протяженность второго типа в том  случае, когда k будет задавать точку на изотропном конусе,
    принадлежащую сечению плоскостью T+Z=1 и протяженность  первого типа не будет конечной.  Определим протяженность вектора К как конечную протяженность
    первого типа, а если такой не существует, то  как протяженность второго типа. Отметим, что вектор $\tilde L$  не является координатно-независимым
    в пространстве $\mathbb R^6_{(2,4)}$, хотя и является инвариантом касательного пространства к гиперболоиду, полученному сечением
    конуса $K_6$ плоскостью V+W=1. Теперь появилась возможность наглядно изобразить твистор в пространстве $\mathbb R^6{}_{(2,4)}$.
    Рассмотрим пару векторов из $\mathbb R^6{}_{(2,4)}$ равной протяженности
\begin{equation}
K^\alpha=\eta^\alpha{}_{ab}\ iT^{\left[\right. a}X^{b\left.\right]}\sps
N^\alpha=\eta^\alpha{}_{ab}\ T^{\left[\right. a}Z^{b\left.\right]}\ \Rightarrow\
K^\alpha K_\alpha=0\sps N^\alpha K_\alpha=0\sps N^\alpha N_\alpha=0\spsd
\end{equation}
    Выберем вектор $Y^a$ таким образом, чтобы были выполнены условия
\begin{equation}

\end{equation}
    Таким образом получится базис из векторов $X^a,Y^a,Z^a,T^a$ (напомним, что $Y_a=s_{aa'}Y^{a'}$)
\begin{equation}
%
\end{equation}
    Вот теперь мы можем построить тривектор
\begin{equation}
    P^{\alpha\beta\gamma}=6K^{\left[\right.\alpha}
                           N^\beta L^{\gamma\left.\right]}\spsd
\end{equation}
    Зная $K^\alpha$, мы знаем $T^a$ и $X^a$  с точностью до
\begin{equation}
    X^a\longmapsto\lambda_1 X^a+\mu_1 T^a\sps
    T^a\longmapsto\nu_1 X^a+\xi_1 T^a\sps
    det\left(
    %
    \right)=1\spsd
\end{equation}
    А если нам известен $N^\alpha$, то произвол в выборе $T^a$ и $Z^a$ таков
\begin{equation}
    Z^a\longmapsto\lambda_2 Z^a+\mu_2 T^a\sps
    T^a\longmapsto\nu_2 Z^a+\xi_2  T^a\sps
    det\left(
    %
    \right)=1\spsd
\end{equation}
    Поэтому $\nu_1=\nu_2=0$ и $\xi_2=\xi_1$. Для $Y^a$ получим
\begin{equation}
    Y^a  \longmapsto  \alpha X^a+\beta Y^a+\gamma Z^a+\delta T^a\spsd
\end{equation}
    Если теперь потребовать, чтобы два таких базиса связаны преобразованием  из группы LU(2,2), то получим
\begin{equation}
    %
\end{equation}
    Таким образом 3-полуплоскость натянутая, на векторы  $K^\alpha,N^\alpha,L^\alpha$, будет координатно-независима в пространстве $\mathbb R^6_{(2,4)}$.
    Итак, нашу конструкцию можно представить в следующем виде.  Протяженность первого типа векторов $K^\alpha$ и $N^\alpha$ должна быть одинакова.
    $K^\alpha$ и $N^\alpha$ определяют 2-плоскость, множество  векторов которой с протяженностью первого типа, равной протяженности вектора
    $K^\alpha$ и началом, совпадающим с началом вектора $K^\alpha$, назовем флагштоком. $K^\alpha$,$N^\alpha$,$L^\alpha$ определят 3-полуплоскость,
    которую назовем полотнищем флага.    Таким образом, зная $K^\alpha$ и $N^\alpha$, мы знаем твистор $T^a$ с точностью до фазы $\Theta$. В свою очередь,
    $2\Theta$ - это угол поворота полотнища флага - 3-полуплоскости  $P^{\alpha\beta\gamma}$ - в 2-плоскости $(L^\alpha,M^\alpha)$
    вокруг флагштока - 2-плоскости $(N^\alpha,K^\alpha)$. Поэтому, поворот флага на $2\pi$ приведет к твистору $-T^a$, и только
    поворот на $4\pi$ вернет нашу конструкцию к исходному состоянию. Кроме того, коллинеарные твисторы определяются протяженностью
    вектора $K^\alpha$ так, что при преобразовании $T^a\longmapsto rT^a,\ Y^a\longmapsto r^{-1}Y^a\ (r\in R\backslash \{0\})$ флагшток умножается на r, а полотнище остается неизменным. Следует, наконец, отметить тот факт, что указанная  геометрическая структура однозначно  восстанавливается по твистору $T^a$.
    В случае бесконечной протяженности первого типа вектора $K^\alpha$ рассматривается конус $K_4\subset K_6$, на котором лежит вектор $N^\alpha$.
    Но ненулевые векторы $K^\alpha$ и $N^\alpha$ имеют конечную протяженность второго типа, давая геометрическую интерпретацию спинора на
    изотропном конусе $K_4$ для $K^\alpha$ и $N^\alpha$.

\newpage
\subsection{О структурных константах алгебры октонионов. Спинорный формализм для n = 8}
\Abstract{
\indent В этом параграфе рассказывается о том, как по связующим операторам, удовлетворяющим уравнению Клиффорда, построить структурные константы
алгебры октонионов. Такие алгебры являются частным случаем нормируемой алгебры с делением (n=8). Рассматривается аксиоматика
таких алгебр. Вывод всех результатов этого параграфа сделан на основе \cite{Andreevr4}.
}

Пусть операторы $\eta_\alpha{}^{ab}\sps(\alpha\ ,\beta\ , ...=\overline{1,6};a\ ,b\ , ...=\overline{1,4})$ - это кососимметричные операторы,
построенные выше для спинорного формализма при n=6. Тогда операторы $\eta_\Lambda{}^{AB}\ (\Lambda\ ,\Psi\ , ...=\overline{1,8};A\ ,B\ , ...=\overline{1,8})$
для спинорного формализма при n=8 строятся согласно схеме (\ref{re6.6}). Поэтому тензор (\ref{re6.5}) будет обладать симметриями
\begin{equation}
\label{re12.6}
\varepsilon_{AB(CD)}=\frac{1}{2}\varepsilon_{AB}\varepsilon_{CD}\sps
\varepsilon_{A(B|C|D)}=\frac{1}{2}\varepsilon_{AC}\varepsilon_{BD}\sps
\end{equation}
где согласно (\ref{re6.3}) $\varepsilon_{AC}=\tilde \varepsilon_{AC}$ является метрическим на спинорном пространстве $\mathbb C^8$.
\begin{equation}
\eta_\Lambda{}_{AB}=\frac{1}{4}\eta_\Lambda{}^{CD}\varepsilon_{ABCD}=\eta_\Lambda{}^{CD}\varepsilon_{AD}\varepsilon_{BC}\sps
\eta_\Lambda:=\frac{1}{4}\eta_\Lambda{}^{CD}\varepsilon_{CD}\spsd
\end{equation}
Второе тождество (\ref{re12.6}) очень важно. Оно обеспечивает невырожденность оператора $\varepsilon_{XA}{}^{YB}X^XX_Y$ и, как следствие,
нормируемость алгебры октав. При этом тождество (\ref{re13.5}) распадается на три: два тождества альтернативности (левое и правое) и центральное тождество эластичности.
Как это происходит.
\begin{theoremr}
\label{rt12.2}
Для любого $r^\Lambda \in \mathbb C^8$ имеет место разложение
\begin{equation}
\label{re12.7}
r^\Lambda=\eta^\Lambda{}_{AB}X^AY^B
\end{equation}
для некоторых $X^A\sps Y^B\in \mathbb C^8$.
\end{theoremr}
\begin{proof}
Из уравнения Клиффорда (\ref{re6.1}) будет следовать, что
\begin{equation}
R^{AB}R_{CB}=\frac{1}{2}r^\Lambda r_\Lambda\delta_C{}^A\sps
R_{KL}\delta_A{}^B=R_{AR}\varepsilon_{KL}{}^{BR}+R^{BR}\varepsilon_{KL}{}_{AR}\spsd
\end{equation}
Свернем это уравнение с $P^K P^A$, выбирая $P^K$ так, что $p:=\frac{1}{2}\varepsilon_{AK}P^K P^A\ne 0$, и полагая, что
$Q_L:=R_{KL}P^K$,
\begin{equation}
Q_LP^B=Q_RP^K\varepsilon_{KL}{}^{BR}+pR^B{}_L\spsd
\end{equation}
Тогда из (\ref{re12.6}) будет следовать
\begin{equation}
R^{BL}=\frac{1}{p}Q^RP^K(\delta_R{}^L\delta_K{}^B-\varepsilon_K{}^{LB}{}_R)=\frac{1}{p}Q^RP^K\varepsilon^{BL}{}_{KR}\spsd
\end{equation}
Остается положить
\begin{equation}
X^A:=\frac{1}{\sqrt p}P^A\sps Y^B:=\frac{1}{\sqrt p}Q^B\sps
\end{equation}
что и докажет теорему.
\end{proof}
Определим
\begin{equation}
P^\Lambda{}_A:=\eta^\Lambda{}_{AB}X^A\sps X^AX_A:=2\sps
\end{equation}
тогда
\begin{equation}
P^\Lambda{}_AP^\Psi{}_B\varepsilon^{AB}=g^{\Lambda\Psi}\sps P^\Lambda{}_AP^\Psi{}_Bg_{\Lambda\Psi}=\varepsilon_{AB}\spsd
\end{equation}
Таким образом  оператор $P^\Lambda{}_A$ определяет изоморфизм между пространствами $\mathbb C^n\cong \mathbb C^N\ (n=N=8)$
\begin{equation}
r^\Lambda=P^\Lambda{}_BY^B\spsd
\end{equation}
Положим
\begin{equation}
\eta_\Lambda{}^{\Psi\Theta}:=\sqrt{2}\eta_\Lambda{}^{AB}P^\Psi{}_AP^\Theta{}_B\sps
\end{equation}
тогда для таких структурных констант выполнены тождества альтернативности алгебры октав
\begin{equation}
\label{re12.8}
\begin{array}{c}
\eta_{\Lambda\Theta}{}^\Phi\eta_{(\Psi\Upsilon)}{}^{\Theta}=
\eta_{\Lambda\left(\right.\Psi}{}^\Theta\eta_{|\Theta|\Upsilon\left.\right)}{}^\Phi\sps
\eta_{(\Lambda\Psi)}{}^\Theta\eta_{\Theta\Phi}{}^{\Upsilon}=
\eta_{\left(\right.\Lambda|\Theta|}{}^\Upsilon\eta_{\Psi\left.\right)\Phi}{}^\Theta
\end{array}
\end{equation}
и центральное тождество Муфанг
\begin{equation}
\label{re12.9}
\begin{array}{c}
r^\Phi\eta_{\Phi\Theta}{}^\Omega\eta_{\Lambda\Psi}{}^{\Theta}\eta_{\Omega\Upsilon}{}^\Gamma r^\Upsilon=
r^\Phi\eta_{\Phi\Lambda}{}^\Theta\eta_{\Theta\Omega}{}^{\Gamma}\eta_{\Psi\Upsilon}{}^\Omega r^\Upsilon\spsd
\end{array}
\end{equation}
Доказательство в приложении. Теперь, для того, чтобы перейти к структурным константам алгебры октав, необходимо воспользоваться оператором вложения
$H_i{}^\Lambda:\ \mathbb R^6\subset\mathbb C^6$. Нормируемость алгебры октав будет следовать из уравнения Клиффорда (\ref{re6.1}), выписанного для
структурных констант
\begin{equation}
(\eta_{\Phi\Theta}{}^\Omega\eta_{\Psi\Xi}{}^\Delta+\eta_{\Psi\Theta}{}^\Omega\eta_{\Phi\Xi}{}^\Delta)g_{\Omega\Delta}=2g_{\Phi\Psi}g_{\Theta\Xi}\spsd
\end{equation}
Операторы же вида (\ref{re10.1})
\begin{equation}
\label{re12.10}
\begin{array}{c}
A_{\Theta\Phi L}{}^X:=\frac{1}{2}\eta_{\left[\right.\Theta}{}^{MX}\eta_{\Phi\left.\right]}{}_{ML}\sps
\tilde A_{\Theta\Phi L}{}^X:=\frac{1}{4}(\eta_{\left[\right.\Theta}{}^{MX}\eta_{\Phi\left.\right]}{}_{ML}+
\eta_{\left[\right.\Theta}{}^{XM}\eta_{\Phi\left.\right]}{}_{LM})
\end{array}
\end{equation}
будут удовлетворять тождествам
\begin{equation}
\label{re12.11}
\begin{array}{cc}
A_{\Lambda\Psi}{}^{AB}A^{\Lambda\Psi}{}_{CD}=\delta_{\left[\right.C}{}^A\delta_{D\left.\right]}{}^B\sps
&\tilde A_{\Lambda\Psi}{}^{AB}\tilde A^{\Lambda\Psi}{}_{CD}=\frac{1}{2}\delta_{\left[\right.C}{}^A\delta_{D\left.\right]}{}^B+
\frac{1}{4}\varepsilon_{\left[\right.C}{}^A{}_{D\left.\right]}{}^B\sps\\
A_{\Lambda\Psi}{}^{AB}A^{\Upsilon\Omega}{}_{AB}=\delta_{\left[\right.\Lambda}{}^\Upsilon\delta_{\Psi\left.\right]}{}^\Omega\sps
&\tilde A_{\Lambda\Psi}{}^{AB}\tilde A^{\Upsilon\Omega}{}_{AB}=\delta_{\left[\right.\Lambda}{}^\Upsilon\delta_{\Psi\left.\right]}{}^\Omega+
\eta^{\left[\right.\Upsilon}\delta_{\left[\right.\Lambda}{}^{\Omega\left.\right]}\eta_{\Psi\left.\right]}\spsd\\
\end{array}
\end{equation}
Доказательство в приложении.

\newpage
\subsection{Геометрия индуктивного перехода от n=6 к n=8}
\Abstract{
\indent В этом параграфе рассказывается о том, как построить две квадрики, связанные принципом тройственности Картана. Построение проводится в явном
виде, обосновывая стандартную схему индуктивного перехода от n=6 к n=8. Вывод всех результатов этого параграфа сделан на основе \cite{Andreevr6}.
}

\subsubsection{Нуль-пары Розенфельда}
\Abstract{
\indent В этом параграфе рассказывается о том, как построить явное решение твисторного уравнения при n=6. Вывод всех результатов этого параграфа сделан
на основе \cite{Andreevr6}, \cite{Rosenfeldr1}.
}

    Обозначим через $\mathbb A^{\mathbb C*}$ пространство, двойственное к спинорному 4-мерному комплексному векторному пространству $\mathbb A^{\mathbb C}$, и образуем
    8-мерное комплексное пространство $\mathbb T^2$ как прямую сумму $\mathbb A^{\mathbb C}\oplus \mathbb A^{\mathbb C*}$. То есть, если $X^a$ $(a,b,...=\overline{1,4})$ - координаты вектора в $A^{\mathbb C}$, а $Y_b$ - координаты ковектора в $A^{\mathbb C*}$, то $X^A:=(X^a,Y_b)\sps (A,B,...=\overline{1,8})$
    будут координатами вектора из $\mathbb T^2$.  Будем рассматривать координаты бивектора $r^{ab}$ из (\ref{re11.3})
\begin{equation}
\label{re14.1}
    X^a=ir^{ab}Y_b
\end{equation}
    как координаты точки комплексного аффинного пространства $\mathbb C\mathbb A_6$. Это есть система из 4 линейных уравнений с 6 неизвестными.
    Для выяснения ее ранга рассмотрим однородное уравнение $r^{ab}Y_b=0$, которое имеет ненулевые решения тогда и только тогда, когда
    бивектор простой: $r_{kc}r^{ac}=\frac{1}{4}r^{ab}r_{ab}\delta_k{}^d=0$ - и, следовательно, представим в виде
\begin{equation}
     r^{ab}_{\mbox{\scriptsize{однородное}}}=P^aQ^b-P^bQ^a\sps
\end{equation}
     причем $P^a$ и $Q^a$ определены с точностью до их линейных комбинаций. Из этого следует, что $P^aY_a=0\sps Q^aY_a=0$.
     Обозначим через $X^a$,$S^a$,$Z^a$ все решения уравнения $X^aY_a=0$, которые образуют базис.  Тогда наше решение примет вид
\begin{equation}
    r^{ab}_{\mbox{\scriptsize{однородное}}}=
    \lambda_1 S^{\left[\right.a} X^{b\left.\right]}+
    \lambda_2 X^{\left[\right.a} Z^{b\left.\right]}+
    \lambda_3 S^{\left[\right.a} Z^{b\left.\right]}
\end{equation}
    и, следовательно, определит в пространстве бивекторов  3-мерное подпространство. Отсюда получается общее решение
\begin{equation}
    r^{ab}=
    r^{ab}_{\mbox{\scriptsize{частное}}}+
    \lambda_1 S^{\left[\right.a} X^{b\left.\right]}+
    \lambda_2 X^{\left[\right.a} Z^{b\left.\right]}+
    \lambda_3 S^{\left[\right.a} Z^{b\left.\right]}\sps
\end{equation}
    где $r^{ab}_{\mbox{\scriptsize{частное}}}$-произвольный бивектор, являющийся частным решением.
\newpage

\subsubsection{\texorpdfstring{Построение квадрик $\mathbb CQ_6$ и $\mathbb C\tilde Q_6$}{Построение квадрик CQ и C'Q}}
\Abstract{
\indent В этом параграфе рассказывается о том, как построить квадрики, удовлетворяющие принципу тройственности Картана. Вывод всех результатов этого
параграфа сделан на основе \cite{Andreevr6}.
}
    Покажем, что определенное ранее пространство $\mathbb T^2$ будет комплексным пространством, в котором скалярный квадрат вектора определится
    квадратичной формой
\begin{equation}
\label{re14.2}
    \varepsilon_{AB}X^AX^B=2X^aY_a
\end{equation}
    так, что матрица тензора $\varepsilon_{AB}$ имеет вид
\begin{equation}
    \parallel \varepsilon_{AB}\parallel=\left(
    \begin{array}{cc}
    0 & \delta_a{}^c \\
    \delta^b{}_d  & 0
    \end{array}
    \right)\spsd
\end{equation}
    При фиксированном $r^{ab}$ уравнение (\ref{re14.1})  определит в $\mathbb T^2$ 4-мерное пространство, которое будет
    являться 4-мерной плоской образующей конуса $\varepsilon_{AB}X^AX^B=0$. Таким образом, мы можем рассматривать квадрику
    $\mathbb CQ_6$,  задаваемую уравнением (\ref{re14.2}), в проективном пространстве $\mathbb C\mathbb P_7$. Ее 4 базисные точки
    будут удовлетворять условию $\varepsilon_{AB}X^A_IX^B_J=0\sps (I,J,...=\overline{1,4})$. Положим
\begin{equation}
    X^A_1:=(X^a,Y_b)\sps
    X^A_2:=(Z^a,T_b)\sps
    X^A_3:=(L^a,N_b)\sps
    X^A_4:=(K^a,M_b)\spsd
\end{equation}
    Исходя из общего решения уравнения (\ref{re14.1}), каждой точке квадрики $\mathbb CQ_6$ можно поставить в соответствие 3-мерную
    изотропную плоскость пространства $\mathbb C\mathbb A_6$. Точку (t,v,w,x,y,z) пространства $\mathbb C\mathbb A_6$ можно представить
    прямой $(\lambda T, \lambda V, \lambda U, \lambda S, \lambda W, \lambda X, \lambda Y, \lambda Z)$ пространства $\mathbb C^8$,
    обладающего метрикой
\begin{equation}
    dL^2=dT^2+dV^2+dU^2+dS^2+dW^2+dX^2+dY^2+dZ^2\spsd
\end{equation}
    Эти прямые будут образующими изотропного конуса $\mathbb CK_8$: $T^2+V^2+U^2+S^2+W^2+X^2+Y^2+Z^2=0$. Пересечение 7-плоскости
    $U-iS=1$ с указанным  конусом $\mathbb CK_8$ обладает индуцированной метрикой
\begin{equation}
    d\tilde L^2=dT^2+dV^2+dW^2+dX^2+dY^2+dZ^2\spsd
\end{equation}
    Это пространство имеет вид параболоида на $\mathbb CK_8$ и тождественно пространству $\mathbb C\mathbb R^6$: $U=1+iS=\frac{1}{2}(1-T^2-V^2-W^2-X^2-Y^2-Z^2)$
    Всякая образующая этого конуса (множество точек, принадлежащих $\mathbb CK_8$ с постоянным отношением T:V:U:S:W:X:Y:Z), не лежащая на
    гиперплоскости $U=iS$, пересекает параболоид в единственной точке. Образующим конуса, лежащим на гиперплоскости $U=iS$, соответствуют точки,
    принадлежащие бесконечности пространства $\mathbb C^6$. Таким образом, прямым $\mathbb C^8$, проходящим через начало $\mathbb C^8$,
    соответствуют точки проективного пространства $\mathbb C\mathbb P_7$.  Стереографическая проекция указанного сечения на плоскость (U=1, S=0)
    с полюсом $N(0,0,\frac{1}{2},\frac{i}{2},0,0,0,0)$ отображает точку P(T,V,U,S,W,X,Y,Z) гиперболоида на точку p(t,v,1,0,w,x,y,z) плоскости (U=1, S=0)
\begin{equation}
    \begin{array}{c}
    \lambda T=t\sps
    \lambda V=v\sps
    \lambda W=w\sps
    \lambda X=x\sps
    \lambda Y=y\sps
    \lambda Z=z\sps
    \lambda =\frac{1}{U+iS}\sps\\
    r^\alpha r_\alpha=-\frac{U-iS}{U+iS}\sps
    \lambda U=\frac{1}{2}(1-t^2-v^2-w^2-x^2-y^2-z^2)=-\lambda iS+1\spsd\\
    \end{array}
\end{equation}
    Образующим же конуса $\mathbb CK_8$ соответствует квадрика $\mathbb C\tilde Q_6$ в проективном пространстве $\mathbb C\mathbb P_7$
\begin{equation}
    g_{\Lambda\Psi}r^\Lambda r^\Psi=0\spsd
\end{equation}

\subsubsection{\texorpdfstring{Соответствие $\mathbb CQ_6\longmapsto \mathbb C\tilde Q_6$}{Соответствие CQ => C'Q}}
\Abstract{
\indent В этом параграфе рассказывается о том, как построить соответствие между образующими квадрики различного ранга согласно принципу тройственности Картана.
Вывод всех результатов этого параграфа сделана основе \cite{Andreevr6}.
}

\begin{enumerate}
\item
    Общее решение уравнения (\ref{re14.1})
\begin{equation}
    r^{ab}=r^{ab}_{\mbox{\scriptsize{частное}}}+r^{ab}_{\mbox{\scriptsize{однородное}}}=
    r^{ab}_{\mbox{\scriptsize{частное}}}+
    \lambda_1S^{\left[\right.a}X^{b\left.\right]}+
    \lambda_2X^{\left[\right.a}Z^{b\left.\right]}+
    \lambda_3S^{\left[\right.a}Z^{b\left.\right]}
\end{equation}
    определит 4-мерную плоскую образующую конуса $\mathbb CK_8$. Тогда для такой образующей определится система
\begin{equation}
\label{re14.3}
    \left\{
    \begin{array}{lcl}
    ir^{ab} Y_b & = &  X^a\sps\\
    ir^{ab} T_b & = &  Z^a\sps\\
    ir^{ab} N_b & = &  L^a\sps\\
    ir^{ab} M_b & = &  K^a
    \end{array}
    \right.
\end{equation}
    с условиями
\begin{equation}
    \begin{array}{c}
     X^a Y_a=0\sps
     Z^a T_a=0\sps
     L^a N_a=0\sps
     K^a M_a=0\sps\\ \\
     X^a T_a=- Z^a Y_a\sps
     X^a N_a=- L^a Y_a\sps
     X^a M_a=- K^a Y_a\sps\\ \\
     Z^a N_a=- L^a T_a\sps
     Z^a M_a=- K^a T_a\sps
     K^a N_a=- L^a M_a\spsd
    \end{array}
\end{equation}
    Таким образом из 16 уравнений с 6 неизвестными  $r^{ab}$ существенными будут только 6 уравнений (10 условий связи),
    что определит точку $\mathbb C\mathbb A_6$, а значит и точку квадрики $\mathbb C\tilde Q_6$.
\item
    Если же нам известно только одно уравнение
\begin{equation}
    ir^{ab} Y_b= X^a
\end{equation}
    с условием
\begin{equation}
    X^aY_a=0\sps
\end{equation}
    то из 4 уравнений существенными будут лишь 3 (одно условие связи).  Это означает, что точке квадрики  $\mathbb CQ_6$
    будет соответствовать плоская 3-мерная образующая $\mathbb C\mathbb P_3$, принадлежащая квадрике $\mathbb C\tilde Q_6$.

\setcounter{mypage}{216}
\begin{figure}\center
\includegraphics[width=0.7\textwidth]{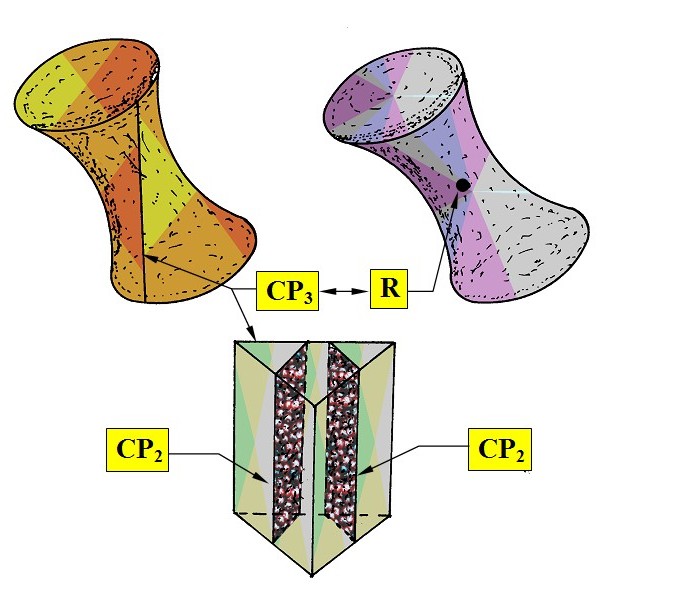}
\caption{Соответствие  $\forall CP_2\subset CP_3 \leftrightarrow R$}
\vspace{1cm}
\includegraphics[width=0.7\textwidth]{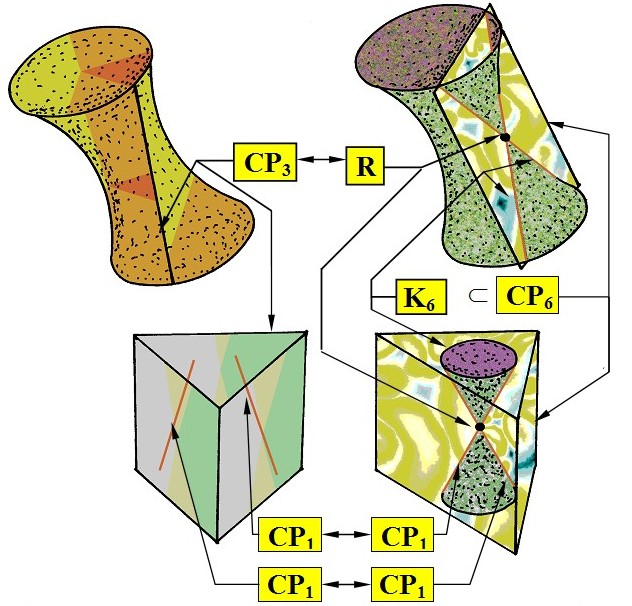}
\caption{Соответствие $CP_3\supset CP_1\leftrightarrow CP_1\subset K_6$}
\end{figure}

\item
    Если  нам известны два уравнения
\begin{equation}
    \left\{
    \begin{array}{lcl}
    ir^{ab}Y_b & = & X^a\sps\\
    ir^{ab}T_b & = & Z^a
    \end{array}
    \right.
\end{equation}
    с условиями
\begin{equation}
    \begin{array}{c}
    X^a Y_a=0\sps
    Z^a T_a=0\sps
    X^a T_a=-Z^a Y_a\sps
    \end{array}
\end{equation}
    то из 8 уравнений существенными будут лишь 5 (неизвестных же 6 и 3 условия связи). Это означает, что прямолинейной образующей
    $\mathbb C\mathbb P_1$ квадрики  $\mathbb CQ_6$ будет соответствовать прямолинейная образующая $\mathbb C\mathbb P_1$,
    принадлежащая квадрике $\mathbb C\tilde Q_6$. При этом многообразие образующих $\mathbb C\mathbb P_1(\mathbb CQ_6)$, принадлежащих
    одной и той же образующей $\mathbb C\mathbb P_3(\mathbb CQ_6)$, определит пучок образующих $\mathbb C\mathbb P_1(\mathbb C\tilde Q_6)$,
    принадлежащий квадрике  $\mathbb C\tilde Q_6$ (этот пучок является на самом деле конусом).  Центр пучка определится системой (\ref{re14.3}).
\item
    Если нам известны три уравнения
\begin{equation}
    \left\{
    \begin{array}{lcl}
    ir^{ab} Y_b & = &  X^a\sps\\
    ir^{ab} T_b & = &  Z^a\sps\\
    ir^{ab} N_b & = &  L^a\sps
    \end{array}
    \right.
\end{equation}
    с условиями
\begin{equation}
    \begin{array}{c}
     X^a Y_a=0\sps
     Z^a T_a=0\sps
     L^a N_a=0\sps\\
     X^a T_a=- Z^a Y_a\sps
     X^a N_a=- L^a Y_a\sps
     Z^a N_a=- L^a T_a\sps
    \end{array}
\end{equation}
    то из 12 уравнений существенными будут лишь 6 (и неизвестных 6 с 6 условиями связи).  Это означает, что 2-мерной образующей
    $\mathbb C\mathbb P_2$ квадрики  $\mathbb CQ_6$ будет соответствовать точка квадрики $\mathbb C\tilde Q_6$. При этом
    многообразие образующих $\mathbb C\mathbb P_2(\mathbb CQ_6)$, принадлежащих одной и той же образующей $\mathbb CP_3(\mathbb CQ_6)$,
    определит единственную точку квадрики $\mathbb C\tilde Q_6$. Эта точка определится системой (\ref{re14.3}).
\end{enumerate}

\setcounter{mypage}{217}
\subsubsection{\texorpdfstring{Геометрия перехода к связующим операторам $\eta_\Lambda{}^{KL}$ для n=8}{Геометрия перехода к связующим операторам при n=8}}
\Abstract{
\indent В этом параграфе рассказывается о том, как построить связующие операторы для спинорного формализма при n=8. Вывод всех результатов этого
параграфа сделана основе \cite{Andreevr6}.
}

    Рассмотрим далее бивектор $\hat R^{AB}$ такой, что его составляющие векторы $X_1{}^A,X_2{}^A$  определены как
\begin{equation}

\end{equation}
     Этим определится цепочка тождеств
\begin{equation}
    %
\end{equation}
    Свернем последнее тождество с $Z^m$, что даст
\begin{equation}
    %
\end{equation}
    Исходя из вышесказанного, рассмотрим прямолинейную образующую квадрики $\mathbb CQ_6$, определенную бивектором $(T_k X^k\ne 0)$
\begin{equation}
    %
    \right)\sps
    \end{array}
\end{equation}
    где $\varepsilon_{AC}$ все тот же метрический симметрический спин-тензор. При этом будет верно уравнение
\begin{equation}
    R_A{}^C\hat R^{AB}=0\sps
\end{equation}
    которое означает, что любой тензор $\hat R^{AB}$, представляющий образующую $\mathbb C\mathbb P_1(\mathbb CQ_6)$, будет содержать один и тот же
    тензор $R^A{}_K$ в своем разложении, при этом второй тензор разложения  $P^{KB}$ будет отвечать за положение $\mathbb C\mathbb P_1$ в
    $\mathbb C\mathbb P_3$. Поэтому есть резон  поставить в соответствие точке квадрики $\mathbb C\tilde Q_6$ биспинор $R^{AB}$,
    которым она определится однозначно. При переходе к пространству $\mathbb C\mathbb R^8$, исходя из
\begin{equation}

\end{equation}
    определим однородные координаты $\mathbb C\mathbb R^8$ следующим образом
\begin{equation}
    \lambda=\left\{%
\end{equation}
    Для того, чтобы $(r_I)^\Lambda$ определяли образующую квадрики $\mathbb C\tilde Q_6$ необходимо и достаточно выполнение условия
\begin{equation}
    g_{\Lambda\Psi}(r_I)^\Lambda (r_J)^\Psi=0\spsd
\end{equation}
    Определим некоторые связующие операторы $\eta_\Lambda{}^{BC}$  так, чтобы выполнялись условия
\begin{equation}
     r_\Lambda=\frac{1}{4}\eta_\Lambda{}^{BC}R_{BC}\sps
     r^\Lambda=\frac{1}{4}\eta^\Lambda{}_{BC}R^{BC}\spsd
\end{equation}
     Тогда эти операторы удовлетворяют уравнению Клиффорда
\begin{equation}
     g_{\Lambda\Psi}\delta_K{}^L=\eta_{\Lambda K}{}^R\eta_\Psi{}^L{}_R+\eta_{\Lambda K}{}^R\eta_\Psi{}^L{}_R\spsd
\end{equation}
     Определим
\begin{equation}
     \varepsilon_{PQRT}:=\eta^\Lambda{}_{PQ}\eta^\Psi{}_{RT}g_{\Lambda\Psi}\sps
     \varepsilon_{PQRT}=\varepsilon_{RTPQ}\sps
\end{equation}
     что даст еще один метрический спинор $\varepsilon_{PQRT}$, с помощью которого можно поднимать и опускать парные индексы
\begin{equation}

\end{equation}
     при  этом результат применения двух метрических тензоров  должен быть одинаков. Если же матрицы тензоров $g_{\Lambda\Psi}$  и
     $\varepsilon_{KL}$ имеют вид
\begin{equation}
    %
\end{equation}
    то существенные координаты операторов $\eta^\Lambda{}_{KL}$ в некотором базисе будут такими
\begin{equation}
    %
\end{equation}
    а существенные координаты операторов $\eta_\Lambda{}^{KL}$  такими
\begin{equation}
    %
\end{equation}
    что несколько отличается от базиса, в котором  выше  были определены однородные координаты $\mathbb C\mathbb R^8$.
    Таким образом, биспинор  $\hat R^{AB}$ определит прямолинейную образующую $\mathbb C\mathbb P_1(\mathbb CQ_6)$, принадлежащую некоторой
    плоской образующей $\mathbb C\mathbb P_3(\mathbb CQ_6)$, которая определит некоторую точку с координатами  $R^{AB}$ квадрики
    $\mathbb C\tilde Q_6$. Определим некоторый тензор $\hat{\hat R}^{AB}$
\begin{equation}
    \hat{\hat R}^{AB}:=\hat R^{AK} \hat P_K{}^B\sps
    \hat P_K{}^B:=\left(
    %
    \right)\spsd
\end{equation}
    Тензор  $\hat{\hat R}^{AB}$ по-прежнему будет представлять  прямолинейную образующую $\mathbb C\mathbb P_1(\mathbb CQ_6)$, принадлежащую
    некоторой плоской образующей $\mathbb C\mathbb P_3(\mathbb CQ_6)$. Применим операторы $\eta^\Lambda{}_{KL}$ к тензору $\hat{\hat R}^{AB}$ и
    получим
\begin{equation}
    \eta^\Lambda{}_{KL}\hat{\hat R}^{KL}=
    \eta^\Lambda{}_{KL} T_m X^m
    \left(
    %
    \right)=
    \eta^\Lambda{}_{KL}R^{KL}\spsd
\end{equation}
    Таким образом, операторы $\eta^\Lambda{}_{KL}$ осуществляют факторизацию прямолинейных образующих
    $\mathbb C\mathbb P_1(\mathbb CQ_6)$  по принадлежности к одной плоской образующей $\mathbb C\mathbb P_3(\mathbb CQ_6)$, и это
    определяет точку квадрики $\mathbb C\tilde Q_6$. В однородных же координатах тензор $R^{KL}$ определит координаты точки R
    пространства $\mathbb C\mathbb R^8$.\\

    Выясним какому семейству принадлежат рассматриваемые выше образующие $\mathbb C\mathbb P_3(\mathbb CQ_6)$. Для этого рассмотрим условия
\begin{equation}
    \begin{array}{c}
    \varepsilon_{AB}(X_I){}^A(X_J){}^B=0\sps\\
    X^{ABCD}:=\varepsilon^{IJKL}(X_I){}^A(X_J){}^B(X_K){}^C(X_L){}^D\sps
    \end{array}
\end{equation}
    где $\varepsilon^{IJKL}$ - квадривектор, кососимметричный по всем  индексам. Кроме того рассмотрим 8-вектор $e_{ABCDKLMN}$, тоже
    кососимметричный по всем индексам. Тогда, если в условии
\begin{equation}
    \frac{1}{24}e_{ABCDKLMN}X^{ABCD}=\rho\varepsilon_{KR}\varepsilon_{LT}\varepsilon_{MU}
    \varepsilon_{NV}X^{RTUV}\sps \rho^2=1
\end{equation}
    $\rho=1$, то будем говорить, что плоские образующие $\mathbb C\mathbb P_3(\mathbb CQ_6)$ принадлежат I семейству, а если $\rho=-1$,
     то - II семейству.  В нашем случае
\begin{equation}
    (X_I){}^A=( (X_I){}^a, (Y_I)_{b})\sps
\end{equation}
    и тогда
\begin{equation}
    \varepsilon^{IJKL}(X_I){}^1(X_J){}^2(X_K){}^3(X_L){}^4=\rho
    \varepsilon^{IJKL}(X_I){}^1(X_J){}^2(X_K){}^3(X_L){}^4\spsd
\end{equation}
    Откуда $\rho=1$. Это означает, что наши образующие необходимо принадлежат  I семейству.
    Кроме того, существует тензор $P_\Lambda{}^B:=\eta_\Lambda{}^{AB}X_A$ для $X_A=(1,0,0,0,1,0,0,0)$
\begin{equation}

\end{equation}
    Отсюда видно, что и образующие $\mathbb C\mathbb P_3(\mathbb C\tilde Q_6)$ необходимо принадлежат I семейству. Для того, чтобы получить
    образующие II семейства, необходимо подействовать элементарным ортогональным отражением (с определителем, равным -1) на оператор
    $P_\Lambda{}^K$.

\setcounter{mypage}{222}
\subsubsection{\texorpdfstring{Соответствие $\mathbb C\tilde Q_6\longmapsto \mathbb CQ_6$}{Соответствие C'Q_6 => CQ}}
\Abstract{
\indent В этом параграфе рассказывается о том, как построить обратное соответствие для квадрик, удовлетворяющие принципу тройственности Картана. Вывод всех результатов этого
параграфа сделан на основе \cite{Andreevr6}.
}

     Применяя операторы $\eta_\Lambda{}^{KL}$ к $g_{\Lambda\Psi}(r_I){}^\Lambda (r_J){}^\Psi=0$, получим
\begin{equation}
      \begin{array}{c}
      (R_I)^{AB}(R_J)_{AB}=0\Leftrightarrow
      ((R_I){}^{AB}-(R_J){}^{AB})((R_K){}_{AB}-(R_L){}_{AB})=0
      \Leftrightarrow\\[2ex]
      ((R_I){}^{ab}-(R_J){}^{ab})((R_K){}_{ab}-(R_L){}_{ab})=0\spsd
      \end{array}
\end{equation}
      Здесь I,J, как обычно, номера базисных точек. Этим определится система
\begin{equation}
\label{re13.7}
    \left\{
    \begin{array}{lcl}
    i(R_1){}^{ab} Y_b & = &  X^a\sps\\
    i(R_2){}^{ab} Y_b & = &  X^a\sps\\
    i(R_3){}^{ab} Y_b & = &  X^a\sps\\
    i(R_4){}^{ab} Y_b & = &  X^a\sps
    \end{array}
    \right.
    \Leftrightarrow
    \left\{
    \begin{array}{lcl}
    i((R_1){}^{ab}-(R_2){}^{ab}) Y_b & = & 0\sps\\
    i((R_1){}^{ab}-(R_3){}^{ab}) Y_b & = & 0\sps\\
    i((R_3){}^{ab}-(R_4){}^{ab}) Y_b & = & 0\sps\\
    i(R_1){}^{ab} Y_b & = &  X^a\spsd\\
    \end{array}
    \right.
\end{equation}
     Далее будем рассматривать только правую систему.  Она строится следующим образом. Всегда существует такой ковектор $Y_a$, который
     обнуляет 3 различных  простых бивектора. Это утверждение сводится к существованию ковектора, ортогонального данным трем векторам,
     поскольку каждый из простых бивекторов  раскладывается по формуле $R^{ab}=2P^{\left[\right.a} Q^{b\left.\right]}$. По четвертому же
     уравнению определится некоторый вектор $X^a$. С другой стороны, на основании того, что все
     $(R_I)^{ab}$ имеют вид
     $R^{ab}_{\mbox{\scriptsize{частное}}}+
    \lambda_1 S^{\left[\right.a} X^{b\left.\right]}+
    \lambda_2 X^{\left[\right.a} Z^{b\left.\right]}+
    \lambda_3 S^{\left[\right.a} Z^{b\left.\right]}$
    при фиксированных $\lambda_1,\lambda_2,\lambda_3$, верно равенство
\begin{equation}
     ((R_I){}^{ab}-(R_J){}^{ab})((R_K){}_{ab}-(R_L){}_{ab})=0\spsd
\end{equation}

\begin{enumerate}
\item
     Итак, пусть нам известно последнее уравнение системы (\ref{re13.7})
\begin{equation}
     i{R_1}^{ab} Y_b  =   X^a\sps
\end{equation}
     то мы имеем 4 уравнения, которые все будут существенными.  Поскольку у нас 8 неизвестных $( X^a, Y_b)$ при фиксированных $(R_I){}^{ab}$,
     то точка квадрики  $\mathbb C\tilde Q_6$ определит 3-мерную плоскую образующую $\mathbb C\mathbb P_3(\mathbb CQ_6)$.
\item
     Если нам известны все уравнения системы (\ref{re13.7}) с условиями
\begin{equation}
     \begin{array}{c}
     ((R_1){}^{ab}-(R_2){}^{ab})((R_1){}_{ab}-(R_2){}_{ab})=0\sps
     ((R_1){}^{ab}-(R_3){}^{ab})((R_1){}_{ab}-(R_3){}_{ab})=0\sps\\ \\
     ((R_3){}^{ab}-(R_4){}^{ab})((R_3){}_{ab}-(R_4){}_{ab})=0\sps\\ \\
      (R_1){}^{ab}(R_2){}_{ab}=0\sps
      (R_1){}^{ab}(R_3){}_{ab}=0\sps
      (R_1){}^{ab}(R_4){}_{ab}=0\sps\\ \\
      (R_2){}^{ab}(R_3){}_{ab}=0\sps
      (R_2){}^{ab}(R_4){}_{ab}=0\sps
      (R_3){}^{ab}(R_4){}_{ab}=0\sps
     \end{array}
\end{equation}
     то из 16 уравнений, существенными будут 7 (8 неизвестных и 9 условий связи).  Поэтому образующей $\mathbb C\mathbb P_3(\mathbb C\tilde Q_6)$
     будет соответствовать точка квадрики $\mathbb CQ_6$.
\item
     Если нам известны 3 уравнения системы (\ref{re13.7})
\begin{equation}
    \left\{
    \begin{array}{lcl}
    i((R_1){}^{ab}-(R_2){}^{ab}) Y_b & = & 0\sps\\
    i((R_1){}^{ab}-(R_3){}^{ab}) Y_b & = & 0\sps\\
    i(R_1){}^{ab} Y_b & = &  X^a\\
    \end{array}
    \right.
\end{equation}
      с условиями
\begin{equation}
     \begin{array}{c}
     ((R_1){}^{ab}-(R_2){}^{ab})((R_1){}_{ab}-(R_2){}_{ab})=0\sps
     ((R_1){}^{ab}-(R_3){}^{ab})((R_1){}_{ab}-(R_3){}_{ab})=0\sps\\ \\
      (R_1){}^{ab}(R_2){}_{ab}=0\sps
      (R_1){}^{ab}(R_3){}_{ab}=0\sps
      (R_2){}^{ab}(R_3){}_{ab}=0\sps
     \end{array}
\end{equation}
     то из 12 уравнений, существенными будут тоже 7 уравнений (8 неизвестных и 5 условий связи). Это означает, что образующей
     $\mathbb C\mathbb P_2(\mathbb C\tilde Q_6)$ будет соответствовать точка квадрики $\mathbb CQ_6$. При этом многообразие образующих
     $\mathbb C\mathbb P_2(\mathbb C\tilde Q_6)$, принадлежащих одной образующей $\mathbb C\mathbb P_3(\mathbb C\tilde Q_6)$, определит
     единственную точку квадрики $\mathbb CQ_6$.
\item
     Если нам известны только 2 уравнения системы (\ref{re13.7})
\begin{equation}
    \left\{
    \begin{array}{lcl}
    i((R_1){}^{ab}-(R_2){}^{ab}) Y_b & = & 0\sps\\
    i(R_1){}^{ab} Y_b & = &  X^a\\
    \end{array}
    \right.
\end{equation}
      с условиями
\begin{equation}
     ((R_1){}^{ab}-(R_2^{ab})((R_1){}_{ab}-(R_2){}_{ab})=0\sps
      (R_1){}^{ab}(R_2){}_{ab}=0\sps
\end{equation}
     то из 8 уравнений, существенными будет только 6 уравнений (8 неизвестных и 2 условия связи). Поэтому образующей
     $\mathbb C\mathbb P_1(\mathbb C\tilde Q_6)$ будет соответствовать прямолинейная образующая  $\mathbb C\mathbb P_1(\mathbb CQ_6)$. При этом
     многообразие образующих $\mathbb C\mathbb P_1(\mathbb C\tilde Q_6)$, принадлежащих одной образующей $\mathbb C\mathbb P_3(\mathbb C\tilde Q_6)$,
     определит пучок прямых $\mathbb C\mathbb P_1(\mathbb CQ_6)$, принадлежащий квадрике $\mathbb CQ_6$. Центр пучка определится системой
     (\ref{re13.7}).
\end{enumerate}

\subsubsection{О майорановских спинорах.}
\Abstract{
\indent В этом параграфе рассказывается о том, что такое условие майорантности спинора с точки зрения инволюции в спинорном пространстве.
Вывод всех результатов этого параграфа сделан на основе \cite{Penroser1}, \cite{Zeer1}.
}

В качестве иллюстрации рассмотрим символы Инфельда - Ван дер Вардена \cite[т. 1, с. 161]{Penroser1}. Пусть уравнение Клиффорда
задано на векторном пространстве $\mathbb C^4$ и в некотором специальном базисе имеет вид
\begin{equation}
\eta_\Lambda\sigma_\Psi+\eta_\Psi\sigma_\Lambda=\delta_{\Lambda\Psi}\sps (\Lambda,\Psi,...=\overline{0,3})\spsd
\end{equation}
Если задано действительное вложение
$H_i{}^\Lambda :\mathbb R^4_{(1,3)}\subset \mathbb C^4$, то этим определена инволюция $S_A{}^{A'}:=\varepsilon^{A'B'}S_{AB'}$
согласно следствию \ref{rc6.3}, случай II).
\begin{equation}
S_\Lambda{}^{\Psi '}\bar \eta_{\Psi '}{}^{A'B'}=\eta_{\Lambda}{}^{BA} S_A{}^{A'}S_B{}^{B'}\sps
(A,A',...=\overline{1,2})\spsd
\end{equation}
Это позволит определить символы Инфельда - Ван дер Вардена следующим образом
\begin{equation}
\begin{array}{c}
g_i{}^{AA'}:=H_i{}^\Lambda \eta_\Lambda{}^{AB}S_B{}^{A'}\sps \\
\overline{g_i{}^{AD'}}=\bar g_i{}^{A'D}:=\bar H_i{}^{\Lambda '} \bar\eta_{\Lambda '}{}^{A'B'}\bar S_{B'}{}^D=
\bar H_i{}^{\Lambda '} \bar S_{\Lambda '}{}^\Lambda\eta_\Lambda{}^{DC}S_C{}^{A'}=
H_i{}^\Lambda\eta_\Lambda{}^{DC}S_C{}^{A'}=g_i{}^{DA'}\sps\\
\bar g_i{}^{A'D}=g_i{}^{DA'}\sps
(i,j,...=\overline{0,3})\spsd
\end{array}
\end{equation}
Этим уравнением инволюция $S_B{}^{A'}$ вводится в определение символов $g_i{}^{AA'}$. Такую операцию позволяет осуществить наличие метрического
кососимметрического тензора $\varepsilon_{AB}$, с помощью которого можно поднимать и опускать одиночные индексы. Построим теперь операторы Дирака
\begin{equation}
\gamma_i=
\left(
\begin{array}{cc}
 0           & (g^T)_i{}_{AA'} \\
 g_i{}^{BB'} & 0
\end{array}
\right)\spsd
\end{equation}
В некотором специальном базисе они будут иметь вид
\begin{equation}
\begin{array}{c}
\gamma_0=
\left(
\begin{array}{rrrr}
 0 & 0 & 1 & 0\\
 0 & 0 & 0 & 1\\
 1 & 0 & 0 & 0\\
 0 & 1 & 0 & 0
\end{array}
\right)\sps
\gamma_1=
\left(
\begin{array}{rrrr}
 0 & 0 & 0 &-1\\
 0 & 0 &-1 & 0\\
 0 & 1 & 0 & 0\\
 1 & 0 & 0 & 0
\end{array}
\right)\sps\\
\gamma_2=
\left(
\begin{array}{rrrr}
 0 & 0 & 0 &-i\\
 0 & 0 & i & 0\\
 0 &-i & 0 & 0\\
 i & 0 & 0 & 0
\end{array}
\right)\sps
\gamma_3=
\left(
\begin{array}{rrrr}
 0 & 0 &-1 & 0\\
 0 & 0 & 0 & 1\\
 1 & 0 & 0 & 0\\
 0 &-1 & 0 & 0
\end{array}
\right)\sps
\end{array}
\end{equation}
Соответственно, инволюция $S_A{}^{A'}$ будет генерировать инволюцию $S$ комплексных операторов Дирака $\bar \gamma_i = S\gamma_i S$, и
в указанном специальном базисе выполнено $||S||=||\gamma^2||$. Таким образом, можно определить майорановский спинор как
\begin{equation}
 \psi^i=\gamma^2 \bar\psi^i\sps
\end{equation}
что соответствует изложению \cite[c. 115, Приложение Е]{Zeer1}. При переходе к пространству $\mathbb R^6_{(2,4)}$
условие майорантности будет таким ($A,A',...=\overline{1,4}$, $a,a',...=\overline{1,2}$)
\begin{equation}
\bar X^{A'}=(\bar X^{a'},\bar Y_{b'})=(\bar S^{a'}{}_aX^a,\bar S_{b'}{}^bY_b)=
\left(
\begin{array}{cc}
0                & \bar S^{a'}{}_a            \\
\bar S_{b'}{}^b  & 0 \\
\end{array}
\right)
\left(
\begin{array}{c}
Y_b\\X^a
\end{array}
\right)=
\bar S^{A'A}X_A\spsd
\end{equation}
Однако, на таком спинорном пространстве согласно (\ref{re6.3}) и (\ref{re6.3a}) не существует метрических тензоров, кроме $S_{AA'}$,
способных поднимать и опускать одиночные индексы. Таким же будет условие майорантности и для пространства $\mathbb R^6$.
В этом случае воспользуемся системой ($a,a',...=\overline{1,4}$)
\begin{equation}
\left\{
\begin{array}{c}
iR^{ab}Y_b=X^a\sps\\
-iR_{ab}Z^b=T_a\spsd
\end{array}
\right.
\end{equation}
Если наше пространство было бы комплексным $\mathbb C^6$, тогда бы эта система устанавливала взаимно-однозначное
соответствие между прямолинейными образующими соответствующих квадрик. Свобода представления диктовалась бы однородным
решением системы, которое было бы изотропным. Но, из-за действительного вложения $\mathbb R^6\subset \mathbb C^6$
изотропные векторы представлены только нулевым вектором. Поэтому возможно взаимно однозначное соответствие
между точками соответствующих квадрик
\begin{equation}
R^{ab}=\frac{-i}{Y^k Y_k}(Y_cX_d \varepsilon^{abcd}+2X^{\left[\right. a}Y^{b\left.\right]})\sps\\
Z^b:=Y^b=S^{bb'}\bar Y_{b'}\sps T_a:=X_a=S_{aa'}\bar X^{a'}\sps X^a Y_a=0\spsd
\end{equation}
\begin{equation}
X^a X_a+Y^a Y_a(-\frac{1}{4} R^{ab}R_{ab})=0\spsd
\end{equation}
(Очевидно, что $R^{ab} R_{ab}>0\sps X^aX_a= S_{aa'}X^a\bar X^{a'}>0\sps Y^aY_a= S_{aa'}Y^a\bar Y^{a'} >0$
для ненулевых векторов.) Последнее условие можно переписать следующим образом
\begin{equation}
(X^a,Y_b)(X_a,\frac{-X^d X_d}{Y^c Y_c} Y^b)=0\spsd
\end{equation}
Таким образом 1-мерной прямолинейной образующей, проходящей через точки $(X^a,Y_b)$ и $(Y^a \frac{-X^d X_d}{Y^c Y_c} ,X_b)$ ставится в соответствие
ортогональная плоскость $(X_a,\frac{-X^d X_d}{Y^c Y_c} Y_b)$, которая высекает на квадрике $(X^aY_a=0)$ 2-мерную плоскую образующую. Совокупность
1-мерной и 2-мерной ортогональных образующих определит 3-мерную плоскую образующую, которая однозначно определит точку на квадрике с координатами $R^{ab}$.
В этом и состоит геометрический смысл майорантности при n=6 для вложения $\mathbb R_6\subset \mathbb C \mathbb R_6$. Кстати говоря, $R^{ab}$  представляет
собой изотропный вектор изотропного конуса пространства $R^8_{(1,7)}$. Соответствующая квадрика высекается сечением этого конуса.
\begin{example}
Итак, пусть $Y^kY_k=2\sps X^aY_a=0$, тогда в специальном базисе  возможно представление
\begin{equation}
\begin{array}{c}
iR^{12}=-i\ \overline{R^{34}}=X^{\left[\right. 1}Y^{2\left.\right]}+\overline{X^{\left[\right. 3} Y^{4\left.\right]}}\sps
iR^{13}=i\ \overline{R^{24}}=X^{\left[\right. 1}Y^{3\left.\right]}-\overline{X^{\left[\right. 2} Y^{4\left.\right]}}\sps\\
iR^{14}=-i\ \overline{R^{23}}=X^{\left[\right. 1}Y^{4\left.\right]}+\overline{X^{\left[\right. 2} Y^{3\left.\right]}}\spsd
\end{array}
\end{equation}
\end{example}

\subsubsection{Теорема о двух квадриках.}
\Abstract{
\indent В этом параграфе рассказывается о том, как можно интерпретировать принцип тройственности Картана на основании предыдущих результатов и
свести их воедино. Вывод всех результатов этого параграфа сделан на основе \cite{Andreevr6}.
}

     Таким образом доказана теорема
\begin{theoremr}
\label{rtheorem9}(принцип тройственности для двух B- цилиндров).\\
     В проективном пространстве $\mathbb C\mathbb P_7$ существуют две квадрики (два B - цилиндра), обладающие
     следующими общими свойствами
     \begin{enumerate}
     \item Плоская образующая $\mathbb C\mathbb P_3$ одной квадрики взаимно однозначно определит точку R другой.
     \item Плоская образующая $\mathbb C\mathbb P_2$ одной квадрики однозначно определит точку R другой. Но точке R можно сопоставить
           многообразие плоских образующих $\mathbb C\mathbb P_2$, принадлежащих одной плоской образующей $\mathbb C\mathbb P_3$
           второй квадрики.
     \item Прямолинейная образующая $\mathbb C\mathbb P_1$ одной квадрики взаимооднозначно определит прямолинейную образующую
           $\mathbb C\mathbb P_1$ из другой. Причем все прямолинейные образующие, принадлежащие одной плоской образующей $\mathbb C\mathbb P_3$
           первой квадрики, определят пучок с центром в точке R, принадлежащий второй квадрике.
     \end{enumerate}
\end{theoremr}
     Эта теорема на самом деле является обобщением соответствия Кляйна. Докажем это.
\begin{proof}
     Рассмотрим на квадрике $\mathbb CQ_6$ только те образующие,  которые имеют вид $X^A=(0,Y_b)$. Многообразие таких образующих диффеоморфно
     $\mathbb C\mathbb P_3$. При этом каждой такой образующей можно поставить в соответствие точку квадрики $\mathbb CQ_4\subset \mathbb C\tilde Q_6$.
     Тогда $R^{ab} Y_b=0$. До конца доказательства положим $\bf{A,B,A',B',...}$$=\overline{1,2}$. Кроме того, рассмотрим спинорное представление
     твисторов согласно \cite[формулы (6.1.24) и (6.2.18)]{Penroser1}
\begin{equation}

\end{equation}
     Поэтому уравнение $R^{ab} Y_b=0$ перепишется в виде системы двух уравнений
\begin{equation}
     \left\{
     %
     \right.
\end{equation}
     из которых будет существенным только одно
\begin{equation}
     ir^{{\mbox{\scriptsize\bf AA}}'}
      \bar\pi_{{\mbox{\scriptsize\bf A}}'}=
      \omega^{\mbox{\scriptsize\bf A}}\spsd
\end{equation}
     Здесь мы воспользовались операцией сопряжения и метрическими спинорами $\varepsilon_{\mbox{\scriptsize\bf A'B'}},
      \varepsilon_{\mbox{\scriptsize\bf AB}}$, с помощью которых подымаются и опускаются спинорные индексы и
      которые при сопряжении переходят друг в друга. Этим  определится система
\begin{equation}
     %
     \right.
      Y_b=( \pi_{\mbox{\scriptsize\bf B}},
      \bar\omega^{{\mbox{\scriptsize\bf B}}'})\sps
      T_b=( \eta_{\mbox{\scriptsize\bf B}},
      \bar\xi^{{\mbox{\scriptsize\bf B}}'})\spsd
     \end{array}
\end{equation}
     Эта система совпадает с системой \cite[формула (6.2.14)]{Penroser1}, которая, в свою очередь, приводит к соответствию Кляйна.
\end{proof}
     Следует в заключении отметить, что из этой теоремы следует принцип тройственности Картана:
     существует 3 диффеоморфных многообразия -  многообразие точек квадрики и 2 многообразия плоских образующих  I  и II семейств.
     Это действительно так, поскольку две построенные  квадрики можно отождествить, например, с помощью тензора $P_\Lambda{}^L$.
     При этом многообразие точек квадрики  будет диффеоморфно многообразию плоских образующих одного из двух семейства. Кроме того,
     поскольку принцип тройственности Картана выполнен, то операторы $\eta^i{}_{KL}$ для вложения $\mathbb R^8\subset\mathbb C^8$
     определят алгебру октав, поскольку удовлетворяют уравнению Клиффорда. Это утверждение основано на результатах,
     приведенных в монографии \cite[т. 2, c. 543-544]{Penroser1}, где рассматриваются структурные константы этой алгебры.

\subsection{Явное построение аналогов операторов Ли при n=4.}
\Abstract{
\indent В этом параграфе рассказывается о том, как можно в некотором базисе получить представление операторов $P$, с помощью которых можно
построить спинорный аналог операторов Ли. Вывод всех результатов этого параграфа сделан на основе \cite{Bilyalovr1}.
}

Пусть на пространстве $\mathbb R^8_{(4,4)}$ задана метрика
\begin{equation}
dS^2=-dU^2+dS^2+dV^2-dW^2+dX^2-dY^2-dT^2+dZ^2\sps
\end{equation}
тогда в некотором базисе связующие операторы $\eta_\Lambda{}^{KL}$ будут иметь вид
\begin{equation}

\end{equation}
а существенные координаты операторов $\eta^\Lambda{}_{KL}$  будут такими
\begin{equation}
    %
\end{equation}
Поэтому в том же базисе
\begin{equation}
    %
\end{equation}
Следовательно, алгоритм \ref{ra6.1} примет вид
\newpage
\begin{algorithmr}$ $\\
\label{ra14.1}
1.  На первом шаге используем в качестве операторов (\ref{re3.12}) и (\ref{re3.12/1}) операторы
\begin{equation}
%
\right)}_{:=\eta_{1_\pm}{}_\myf\Lambda{}^{AB}}\sps\\
\alpha_1=\frac{1}{2}(\eta_3+\eta_4)\sps
\beta_1=\frac{1}{2}(\eta_3-\eta_4)\sps
\gamma_1=\frac{1}{2}(\eta_2-\eta_1)\sps
\delta_1=\frac{1}{2}(\eta_2+\eta_1)\spsd
\end{array}
\end{equation}
Здесь $\varepsilon^{\myff a\myff b}$ - кососимметрический метрический спинор для связующих операторов $\eta_\alpha{}_{\myff p}{}^{\myff l}$
($a,b,...,\myff a,\myff b,...=\overline{1,2},\ \myf\Lambda,...,\myff A,\myff B,... =\overline{1,8},\ A,B,...=\overline{1,4}$).\\
\newpage
2.  На втором шаге используем в качестве операторов (\ref{re3.12}) и (\ref{re3.12/1}) операторы
\begin{equation}

\end{equation}
($\myf\Lambda,...=\overline{1,8},\ \myff A,\myff B,...=\overline{1,4},\ A,B,...=\overline{1,2}$).\\
3. Для касательного пространства операторы $m_\Lambda{}^\myf\Lambda$ будут иметь вид
\begin{equation}
%
\right)}_{:=(\eta_{1,2,3,4}){}_\Lambda{}^{AB}}\\
\end{equation}
будут связующими операторами для комплексной реализации при n=4. Таким образом построен переход от вещественной реализации связующих
операторов при n=8 к их комплексным аналогам при n=4.
\end{algorithmr}
Таким образом, операторы $2((\tilde M^*_K)^T)_\myff A{}^A$ согласно (\ref{re3.13/1}) и алгоритму \ref{ra6.1} будут иметь вид
\begin{equation}
\label{re14.6}
\left(
%
\right)\spsd
\end{equation}
При этом $p_1,p_2$ принимают два значения $\pm 1$, обеспечивая 4 различных варианта операторов $\tilde M^*_K$. Осталось применить эти операторы и
операторы $m_\Lambda{}^\myf\Lambda$, $\bar m_{\Lambda '}{}^\myf\Lambda$ к операторам $P_\myf\Lambda{}^\myff A$
\begin{equation}
\label{re14.5}
(P^*_K)_\Lambda{}^A:=
\frac{1+i}{4}\left(
%
\right)\spsd
\end{equation}
Очевидно, что из 4 вариантов для $K=\overline{1,4}$, существенными будут только два $(P^*_2)_\Lambda{}^A$, $(P^*_3)_\Lambda{}^A$ и
$(\bar P^*_1)_{\Lambda '}{}^A$, $(\bar P^*_4)_{\Lambda '}{}^A$ ($\Lambda,...=\overline{1,4}$). При условии постоянства таких операторов при обычном и ковариантном
дифференцировании коэффициенты связности $\Gamma_{\myf\Lambda\Psi}{}^\Phi$ распадаются на 8 существенных компонент
\begin{equation}
\Gamma_{\myf\Lambda\Psi}{}^\Phi\rightarrow
\left\{
%
\right.
\end{equation}
В свою очередь, коэффициенты $\Gamma_{\myf\Lambda\Psi}{}^\Phi$ определяются из уравнения $\nabla_\myf\Lambda m_\Psi{}^\myf\Psi=0$.
Таким образом теоремы \ref{rtheorem9.1}, \ref{rtheorem9.2} остаются верными и в этом случае. В случае вещественного вложения с помощью операторов
$H_i{}^\Lambda$ при условии ковариантного и обычного постоянства операторов $(P^*_2)_i{}^A$, $(P^*_3)_i{}^A$ и $(P^*_1)_i{}^A$, $(P^*_4)_i{}^A$
коэффициенты связности $\Gamma_{\myf\Lambda i}{}^j$ распадаются на 4 существенные компоненты
\begin{equation}
\Gamma_{\myf\Lambda i}{}^j\rightarrow
\left\{
\begin{array}{l}
(\Gamma_K)_{kA}{}^B:=\Gamma_{ki}{}^j (P_K)^i{}_A(P^*_K)_j{}^B\sps\\
(\bar \Gamma_K)_{kA'}{}^{B'}:=\Gamma_{ki}{}^j (\bar P_K)^i{}_{A'}(\bar P^*_K)_j{}^{B'}\spsd\\
\end{array}
\right.
\end{equation}
Однако, условие действительности ограничивает выбор спинорa $X^\myff A$ при определении операторов $P_\myf\Lambda{}^\myff A$
\begin{equation}
X^\myff A=\tilde S_\myff B{}^\myff A X^\myff B\sps
X^\myff A=\tilde {\tilde S}^{\myff B\myff A} X_\myff B\sps
\end{equation}
где $\tilde S_\myff B{}^\myff A$ ($S^{\myff B\myff A}$) есть действительная реализация комплексной инволюции $\tilde S_B{}^{A'}$ ($\tilde S^{BA'}$),
которая является образом инволюции $S_\Lambda{}^{\Psi '}$ в касательном пространстве согласно следствию \ref{rc6.3}. Необходимо добавить, что
существует несобственное ортогональное преобразование
\begin{equation}
(-\eta_\Lambda\eta^\Psi+\delta_\Lambda{}^\Psi)\eta_\Psi{}^{AB}=\eta_\Lambda{}^{BA}\sps
\end{equation}
осуществляющее транспонирование связующих операторов (\ref{re12.6}).

\section{Благодарности}$ $\\
\indent Эта статья является ответами на часто задаваемые вопросы. Поэтому автор благодарен всем, кто эти вопросы верно сформулировал, что
дало возможность найти на них ответ. Отдельная благодарность К. Шарнхорсту за копии статей, без которого этой публикации не было.

\newpage

\section{Приложение}
\subsection{Доказательство формул о метрических тензорах и инволюции}
$\phantom{fff}$Необходимо доказать тождества (\ref{re6.3}).\\
Случай I)., метрический тензор $\varepsilon_{AB}=-\varepsilon_{BA}$.
\begin{equation}

\end{equation}

\newpage
\subsection{Доказательство формул о связующих операторах}
$\phantom{fff}$ Необходимо доказать тождество (\ref{re6.4})
\begin{equation}
\eta_\Lambda{}^{AB}\eta_\Psi{}_{AD}\eta^\Omega{}^{CD}\eta^\Theta{}_{CB}=
\frac{N}{4}(g_{\Lambda\Psi}g^{\Omega\Theta}+\delta_\Lambda{}^\Theta\delta_\Psi{}^\Omega-\delta_\Psi{}^\Theta\delta_\Lambda{}^\Omega)\spsd
\end{equation}
\begin{enumerate}
\item
Первый шаг.
\begin{equation}
%
\end{equation}
\end{proof}
\end{enumerate}

\subsection{Доказательство формул об инфинитезимальных преобразованиях}
$\phantom{fff}$Необходимо доказать тождество (\ref{re8.2}).\\
\begin{enumerate}
\item Случай 1).
\begin{equation}
%
\end{equation}

\subsection{Доказательство тождества о скобке Ли}
$\phantom{fff}$Необходимо доказать тождество (\ref{re9.7}).
\begin{equation}
%
\end{equation}
\newpage

\subsection{Доказательство формул о тензоре кривизны}
$\phantom{fff}$Необходимо доказать тождество (\ref{re10.2}).\\
Тензор кривизны  в спинорном расслоении вычисляется по формуле
\begin{equation}
(\mathcal R_K)_{\Lambda\Psi C}{}^A=2(\partial_{\left[\right.\Lambda}(\Gamma_K)_{\Psi\left.\right] C}{}^A+\Gamma_{\left[\right.\Lambda |L|}{}^A(\Gamma_K)_{\Psi\left.\right] C}{}^L)\spsd
\end{equation}
Тогда $\partial_\Lambda(\varepsilon_K)_{AB}{}^{CD}=0$ при выполнении условия (\ref{re9.9}) означает
\begin{equation}
\partial_{\left[\right.\Lambda}(\Gamma_K)_{\Psi\left.\right] A}{}^C=
\underbrace{-\frac{2}{n-4}((\varepsilon_K)_{AB}{}^{CD}-\frac{n-2}{N}\delta_A{}^C\delta_B{}^D)}_{:=(\tilde\varepsilon_K)_{AB}{}^{CD}}\partial_{\left[\right.\Lambda}(\Gamma_K)_{\Psi\left.\right] D}{}^B\spsd
\end{equation}
В свою очередь,  $\nabla_\Lambda(\varepsilon_K)_{AB}{}^{CD}=0$ при выполнении условия (\ref{re9.9}) означает
\begin{equation}

\end{equation}
Откуда
\begin{equation}
(\mathcal R_K)_{\Lambda\Psi C}{}^M=(\mathcal R_K)_{\Lambda\Psi L}{}^D\tilde\varepsilon_{CD}{}^{ML}\spsd
\end{equation}

\subsection{Доказательство вспомогательных тождеств для теоремы о спинорных аналогах тензора кривизны}
$\phantom{fff}$Необходимо доказать тождество (\ref{re10.3}).\\
\begin{equation}
%
\end{equation}
откуда и следует требуемое.
\end{proof}

Необходимо доказать тождество (\ref{re10.4}).\\
\begin{equation}
%
\right|\\
=(-1+\frac{4}{n-4})
\eta_{\left[\right. \Lambda_1}{}^{A_1A_2}\eta_{\Lambda_2\left.\right]}{}_{A_1A_3}
\eta_{\left[\right. \Lambda_3}{}^{A_4A_3}\eta_{\Lambda_4\left.\right]}{}_{A_4A_5}
\eta_{\left[\right. \Lambda_5}{}^{A_6A_5}\eta_{\Lambda_6\left.\right]}{}_{A_6A_2}+\\
+(\frac{8}{n-4}-4)\frac{N}{2}
  g_{\left[\right. \Lambda_3 | \left[\right. \Lambda_2}
  g_{\Lambda_1 \left.\right] | \left[\right. \Lambda_5}
  g_{\Lambda_6 \left.\right] | \Lambda_4 \left]\right.}\spsd
\end{array}
\end{equation}
откуда и следует требуемое.
\end{proof}
\newpage
Необходимо доказать тождество (\ref{re10.5}).\\
\begin{equation}

\end{equation}
откуда и следует требуемое.
\end{proof}
\begin{noter} Очевидно, что данные выкладки не применимы для n<8, поскольку $\frac{8}{n-4}-4$ не будет иметь смысла в этом случае.
\end{noter}

\subsection{Доказательство формул о твисторном уравнении.}
$\phantom{fff}$Необходимо доказать  (\ref{re11.2}).\\
\begin{equation}
%
\end{equation}
Таким образом условия интегрируемости примут вид
\begin{equation}
%
\end{equation}
Будем считать, что данное условие выполнено для любого $X^K$. Воспользуемся тождеством Бианки $2R^{\Delta[\Phi\Psi]\Lambda}=R^{\Psi\Phi\Delta\Lambda}$
\begin{equation}
%
\end{equation}
Окончательно
\begin{equation}
R^{\Phi\Psi\Lambda\Delta}-\frac{4}{n-2}R^{\left[\Lambda \right.|\left[ \Phi \right.}g^{\left.\Psi \right]|\left. \Delta \right]}+
\frac{2}{(n-1)(n-2)}Rg^{\left[\Delta \right.|\left[ \Psi \right.}g^{\left.\Phi \right]|\left. \Lambda \right]}=C^{\Phi\Psi\Lambda\Delta}=0\spsd
\end{equation}
\newpage

\subsection{Доказательство тождеств для n=6}
\subsubsection{Доказательство основных тождеств для n=6}
$\phantom{fff}$Необходимо доказать  (\ref{re12.3}).\\
\begin{equation}
       A_{\alpha\beta d}{}^c=
       \eta_{\left[\right.\alpha}{}^{ca}
       \eta_{\beta\left.\right]}{}_{da}\sps
\end{equation}
\begin{equation}

\end{equation}
\newpage

\subsubsection{Доказательство тождеств о 4-векторе}
$\phantom{fff}$Необходимо доказать  (\ref{re12.4}).\\
\begin{equation}
       %
\end{equation}

\subsubsection{Доказательство формул о 6-векторе}
$\phantom{fff}$Необходимо доказать  (\ref{re12.5}).\\
\begin{equation}
       %
\end{equation}
       Откуда
\begin{equation}
       \tilde e=\frac{1}{8}i\spsd
\end{equation}

\subsubsection{Доказательство тождества Бианки}
       $\phantom{fff}$Тождество Бианки имеет вид
\begin{equation}
       %
\end{equation}
       что и завершает доказательство.

\subsubsection{Доказательство тождеств, касающихся тензора Вейля.}
       Поскольку
\begin{equation}
    %
\end{equation}
    то  будет верным следующее соотношение
\begin{equation}
       %
\end{equation}
       Таким образом, из
\begin{equation}
       %
\end{equation}
        что завершает доказательство разложения.

\subsubsection{\texorpdfstring{Доказательство формул, связанных с метрикой, индуцированной в сечении конуса $K_6$}{Доказательство формул, связанных с метрикой, индуцированной в сечении конуса}}
     $\phantom{fff}$Пусть задано сечение конуса $K_6$
\begin{equation}
     T^2+V^2-W^2-X^2-Y^2-Z^2=0
\end{equation}
     плоскостью V+W=1. Сделаем стереографическую проекцию  полученного гиперболоида на плоскость (V=0,W=1).
\begin{equation}
%
\end{equation}
     Поэтому есть резон положить
\begin{equation}
    X:=
    \left(
    %
    \right)\spsd
\end{equation}
    Тогда
\begin{equation}
    d s^2=-\frac{det(dX)}{(det(X))^2}\sps \bar X^T+X=0\spsd
\end{equation}

\subsubsection{Доказательство формул о первом инварианте}
    $\phantom{fff}$Рассмотрим группу дробно-линейных преобразований L
\begin{equation}
    {\tilde X}=(AX+B)(CX+D)^{-1} \sps
    S:=
    \left(
    %
    \right) \sps
    detS=1\spsd
\end{equation}
    Пусть имеется два последовательных преобразования
\begin{equation}
    \tilde X=(AX+B)(CX+D)^{-1} \sps
    \tilde{\tilde X}=(\tilde A\tilde X+\tilde B)(\tilde C\tilde X+\tilde D)^{-1}\sps
\end{equation}
    то
\begin{equation}
%
\end{equation}
    Для унитарных дробно-линейных преобразований имеем
\begin{equation}
%
\end{equation}
    Поэтому верны тождества
\begin{equation}
%
\end{equation}
    Доказательство второго таково
\begin{equation}
%
\end{equation}
    Тогда получим цепочку тождеств
\begin{equation}
%
\end{equation}
    Домножим обе части на $CdX$
\begin{equation}
%
\end{equation}
    Таким образом получается первый инвариант.\\

\subsubsection{Доказательство формул о втором инварианте}
    $\phantom{fff}$Пусть
\begin{equation}
    \hat X=AX+B\sps
\end{equation}
    или в покомпонентной записи
\begin{equation}
    %
\end{equation}
       Если все-таки $det(A)=0$, то можно положить
\begin{equation}
%
\end{equation}
        Очевидно, что $\tilde D$ всегда можно выбрать так, что $det(\tilde{\tilde{\tilde C}})\ne 0$. Поэтому
\begin{equation}
%
\end{equation}
        что и даст второй инвариант.

\subsubsection{Доказательство тождества альтернативности для алгебры октав}
$\phantom{fff}$Докажем тождество (\ref{re12.8}). В случае $P_{\Psi A}=\eta_{\Psi XA}X^X$
\begin{equation}
%
\end{equation}
\newpage

\subsubsection{Доказательство центрального тождества Муфанг}
$\phantom{fff}$Докажем тождество (\ref{re12.9}). Отметим, что $r^\Phi\eta_{\Phi\Theta}{}^\Omega$ и $\eta_{\Omega\Upsilon}{}^\Gamma r^\Upsilon$ являются
ортогональными преобразованиями.
\begin{equation}
%
\end{equation}
\end{enumerate}
Это прямое доказательство. Более простое доказательство основано на формуле (\ref{re8.1}, случай 2).

\subsubsection{Доказательство тождеств для оператора A для n=8}
$\phantom{fff}$Докажем тождество (\ref{re12.11}).
\begin{enumerate}
\item Во-первых.\\
\begin{equation}
%
\end{equation}
\item В-третьих.\\
На основании тождества (\ref{re6.4})
\begin{equation}
\eta_\Lambda{}^{AB}\eta_\Psi{}_{AD}\eta^\Omega{}^{CD}\eta^\Theta{}_{CB}=
2(g_{\Lambda\Psi}g^{\Omega\Theta}+\delta_\Lambda{}^\Theta\delta_\Psi{}^\Omega-\delta_\Psi{}^\Theta\delta_\Lambda{}^\Omega)\sps
\end{equation}
тогда получим
\begin{equation}
%
\end{equation}
\end{enumerate}

}
\newpage
\References{
\bibitem{Andreevr1}
{\sc Андреев К. В.}
Спинорный формализм и геометрия шестимерных римановых пространств. Кандидатская диссертация, Уфа, 1997. [K.V. Andreev. Spinor formalism and the geometry of six-dimensional Riemannian spaces / Ph. D. Thesis, Ufa, 1997], \href{http://arxiv.org/abs/1204.0194}{arXiv:1204.0194v1}.
\bibitem{Andreevr2}
{\sc Андреев К. В.}
О структуре тензора кривизны 6-мерных римановых пространств.  Вестик Башкирского унверситета, (1996) \No 2, с. 44-47. [K. V. Andreev. On the
structure of the curvature tensor of 6-dimensional Riemannian spaces. Vestnik Bashkirskogo Universiteta, (1996) \No 2, 44-47].
\bibitem{Andreevr3}
{\sc Андреев К. В.}
О спинорном формализме при n=6. Известия Высших учебных заведений, 2001, \No 1 (\No 464), 11-23.[K. V. Andreev. On spinor formalism for n=6. Izvestiya Vysshikh Uchebnykh Zavedeni., Matematika, 2001, \No 1 (\No 464), 11-23. [(The article is freely available online at the Mathnet URL:\url{http://mi.mathnet.ru/eng/ivm838}, the two figures of the article are missing in the electronic version, however). English translation: K. V. Andreyev: On spinor formalism for n = 6. Russian Mathematics (Iz. VUZ) 45:1(2001)9-20.]
\bibitem{Andreevr4}
{\sc Андреев К. В.}
О структурных константах алгебры октав. Уравнение Клиффорда , Известия вузов, Математика, 2001,\No 3, 3–6. [K. V. Andreev, Structure constants of the algebra of octaves. The Clifford equation, Izv. Vyssh. Uchebn. Zaved. Mat., 2001, \No 3, 1-4, Mathnet URL:\url{www.ksu.ru/journals/izv\_vuz/arch/2001/03/01-3.PDF}]
\bibitem{Andreevr5}
{\sc Андреев К. В.}
О твисторах 6-мерного пространства. ВИНИТИ -2469-В-98, август 1998. 23 с. [K.V. Andreev. On the twistors of six-dimensional space. VINITI-2469-V-98, Aug 1998. 23pp. Paper deponed on Aug 3, 1998 at VINITI (Moscow), ref. \No 2469-V 98]
\bibitem{Andreevr6}
{\sc Андреев К. В.}
Принцип тройственности для двух квадрик. ВИНИТИ -2470-В-98, август 1998. 19 с. [K.V. Andreev. Triality principle for two quadrics. VINITI-2470-V-98, Aug 1998. 19pp. Paper deponed on Aug 3, 1998 at VINITI (Moscow), ref. \No 2470-V 98.]
\bibitem{Andreevr7}
{\sc Андреев К. В.}
О внутренних геометриях многообразия плоских образующих 6-мерной квадрики, Изв. вузов. Матем., 1998,\No 6, 3-8. [K. V. Andreev. On intrinsic geometries of the manifold of plane generators of a 6-dimensional quadric, Izv. Vyssh. Uchebn. Zaved. Mat., 1998, \No 6, 1-6, Mathnet URL:\url{http://www.mathnet.ru/links/31e5d8f5c91e320943c14afce86b45aa/ivm436.pdf}]
\bibitem{Baezr1}
{\sc Баэз Джон С.}
Октонионы. Гиперкомплексные числа в геометрии и физике. \No 1(5), Vol 3 (2006), c.120-177. [John C. Baez The Octonions. \href{http://arxiv.org/abs/math/0105155}{arXiv:math.RA/0105155v4}]
\bibitem{Bilyalovr1}
{\sc Билялов Р.Ф., Никитин Б.С.}
Спиноры в произвольных реперах. Ковариантная производная и производная Ли спиноров. Известия Высших учебных заведений, 1998, N 6 (N 433), 9-19. [R. F. Bilyalov, B. S. Nikitin. Spinors in arbitrary frames. The covariant derivative and the Lie derivative of spinors. Izv. Vyssh. Uchebn. Zaved. Mat., 1998, \No. 6, 7–15.]
\bibitem{Bergerr1}
{\sc Берже М.}
Геометрия т.1,2. Москва, Мир, 1984.
[Berger, Marcel: Geometry. II. Translated from the French by M. Cole and S. Levy. Universitext. Springer-Verlag, Berlin, 1987.
Berger, Marcel: Geometry. I. Translated from the French by M. Cole and S. Levy. Universitext. Springer-Verlag, Berlin, 1987. ]
\bibitem{Buchdahlr1}
{\sc Бухдаль Х.А.}
H.A. Buchdahl: On the calculus of for-spinors. Proceedings of the Royal Society of London, Series A, Mathematical and Physical Sciences
303(1968)355-378.
\bibitem{Dietmar1}
{\sc Дитмар С.}
Dietmar Salamon: Spin Geometry and Seiberg-Witten invariants. Warwick, 1996.
\bibitem{Klotzr1}
{\sc  Клотц Ф.C.}
F. S. Klotz. Twistors and the conformal group. Journal of Mathematical Phisics, 15(1974) 2242-2247.
\bibitem{Kotelnikovr1}
{\sc Котельников А.П.}
Винтовое счисление и некоторые приложения его к геометрии механики. Казань, 1895 г.
[Screw calculation and some of its applications to geometry, mechanics. Kazan, 1895.]
\bibitem{Neifeldr1}
{\sc Нейфельд Э.Г.}
Об инволюциях в комплексных пространствах. ТГС, Казань, 1989, Выпуск 19, с.~71-82. [E. G. Neifel'd. On the involutions in complex spaces. Tr. Geom. Semin, Kazan, 1989, \No~19, pp.~71-82.]
\bibitem{Neifeldr2}
{\sc Нейфельд Э.Г.}
Нормализация комплексных грассманианов и квадрик. ТГС, Казань, 1990, Выпуск 20, с.~58-69 [E. G. Neifel'd. Normalization of complex Grassmannians and quadrics. Tr. Geom. Semin, Kazan, 1990, N\o~20, pp.~58-69.]
\bibitem{Nordenr1}
{\sc Норден А.П.}
О комплексном представлении тензоров пространства Лоренца. Известия ВУЗов, Казань, 1959, т.~8, \No~1, с.~156-164, Mathnet URL:\url{http://www.mathnet.ru/links/456d047e23563a194e6ea7d5c013a792/ivm2415.pdf}.
\bibitem{Postnikovr1}
{\sc Постников М.М.} Лекции по геометрии. V семестр. Группы и алгебры Ли. Москва, Наука, 1982 г.
\bibitem{Rosenfeldr1}
{\sc Розенфельд Б.А.}
Проективная геометрия как метрическая геометрия. Труды семинара по векторному и тензорному анализу с их приложениями
к геометрии, механике и физике.- М., 1950.-Т8.-с.~328-354. [B. A. Rosenfeld. Projective geometry as the metric geometry. Proceedings of the Seminar on Vector and tensor analysis with their applications.- Moscow, 1950.-v8.-p.~328-354.]
\bibitem{Rosenfeldr2}
{\sc Розенфельд Б.А.}
Неевклидовы геометрии. Москва, ГИТО, 1955.
[B. A. Rosenfeld. Non-euclidian geometry. Moscow, Gosudarstv. Izdat. Tehn., 1955]
\bibitem{Zeer1}
{\sc Энтони Зи}
Квантовая теория в двух словах. РХД, Москва, Ижевск, 2009.
[A. Zee. Quantum field theory in a nutshell. Princeton university press, 2003]
\bibitem{Penroser1}
{\sc Пенроуз Р. Риндлер В.}
Спиноры и пространство-время.  Т.1, Мир, Москва, 1987.
Спиноры и пространство-время.  Т.2, Мир, Москва, 1988.
\bibitem{Sintcovr1}
{\sc Синцов Д.М.}
Теория коннексов в пространстве в связи с теорией дифференциальных уравнений в частных производных первого порядка. Казань, 1894 г.
[Theory of connex in space in relation to the theory of partial differential equations of first order. Kazan. 1894]
\bibitem{Stepanovskiir1}
{\sc  Степановский Ю.П.}
Спиноры 6-мерного пространства и их применение к описанию поляризованных частиц со спином 1/2. Проблемы Ядерной Физики и Космических Лучей,
Харьков, 4(1976)9-21.
[Yu. P. Stepanovskii]: Spinors of a sixdimensional space and their application to the description of polarized particles with spin 1/2.
Problems of Nuclear Physics and Cosmic Rays, Kharkiv, 4(1976)9-21.
\bibitem{Stepanovskiir2}
{\sc Степановський Ю.П.}
Алгебра матриць Дiрака у шестiмiрному виглядi. Украiнський Фiзичний Журнал. Kиiв, т. XI, 8(1968) 813-824.
[Yu. P. Stepanovskii: Algebra of Dirac matrices in six-dimensional form. Ukrainian Journal of Physics, Kiev, v.XI, 8(1968) 813-824].
\bibitem{Schoutenr1}
{\sc  Схоутен Я.А.}
J.A. Schouten, J. Haantjes: Konforme Feldtheorie II; $R_6$ und Spinuraum [Conformal field theory II and spin space]. Annali della R.Scuola
Normale Superiore di Piza, Scienze Fisiche e Matematiche,2. Series, 4(1935)175-189.
\bibitem{Husemollerr1}
{\sc Хьюзмоллер Д.}
Расслоенные пространства. Москва, Мир, 1970 г.
[Dale Husemoller. Fibre bundles. McGraw-Hill Book Company,1966]
\bibitem{Scharnhorstr1}
{\sc К. Шарнхорст, Ж.-В. ван Хольтен}
K. Scharnhorst, J.-W. van Holten: Nonlinear Bogolyubov-Valatin transformations: 2 modes. NIKHEF preprint NIKHEF/2010-005, \href{http://arxiv.org/abs/1002.2737}{arXiv:1002.2737v3}, 114 pp.
}
  
\label{origine}
\newpage
\makeatletter
\def\addcontentsline#1#2#3{
  \begingroup
    \let\label\@gobble
    \ifx\@currentHref\@empty
      \Hy@Warning{%
        No destination for bookmark of \string\addcontentsline,%
        \MessageBreak destination is added%
      }%
      \phantomsection
    \fi
    \expandafter\ifx\csname toclevel@#2\endcsname\relax
      \begingroup
        \def\Hy@tempa{#1}%
        \ifx\Hy@tempa\Hy@bookmarkstype
          \Hy@WarningNoLine{%
            bookmark level for unknown #2 defaults to 0%
          }%
        \else
          \Hy@Info{bookmark level for unknown #2 defaults to 0}%
        \fi
      \endgroup
      \expandafter\gdef\csname toclevel@#2\endcsname{0}%
    \fi
    \edef\Hy@toclevel{\csname toclevel@#2\endcsname}%
    \Hy@writebookmark{\csname the#2\endcsname}%
      {#3}%
      {\@currentHref}%
      {\Hy@toclevel}%
      {#1}%
    \ifHy@verbose
      \begingroup
        \def\Hy@tempa{#3}%
        \@onelevel@sanitize\Hy@tempa
        \let\temp@online\on@line
        \let\on@line\@empty
        \Hy@Info{%
          bookmark\temp@online:\MessageBreak
          thecounter {\csname the#2\endcsname}\MessageBreak
          text {\Hy@tempa}\MessageBreak
          reference {\@currentHref}\MessageBreak
          toclevel {\Hy@toclevel}\MessageBreak
          type {#1}%
        }%
      \endgroup
    \fi
    \addtocontents{#1}{\protect\contentsline{#2}{#3}{\thepage}{\@currentHref}}
  \endgroup
}
\pagestyle{plain}
\foreignlanguage{russian}{
\section{Добавление}
$\phantom{fff}$Рассмотрим вложение $\mathbb R^4_{(1,3)}\subset \mathbb C^4$ с операторами \footnote{Этот пример не вошел в опубликованную статью.}
\begin{equation}
\bar H_i{}^{\Lambda '}=\left(
\begin{array}{cccc}
1 & 0 & 0 & 0\\
0 & 1 & 0 & 0\\
0 & 0 & 1 & 0\\
0 & 0 & 0 &-i\\
\end{array}
\right)\sps
H^i{}_\Lambda =\left(
\begin{array}{cccc}
1 & 0 & 0 & 0\\
0 & 1 & 0 & 0\\
0 & 0 & 1 & 0\\
0 & 0 & 0 &-i\\
\end{array}
\right)\sps
S_\Lambda{}^{\Lambda '}=\left(
\begin{array}{cccc}
1 & 0 & 0 & 0\\
0 & 1 & 0 & 0\\
0 & 0 & 1 & 0\\
0 & 0 & 0 &-1\\
\end{array}
\right)\spsd
\end{equation}
Для значений комплексных операторов из алгоритма \ref{ra14.1} метрика имеет вид
\begin{equation}
g_{\Lambda\Psi}=\left(
\begin{array}{cccc}
1 & 0 & 0 & 0\\
0 &-1 & 0 & 0\\
0 & 0 &-1 & 0\\
0 & 0 & 0 & 1\\
\end{array}
\right)\sps
g_{ij}=\left(
\begin{array}{cccc}
1 & 0 & 0 & 0\\
0 &-1 & 0 & 0\\
0 & 0 &-1 & 0\\
0 & 0 & 0 &-1\\
\end{array}
\right)\spsd
\end{equation}
Тогда определение (\ref{re14.5}) станет таким
\begin{equation}
\begin{array}{c}
(P^*_K)_i{}^A:=
\frac{1+i}{4}\left(
\begin{array}{cc}
 0          &  p_1-p_2   \\
 0          &  1-p_1p_2  \\
 p_1-p_2    &  0         \\
 i(1-p_1p_2)&  0         \\
\end{array}
\right)\sps
(P_K)^i{}_A:=
\frac{1-i}{4}\left(
\begin{array}{cc}
  0          &  p_1-p_2   \\
  0          &  1-p_1p_2  \\
  p_1-p_2    &  0         \\
-i(1-p_1p_2) &  0         \\
\end{array}
\right)\spsd
\end{array}
\end{equation}
Поскольку$x^i:=(P_2)^i{}_A(X_2){}^A+(P_3)^i{}_A(X_3){}^A$ то из (\ref{re3.13/1}), (\ref{re12.6}) и следствия (\ref{rc6.2}) следует, что $(X_K){}^A:=x^i(P^*_K)_i{}^A$, $(X^*_K){}_A:=x^i(P_K)_i{}_A$, а из (\ref{re3.13/1}), (\ref{re12.6}), следствия (\ref{rc6.2}) и соотношения $(M_K)^i{}^A(M_K)_i{}_A=0,\ (M^*_K)^i{}^A(M^*_K)_i{}_A=0$ (смотри (\ref{re14.6})) следует, что $(X^*_2){}_C=(P_2)^i{}_C(P_3)_i{}_B(X_3){}^B$. Поэтому мы можем построить взаимно однозначное соответствие ($a,b,...=\overline{1,4}$)
\begin{equation}
x^i\cdot\underbrace{((P^*_2)_i{}^A,(P^*_3)_i{}^B)}_{:=P_i{}^a}=\underbrace{((X_2){}^A,(X_3){}^B)}_{:=X^a}\sps
x^i\cdot\underbrace{((P_2)_i{}_A,(P_3)_i{}_B)}_{:=P_i{}_a}=\underbrace{((X^*_2){}_A,(X^*_3){}_B)}_{:=X_a}\spsd
\end{equation}
Тогда на спин-парах $X^a=((X_2){}^A,(X_3){}^B)$ индуцируется метрика и инволюция вида
\begin{equation}
\begin{array}{c}
\varepsilon_{ab}=
\left(
\begin{array}{cc}
                      0 & (P_2)^i{}_A(P_3)_i{}_B \\
 (P_3)^i{}_C(P_2)_i{}_D & 0 \\
\end{array}
\right)=
\left(
\begin{array}{cccc}
 0 & 0 &-i & 0\\
 0 & 0 & 0 & i\\
-i & 0 & 0 & 0\\
 0 & i & 0 & 0\\
\end{array}
\right)\sps\\
S_a{}^{b'}=
\left(
\begin{array}{cc}
 (P_2)^i{}_A(\bar P^*_2)_i{}^{C'} & (P_2)^i{}_A(\bar P^*_3)_i{}^{D'} \\
 (P_3)^i{}_B(\bar P^*_2)_i{}^{C'} & (P_3)^i{}_A(\bar P^*_3)_i{}^{D'} \\
\end{array}
\right)=
\left(
\begin{array}{cccc}
 0 & 0 & i & 0\\
 0 &-i & 0 & 0\\
 i & 0 & 0 & 0\\
 0 & 0 & 0 &-i\\
\end{array}
\right)\spsd
\end{array}
\end{equation}
Поэтому возможно построить на спин-парах лиевские аналоги
\begin{equation}
L_x(Y)^a=x^i\nabla_iY^a-P^j{}_bY^bP_i{}^a\nabla_j x^i\sps
L_X(Y)^a=X^c\nabla_cY^a-Y^c\nabla_c X^a\spsd
\end{equation}
Заметим, что разложение $x^i:=(P_2)^i{}_A(X_2){}^A+(P_3)^i{}_A(X_3){}^A$ является разложением по двум изотропным векторам. А для таких векторов лиевские аналоги были построены в \cite[т. 2, с. 127]{Penroser1}.

Ограниченное преобразование Лоренца переводит вектор $x^i$ в вектор $\tilde x^i$. Это индуцирует спинорные преобразования $(X_K)^A\longmapsto (\tilde X_K)^A$. Они воздействуют на оператор $(P^*_K)_i{}^A,\ (P_K)^i{}_A$, или более точно, на управляющий спинор $X^\myff A$ из определения (\ref{re9.10}). В свою очередь, это индуцирует преобразования на спин-парах $X^a=((X_2){}^A,(X_3){}^B)\longmapsto \tilde X^a$.

\newpage
}
\tableofcontents
\listoftables
\listoffigures
\end{document}